\documentclass[%
reprint,
 showpacs,
 showkeys,
 amsmath,amssymb,
 aps,
 prd,
]{revtex4-1}

\usepackage[T1]{fontenc}
\usepackage[latin2]{inputenc} 
\usepackage[czech,english]{babel}
\usepackage{graphicx}
\usepackage{dcolumn}
\usepackage{bm}
\usepackage[mathlines]{lineno}

\newcommand*{\Schw}{Schwarzschild}
\newcommand*{\SdS}{\Schw\discretionary{--}{--}{--}de~Sitter}
\newcommand*{\SaodS}{\Schw\discretionary{--}{--}{--}\mbox{(anti-)de}~Sitter}
\newcommand*{\dif}{\ensuremath{\mathrm{d}}}
\newcommand*{\oder}[2]{\ensuremath{\frac{\dif #1}{\dif #2}}}
\newcommand*{\loder}[2]{\ensuremath{\dif #1/\dif #2}}
\newcommand*{\pder}[2]{\ensuremath{\frac{\partial #1}{\partial #2}}}
\newcommand*{\eto}[1]{\ensuremath{\mathrm{e}^{#1}}}
\newcommand*{\rhocent}{\ensuremath{\rho_{\mathrm{c}}}}
\newcommand*{\pcent}{\ensuremath{p_{\mathrm{c}}}}
\newcommand*{\msun}{\ensuremath{\mathrm{M}_{\odot}}}
\renewenvironment{equation*}%
  {\begin{equation}}{\end{equation}\ignorespacesafterend}
\let\nobadge\relax

\begin{document}

\preprint{APS/123-QED}

\title{General relativistic polytropes with a~repulsive\\ cosmological constant}

\author{Zden\v{e}k Stuchl\'{\i}k}
  \altaffiliation[Also at ]{Research Centre of Theoretical Physics
    and Astrophysics,
    Faculty of Philosophy and Science, Silesian University in Opava,
    Bezru\v{c}ovo n\'am.~13, CZ-746\,01 Opava, Czech Republic}
  \email{zdenek.stuchlik@fpf.slu.cz}
\author{Stanislav Hled\'{\i}k}%
  \altaffiliation[Also at ]{Research Centre of Computational Physica
    and Data Processing,
    Faculty of Philosophy and Science, Silesian University in Opava,
    Bezru\v{c}ovo n\'am.~13, CZ-746\,01 Opava, Czech Republic}
  \email{stanislav.hledik@fpf.slu.cz}
\author{Jan Novotn\'y}
  \email{jan.novotny@fpf.slu.cz}
\affiliation{%
  Institute of Physics, Faculty of Philosophy and Science,
  Silesian University in Opava,\\
  Bezru\v{c}ovo n\'am.~13, CZ-746\,01 Opava, Czech Republic%
}

\date{\today}

\begin{abstract}
  Spherically symmetric equilibrium configurations of perfect fluid obeying a
  polytropic equation of state are studied in spacetimes with a repulsive
  cosmological constant. The configurations are specified in terms of three
  parameters---the polytropic index $n$, the ratio of central pressure and
  central energy density of matter $\sigma$, and the ratio of energy density
  of vacuum and central density of matter $\lambda$. The static equilibrium
  configurations are determined by two coupled first-order nonlinear
  differential equations that are solved by numerical methods with the
  exception of polytropes with $n=0$ corresponding to the configurations with
  uniform distribution of energy density, when the solution is given in terms
  of elementary functions. The geometry of the polytropes is conveniently
  represented by embedding diagrams of both the ordinary space geometry and
  the optical reference geometry reflecting some dynamical properties of the
  geodesic motion. The polytropes are represented by radial profiles of energy
  density, pressure, mass, and metric coefficients. For all tested values of
  $n>0$, the static equilibrium configurations with fixed parameters $n$,
  $\sigma$, are allowed only up to a critical value of the cosmological
  parameter $\lambda_{\mathrm{c}}=\lambda_{\mathrm{c}}(n,\sigma)$. In the case
  of $n>3$, the critical value $\lambda_{\mathrm{c}}$ tends to zero for
  special values of $\sigma$.
  The gravitational potential energy and the binding energy of the polytropes
  are determined and studied by numerical methods.
  We discuss in detail the polytropes with extension comparable to those of
  the dark matter halos related to galaxies, i.e., with extension
  $\ell > 100\,\mathrm{kpc}$ and mass $M > 10^{12}\,\msun$. For such largely
  extended polytropes the cosmological parameter relating the vacuum energy to
  the central density has to be larger than
  $\lambda = \rho_{\mathrm{vac}}/\rhocent \sim 10^{-9}$. We demonstrate that
  extension of the static general relativistic polytropic configurations
  cannot exceed the so called static radius related to their external
  spacetime, supporting the idea that the static radius represents a natural
  limit on extension of gravitationally bound configurations in an expanding
  universe dominated by the vacuum energy.
\end{abstract}

\maketitle

\section{\label{intro}Introduction}

Data from cosmological observations indicate that in the framework of the
inflationary paradigm~\cite{Lin:1990:InfCos:}, a very small relict repulsive
cosmological constant $\Lambda > 0$, i.e., vacuum energy, or, generally, a
dark energy demonstrating repulsive gravitational effect, has to be invoked in
order to explain the dynamics of the recent
Universe~\cite{Kra-Tur:1995:GENRG2:,Ost-Ste:1995:NATURE:,Kra:1998:ASTRJ2:,Bah-etal:1999:SCIEN:,Cal-Dav-Ste:1998:PHYRL:,ArP-Muk-Ste:2000:PHYRL:,Wan-etal:2000:ASTRJ2:}. The
total energy density of the Universe is very close to the critical energy
density $\rho_{\mathrm{crit}}$ corresponding to almost flat universe predicted
by the inflationary scenario~\cite{Spe-etal:2007:ASTJS:3yrWMAP}. Observations
of distant Ia-type supernova explosions indicate that starting at the
cosmological redshift $z \approx 1$ expansion of the Universe is
accelerated~\cite{Rie-etal:2004:ASTRJ2:}. The cosmological tests demonstrate
convincingly that the dark energy represents about $70\%$ of the energy
content of the observable
universe~\cite{Spe-etal:2007:ASTJS:3yrWMAP,Cal-Kam:2009:NATURE:CosDarkMat}. These
results are confirmed by recent measurements of cosmic microwave background
anisotropies obtained by the space satellite observatory
PLANCK~\cite{Ada-etal:2013:ASTRA:DifLiYoClG,Ade-etal:2014:ASTRA:PlanckXII}.

There are strong indications that the dark energy equation of state is very
close to those corresponding to the vacuum energy, i.e., to the repulsive
cosmological constant~\cite{Cal-Kam:2009:NATURE:CosDarkMat}. Therefore, it is
important to study the cosmological and astrophysical consequences of the
effect of the observed cosmological constant implied by the cosmological tests
to be $\Lambda \approx 1.3\times 10^{-56}\,\mathrm{cm^{-2}}$, and the related
vacuum energy $\rho_{\mathrm{vac}} \sim 10^{-29}\,\mathrm{g/cm^{3}}$ that is
comparable to the critical density of the universe. The presence of a
repulsive cosmological constant changes dramatically the asymptotic structure
of \mbox{black-hole}, naked singularity, or any compact-body backgrounds as
such backgrounds become asymptotically de~Sitter spacetimes, not flat
spacetimes. In such spacetimes, an event horizon (cosmological horizon) always
exists, behind which the geometry is dynamic.

The repulsive cosmological constant was discussed mainly in the scope of the
cosmological models~\cite{Mis-Tho-Whe:1973:Gra:}. Its role in the vacuola
models of mass concentrations immersed in the expanding universe has been
considered
in~\cite{Stu:1983:BULAI:,Stu:1984:BULAI:,Uza-Ell-Lar:2011:GENRG2:2MassExp,Gre-Lak:2010:PHYSR4:SwissCheeseMass,Fle-Dup-Uza:2013:PHYSR4:HubbleNonhomUn,Fir-Feg:2016:arXiv160805491:PtMassCBH}. Recently,
relevance of the repulsive cosmological constant has been found in the
McVittie model~\cite{McV:1933:MONNR:MassPartExpUn} of mass concentrations
immersed in the expanding
universe~\cite{Nol:1998:PHYSR4:PtMassIsotUn,Nan-Las-Hob:2012:MONNR:ExpUnMassOb,Nan-Las-Hob:2012:MONNR:MassObExpUn,Kal-Kle-Mar:2010:PHYSR4:McVittieLegacy,Lak-Abd:2011:PHYSR4:McVittieLegacy,Sil-Fon-Gua:2013:PHYSR4:McVittieST,Nol:2014:1408.0044:MvVitOrb}. Significant
role of the repulsive cosmological constant has been demonstrated also for
astrophysical situations related to active galactic nuclei and their central
supermassive black holes~\cite{Stu:2005:MODPLA:}. The black hole spacetimes
with the $\Lambda$ term are described in the spherically symmetric case by the
vacuum \SaodS{} (SdS) geometry
\cite{Kot:1918:ANNPH2:PhyBasEinsGr,Stu-Hle:1999:PHYSR4:}, while the internal,
uniform density SdS spacetimes are given
in~\cite{Stu:2000:ACTPS2:,Boh:2004:GENRG2:}. In axially symmetric, rotating
case, the vacuum spacetime is determined by the Kerr--de~Sitter (KdS) geometry
\cite{Car:1973:BlaHol:}. In the spacetimes with the repulsive cosmological
term (and the related solutions of the f(R) gravity), motion of photons is
treated in a series of
papers~\cite{Stu:1990:BULAI:,Stu-Cal:1991:GENRG2:,Stu-Hle:2000:CLAQG:,Lak:2002:PHYSR4:BendLiCC,Bak-etal:2007:CEURJP:,Ser:2008:PHYSR4:CCLens,Ser:2008:PHYSR4:CCLens,Mul:2008:GENRG2:FallSchBH,Sch-Zai:2008:0801.3776:CCTimeDelay,Vil-etal:2012:ASTSS:PhMoChgAdS,Zha-Tan:2015:PHYSR4:GraLensSdS,Zha-Tan-He:2016:PHYSR4:GraLensRNdS},
while motion of test particles was studied
in~\cite{Stu:1983:BULAI:,Stu-Hle:1999:PHYSR4:,Stu-Hle:2002:ACTPS2:,Stu-Sla:2004:PHYSR4:,Kra:2004:CLAQG:,Kra:2005:DARK:CCPerPrec,Kra:2007:CLAQG:Periapsis,Cru-Oli-Vil:2005:CLAQG:GeoSdSBH,Kag-Kun-Lam:2006:PHYLB:SolarSdS,Ali:2007:PHYSR4:EMPropKadS,Che:2008:CHINPB:DkEnGeoMorSchw,Ior:2009:NEWASTR:CCDGPGrav,Hac-etal:2010:PHYSR4:KerrBHCoStr,Oli-etal:2011:MODPLA:ChaParRNadS,Cha-Har:2012:PHYSR4:BEConGRStar,Cha-Reg:2012:PHYSR4:LocEffCC,Zou-Li-Li:2014:INTJMD:TOVSdS,Sar-Gho-Bha:2014:PHYSR4:NewtAnalogSdS,Gon-Oli-Vas:2015:EPHYJC:4DAdS,Mac-Gua-Mol:2015:PHYSR4:CBHandWH,Kun-Per-Lam:2015:PHYSR4:IsofreqPairSdS,Zak:2015:JASPASN:AltToSmBH,Arr:2015:IJMPD:OnBHAltThGraNonlin,Arr:2015:EURLE:OnApparLoss,Spo-Bor:2016:INTJMD:GBfactorsSdS}. Oscillatory
motion of current carrying string loops in SdS and KdS spacetimes was treated
in~\cite{Jac-Sot:2009:PHYSR4:StrDynSpiBH,Kol-Stu:2010:PHYSR4:CurCarStrLoops,Gu-Cheng:2007:GENRG2:CircLoopKdS,Wan-Che:2012:PHYLB:CirLoopPerTens,Stu-Kol:2012:PHYSR4:AccStringLoops,Stu-Kol:2012:JCAP:StringLoops,Stu-Kol:2014:PHYSR4:KerrStrLoopOsc}.

The cosmological constant can be relevant in both the geometrically thin
Keplerian accretion
disks~\cite{Stu:2005:MODPLA:,Stu-Hle:1999:PHYSR4:,Stu-Sla:2004:PHYSR4:,Mul-Asc:2007:CLAQG:VelProKadSBH,Sla-Stu:2008:CLAQG:CmtNoMonKadS}
and geometrically thick toroidal accretion
disks~\cite{Stu-Sla-Hle:2000:ASTRA:,Sla-Stu:2005:CLAQG:,Rez-Zan-Fon:2003:ASTRA:,Asc:2008:CHIAA:MassSpinBHQPO,Stu-etal:2005:PHYSR4:AschenUnexpTopo,Mul-Asc:2007:CLAQG:VelProKadSBH,Sla-Stu:2008:CLAQG:CmtNoMonKadS,Kuc-Sla-Stu:2011:JCAP:ToroPerFlRNadS,Per-Rom-PeB:2013:ASTRA:AccDiBHModGra,Cha:2015:CLAQG:EqConfArndBH}
orbiting supermassive black holes in the central parts of giant galaxies, or
in the recently discussed ringed accretion
disks~\cite{Pug-Stu:2015:ASTJS:RingedAccDiskEq,Pug-Stu:2016:ASTJS:RingAccDiskInstab}. Spherically
symmetric, stationary polytropic accretion in the spacetimes with the
repulsive cosmological constant has been studied
in~\cite{Kar-Mal:2013:PHYSR4:BondiAccCBH,Mac-Mal:2013:PHYSR4:StabBondiAccSadS,Mac-Mal-Kar:2013:PHYSR4:SphSteadAccFlow,Mac:2015:PHYSR4:HomoclinAccrSadS,Fic:2015:CLAQG:BondiTypeAcc}.

In spherically symmetric spacetimes Keplerian and toroidal disk structures can
be described with high precision by an appropriately chosen Pseudo-Newtonian
potential~\cite{Stu-Kov:2008:INTJMD:PsNewtSdS,Stu-Sla-Kov:2009:CLAQG:PseNewSdS}
that appears to be useful also in studies of motion of interacting
galaxies~\cite{Stu-Sch:2011:JCAP:CCMagOnCloud,Sch-Stu-Pet:2013:JCAP:MOND,Stu-Sch:2012:INTJMD:GRvsPsNewtMagClou}. It
should be mentioned that the KdS geometry can be relevant also in the case of
Kerr superspinars representing an alternate explanation of active galactic
nuclei~\cite{Gim-Hor:2004:hep-th0405019:GodHolo,Boy-etal:2003:PHYSR4:HoloProtChron,Gim-Hor:2009:PHYLB:AstVioSignStr,Stu-Sch:2012:CLAQG:ObsPrimKerrSS}. The
superspinars breaking the black hole bound on the spin exhibit a variety of
unusual physical
phenomena~\cite{deFel:1974:ASTRA:,Cun:1975:ASTRJ2:,deFel:1978:NATURE:InstabNS,Stu:1980:BULAI:,Stu-Hle-Tru:2011:CLAQG:,Hio-Mae:2009:PHYSR4:KerrSpinMeas,Stu-Sch:2010:CLAQG:AppKepDiOrKerrSSp,Stu-Sch:2012:CLAQG:CntRKerrSSpi,Stu-Sch:2013:CLAQG:UHEKerrGeo}.

Besides the vacuum \mbox{black-hole} (\mbox{naked-singularity}) spacetimes, we
have to study the role of a repulsive cosmological constant also in non-vacuum
spacetimes representing static mass configurations. Such general relativistic
non-vacuum solutions can be interesting, e.g., in connection to the cold dark
matter (CDM) halos that are recently widely discussed as an explanation of
hidden structure of galaxies enabling correct treatment of the motion in the
external parts of
galaxies~\cite{Bos:1981:ASTRJ1:21cmSpiGal,Rub:1982:HIContNorGal:SaSbScGal} and
are at present assumed usually in the Newtonian
approximation~\cite{Bin-Tre:1988:GalacDynam:,Ior:2010:MONNR:GalOrbMoDarkMat,Nav-Fre-Whi:1997:ASTRJ2:UniDeProHiCl,Stu-Sch:2011:JCAP:CCMagOnCloud,Cre-Stu:2013:INTJMD:KinThGravSys}. There
is a variety of candidates for the CDM~\cite{Wei:2008:Cosmology:1},
nevertheless none of these candidates is considered to be confirmed in the
present state of knowledge. Therefore, it is important to test the possibility
to represent such CDM halos in a relatively simple manner that enables to
estimate easily the role of the cosmological constant. We shall discuss the
most simple case of spherically symmetric static configurations of perfect
fluid with a polytropic equation of state generalizing thus the standard
discussion of Tooper~\cite{Too:1964:ASTRJ2:} by introducing the vacuum energy
represented by the repulsive cosmological constant. Outside of these
polytropic spheres the spacetime is described by the vacuum \SdS{} geometry.

Choosing the polytropic equation of state means that details of the processes
inside the polytropic spheres are not considered, and a simple power law
relating the total pressure to the total energy density of matter is
assumed. Such an approximation seems to be applicable in the dark matter
models that assume weakly interacting particles (see,
e.g.,~\cite{Bor:1993:EarlyUniv:,Kol-Tur:1990:EarUni:,Cre-Stu:2013:INTJMD:KinThGravSys}). In
fact, such a simple assumption enables to obtain basic properties of the
non-vacuum configurations governed by the relativistic laws. For example, the
equation of state of the ultrarelativistic degenerate Fermi gas is determined
by the polytropic equation with the adiabatic index $\Gamma = 4/3$
corresponding to the polytropic index $n = 3$, while the non-relativistic
degenerate Fermi gas is determined by the polytropic equation of state with
$\Gamma = 5/3$, and $n = 3/2$~\cite{Sha-Teu:1983:BHWDNS:}. It should be noted
that a similar case of the adiabatic equation of state can be used in the case
of a general ideal gas. This case was appropriately applied to describe the
(test) perfect fluid toroidal configurations orbiting black
holes~\cite{Stu-Sla-Kov:2009:CLAQG:PseNewSdS} and can be, in principle,
applied for modeling of self-gravitating adiabatic spherically symmetric
general relativistic configurations. The special case of polytropes with
polytropic index $n = 0$ corresponding to the simplest, although rather
unphysical and artificial case of spheres with uniform distribution of energy
density (but radius dependent distribution of pressure) can be treated as a
very useful model---it can serve as a test bed for properties of general
relativistic polytropes (GRP hereinafter) because its structure equations can
be solved in terms of elementary
functions~\cite{Stu:2000:ACTPS2:,Stu-etal:2001:PHYSR4:,Boh:2004:GENRG2:,Nil-Ugg:2000:ANNPH1:GRStarPoEqSt,Boe-Fod:2008:PHYSR4:}. For
non-zero values of the polytropic index, the structure equations have to be
solved by numerical methods.

The Einstein equations with a non-zero cosmological constant lead in the case
of spherically symmetric, static equilibrium configurations to generalized
Tolman--Oppenheimer--Volkoff (TOV) equation. By using the standard ansatz for
the polytropic equation of state, the equations are transferred into
dimensionless form of two coupled first-order nonlinear differential equations
that are solved by numerical methods under boundary conditions requiring
regularity of the solution at the center of the polytrope, and smooth matching
of the internal spacetime at the surface of the polytrope to the external
\SdS{} spacetime characterized by the same mass parameter (and the
cosmological constant) as the internal spacetime. The configurations are
specified in terms of three parameters---the polytropic index $n$, the ratio
of central pressure and central energy density of matter $\sigma$, and the
ratio of energy density of vacuum and central density of matter $\lambda$. By
simultaneously solving the coupled equations, the structure of the polytrope
is obtained; it is characterized by the profiles of the energy density,
pressure, mass, and two metric coefficients ($g_{tt}, g_{rr}$) giving the
geometry of the internal spacetime of the polytropic sphere. The spacetime
structure can be reflected by the embedding diagrams of the ordinary space and
the optical reference geometry reflecting some hidden properties of the
spacetime~\cite{Abr:1990:MONNR:,Stu-Hle-Jur:2000:CLAQG:}. The other relevant
characteristics of the polytropes are the gravitational potential energy and
the binding energy \cite{Too:1964:ASTRJ2:}.

\section{\label{eost}Equations of structure}

In terms of the standard Schwarzschild coordinates, the line element of a
spherically symmetric, static spacetime is given in the form
\begin{equation*}
  \dif s^{2} = - \eto{2\Phi}\, c^{2} \dif t^{2} + \eto{2\Psi}\,\dif r^{2} + r^{2}
    (\dif\theta^{2} + \sin^{2} \theta\,\dif\phi^{2})            \label{tpa007e1}
\end{equation*}
with just two unknown functions of the radial coordinate, $\Phi(r)$ and
$\Psi(r)$. Matter inside the configuration is assumed to be a perfect fluid
with $\rho = \rho (r)$ being the density of mass-energy in the rest-frame of
the fluid and $p = p(r)$ being the isotropic pressure. The stress-energy
tensor of the perfect fluid reads
\begin{equation*}
  T^\mu_{\hphantom{\mu}\nu} =
    (p+\rho c^{2}) U^{\mu} U_{\nu} + p\,\delta^{\mu}_{\nu}               \label{e3}
\end{equation*}
where $U^{\mu}$ denotes the four-velocity of the fluid.  We consider here the
simplest direct relation between the energy density and pressure of the fluid
given by the polytropic equation of state
\begin{equation*}
  p = K \rho^{1+\frac{1}{n}}                                          \label{grp6}
\end{equation*}
where $n$ is the `polytropic index' assumed to be a given constant (not
necessarily an integer) and $K$ is a constant that has to be determined by the
thermal characteristics of a given fluid sphere, by specifying the density
$\rhocent$ and pressure $\pcent$ at the center of the polytrope. Since the
density is a function of temperature for a given pressure, $K$ contains the
temperature implicitly. It can be shown that $K$ is determined by the total
mass, radius, and $\pcent/\rhocent c^{2}$ ratio. (The polytropic equation
represents a limiting form of the parametric equations of state for a
completely degenerate gas at zero temperature, relevant, e.g., for neutron
stars. Then both $n$ and $K$ are universal physical
constants~\protect\cite{Too:1964:ASTRJ2:,Too:1964:ASTRJ2:}.)

In a static configuration, each element of the fluid must remain at rest in
the static coordinate system where the spatial components of 4-velocity field
$\loder{r}{\tau}$, $\loder{\theta}{\tau}$, $\loder{\phi}{\tau}$ vanish,
leaving the temporal component
\begin{equation*}
  u^{t} = \oder{t}{\tau} = \eto{-\Phi}                          \label{tpa007e3}
\end{equation*}
the only nonvanishing one.  The structure of a relativistic star is determined
by the Einstein field equations
\begin{equation*}
  G_{\mu\nu} \equiv R_{\mu\nu} - \frac{1}{2}Rg_{\mu\nu}
    + \Lambda g_{\mu\nu} = \frac{8\pi G}{c^4} T_{\mu\nu}
\end{equation*}
and by the law of local energy-momentum conservation 
\begin{equation*}
  T^{\mu\nu}_{\hphantom{\mu\nu};\nu} = 0\,.
\end{equation*}
It is convenient to express the equations in terms of the orthonormal tetrad
components using the 4-vectors carried by the fluid elements:
\begin{align}
  \vec{e}_{(t)} &= \frac{1}{\eto{\Phi}} \pder{}{t}\,,\quad
  \vec{e}_{(r)} = \frac{1}{\eto{\Psi}} \pder{}{r}\,,\notag\\
  \vec{e}_{(\theta)} &= \frac{1}{r} \pder{}{\theta}\,,\quad
  \vec{e}_{(\phi)}  = \frac{1}{r\sin\theta}\pder{}{\phi}\,.\nobadge
\end{align}
Projection of $T^{\mu\nu}_{\hphantom{\mu\nu};\nu} = 0$ orthogonal to $u^{\mu}$
(by the projection tensor $P^{\mu\nu} = g^{\mu\nu} + u^{\mu} u^{\nu}$) gives
the relevant equation
\begin{equation}
  (\rho c^{2} + p) \frac{\dif\Phi}{\dif r}
    = - \frac{\dif p}{\dif r}                                   \label{tpa007e9}
\end{equation}
which is the equation of hydrostatic equilibrium describing the balance
between the gravitational force and pressure gradient.

There are two relevant structure equations following from the Einstein
equations. These are determined by the $(t)(t)$ and $(r)(r)$ tetrad components
of the field equations (the $(\theta)(\theta)$ and $(\phi)(\phi)$ components
give dependent equations).  First we shall discuss the $(t)(t)$ component:
\begin{equation*}
  G_{(t)(t)} = \frac{1}{r^{2}} - \frac{\eto{-2 \Psi}}{r^{2}} - \frac{1}{r}
  \frac{\dif}{\dif r}\,\eto{-2\Psi} - \Lambda
    = \frac{8 \pi G}{c^{2}} \rho\,.                            \label{tpa007e10}
\end{equation*}
This can be transferred into the form
\begin{equation*}
  \frac{\dif}{\dif r}\left[r \left(1-\eto{-2 \Psi} \right)
    - \frac{1}{3}\Lambda r^{3}
    \right] = \frac{\dif}{\dif r} \frac{2G}{c^{2}} m (r)       \label{tpa007e11}
\end{equation*}
where
\begin{equation}
  m(r) = \int^{r}_{0} 4 \pi r'^{2} \rho \, \dif r'\,.           \label{tpa007e12}
\end{equation}
The integration constant in~(\ref{tpa007e12}) is chosen to be $m(0) = 0$
because then the spacetime geometry is smooth at the origin
(see~\cite{Mis-Tho-Whe:1973:Gra:}) and we arrive to the relation
\begin{equation}
  \eto{2 \Psi} = \left[1 - \frac{2Gm(r)}{c^{2} r}
    - \frac{1}{3}\Lambda  r^{2} \right]^{-1}\,.                 \label{tpa007e13}
\end{equation}
The $(r)(r)$ component of the field equations reads 
\begin{equation*}
  G_{(r)(r)} = - \frac{1}{r^{2}} + \frac{\eto{-2\Psi}}{r^{2}} + \frac{2
    \eto{-2 \Psi}}{r}\,\frac{\dif\Phi}{\dif r} + \Lambda
    = \frac{8 \pi G}{c^4} p\,.                                 \label{tpa007e14}
\end{equation*}
Using Eq.\,(\ref{tpa007e13}), we obtain the relation
\begin{equation*}
  \frac{\dif\Phi}{\dif r} = \frac{\frac{G}{c^{2}}m(r)-\frac{1}{3}\Lambda  r^{3}
    + \frac{4\pi G}{c^4}p r^{3}}{r
    \left[r-\frac{2G}{c^{2}}m(r)
    - \frac{1}{3}\Lambda  r^{3} \right]}                       \label{tpa007e15}
\end{equation*}
which enables us to put the equation of hydrostatic
equilibrium~(\ref{tpa007e9}) into the Tolman--Oppenheimer--Volkoff (TOV) form
modified by the presence of a non-zero cosmological
constant~\cite{Stu:2000:ACTPS2:}:
\begin{equation*}
  \frac{\dif p}{\dif r} = - (\rho c^{2} + p) \frac{\frac{G}{c^{2}}m(r)
    - \frac{1}{3}\Lambda  r^{3} + \frac{4\pi G}{c^4} p r^{3}}
    {r \left[r-\frac{2G}{c^{2}}m(r)
    - \frac{1}{3}\Lambda  r^{3} \right]}\,.                    \label{tpa007e16}
\end{equation*}

The $(t)(t)$ component of the Einstein equations can be expressed and applied
in the form
\begin{equation}
  \oder{m(r)}{r} = 4\pi\rho(r)r^{2}\,.                             \label{grp11}
\end{equation}

For integration of the structure equations it is convenient to introduce,
following the approach of~\cite{Too:1964:ASTRJ2:}, a new variable $\theta$
related to the density radial profile $\rho(r)$ and the central density
$\rhocent$, by
\begin{equation*}
  \rho = \rhocent\theta^{n}                                        \label{grp14}
\end{equation*}
with $n$ being the polytropic index. The boundary condition on $\theta(r)$
reads $\theta(r=0)=1$. The pressure dependence is given by the relation
\begin{equation*}
  p = K\rhocent^{1+\frac{1}{n}}\theta^{n+1}\,.                         \label{grp15}
\end{equation*}
The conservation law~(\ref{grp11}) can be expressed in the form
\begin{equation}
  \sigma (n+1)\,\dif\theta + (\sigma\theta + 1)\,\dif\Phi = 0     \label{grp16}
\end{equation}
where the parameter $\sigma$ is given by the relation
\begin{equation*}
  \sigma = \frac{K}{c^{2}}\rhocent^{1/n}
    = \frac{\pcent}{\rhocent c^{2}}\,.                          \label{grp17-18}
\end{equation*}

At the edge of the configuration, $r=R$, there is $\rho(R)= p(R)=0$.  Outside
the mass configuration with mass parameter $M$ related to the mass of the
polytrope by $M=m(R)$, the spacetime is described by the vacuum \SaodS{}
metric. Solving Eq.\,(\ref{grp16}) and using the boundary condition that the
internal and external metric coefficients are smoothly matched at $r=R$, we
obtain
\begin{equation*}
  \eto{2\Phi} = (1+\sigma\theta)^{-2(n+1)}
    \left(1-\frac{2GM}{c^{2}R}-\frac{1}{3}\Lambda R^{2}\right)\,.   \label{grp21}
\end{equation*}
Thus, the internal metric coefficient $g_{tt}$ is determined by the function
$\theta(r)$ and the parameter $\sigma$. The function $\eto{2\Psi}$ remains to
be expressed in terms of $\theta$, and we need to find the function
$\theta=\theta(r)$ using the structure equations. First, we rewrite
Eq.\,(\ref{grp16}) in the form
\begin{equation*}
  \oder{\Phi}{r}
    = \frac{\sigma(n+1)}{1+\sigma\theta}\oder{\theta}{r}\,.        \label{grp22}
\end{equation*}
Then we can express the $(r)(r)$ component of the Einstein equations and
Eq.\,(\ref{grp11}) in the form
\begin{widetext}
\begin{subequations}
\label{grp24and25}
\begin{align}
  \frac{\sigma(n+1)}{1+\sigma\theta}\,r\,\oder{\theta}{r}
  \left(1-\frac{2Gm(r)}{c^{2}r}-\frac{1}{3}\Lambda r^{2}\right)
  + \frac{Gm(r)}{c^{2}r} - \frac{1}{3}\Lambda r^{2}
  &= -\frac{G}{c^{2}}\sigma\theta\oder{m}{r}\,,                  \label{grp24}\\
  \oder{m}{r} &= 4\pi r^{2} \rhocent\theta^{n}\,.                 \label{grp25}
\end{align}
\end{subequations}
\end{widetext}
Introducing factor $\mathcal{L}$ giving a characteristic length scale of the
polytrope
\begin{equation}
  \mathcal{L} = \left[\frac{(n+1)K\rhocent^{1/n}}%
                 {4\pi G\rhocent}\right]^{1/2}
    = \left[\frac{\sigma(n+1)c^{2}}%
                 {4\pi G\rhocent}\right]^{1/2}                      \label{grp26}
\end{equation}
and factor $\mathcal{M}$ giving a characteristic mass scale of the polytrope
\begin{equation}
   \mathcal{M} = 4\pi \mathcal{L}^3 \rhocent
     = \frac{c^2}{G}\sigma (n+1)\mathcal{L}\,,                    \label{grp26a}
\end{equation}
Eqs~(\ref{grp24and25}) can be transformed into dimensionless form
by introducing a dimensionless radial coordinate
\begin{equation*}
  \xi = \frac{r}{\mathcal{L}}                                      \label{grp27}
\end{equation*}
and dimensionless quantities
\begin{subequations}
\label{grp28and29}
\begin{align}
  v(\xi) &= \frac{m(r)}{4\pi \mathcal{L}^{3}\rhocent}
          = \frac{m(r)}{\mathcal{M}}\,,                         \label{grp28}\\
  \lambda &= \frac{\rho_{\mathrm{vac}}}{\rhocent}\,,               \label{grp29}
\end{align}
\end{subequations}
where $v(\xi)$ represents a dimensionless mass parameter and $\lambda$
represents a dimensionless cosmological constant related to the polytrope. The
vacuum energy density is related to the cosmological constant by
\begin{equation*}
  \rho_{\mathrm{vac}}c^{2} = \frac{\Lambda c^4}{8\pi G}
    = \frac{8\pi G}{c^2} \rhocent \lambda\,.                \label{grp30}
\end{equation*}

The dimensionless form of Eqs~(\ref{grp24and25}) determining the
polytrope structure then can be written down as
\begin{subequations}                                         \label{grp31and32}
\begin{align}
  \oder{\theta}{\xi} &=
    \frac{\frac23\lambda\xi^3-\sigma\xi^3\theta^{n+1}-v}%
      {\xi^2(1 + \sigma\theta)^{-1}}\,
    g_{rr}(\xi,v;n,\sigma,\lambda)\,,                            \label{grp31}\\
  \oder{v}{\xi} &= \xi^2\theta^n                                 \label{grp32}
\end{align}
\end{subequations}
where
\begin{equation}
  g_{rr}(\xi,v;n,\sigma,\lambda) \equiv
    \frac{1}{1 - 2\sigma(n+1)\left( \frac{v}{\xi}
    + \frac{1}{3}\lambda\xi^2 \right)}\,.                           \label{grr}
\end{equation}
coincides with the radial metric coefficient~(\ref{tpa007e13}).  For given
$n$, $\sigma$ and $\lambda$, Eqs~(\ref{grp31and32}) have to be simultaneously
solved under the boundary conditions
\begin{equation}
  \theta(0) = 1\,, \quad v(0) = 0\,.                               \label{grp33}
\end{equation}
It follows from~(\ref{grp32}) and~(\ref{grp33}) that $v(\xi) \sim \xi^{3}$ for
$\xi \to 0$ and, according to Eq.\,(\ref{grp31}),
\begin{equation*}
  \lim_{\xi\to 0_+}\oder{\theta}{\xi} = 0\,.                         \label{grp34}
\end{equation*}
The boundary of the fluid sphere ($r=R$) is represented by the first zero
point of $\theta(\xi)$, say at $\xi_{1}$:
\begin{equation*}
  \theta(\xi_{1}) = 0\,.                                           \label{grp35}
\end{equation*}
The solution $\xi_{1}$ determines the surface radius of the polytrope and the
solution $v(\xi_{1})$ determines its gravitational mass.

In the Newtonian limit ($\sigma\ll 1$), the structure equations can be
transformed to one differential equation of the second order
\begin{equation*}
  \frac{1}{\xi^{2}} \oder{}{\xi} \left(\xi^{2} \oder{\theta}{\xi}
    \right)+ \theta^{n} - 2 \lambda = 0                            \label{grp-0}
\end{equation*}
that is reduced to the Lane--Emden equation, if the cosmological term vanishes
($\lambda=0$)
\begin{equation*}
  \oder{}{\xi}\left(\xi^{2} \oder{\theta}{\xi}      \right)
    + \xi^{2} \theta^{n} = 0\,.                                    \label{grp-00}
\end{equation*}
The differential equations governing the structure of GRPs have to be solved
by numerical methods (even in the Newtonian limit). Only polytropes with the
polytropic index $n=0$, corresponding to configurations having uniform
distribution of the energy density but non-uniform pressure profile, allow for
solutions of the differential equations in terms of elementary functions.

\section{\label{properties}Properties of the polytropes}

The general relativistic polytropic spheres with given polytropic index $n$
are determined by the functions $\theta(\xi)$ and $v(\xi)$ of the
dimensionless coordinate $\xi$ and by the length and mass scales,
$\mathcal{L}$~(\ref{grp26}) and $\mathcal{M}$~(\ref{grp26a}). The functions
$\theta(\xi)$ and $v(\xi)$ are governed by the structure equations, the values
of the central energy density $\rhocent$, and the parameters $\sigma$ and
$\lambda$. A concrete polytropic sphere is then given by the first (lowest)
solution $\xi_{1}$ of the equation $\theta(\xi)=0$ that determines all the
characteristics of the polytropic configuration and the radial profiles of its
energy density, pressure, metric coefficients, or gravitational and binding
energy.

Assuming $\Lambda$, $n$, $\sigma$, and $\rhocent$ are given, then mass $M$,
radius $R$, and the internal structure of the polytropes can be easily
determined. First, the length scale $\mathcal{L}$ given by Eq.\,(\ref{grp26})
has to be found. By numerical integration of Eqs~(\ref{grp31and32}), functions
$\theta(\xi)$ and $v(\xi)$ are found and $\xi_{1}$, where $\theta(\xi_{1})=0$,
is determined together with $v(\xi_{1})$. The radius of the sphere is
\begin{equation*}
  R = \mathcal{L}\xi_{1}                                           \label{grp47}
\end{equation*}
and the mass of the sphere is given by
\begin{equation*}
  M = \mathcal{M}v(\xi_{1})
    = \frac{c^{2}}{G}\mathcal{L}\sigma(n+1) v(\xi_{1})\,.           \label{grp48}
\end{equation*}
The density, pressure, and mass-distribution profiles are determined by the
relations
\begin{subequations}                                           \label{grp49to51}
\begin{align}
  \rho(\xi) &= \rhocent\theta^{n}(\xi)\,,^{ }                     \label{grp49}\\
  p(\xi) &= \sigma\rhocent\theta^{n+1}(\xi)\,,                   \label{grp50}\\
  M(\xi) &= M\frac{v(\xi)}{v(\xi_{1})}\,.                        \label{grp51}
\end{align}
\end{subequations}
The temporal and radial metric coefficients can be expressed in the form
\begin{subequations}                                          \label{grp52and43}
\begin{align}
  \eto{2\Phi} &=
    \frac{1-2\sigma(n+1)
           \left[
             \frac{v(\xi_{1})}{\xi_{1}}+\frac{\lambda\xi_{1}^{2}}{3}
           \right]}%
         {(1+\sigma\theta)^{2(n+1)}}\,,                           \label{grp52}\\
  \eto{-2\Psi} &= 1 - 2\sigma(n+1)
    \left[\frac{v(\xi)}{\xi}+\frac{1}{3}\lambda\xi^{2}\right]\,.  \label{grp43}
\end{align}
\end{subequations}
(see also~(\ref{grr})).

One of the basic characteristics of the polytropes is the mass-radius
($M$-$R$) relation. Using Eq.\,(\ref{grp28and29}), we obtain
\begin{equation*}
  v(\xi) = \frac{m(r)}{4\pi\rhocent \mathcal{L}^{2}}\frac{\xi}{r}
         = \frac{G}{c^{2} \sigma(n+1)}\xi\frac{m(r)}{r}            \label{grp41}
\end{equation*}
and the $M$-$R$ relation can be expressed by the formula 
\begin{equation*}
  \mathcal{C} \equiv \frac{GM}{c^{2}R} = \frac{1}{2}\frac{r_{\mathrm{g}}}{R}
              = \frac{\sigma(n+1)v(\xi_{1})}{\xi_{1}}               \label{grp42}
\end{equation*}
where 
\begin{equation*}
  r_{\mathrm{g}} = \frac{2GM}{c^{2}}
\end{equation*}
is the gravitational radius of the polytropic configuration determined by its
total gravitational mass $M$. The quantity $\mathcal{C}$ determines
compactness of the sphere, i.e., effectiveness of the gravitational binding,
and it can be represented by the gravitational redshift of radiation emitted
from the surface of the polytropic sphere~\cite{Hla-Stu:2011:JCAP:}.

The external vacuum \SdS{} spacetime, with the same mass parameter $M$ and the
cosmological constant $\Lambda$ as those characterizing the internal spacetime
of the polytrope, has the metric coefficients
\begin{equation*}
  \eto{2\Phi} = \eto{-2\Psi}
    = 1 - \frac{2GM}{c^{2} r} -\frac{1}{3}\Lambda r^{2}\,.
\end{equation*}
There are two pseudosingularities of the external vacuum geometry that give
two length scales related to the polytropic spheres. The first one is
determined by the radius of the black hole horizon
\begin{equation*}
  r_{\mathrm{h}} = \frac{2}{\sqrt{\Lambda}}\,\cos\frac{\pi+\alpha}{3}	
\end{equation*}
and the second one is given by the cosmological horizon
\begin{equation*}
  r_{\mathrm{c}} = \frac{2}{\sqrt{\Lambda}}\,\cos\frac{\pi-\alpha}{3}\,;	
\end{equation*}
there is
\begin{equation*}
  \alpha=\arccos\left(\frac{3}{2}\sqrt{\Lambda r_{\mathrm{g}}^{2}}\right)\,.
\end{equation*}
In astrophysically realistic situations, even for the most massive black holes
in the central part of giant galaxies, such as the one observed in the quasar
TON~618 with the mass
$M \sim 6.6\times10^{10}\,\msun$~\cite{Zio:2008:CHIAA:MassBHU}, or for whole
giant galaxies containing an extended CDM halo and having mass up to
$M \sim 10^{14}\,\msun$, the black hole horizon and the cosmological horizon
radii are given with very high precision by the simplified formulae
\begin{equation*}
  r_{\mathrm{h}} = r_{\mathrm{g}}\,,\quad
  r_{\mathrm{c}} = \left(\tfrac{1}{3}\Lambda\right)^{1/2}\,.  
\end{equation*}
The horizons (black-hole and cosmological) thus give two characteristic length
scales of the SdS spacetimes. Clearly, the radius corresponding to the black
hole horizon is located inside the polytropic spheres, while the cosmological
horizon is located outside the polytropic sphere, usually at extremely large
distance from the polytrope for the observationally given value of the relict
cosmological constant.

The \SdS{} geometry can be characterized by a dimensionless
parameter~\cite{Stu-Hle:1999:PHYSR4:}
\begin{equation*}
  y = \tfrac{1}{12}\Lambda r_{\mathrm{g}}^{2}\,.
\end{equation*}
Considering the observationally given repulsive cosmological constant
$\Lambda = 1.3\times 10^{-56}\,\mathrm{cm^{-2}}$, the cosmological parameter
$y$ takes extremely small values for astrophysically relevant objects such as
the stellar mass black holes and galactic center black holes, and even for the
largest compact objects of the universe, i.e., the central supermassive black
holes in the active galactic nuclei or for the related giant
galaxies~\cite{Stu-Sla-Hle:2000:ASTRA:,Stu:2005:MODPLA:}. However, we can
introduce a third characteristic length scale determining the boundary of the
gravitationally bound system, where cosmic repulsive effects start to be
decisive. This is the so called static
radius~\cite{Stu-Hle:1999:PHYSR4:,Stu-Sla:2004:PHYSR4:,Stu-Sch:2011:JCAP:CCMagOnCloud,Stu-Sch:2012:INTJMD:GRvsPsNewtMagClou,Ara:2013:MODPLA:PropGwAsymdS,Ara:2014:1406.2571:KomarMass}
defined as
\begin{equation*}
  r_{\mathrm{s}} = \frac{r_{\mathrm{g}}}{2y^{1/3}} . 
\end{equation*}
At the static radius, the gravitational attraction of the central mass source
is just balanced by the cosmic repulsion and behind the static radius the
cosmic repulsive acceleration prevails~\cite{Stu:2005:MODPLA:}.

It is relevant and instructive to relate the three characteristic length
scales of the external vacuum spacetime to the length scale of the general
relativistic polytrope $\mathcal{L}$ and its radius $R=\mathcal{L}\xi_{1}$. In
the case of polytropes with very large central density, related to the central
densities of neutron stars, quark stars, or other very compact objects, the
polytrope length scale is comparable to the scale of the black hole horizon,
while with decreasing central density the polytrope length scale increases in
comparison to the black hole horizon scale. In the case of extremely low
central densities related to extremely extended polytropes that could
represent, e.g., the CDM halos, their length scale is comparable to the static
radius of the external spacetime. We shall see that the static radius cannot
be exceeded by the polytrope extension. For observationally given cosmological
constant, the length scale (extension) of all astrophysically relevant
polytropes is much lower than the length scale of the cosmological horizon.

\section{\label{energies}Gravitational energy and binding energy of the polytropic spheres}

Properties of the GRPs are well characterized by their gravitational potential
energy and binding energy. The later reflects amount of the microscopic
kinetic energy bounded in the relativistic polytropes. Both the (negative)
gravitational potential energy and the binding energy are related to the total
energy given by the mass parameter of the polytropes, and are expressed in
terms of the parameters characterizing the polytropes that can be determined
numerically. In the case of the $n=0$ polytropes, the binding energy must be
just negatively valued gravitational potential energy, because the polytropic
configurations with uniform distribution of energy density have to be
considered as incompressible.

\subsection{Gravitational potential energy}

Due to the equivalence of matter and energy, the total energy $E$ of the mass
configuration, including the internal energy and gravitational potential
energy, is given by the gravitational mass $M$ generating the external
gravitational field:
\begin{equation*}
  E = Mc^{2} = 4\pi c^{2} \mathcal{L}^{3}\rhocent v(\xi_{1})
        = 4\pi c^{2} \int_{0}^{R}\rho r^{2}\,\dif r\,.               \label{grp55}
\end{equation*}
The proper energy $E_{0}$ is defined as the integral of the energy density
over the proper volume of the fluid sphere
\begin{equation*}
  E_{0} = 4\pi c^{2} \mathcal{L}^{3}\rhocent
         \int_{0}^{\xi_{1}}\!
           g_{rr}^{1/2}
         \theta^{n}\xi^{2}\,\dif\xi                                 \label{grp57}
\end{equation*}
with $g_{rr}(\xi,v;n,\sigma,\lambda)$ given by~(\ref{grr}). The gravitational
potential energy is thus given by
\begin{equation*}
  \mathcal{G} = E - E_{0}\,.                                       \label{grp58}
\end{equation*}
Since $\eto{\Psi}\geq 1$, there is $E_{0} \geq E$ and
$\mathcal{G}\leq 0$---the gravitational potential energy is always
negative. Following the basical work of Tooper~\cite{Too:1964:ASTRJ2:}, we can
consider the negatively valued gravitational potential energy,
$(-\mathcal{G})$, as the gravitational binding energy, i.e., the energy
representing the work that has to be applied to the system in order to
disperse the matter against the gravitational forces. The intensity of the
gravitational binding of the polytropic spheres can be represented by the
ratio
\begin{equation*}
  g = \frac{\mathcal{G}}{E}
    = 1-\frac{1}{v(\xi_{1})}
      \int_{0}^{\xi_{1}}
        g_{rr}^{1/2}\theta^{n}\xi^{2}\,\dif\xi\,.                    \label{grp59}
\end{equation*}

The proper energy of a relativistic polytrope consists of the rest energy of
gas, the kinetic energy of microscopic motion of the gas, and the radiation
energy. The simple polytropic law relates the total energy density and the
total pressure which consists of gas pressure related to the kinetic energy of
the microscopic motion, and the radiation pressure. Therefore, we have to
determine the gas density of the polytropic matter.

\subsection{Adiabatic processes and speed of sound}

In the relativistic polytropes, the special case of adiabatic processes
implies a unique relation between the gas density $\rho_{\mathrm{g}}$ and the
total mass density $\rho$, or between $\rho_{\mathrm{g}}$ and
$\theta$~\cite{Too:1964:ASTRJ2:}. The assumption of an adiabatic process is
consistent with the absence of heat terms in the energy-momentum tensor. For
an adiabatic process the relativistic generalization of the first law of
thermodynamics takes the form
\begin{equation*}
  \dif\epsilon + (p+ \epsilon)\frac{\dif V}{V} = 0                 \label{grp60}
\end{equation*}
where $\dif\epsilon$ is the change in the energy density due to a change
$\dif V$ in the specific volume. Since
\begin{equation*}
  \frac{\dif V}{V}
    = - \frac{\dif\rho_{\mathrm{g}}}{\rho_{\mathrm{g}}}\,,              \label{grp61}
\end{equation*}
we arrive at
\begin{equation*}
  \frac{\dif \rho_{\mathrm{g}}}{\rho_{\mathrm{g}}} = \frac{\dif \epsilon}
        {p + \epsilon}                                             \label{grp62}
\end{equation*}
and using the variable $\theta$, we find equation
\begin{equation*}
  \frac{\dif \rho_{\mathrm{g}}}{\rho_{\mathrm{g}}}
    = \frac{n \dif \theta}{\theta(1+ \sigma \theta)}\,.            \label{grp63}
\end{equation*}
Because the internal energy density is small being compared to the rest energy
density near the boundary of the polytropic sphere, we obtain the profile of
the rest mass density in the form
\begin{equation*}
  \rho_{\mathrm{g}}
    = \rhocent \left(\frac{\theta}{1+ \sigma \theta}\right)^{2}
    = \frac{\rho}{(1+ \sigma \theta)^{n}}\,.                       \label{grp64}
\end{equation*}
In the nonrelativistic limit ($\sigma \ll 1$), the gas density and the total
density are nearly equal.

The standard relativistic (Landau--Lifshitz) formula for the phase velocity of
sound in an adiabatic process~\cite{Lan-Lif:1987:FluidMech:}
\begin{equation*}
  v_{\mathrm{s}}^{2}
    = \left(\oder{p}{\rho}\right)_{\mathrm{adiabatic}}                 \label{grp65}
\end{equation*}
yields the phase sound speed at the center of the polytrope to be given by
\begin{equation*}
  v_{\mathrm{sc}}
    = c \left(\frac{n+1}{n} \sigma \right)^{1/2}\,.                 \label{grp66}
\end{equation*}
For a given $n$ there is a maximum value of the parameter $\sigma$ that
guarantees $v_{\mathrm{sc}} < c$:
\begin{equation*}
  \sigma \leq \frac{n}{n+1}\,.                                     \label{grp67}
\end{equation*}
For the nonrelativistic Fermi gas $n=3/2$ we have
$\sigma_{\mathrm{N}} \leq 3/5$, while for the ultrarelativistic Fermi gas
$n=3$ we have $\sigma_{\mathrm{U}} \leq 3/4$; for the case of $n=4$ there is
$\sigma \leq 4/5$. However, these limits hold for the phase sound velocity,
not the group velocity, so they should not be taken too
literally~\cite{Too:1964:ASTRJ2:}.

\subsection{Binding energy}

The proper mass and the total rest energy of gas in a polytropic sphere are
determined by the relation
\begin{equation*}
  E_{0\mathrm{g}} = M_{0\mathrm{g}} c^{2}
    = 4 \pi c^{2} \int_{0}^R \rho_{\mathrm{g}} 
                              \eto{\Psi} r^{2}\,\dif r\,.          \label{grp68}
\end{equation*}
The energy $E_{0\mathrm{g}}$ represents the sum of the rest masses of the
elementary particles in the polytrope in units of energy, and
$M_{0\mathrm{g}}$ gives number of nucleons in the polytrope multiplied by the
nucleon rest mass. The proper rest energy of the gas in the polytropic
configuration is given by integration over the proper volume and is determined
by the relation
\begin{equation*}
  E_{0\mathrm{g}}
    = 4 \pi c^{2} \mathcal{L}^{3} \rhocent \int_{0}^{\xi_{1}}
      \frac{g_{rr}^{1/2}\theta^{n}\xi^{2} \dif \xi}%
        {(1+ \sigma \theta)^{n}}                                   \label{grp69}
\end{equation*}
with $g_{rr}(\xi,v;n,\sigma,\lambda)$ given by~(\ref{grr}). The binding energy
$E_{\mathrm{b}}$ of the gas of the polytropic sphere is then given by the
formula
\begin{equation*}
  E_{\mathrm{b}} = E_{0\mathrm{g}} - E\,.                               \label{grp70}
\end{equation*}
Considering an `initial' state where the particles are widely dispersed and
the system has zero internal energy, and assuming conservation of the number
of nucleons, the binding energy represents the difference in energy between
the `initial' state and the `final' state in which the particles with given
internal energy are bounded by gravitational forces.

We can consider the quantity giving the difference of the proper energy
$E_{0}$ and the proper rest energy $E_{0\mathrm{g}}$, describing the internal
`kinetic' energy of the polytropic sphere (more precisely of particles
constituting the polytrope)
\begin{equation*}
  E_{\mathrm{k}} = E_{0} - E_{0\mathrm{g}}\,.                           \label{grp71}
\end{equation*}
Polytropic fluid spheres can be characterized by relating the gravitational
potential energy, the binding energy, and the kinetic energy, to the total
energy, introducing the following parameters. The internal energy parameter
\begin{equation*}
  i \equiv \frac{E_{0\mathrm{g}}}{E} = \frac{1}{v(\xi_{1})}
    \int_{0}^{\xi_{1}}
    \frac{g_{rr}^{1/2}\theta^{n} \xi^{2} \dif \xi}%
        {(1+ \sigma \theta)^{n}}\,,                               \label{grp72}
\end{equation*}
binding energy parameter
\begin{equation*}
  b \equiv \frac{E_{\mathrm{b}}}{E}
    = \frac{E_{0\mathrm{g}}}{E} - 1= i- 1\,,                         \label{grp73}
\end{equation*}
and the kinetic energy parameter
\begin{equation*}
  k \equiv \frac{E_{\mathrm{k}}}{E}
    = \frac{E_{0}}{E}- \frac{E_{0\mathrm{g}}}{E}\,.                   \label{grp74}
\end{equation*}
Clearly, the parameters are not independent. They are related by
\begin{equation*}
  k = 1 - g - i = - g - b\,.                                      \label{grp75}
\end{equation*}
It is not apparent, if the binding energy is positive, or negative. The gas
density $\rho_{\mathrm{g}}$ is smaller than the total density $\rho$, but the
radial metric coefficient is in general greater than unity. Recall that in the
Newtonian limit (with $\lambda = 0$), we obtain in the first approximation
\begin{align}
  E_{0} &\approx 4 \pi c^{2} \int_{0}^R \rho(r)
          \left[1+ \frac{Gm(r)}{c^{2} r}\right] r^{2} \dif r  \notag\\
       &= E + \int_{0}^R \frac{Gm(r)\,\dif m(r)}{r}        \nobadge\label{grp76}
\end{align}
and the binding energy is determined by the well known
formula~\cite{Sha-Teu:1983:BHWDNS:}
\begin{equation*}
  E_{\mathrm{b}} \approx \frac{3-n}{5-n} \frac{GM^{2}}{R} 
    \approx \frac{n-3}{3} G\,.                                     \label{grp77}
\end{equation*}
The Newtonian limit demonstrates immediately that the binding energy can be
positive or negative, in dependence on the polytropic index $n$. Since the
gravitational energy is always negative, we can conclude that in this limit
the binding energy is positive (negative) for $n<3$ ($n>3$). In the fully
GRPs, the situations is clearly more complex. The fully general relativistic
polytropic spheres are characterized by the most important quantity relating
the binding energy and the gravitational potential energy through the formula
\begin{equation*}
  \frac{E_{\mathrm{b}}}{G}
    = \frac{%
    \int_{0}^{\xi_{1}}\frac{g_{rr}^{1/2}\theta^{n} \xi^{2} \,\dif \xi}
    {(1+ \sigma \theta)^{n}}
    - v(\xi_{1})}{v(\xi_{1})- \int_{0}^{\xi_{1}}
      g_{rr}^{1/2}\theta^{n} \xi^{2} \,\dif \xi}                     \label{grp78}
\end{equation*}
that enables to find easily the regions of positively valued binding energy
since the gravitational energy is again always negative.

\section{\label{embd}Embeddings of the ordinary and optical space}

We concentrate our attention into visualization of the structure of the
internal spacetime of the GRPs, considering both the ordinary and the optical
geometry of the spacetime.

The curvature of the internal spacetime of the polytropes can conveniently be
represented by the standard embedding of 2D, appropriately chosen, spacelike
surfaces of the ordinary 3-space of the geometry (here, these are
$t=\mathrm{const}$ sections of the central planes) into 3D Euclidean
space~\cite{Mis-Tho-Whe:1973:Gra:}.

The 3D optical reference geometry~\cite{Abr-Car-Las:1988:GENRG2:}, associated
with the spacetime under consideration, enables introduction of a natural
`Newtonian' concept of gravitational and inertial forces and reflects some
hidden properties of the test particle
motion~\cite{Abr:1990:MONNR:,Abr-Mil-Stu:1993:PHYSR4:,Stu-Hle-Jur:2000:CLAQG:,Kov-Stu:2007:CLAQG:}. (In
accord with the spirit of general relativity, alternative approaches to the
concept of inertial forces are possible, e.g., the `special relativistic'
one~\cite{Sem:1995:NUOC2:}.)  Properties of the inertial forces can be
reflected by the embedding diagrams of appropriately chosen 2D sections of the
optical geometry, as reviewed, e.g.,
in~\cite{Hle:2002:JB60:,Stu-Hle-Jur:2000:CLAQG:}. The embedding diagrams of
the $n=0$ polytropes with the uniform distribution of the energy density were
presented in~\cite{Stu-etal:2001:PHYSR4:}, here they are constructed for
typical GRPs with $n>0$. Note that it can be directly shown by using the
optical reference geometry that extremely compact configurations, allowing the
existence of bound null geodesics, can
exist~\cite{Stu-etal:2001:PHYSR4:,Stu-etal:2012:GENRG2:NeutrinoTrap}. Such
extremely compact relativistic polytropes have a turning point of the
embedding diagram of the optical geometry as shown
in~\cite{Stu-Hle-Jur:2000:CLAQG:}. However, as we show later, such
configurations can have compactness parameter $\mathcal{C} > 1/3$.

We embed the equatorial plane of the ordinary space geometry and optical
reference geometry into the 3D Euclidean space with the line element
\begin{equation*}
  \dif\tilde{\sigma}^{2} = \dif \rho^{2} + \rho^{2} \,\dif \alpha^{2}
    + \dif z^{2}\,.                                                   \label{e1}
\end{equation*}
The embedding is a rotationally symmetric surface $z=z(\rho)$ with the line
element (2D):
\begin{equation*}
  \dif \ell_{\mathrm{(E)}}^{2}
    = \left[1+ \left(\oder{z}{\rho}\right)^{2} \right]
    \,\dif \rho^{2} + \rho^{2} \,\dif\alpha^{2}\,.                      \label{e2}
\end{equation*}

\subsection{Ordinary space}

Its equatorial plane has the line element
\begin{equation*}
  \dif \ell^{2}_{\mathrm{(ord)}}
    = g_{rr} \,\dif r^{2} + r^{2} \,\dif \phi^{2}                      \label{ee3}
\end{equation*}
where
\begin{equation}
  g_{rr}= \eto{2\Psi(r)}= \left\{1- 2 \sigma(n+1) \left[
    \frac{v(\xi)}{\xi}+ \frac{\lambda}{3} \xi^{2} 
    \right]\right\}^{-1}                                              \label{e4}
\end{equation}
with $v(\xi)$ being the solution of the TOV for the GRP\@. We have to identify
$\dif\ell^{2}_{\mathrm{(E)}}$ and $\dif \ell^{2}_{\mathrm{(ord)}}$. Clearly,
$\alpha\equiv \phi$, and $\rho \equiv r$. The embedding formula then takes the
form
\begin{equation*}
  \oder{z}{r} = \pm \sqrt{g_{rr} - 1}\,;                              \label{e5}
\end{equation*}
different signs give isometric surfaces. We take `+' sign. Using
Eq.\,(\ref{e4}), we arrive at the dimensionless embedding formula, if we
introduce
\begin{equation*}
  z = \frac{z}{\mathcal{L}}\,, \qquad
  \xi = \frac{r}{\mathcal{L}}                                         \label{e6}
\end{equation*}
in the form
\begin{equation*}
  \oder{z}{\xi} = \left\{\frac{2 \sigma (n+1)\left[\frac{v(\xi)}{\xi}
    + \frac{\lambda}{3} \xi^{2} \right]}
    {1- 2\sigma (n+1) \left[\frac{v(\xi)}{\xi}+ \frac{\lambda}{3}
    \xi^{2} \right]}  \right\}^{1/2}\,.                                \label{e7}
\end{equation*}
This must be integrated numerically using computer code for $v(\xi)$. Clearly,
the embedding is well defined in whole the range of allowed
$\xi \in (0, \xi_{1})$, as $g_{rr} > 1$ there.

\subsection{Optical space (optical reference geometry)}

In the static spacetimes, the optical 3D space has its metric coefficients
determined by~\cite{Abr-Car-Las:1988:GENRG2:,Abr-Nur-Wex:1995:CLAQG:}
\begin{equation*}
  h_{ik}= \frac{g_{ik}}{-g_{tt}}\,.                                     \label{e8}
\end{equation*}
Its equatorial plane has the line element
\begin{equation*}
  \dif \ell^{2}_{(\mathrm{opt})}
    = h_{rr} \,\dif r^{2} + h_{\phi\phi}\,\dif \phi^{2}                   \label{e9}
\end{equation*}
that has to be identified with $\dif \ell^{2}_{\mathrm{(E)}}$. Now, the
azimuthal coordinates still can be identified ($\alpha \equiv \phi$), however
the radial coordinates are related via
\begin{equation*}
  \rho^{2} = h_{\phi \phi}                                              \label{e10}
\end{equation*}
and the embedding formula is given by
\begin{equation*}
  \oder{z}{\rho}
    = h_{rr} \left(\oder{r}{\rho}\right)^{2} -1\,.                    \label{e11}
\end{equation*}
It is convenient to cast the embedding formula into a parametric form
$z(\rho)= z(r(\rho))$. Then
\begin{equation*}
  \oder{z}{r} = \sqrt{h_{rr}- \left(\oder{\rho}{r}\right)^{2}}\,.     \label{e12}
\end{equation*}
Because
\begin{equation*}
  \oder{z}{\rho} = \oder{z}{r} \oder{r}{\rho}\,,                     \label{e13}
\end{equation*}
the turning points of the embedding diagrams are given by the condition
\begin{equation*}
  \oder{\rho}{r} = 0\,.                                              \label{e14}
\end{equation*}
The reality condition, determining the limits of embeddability, reads
\begin{equation*}
  h_{rr} - \left(\oder{\rho}{r}\right)^{2} \geq 0\,.                  \label{e15}
\end{equation*}
For the GRPs, the metric coefficients of the optical geometry are given by the
formulae
\begin{align}
   h_{rr} &= \frac{\eto{2\Psi}}{\eto{2\Phi}}=
     \frac{[1+ \sigma \theta (\xi)]^{2(n+1)}}%
     {1-2\sigma(n+1)
       \left[
         \frac{v(\xi_{1})}{\xi_{1}} + \frac{\lambda}{3}\xi_{1}^{2}
       \right]}  \notag\\
     &\hphantom{==}\times\left\{
       1- 2\sigma (n+1)
       \left[
         \frac{v(\xi)}{\xi} +\frac{\lambda}{3} \xi^{2}
       \right]
     \right\}\,,\nobadge\\                                       
  h_{\phi \phi} &= \frac{r^{2}}{\eto{2 \Phi}} =
    \frac{r^{2}[1+ \sigma \theta (\xi)]^{2(n+1)}}
    {1-2 \sigma(n+1) \left[\frac{v(\xi_{1})}{\xi_{1}} + 
    \frac{\lambda}{3}\xi_{1}^{2}\right]}\,.\nobadge                 
\end{align}
Introducing a dimensionless coordinate $\eta$ by
\begin{equation*}
  \eta = \frac{\rho}{\mathcal{L}}\,,                                 \label{e18}
\end{equation*}
we can write
\begin{equation*}
  \eta = \frac{\xi[1+ \sigma \theta(\xi)]^{n+1}}
    {\left\{1- 2 \sigma(n+1) \left[\frac{v(\xi_{1})}{\xi_{1}}
    + \frac{\lambda}{3} \xi^{2}_{1}\right]\right\}^{1/2}}              \label{e19}
\end{equation*}
and
\begin{equation*}
  \oder{\eta}{\xi} = \frac{\xi[1+ \sigma \theta(\xi)]^n
    \left\{1+ \sigma \left[\theta(\xi) + (n+1) \xi \oder{\theta}{\xi}
                                        \right]\right\}}
    {\left\{1- 2 \sigma(n+1) \left[\frac{v(\xi_{1})}{\xi_{1}}
    + \frac{\lambda}{3} \xi^{2}_{1}\right]\right\}^{1/2}}\,.           \label{e20}
\end{equation*}
The condition for the turning points of the embedding diagrams thus reads
\begin{equation*}
  \theta(\xi) + (n+1) \xi \oder{\theta}{\xi}= 
    -\frac{1}{\sigma}\,.                                             \label{e21}
\end{equation*}
The embedding formula takes the form 
\begin{widetext}
\begin{align}
   \left(\oder{z}{\xi}\right)^{2} &=
     \left\{
       1- 2 \sigma (n+1)
       \left[
         \frac{v(\xi_{1})}{\xi_{1}} + \frac{\lambda}{3} \xi^{2}_{1}
       \right]
     \right\}
     \left\{
       1-2\sigma (n+1)\left[\frac{v(\xi)}{\xi} + \frac{\lambda}{3}\xi^{2}\right]
     \right\}
     2\sigma (n+1)[1+ \sigma \theta(\xi)]^{2n}\notag\\
     & {}\times
     \left\{
       \left\{1+ \sigma
         \left[\theta(\xi) + (n+1)\xi\oder{\theta}{\xi}\right]
       \right\}
     \left[\frac{v(\xi)}{\xi} + \frac{\lambda}{3}\xi^{2}\right]
     - \xi \oder{\theta}{\xi}
     \left[
       1+ \sigma\theta(\xi)+\frac{\sigma}{2}(n+1)\xi \oder{\theta}{\xi}
     \right]
     \right\}\,.                                            \nobadge\label{e22}
\end{align}
\end{widetext}
This has to be solved numerically, together with the condition on the limits
of embeddability given in the form
\begin{gather}
  \left\{1+ \sigma \left[\theta(\xi) + (n+1) \xi \oder{\theta}{\xi}
    \right]\right\}
      \left[\frac{v(\xi)}{\xi}+\frac{\lambda}{3}\xi^{2}\right]       \notag\\
   {} - \xi \oder{\theta}{\xi} \left[1+ \sigma \theta (\xi) + 
    \frac{\sigma}{2}(n+1)
    \xi\oder{\theta}{\xi}\right]\geq 0\,.                   \nobadge\label{e23}
\end{gather}

\section{\label{cud}Configurations of uniform density}

There is a special class of GRPs of the index $n = 0$ where the structure
equations can be integrated in terms of elementary functions. We shall discuss
these polytropes in detail because they can give an intuitive insight into the
role of the cosmological constant and can serve as a test bed for the general
case of polytropes with $n>0$.

\begin{figure}[t]
\begin{center}
\includegraphics[width=\linewidth]{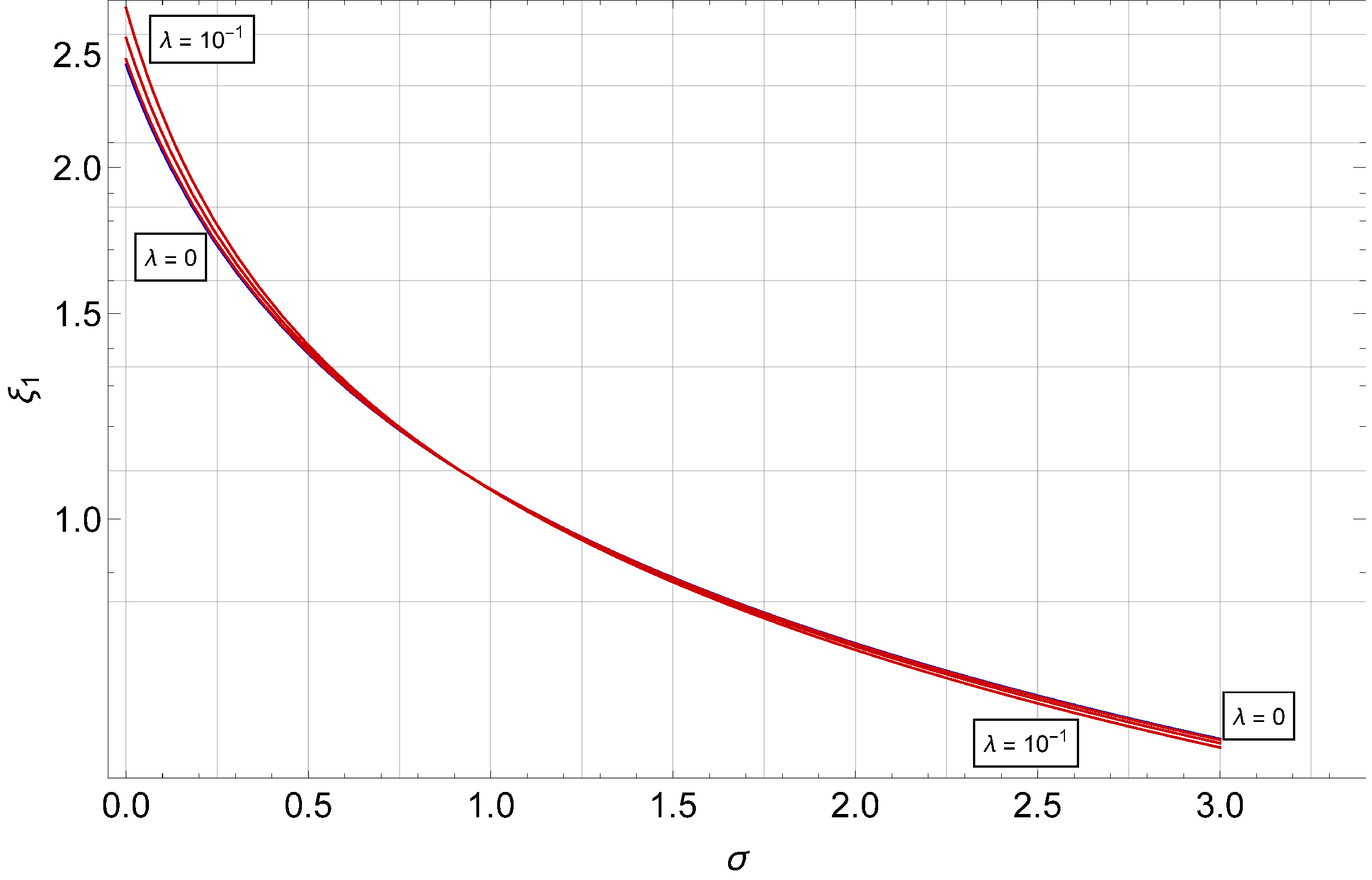}
\end{center}
\caption{\label{UniSigXi1}Dependence of the dimensionless radius
  $\xi_{1}$~(\ref{grp90}) for configurations of uniform density on relativity
  parameter $\sigma$ for cosmological parameter
  $\lambda = 0, 10^{-2}, 5\times10^{-2}, 10^{-1}$, respectively.}
\end{figure}

The $n=0$ polytropes correspond to the special class of the internal \SaodS{}
spacetimes~\cite{Stu:2000:ACTPS2:} where the distribution of density $\rho$ is
uniform although the pressure grows monotonically from its zero value on the
surface of the configuration to a maximum value at its center. Recall that in
the configurations with $\rho = \mathrm {const}$, it is not necessary to use
the unrealistic notion of an incompressible fluid---one can consider fluids
with pressure growing as radius decreases, being `hand
tailored'~\cite{Mis-Tho-Whe:1973:Gra:}.  Assuming $n=0$, Eq.\,(\ref{grp32})
can be integrated to give
\begin{equation*}
  v(\xi) = \tfrac{1}{3}\xi^{3}\,,                                  \label{grp87}
\end{equation*}
while the Eq.\,(\ref{grp31}) takes the form
\begin{equation}
  \left[3-2\sigma(1+\lambda)\xi^{2}\right]\oder{\theta}{\xi}
  + (1-2\lambda+3\sigma\theta)(1+\sigma\theta) = 0                 \label{grp88}
\end{equation}
and can be integrated directly after separation of variables. Using the
boundary condition $\theta(0)=1$, we obtain
\begin{equation*}
  \sigma\theta = \frac{1
    -\frac{(1-2\lambda)(1+\sigma)}%
      {(1-2\lambda+3\sigma)
        \left[1-\frac{2}{3}\sigma(1+\lambda)\xi^{2}\right]^{1/2}}}%
    {\frac{3(1+\sigma)}%
      {(1-2\lambda+3\sigma)\left[1-\frac{2}{3}
      \sigma(1+\lambda)\xi^{2}\right]^{1/2}}-1}\,.                  \label{grp89}
\end{equation*}
This solution determines the dependence of pressure on the radial coordinate,
since for $n=0$ there is $\theta = p(r)/\pcent$. The dependence is
given in units of the energy density since $\sigma = \pcent/\rhocent$.
From the condition $\theta(\xi_{1})= 0$, we find the radius of the
configuration to be determined by
\begin{equation}
  \xi_{1}^{2}(\sigma,\lambda) = \frac{6[1+2\sigma-\lambda(2+\sigma)]}%
                 {(1-2\lambda+3\sigma)^{2}}\,.                     \label{grp90}
\end{equation}
We illustrate behavior of the function $\xi_{1}^{2}(\sigma,\lambda)$ in
Fig.\,\ref{UniSigXi1}. Clearly, the parameters $\sigma$ and $\lambda$ have to
be restricted by the condition
\begin{equation*}
  \lambda \leq \frac{1+2\sigma}{2+\sigma}\,.
\end{equation*}
However, the $n=0$ polytropic configurations should behave regularly for all
allowed values of the relativistic parameter $\sigma > 0$, but
$\xi_{1}^{2}(\sigma,\lambda)$ always diverges for $\sigma$ low enough, if
$\lambda > 1/2$. Therefore, it is natural to put the restriction of
\begin{equation*}
  \lambda \leq \tfrac{1}{2}\,.
\end{equation*}

We can express the pressure profile in terms of $r$, $R$, instead of $\xi$,
$\xi_{1}$ and $\sigma, \lambda$, obtaining thus the form of expression of the
$n=0$ polytrope as being discussed in~\cite{Stu:2000:ACTPS2:}. Introducing a
new parameter $a$, having dimension of length, by the relation
\begin{equation*}
  \frac{1}{a^{2}}
    = \frac{8\pi}{3}(\rhocent + \rho_{\mathrm{vac}})\,,               \label{grp91}
\end{equation*}
we find the relation of $R$, $\sigma$, and $\lambda$ to be given by
\begin{equation*}
  \left(1-\frac{R^{2}}{a^{2}}\right)^{1/2}
  = \frac{(1-2\lambda)(1+\sigma)}{1-2\lambda+3\sigma}\,.           \label{grp93}
\end{equation*}
The central pressure $\pcent$ and the pressure profile $p(r)$ can be then
expressed in the known
form~\cite{Stu:2000:ACTPS2:,Boh:2004:GENRQC:}
\begin{subequations}                                          \label{grp94and95}
\begin{align}
  \pcent &= \rhocent\frac%
    {(1-2\lambda)\left[1-\left(1-\frac{R^{2}}{a^{2}}\right)^{1/2}\right]}%
    {3\left(1-\frac{R^{2}}{a^{2}}\right)^{1/2}-(1-2\lambda)}\,,    \label{grp94}\\
  p(r) &= \rhocent\frac%
    {(1-2\lambda)
      \left[
        \left(1-\frac{r^{2}}{a^{2}}\right)^{1/2}
        -\left(1-\frac{R^{2}}{a^{2}}\right)^{1/2}
      \right]}%
    {3\left(1-\frac{R^{2}}{a^{2}}\right)^{1/2}
      -(1-2\lambda)\left(1-\frac{r^{2}}{a^{2}}\right)^{1/2}}\,.      \label{grp95}
\end{align}
\end{subequations} 
The presented results for the $n=0$ polytropes are relevant for both the
positive and negative values of the cosmological parameter $\lambda$
describing thus also effects of the attractive cosmological constant when
$\lambda < 0$.

The radial metric coefficient is given by the relation
\begin{equation}
  \eto{2\Psi(r)} = \left(1-\frac{r^{2}}{a^{2}}\right)^{-1}\,,        \label{grp96}
\end{equation}
and the total mass reads 
\begin{equation*}
  M = \frac{1}{3}\mathcal{L}\sigma\xi_{1}^{3}
    = \frac{4\pi}{3}\rhocent r^{3}\,.                              \label{grp97}
\end{equation*}
The temporal metric coefficient is determined by the relations  
\begin{subequations}                                          \label{grp98and99}
\begin{align}
   \eto{\Phi(r)} &= \frac{[1 - \frac{2}{3}\sigma(1+\lambda)\xi^{2}]^{1/2}}
                     {1 + \sigma\theta}\,,                       \label{grp99}\\
  \eto{\Phi(r)} &= \frac%
    {3\left(1-\frac{R^{2}}{a^{2}}\right)^{1/2}
      -(1-2\lambda)\left(1-\frac{r^{2}}{a^{2}}\right)^{1/2}}%
    {2(1+\lambda)}\,.                                              \label{grp98}
\end{align}
\end{subequations}

The special case of the attractive cosmological constant corresponding to
$\lambda = -1$ has to be treated separately as $1/a^{2}=0$. In such a case,
Eq.\,(\ref{grp88}) reduces to
\begin{equation*}
  \frac{\dif\theta}{(1+\sigma\theta)^{2}} = -\xi\,\dif\xi         \label{grp100}
\end{equation*}
which leads, after integration with the boundary condition $\theta(0)=1$, to
the formula
\begin{equation*}
  \theta = \frac%
    {1-\frac{1}{2}(1+\sigma)\xi^{2}}%
    {1+\frac{1}{2}(1+\sigma)\xi^{2}}\,.                           \label{grp101}
\end{equation*}
The boundary of the configuration is at 
\begin{equation*}
  \xi_{1}^{2} = \frac{2}{1+\sigma}\,.                            \label{grp102}
\end{equation*}
The central pressure and the pressure profile can be expressed in terms of the
radial coordinates $r$, $R$ in the form
\begin{subequations}
\begin{align}
  \pcent &=
    \frac{\rhocent}{\left(\frac{2R}{3M}-1\right)}\,,\nobadge\\
  p(r) &= \pcent\frac{1-\frac{r^{2}}{R^{2}}}%
    {1+\frac{1}{\frac{2R}{3M}-1}\frac{r^{2}}{R^{2}}}\,.\nobadge 
\end{align}
\end{subequations}
The metric coefficients have a special form, too. The radial $g_{rr}$
component corresponds to the flat $t = \mathrm{const}$ sections
\begin{equation*}
  \eto{2\Psi(r)} = 1                                              \label{grp105}
\end{equation*}
while the temporal $g_{tt}$ component takes the form~\cite{Stu:2000:ACTPS2:}
\begin{equation*}
  \eto{\Phi(r)} = 1
    + \frac{3M}{2R}\left(\frac{r^{2}}{R^{2}}-1\right)\,.           \label{grp107}
\end{equation*}
Properties of the $n=0$ GRPs were discussed
in~\cite{Stu:2000:ACTPS2:,Stu-etal:2001:PHYSR4:}. In the following we
concentrate on the existence of these polytropes in dependence on the
repulsive cosmological constant.

\subsection{Existence conditions of the $\bm{n=0}$ polytropes and their
  compactness}

The reality conditions on the general solution, given by Eqs~(\ref{grp94and95}),
(\ref{grp96}), and~(\ref{grp98}), must guarantee that the pressure is positive
and non-divergent, and the metric coefficients have to be regular at
$r \leq R$.  Therefore, two conditions have to be satisfied:
\begin{equation}
  1-2\lambda > 0                                                  \label{grp108}
\end{equation}
and
\begin{equation}
  3\left(1-\frac{R^{2}}{a^{2}}\right)^{1/2} - (1-2\lambda) > 0\,.   \label{grp109}
\end{equation}
Considering the limiting case of $\lambda = 1/2$, we find
\begin{equation*}
  \theta(\xi) = \frac{(1-\sigma\xi^{2})^{1/2}}%
                     {1+\sigma + \sigma(1-\sigma\xi)^{1/2}}\,.     \label{grp110}
\end{equation*}
The boundary of such a polytrope configuration is at
\begin{equation}
  \xi_{1}^{2} = \frac{1}{\sigma}\,,                                \label{grp111}
\end{equation}
and we can show that
\begin{equation*}
  R^{2} = a^{2}\,.                                                 \label{grp112}
\end{equation*}
Therefore, the metric coefficients are singular because there is
\begin{equation*}
  \eto{-2\Psi(R)} = \eto{2\Phi(R)} = \eto{2\Phi(r)}
                   = 1-\frac{R^{2}}{a^{2}}\,.                      \label{grp113}
\end{equation*}
For polytrope configurations of a given $M$ and $\Lambda > 0$,
condition~(\ref{grp108}) gives an upper limit on admissible values of the
external radius $R$. Configurations with $\lambda \sim 1/2$ can be considered
as nearly `geodetical' since the pressure gradient almost vanishes on the
surface, which is close to the static radius of the external geometry. (For
$\lambda = 1/2$, the surface of the static configuration has to be located at
the horizon of the external spacetime, with $R=3r_{\mathrm{g}}/3$; however, no
static configuration can have its boundary at a \mbox{black-hole} horizon, and
such configurations are forbidden.)

The lower limit on the external radius of the $n=0$ polytropes is determined
by the condition~(\ref{grp109}) that can be transformed into the relations
\begin{equation*}
  \frac{R^{2}}{a^{2}} < \frac{4(1+\lambda)(2-\lambda)}{9}          \label{grp114}
\end{equation*}
and
\begin{equation*}
  R > 2M\frac{9}{4(2-\lambda)}\,.                                 \label{grp115}
\end{equation*}
For $\lambda = 0$, we obtain the well known limit $R> (9/4)(GM/c^{2})$. The
restrictions on physically realistic $n=0$ polytropes can also be transformed
into a form containing dimensionless quantities $x \equiv R/M$,
$y \equiv \Lambda M^{2}/3$ (see
Refs~\cite{Stu:2000:ACTPS2:,Stu-etal:2001:PHYSR4:}).

Compactness of the polytropic spheres of the uniform density is given by the
relation
\begin{equation*}
  \mathcal{C}(\sigma, \lambda)= \frac{2 \sigma [1- 2 \lambda + \sigma
    (2- \lambda)]}{(1- 2 \lambda + 3 \sigma)^{2}}                 \label{grp-c1}
\end{equation*}
that is reduced for $\lambda = 0$ to the formula
\begin{equation*}
  \mathcal{C}(\sigma)
    = \frac{2 \sigma (1+ 2 \sigma)}{(1+ 3 \sigma)^{2}}\,.         \label{grp-c2}
\end{equation*}
The extremely compact configurations with $\mathcal{C} > 1/3$ can exist, if
the central parameter satisfies the relation
\begin{equation*}
  \sigma^{2} \geq \sigma^{2}_{\mathrm{ext}} \equiv
    \tfrac{2}{3}\lambda\,.                                        \label{grp-c3}
\end{equation*}
As extremely compact spherical configurations are denoted those having their
surface located under the photon circular orbit of the external
spacetime~\cite{Stu-etal:2009:CLAQG:NeuTrap1Eff}. It can be
shown~\cite{Stu-etal:2001:PHYSR4:} that in extremely compact configurations
(with $R < 3M$), a stable circular null geodesic exists around which null
geodesics captured by the strong gravitational field are
concentrated. Neutrinos, moving along these bound null geodesics, can
influence cooling of extremely compact neutron stars.  The potential well of
the captured geodesics becomes deeper with the repulsive cosmological constant
increasing, while it gets flatter with the attractive cosmological constant
decreasing~\cite{Stu-etal:2012:GENRG2:NeutrinoTrap}.

\subsection{Gravitational binding of the $\bm{n=0}$ polytropes}

\begin{figure}[t]
\begin{center}
\includegraphics[width=\linewidth]{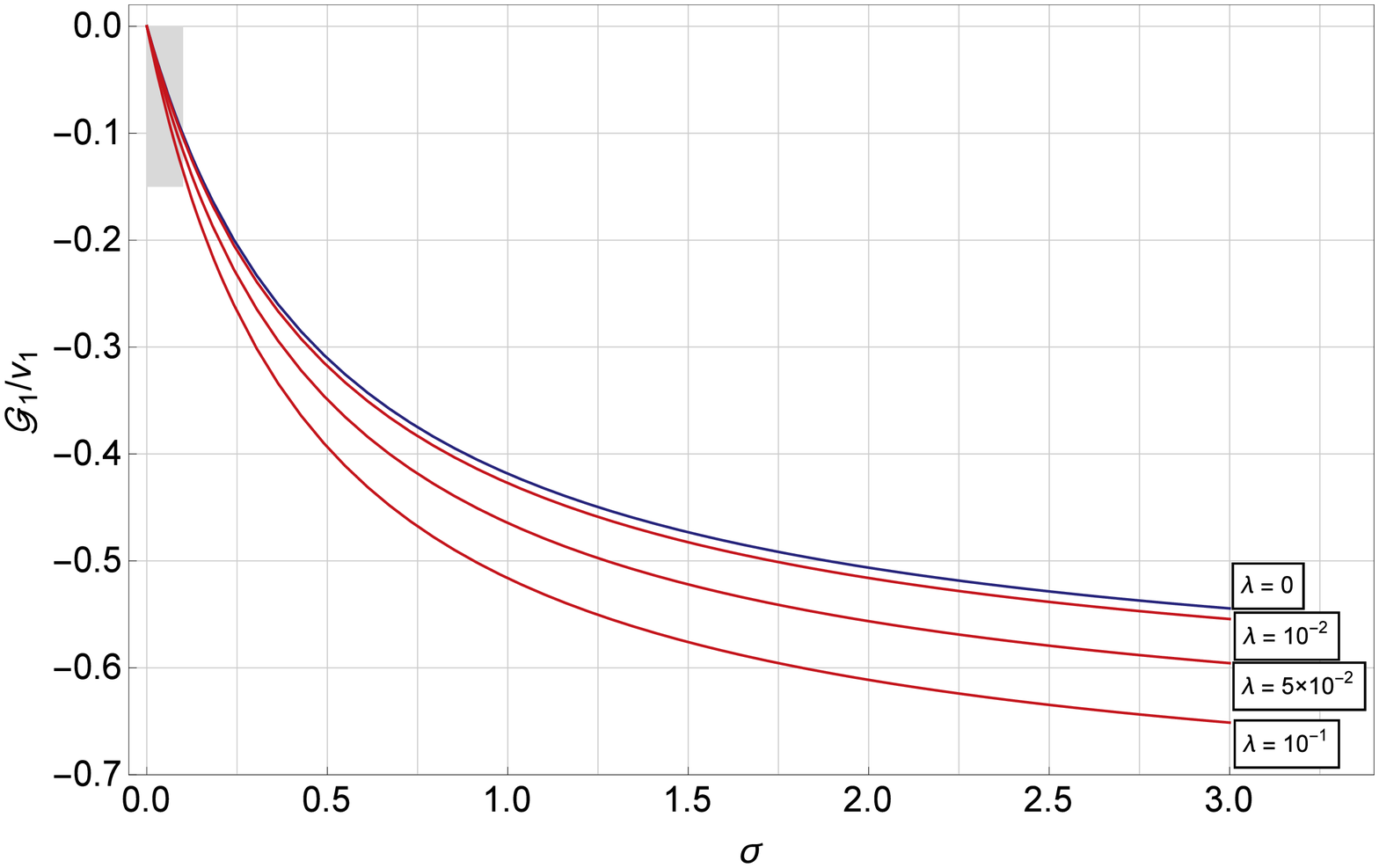}
\par\vspace{0.8\baselineskip}\par
\includegraphics[width=\linewidth]{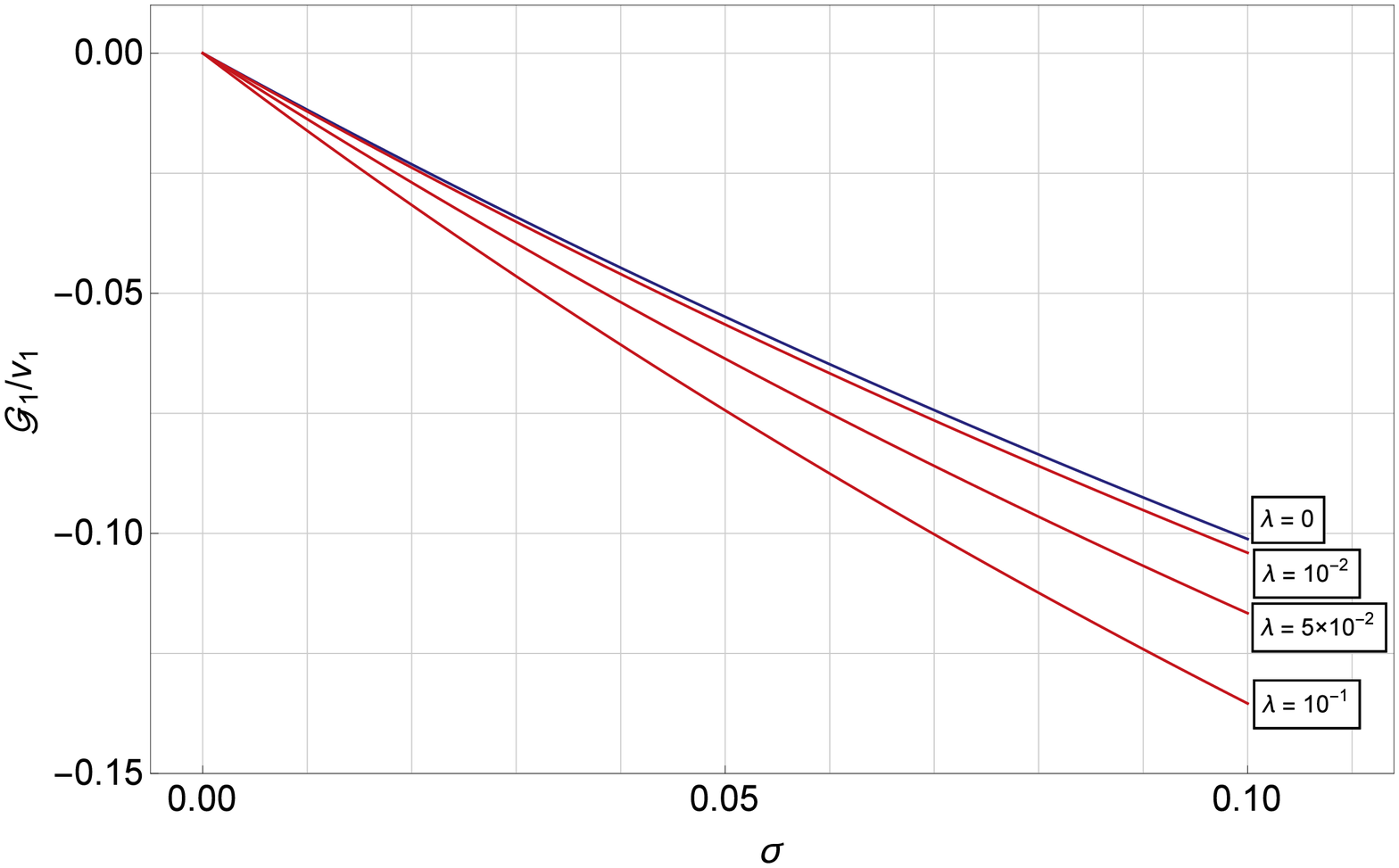}
\end{center}
\caption{\label{UniSigGBEn}Dependence of the gravitational binding
  factor~(\ref{grp121}) for configurations of uniform density on relativity
  parameter $\sigma$ for cosmological parameter
  $\lambda = 0, 10^{-2}, 5\times10^{-2}, 10^{-1}$, respectively. The ``1''
  index emphasizes the energies are related to the whole
  configuration. \emph{Top:}~Wide range of relativity
  parameter. \emph{Bottom:}~Zoom of the shaded region in the top plot.}
\end{figure}

It is instructive to give the gravitational energy of the $n=0$ polytropes and
their gravitational binding factor $g$. The total energy takes the simple form
\begin{equation*}
  E = M = \frac{4\pi}{3}\rhocent R^{3}\,.                         \label{grp116}
\end{equation*}
The formula for the total proper energy reads 
\begin{equation*}
  E_{0} = 4\pi\rhocent\int_{0}^R
    \left(1-\frac{r^{2}}{a^{2}}\right)^{-1/2}\,r^{2}\,\dif r\,;      \label{grp117}
\end{equation*}
after integration we obtain 
\begin{equation*}
  E_{0} = \frac{3M}{2}\left(\frac{a}{R}\right)^{3}
    \left[
      \arcsin\left(\frac{R}{a}\right)
      -\frac{R}{a}\left(1-\frac{R^{2}}{a^{2}}\right)^{1/2}
    \right] ,                                                     \label{grp118}
\end{equation*}
where
\begin{equation*}
  \frac{R}{a} = \frac{2(1+\lambda)^{1/2}}{1-2\lambda+3\sigma}
    \left\{\sigma[1+2\sigma-\lambda(2+\sigma)]\right\}^{1/2}\,.    \label{grp119}
\end{equation*}
The gravitational potential energy can be given in the form 
\begin{equation*}
  \mathcal{G} = M c^2 g                                           \label{grp120}
\end{equation*}
where the negative gravitational binding factor $g=\mathcal{G}/E$ is given in
terms of the $R/a$ ratio
\begin{equation*}
  g = 1 - \frac{3}{2}\left(\frac{a}{R}\right)^{3}
      \left[
        \arcsin\left(\frac{R}{a}\right)
        -\frac{R}{a}\left(1-\frac{R^{2}}{a^{2}}\right)^{1/2}
      \right]                                                     \label{grp121}
\end{equation*}
It expresses the gravitational binding in a `pure' form as there in no
internal energy in `incompressible' configurations. We can directly conclude
that the relation of the binding energy of gas $E_{\mathrm{b}}$, and the
gravitational energy is given by
\begin{equation}
  \frac{E_{\mathrm{b}}}{\mathcal{G}} = -1\,.                        \label{grp122}
\end{equation}
The gravitational potential energy of the $n=0$ GRP is represented in
Fig.\,\ref{UniSigGBEn}. The role of the cosmological constant is illustrated
by the sequence of lines constructed for appropriately chosen values of the
cosmological parameter $\lambda$.

In polytropes with $n>0$, the binding energy $E_{\mathrm{b}}$ differs from the
gravitational binding energy $(-\mathcal{G})$, as some of the work of the
gravitational field is converted into the kinetic energy of microscopic motion
in these polytropes. The role of the cosmological parameter $\lambda > 0$ in
the effect of gravitational binding will be discussed in the next section.

\section{\label{GRePoCC}GRPs with the cosmological constant}

We construct models of the GRPs with $n>0$ and discuss their dependence on the
cosmological parameter $\lambda$. We present the length and mass scales of the
polytropes, and determine the existence restrictions put on the parameters
characterizing their structure. Then we discuss the polytrope global
characteristics as the dimensionless mass and dimensionless radius,
compactness and gravitational and binding energy of the polytrope
configuration. Finally, we study the radial profiles of the energy density,
pressure and metric coefficients, and illustrate the polytrope curvature by
embedding diagrams of the ordinary projected geometry and the optical
geometry.

\subsection{Length scale and mass scale}

The polytropic spheres are determined by the dimensionless structure equations
that are governed by three parameters---the polytropic index $n$, the
relativistic parameter $\sigma$, the cosmological parameter $\lambda$---and by
the central density $\rhocent$ governing, simultaneously with the parameters
$n, \sigma$, the dimensional length and mass scale of the polytropic
spheres. The dimensional length and mass scales are given by the respective
relation
\begin{subequations}                                           \label{scalesLM}
\begin{align}
  \mathcal{L} &= 3.27\,
    \frac{[\sigma (n+1)]^{1/2}}{\rhocent^{1/2}}\,
      (10^{13}\,\mathrm{cm})\,,                                 \label{scaleL}\\
  \mathcal{M} &= 4.41\,
    \frac{[\sigma (n+1)]^{3/2}}{\rhocent^{1/2}}\,
                (10^{41}\,\mathrm{g})\,;                        \label{scaleM}
\end{align}
\end{subequations}
$\rhocent$ has to be substituted in units of $\mathrm{g/cm^{3}}$.  The
polytropic spheres with given mass and length scales are determined by
solutions of the structure equations that are governed by the dimensionless
radial coordinate $\xi_{1}(n, \sigma, \lambda)$ and the related dimensionless
mass parameter $v(\xi_{1})(n,\sigma, \lambda)$.

\subsection{Integration of the structure equations}

The differential structure equations have to be solved numerically for any
polytropic index $n>0$. For each fixed value of $n$ we obtain a sequence of
polytropic spheres determined by the central density $\rhocent$, the
relativistic parameter $\sigma$ and the cosmological parameter $\lambda$. For
the observationally fixed value of the repulsive cosmological constant,
$\Lambda=1.3 \times 10^{-56}\,\mathrm{cm^{-2}}$, and the related vacuum energy
density, $\rho_{\mathrm{vac}}$, the central density of the polytrope,
$\rhocent$, governs the value of the cosmological parameter $\lambda$ and it
is not a free parameter in such a situation. The first (lowest) solution
$\xi_{1}$ of the equation $\theta(\xi)=0$ determines extension of the
polytropic spheres in terms of the dimensionless radius $\xi$; their
dimensional radius reads $R=\mathcal{L} \xi_{1}$. The dimensionless mass
parameter is given by $v(\xi_{1})=v_{1}$, and the polytrope gravitational mass
is then given by $M = \mathcal{M} v(\xi_{1})$. The radial profiles of the
energy density, pressure, gravitational mass parameter, and the metric
coefficients are determined by the functions $\rho(\xi;n,\sigma,\lambda)$,
$p(\xi;n,\sigma,\lambda)$, $v(\xi;n,\sigma,\lambda)$,
$g_{tt}(\xi;n,\sigma,\lambda)$, $g_{rr}(\xi;n,\sigma,\lambda)$---note that the
metric coefficients depend on $\xi$ also through the mass parameter
$v(\xi)$. These functions are given by Eqs~(\ref{grp49to51})
and~(\ref{grp52and43}). In a similar way, the embedding diagrams of the
ordinary and optical geometry are given by the functions
$z_{\mathrm{ord}}(\xi;n,\sigma,\lambda)$ and
$z_{\mathrm{opt}}(\xi g_{tt}^{1/2};n,\sigma,\lambda)$. It is also instructive
to illustrate the dependence of the other global characteristics of the
polytropic spheres on the basic parameters, i.e., the functions of compactness
$\mathcal{C}(\xi=\xi_{1};n,\sigma,\lambda)$, gravitational potential energy
$\mathcal{G}(\xi=\xi_{1};n,\sigma,\lambda)$ and the binding energy of the
polytropic gas $E_{\mathrm{b}}(\xi=\xi_{1};n,\sigma,\lambda)$.

\subsection{Limit on existence of the GRPs}

\begin{figure*}[t]
\begin{center}
\includegraphics[width=0.8\linewidth]{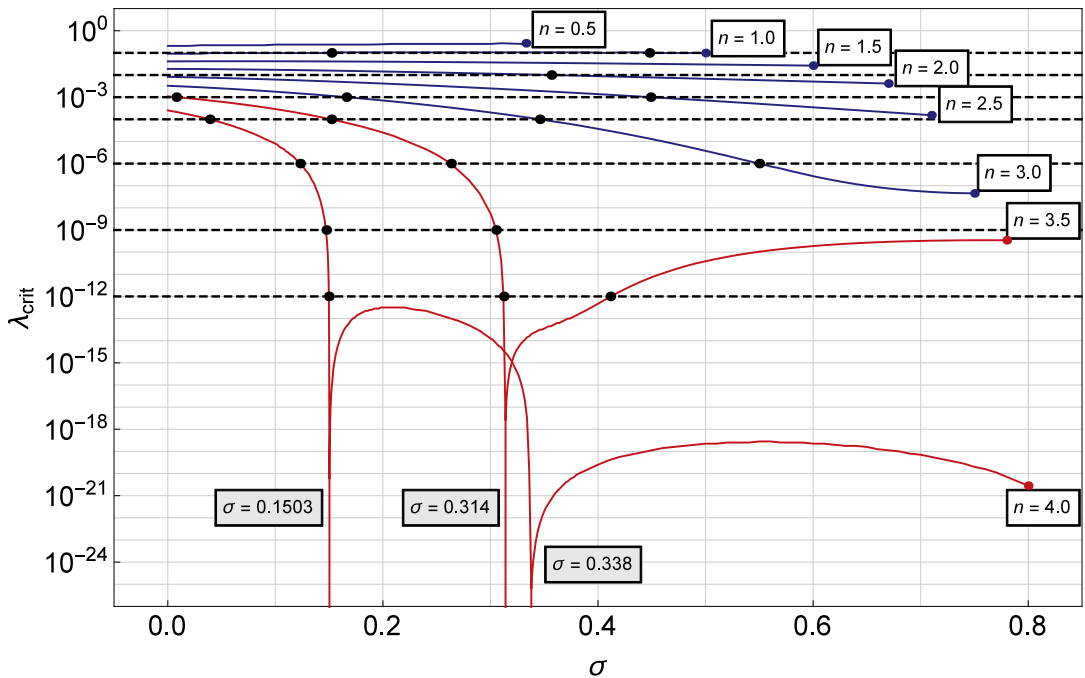}
\end{center}
\caption{\label{SigLamCrit}Dependence of the critical value of the
  cosmological parameter on the relativity parameter $\sigma$ for the
  polytropic index taken from 0.5 to 4.0 with step of 0.5. For a particular
  polytropic index, the polytropic configurations can only exist for parameter
  points $(\sigma, \lambda)$ located below the corresponding curve. For
  polytropic configurations with $n \leq 3.34$, critical values of the
  relativistic parameter, $\sigma_f$, exist, giving infinitely extended
  configurations~\cite{Nil-Ugl:2000:ANNPH1:GRStarsPolyEoS}. For polytropes
  with $n=3.5$ one such critical point is relevant, while for polytropes with
  $n=4$ two such points are relevant.}
\end{figure*}

We have found the role of the cosmological parameter $\lambda$ as being
concentrated in putting strong limits on the existence of polytropic spheres
in dependence on both the polytropic index $n$ and the relativistic parameter
$\sigma$. The critical, limiting values of the cosmological parameter, given
by the function $\lambda_{\mathrm{crit}}=\lambda_{\mathrm{crit}}(n;\sigma)$,
have been determined by numerical calculations and are represented for
selected representative values of $n$ in Fig.\,\ref{SigLamCrit}; the
polytropes are allowed at regions of the parameter space below the critical
curves. We cover the range of standard values of the polytropic index,
starting at the non-relativistic fluid with $n=3/2$ and finishing at $n=3$ for
the ultrarelativistic fluid, and we add both some values of $n < 3/2$, and
some values of $n>3$ when a special character of the polytrope properties
occurs.  Extension of the critical curves is restricted by the value of the
relativistic parameter $\sigma$ corresponding to the equality of the velocity
of sound and the velocity of light (so called causality limit).

We can see that character of the critical function 
$\lambda_{\mathrm{crit}}(n,\sigma)$ strongly depends on the value of the
polytropic index $n$. Generally, it increases with $n$ decreasing. For $n<3$
the function $\lambda_{\mathrm{crit}}(n;\sigma)$ slightly monotonically
decreases with $\sigma$ increasing; it is limited by the value of
$\lambda_{\mathrm{crit}}=10^{-7}$ even in the limit of $\sigma \rightarrow 1$.
In the special case of $n=3$ it decreases from the starting point
$\lambda_{\mathrm{crit}}(n=3,\sigma=0)=3 \times 10^{-3}$ down to
$\lambda_{\mathrm{crit}}(n=3;\sigma=0.7) = 10^{-7}$ and remains constant with
increasing values of $\sigma$.

For $n>3$, the function $\lambda_{\mathrm{crit}}(n;\sigma)$ looses its
monotonic character, and there are forbidden polytropes for some special
values of the relativistic parameter $\sigma$ in dependence on the polytrope
index, since such polytropes should have infinite extension. For example, for
$n=3.5$ the polytropes are forbidden for one specific value of
$\sigma_{\mathrm{f}}=0.314$, while for $n=4$, there are two specific forbidden
values of $\sigma_{\mathrm{f}1}=0.1503$, $\sigma_{\mathrm{f}2}=0.338$. Third
forbidden configuration with $n=4$ corresponds to $\sigma$ breaking the
causality limit and reads $\sigma_{\mathrm{f}3}=0.834$. These forbidden
configurations occur for the general relativistic polytropic configurations in
the spacetimes with $\Lambda=0$, and have been discussed for the first time
in~\cite{Nil-Ugg:2000:ANNPH1:GRStarPoEqSt}.

Notice that in the region of $\sigma > \sigma_{\mathrm{f}1}$ there is
$\lambda_{\mathrm{crit}}(n;\sigma) < 10^{-10}$. On the other hand, for
non-relativistic polytrope spheres with $\sigma < 0.1$, there is
$\lambda_{\mathrm{crit}}(n;\sigma) > 10^{-5}$ for all polytropic indexes
$n<4$. In such situations we can see that the polytropic spheres with very
small central density have their structure strongly influenced by the
repulsive cosmological constant.

\section{\label{globGRP}Global GRP characteristics}

\subsection{Extension and mass}

The basical global characteristics of the GRPs are given by the dimensionless
extension and dimensionless mass. We thus give dependences of the
polytrope extension parameter $\xi_{1}$ and the polytrope mass parameter
$v_{1}=v(\xi_{1})$ on the parameters $n$, $\sigma$ and $\lambda$ and discuss
their properties.

\begin{figure*}[t]
\begin{minipage}{0.48\linewidth}
\centering
\includegraphics[width=\linewidth]{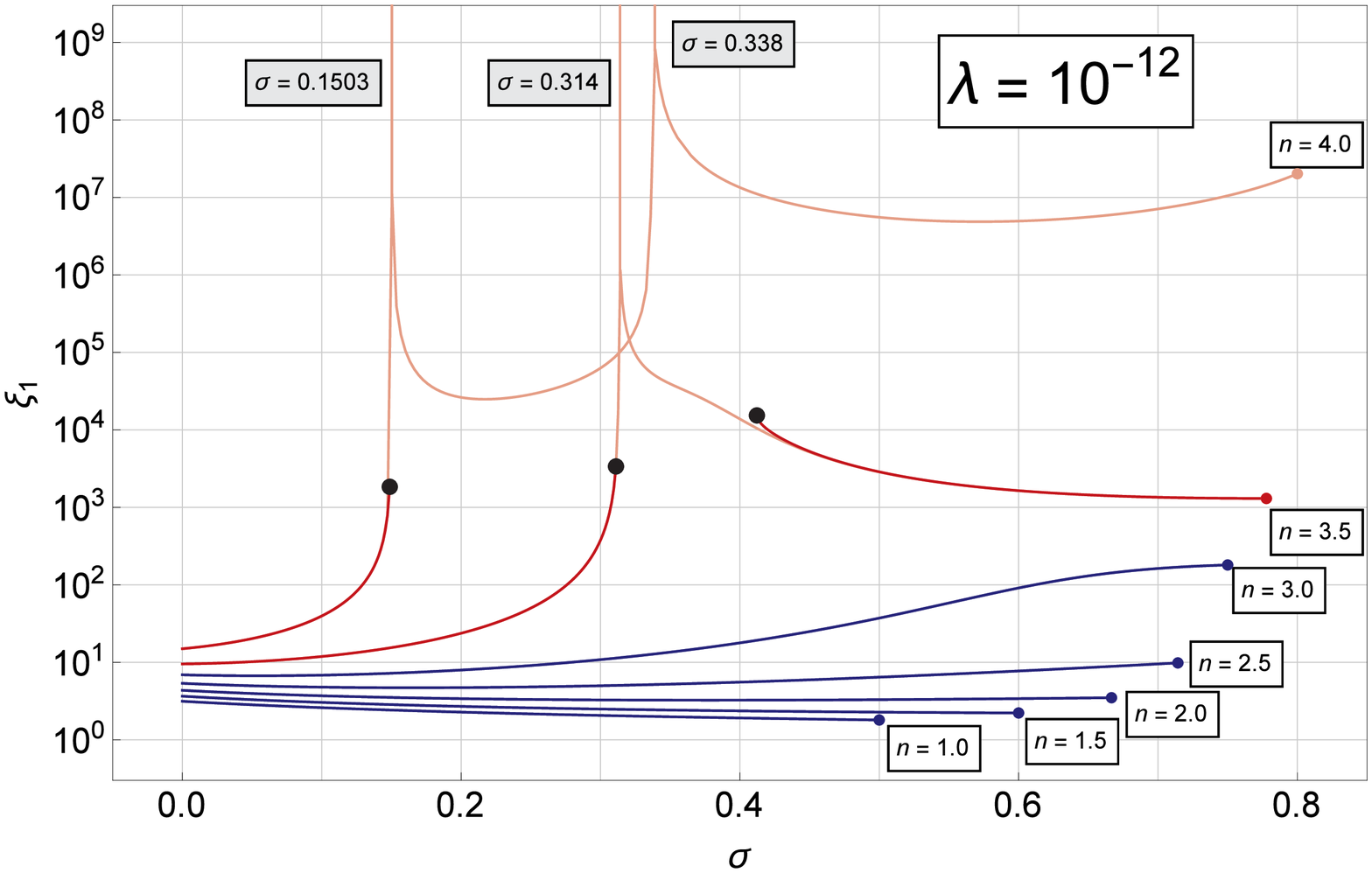}
\end{minipage}\hfill%
\begin{minipage}{0.48\linewidth}
\centering
\includegraphics[width=\linewidth]{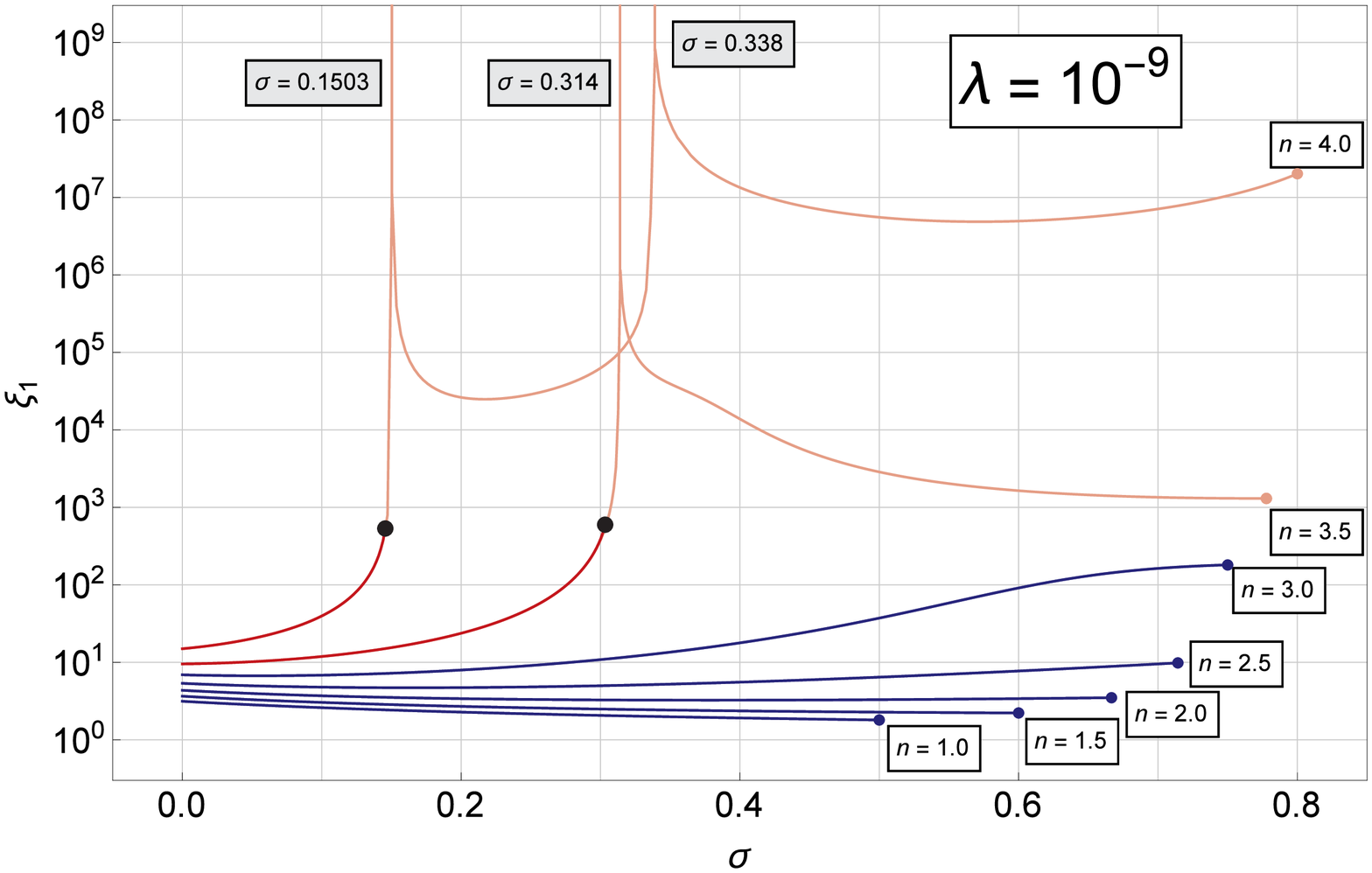}
\end{minipage}
\par\vspace{1.5\baselineskip}\par
\begin{minipage}{0.48\linewidth}
\centering
\includegraphics[width=\linewidth]{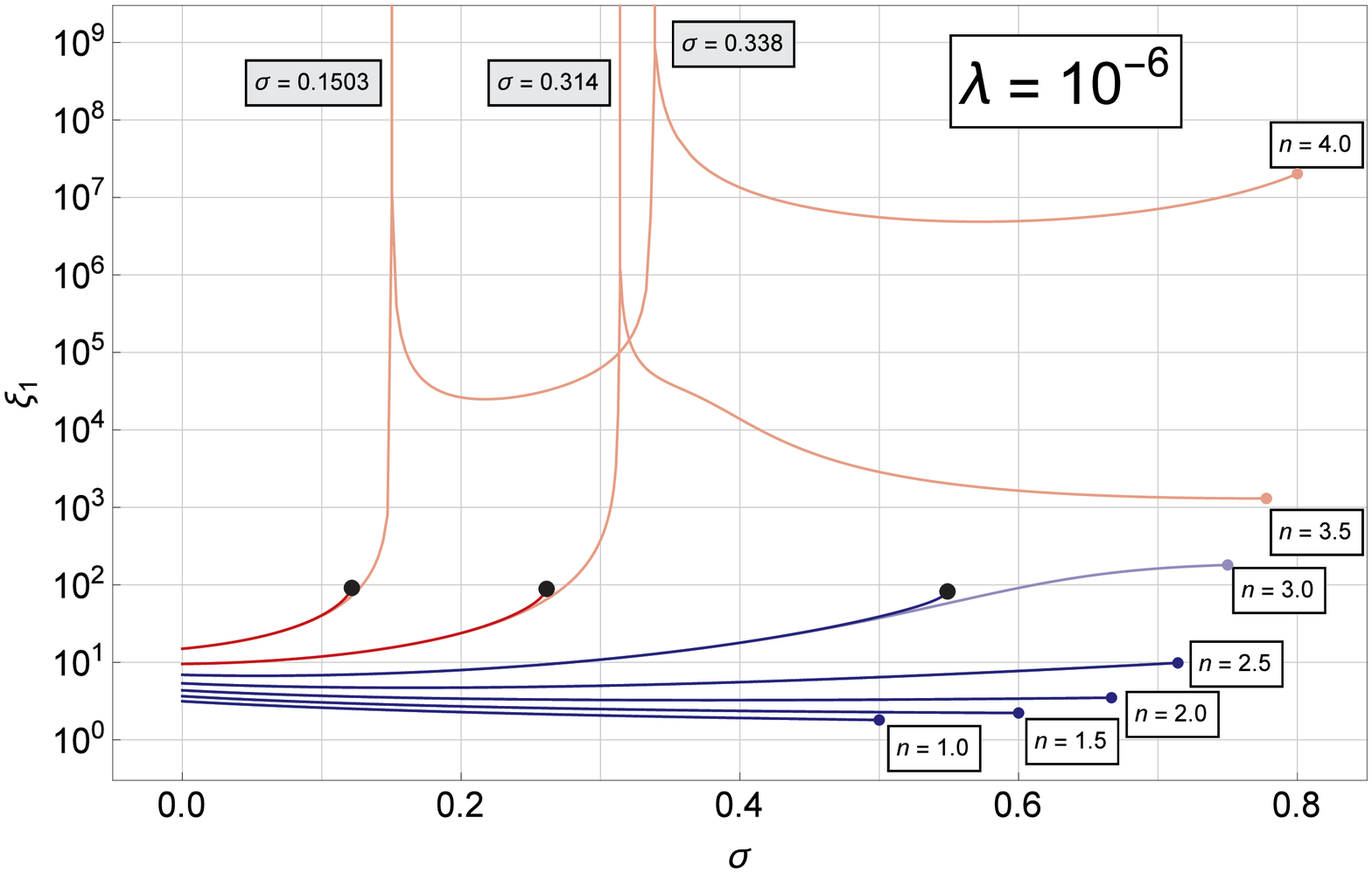}
\end{minipage}\hfill%
\begin{minipage}{0.48\linewidth}
\centering
\includegraphics[width=\linewidth]{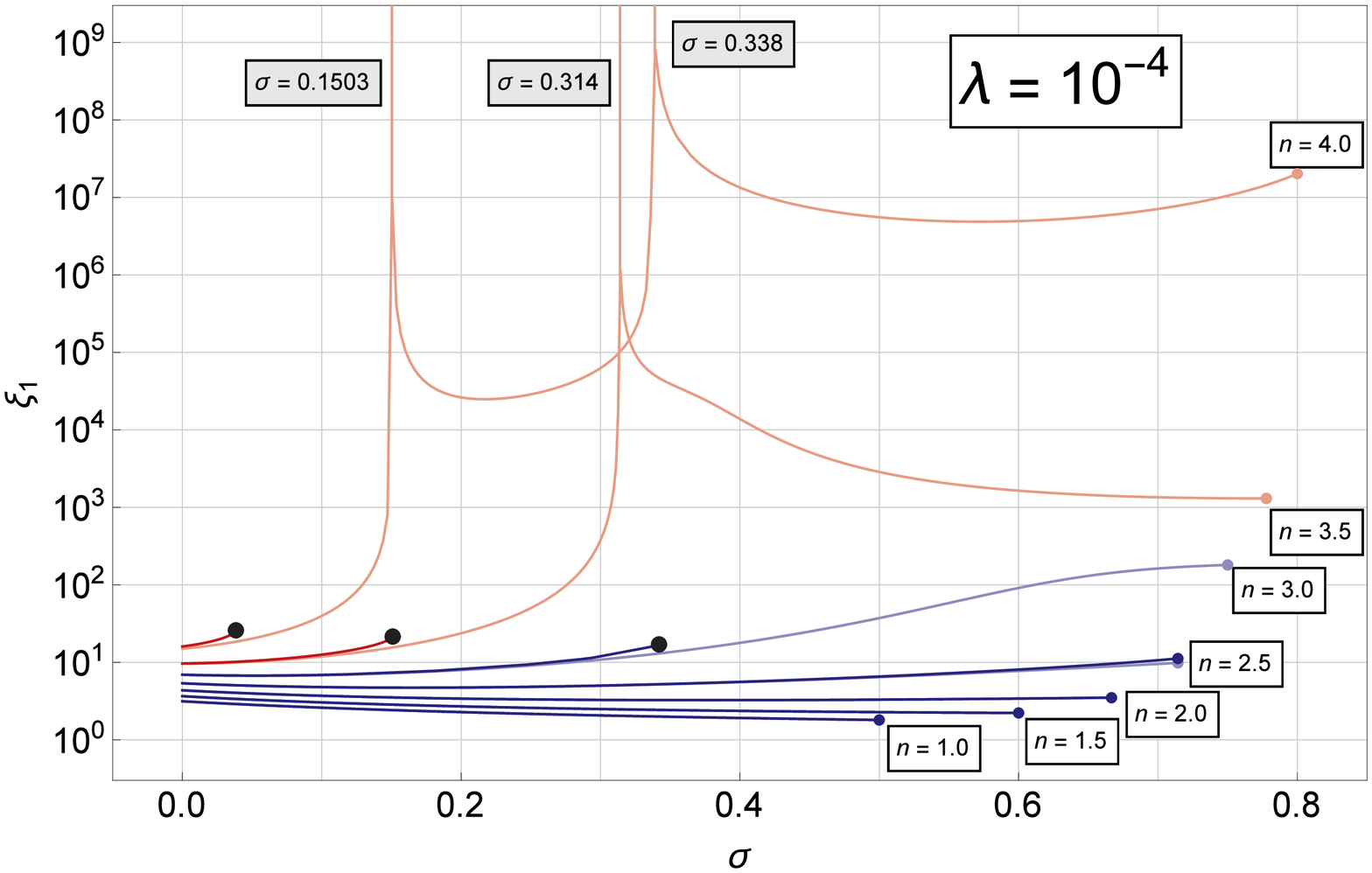}
\end{minipage}
\par\vspace{1.5\baselineskip}\par
\begin{minipage}{0.48\linewidth}
\centering
\includegraphics[width=\linewidth]{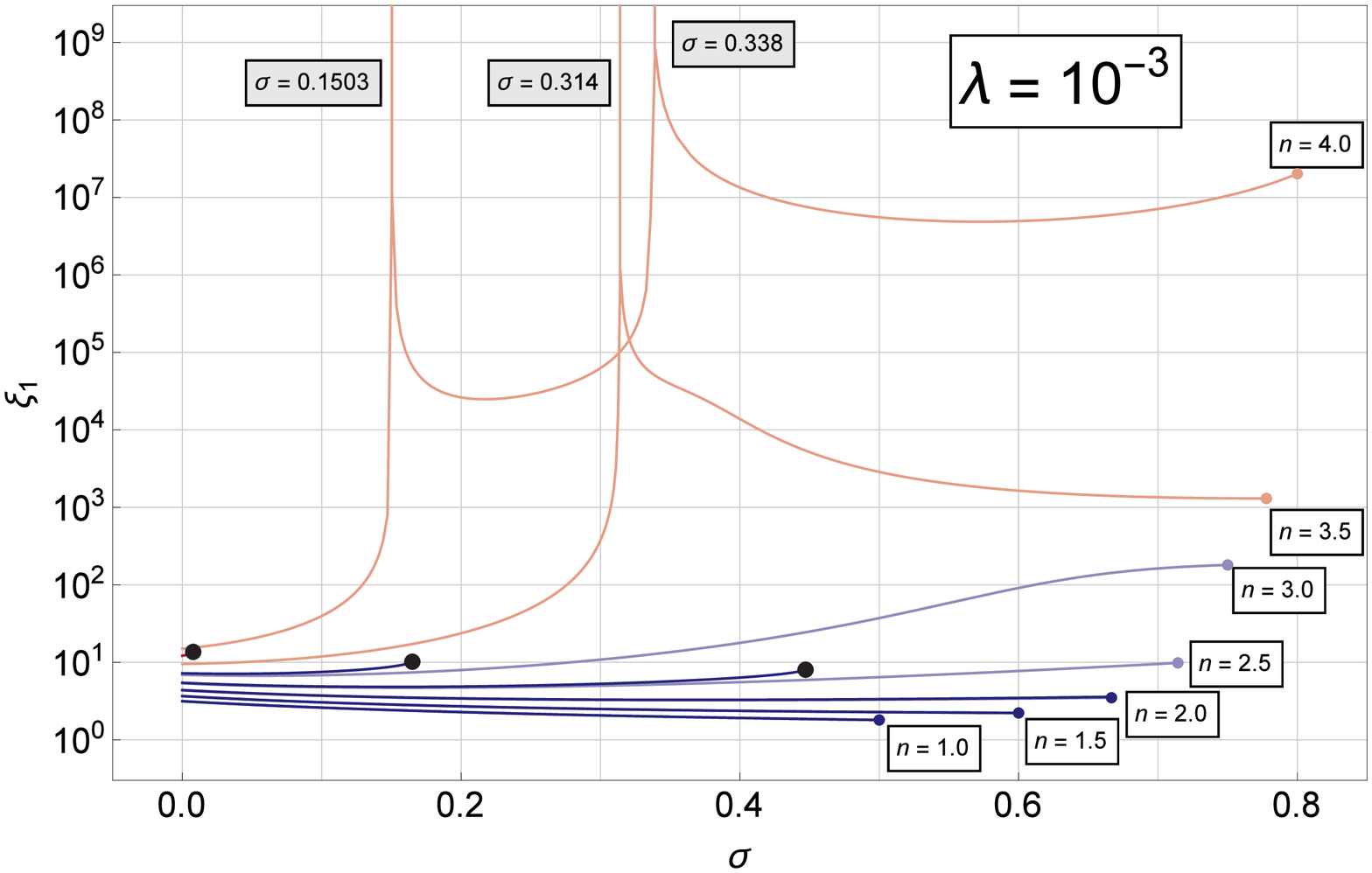}
\end{minipage}\hfill%
\begin{minipage}{0.48\linewidth}
\centering
\includegraphics[width=\linewidth]{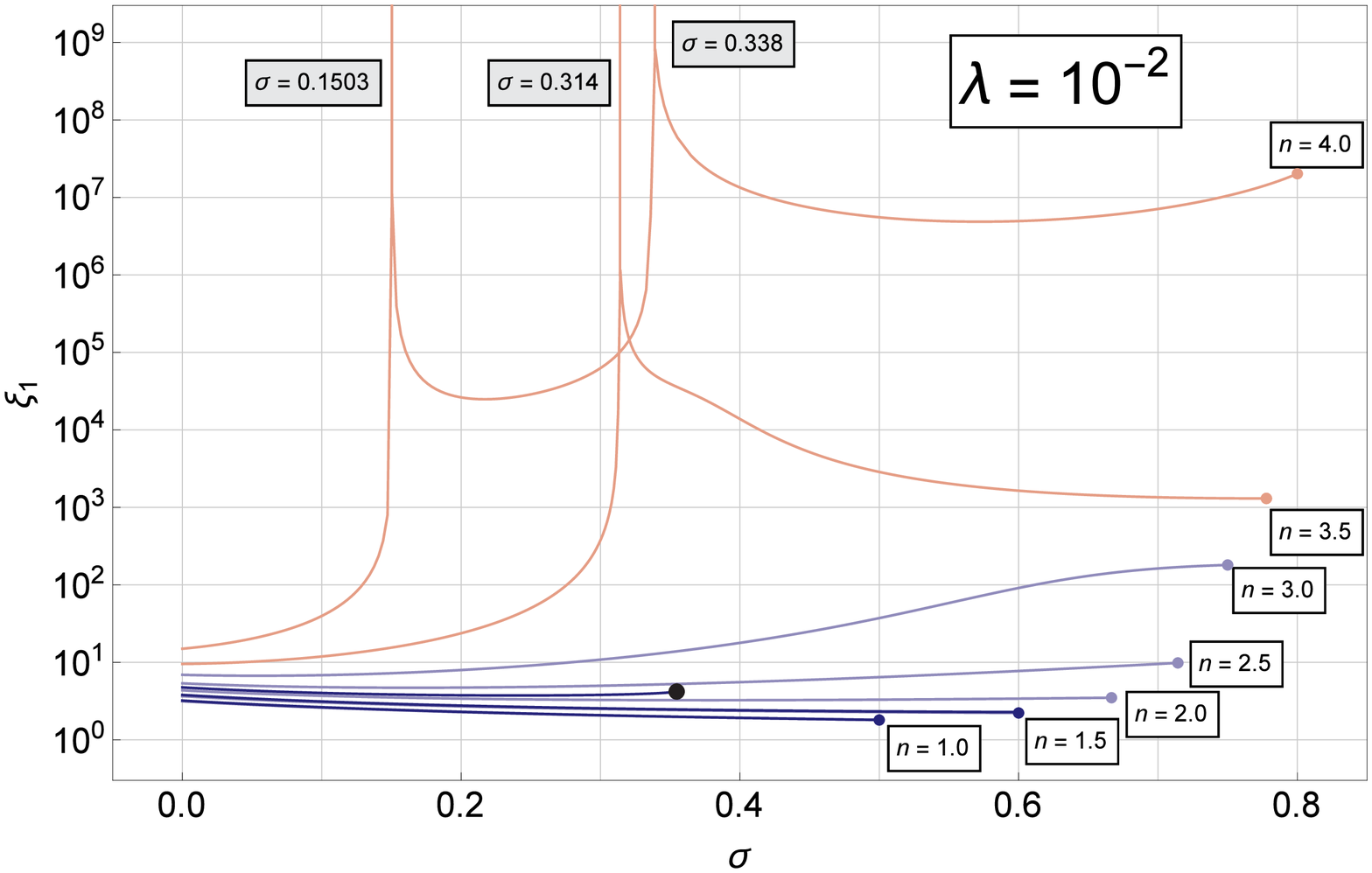}
\end{minipage}
\par\vspace{.8\baselineskip}\par
\caption{\label{SigXi1}Dependences of the extension parameter $\xi_{1}$ for
  the characteristic values of the polytropic index
  $n\in\{1.0, 1.5, 2, 2.5, 3, 3.5, 4\}$ with $\sigma$ varying up to the causal
  limit for $\lambda = 10^{-12}$, $\lambda = 10^{-9}$ (top),
  $\lambda = 10^{-6}$, $\lambda = 10^{-4}$ (middle) and $\lambda = 10^{-3}$,
  $\lambda = 10^{-2}$ (bottom).}
\end{figure*}

We first illustrate dependences of the extension parameter $\xi_{1}$ for the
characteristic values of the polytropic index $n = 1, 1.5, 2, 2.5, 3, 3.5, 4$,
with $\sigma$ varying up to the causal limit. The dependence of the
dimensionless radius of the polytropes on the cosmological constant parameter
$\lambda$ is presented in Fig.\,\ref{SigXi1} where we vary the cosmological
parameter for the characteristic values of
$\lambda = 10^{-12}, 10^{-9}, 10^{-6}, 10^{-4}, 10^{-3}, 10^{-2}$. The curves
$\xi_{1}(\sigma,n,\lambda)$ are compared to the curves
$\xi_{1}(\sigma,n,\lambda=0)$---we can see that at the critical points of
$\sigma_{\mathrm{f}}$, the dimensionless parameter $\xi_{1}$ diverges for
$\lambda=0$, indicating that the critical polytrope cannot be limited and is
not well defined. The validity restriction of the curves
$\xi_{1}(\sigma,n,\lambda)$ at the causal limit is depicted by the shaded
points. The black points depict the limit of validity of the curves
$\xi_{1}(\sigma,n,\lambda)$ meaning that the polytrope radius cannot exceed
the static radius of the external spacetime.

\begin{figure*}[t]
\begin{minipage}{0.48\linewidth}
\centering
\includegraphics[width=\linewidth]{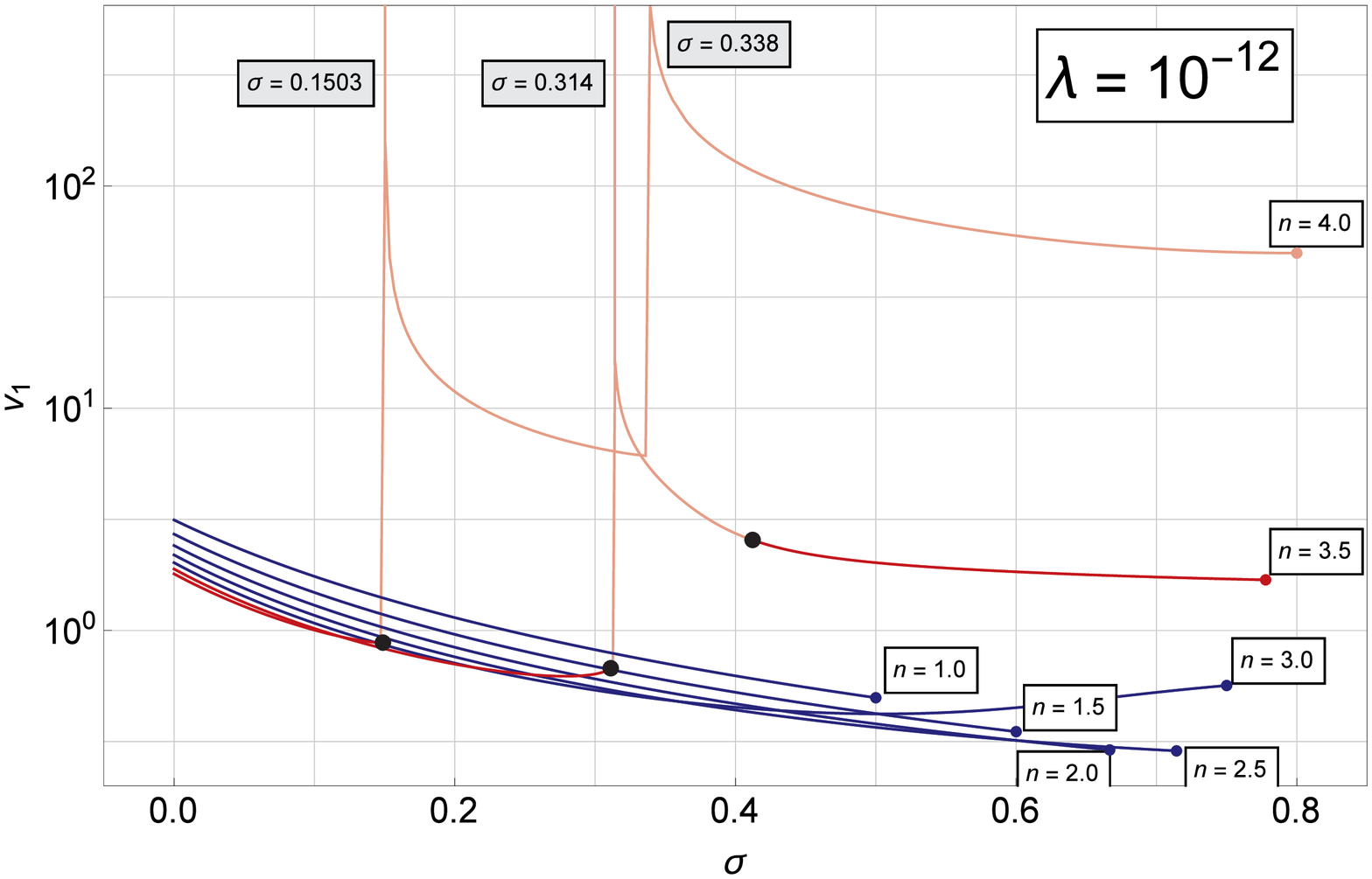}
\end{minipage}\hfill%
\begin{minipage}{0.48\linewidth}
\centering
\includegraphics[width=\linewidth]{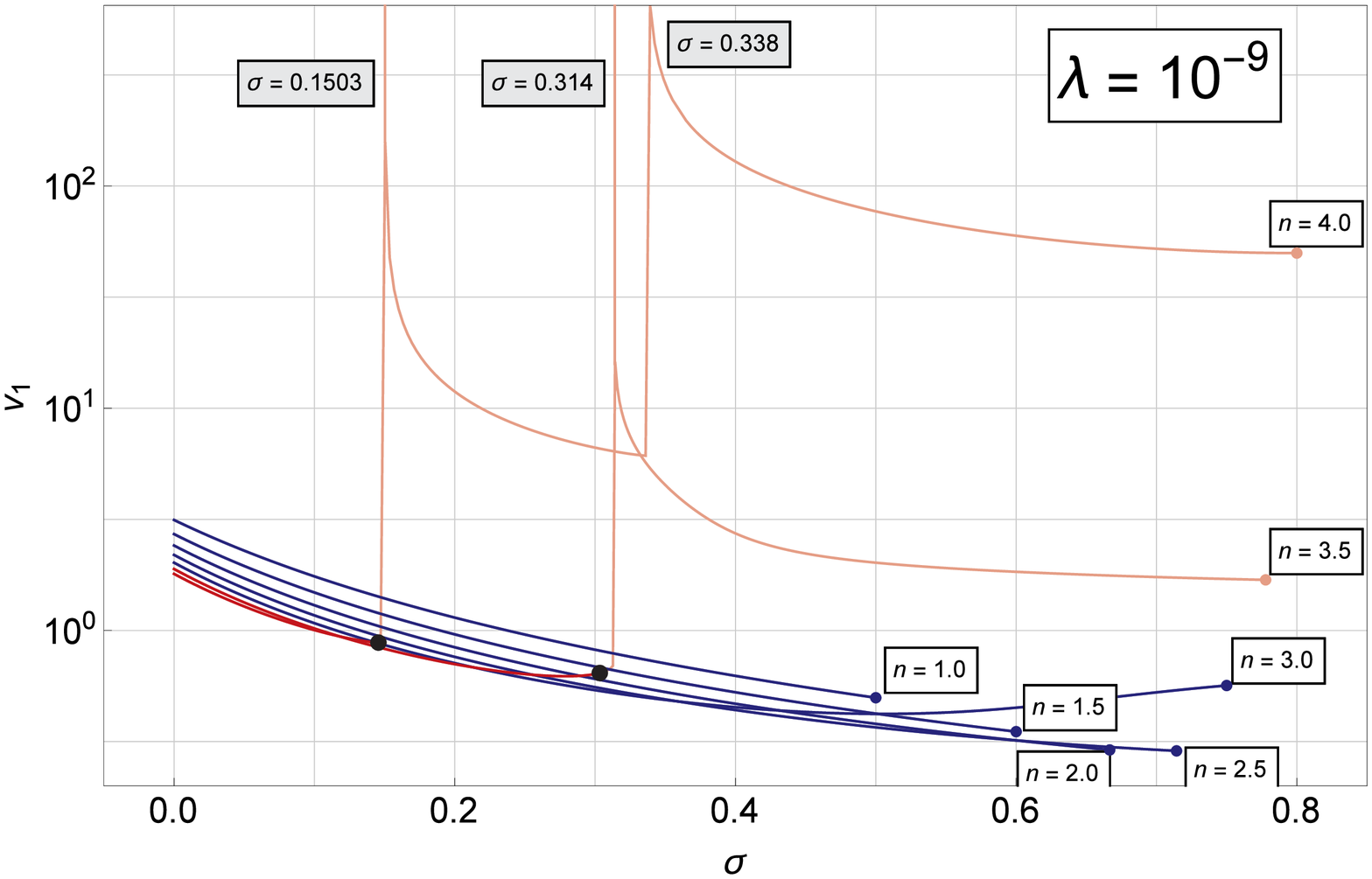}
\end{minipage}
\par\vspace{1.5\baselineskip}\par
\begin{minipage}{0.48\linewidth}
\centering
\includegraphics[width=\linewidth]{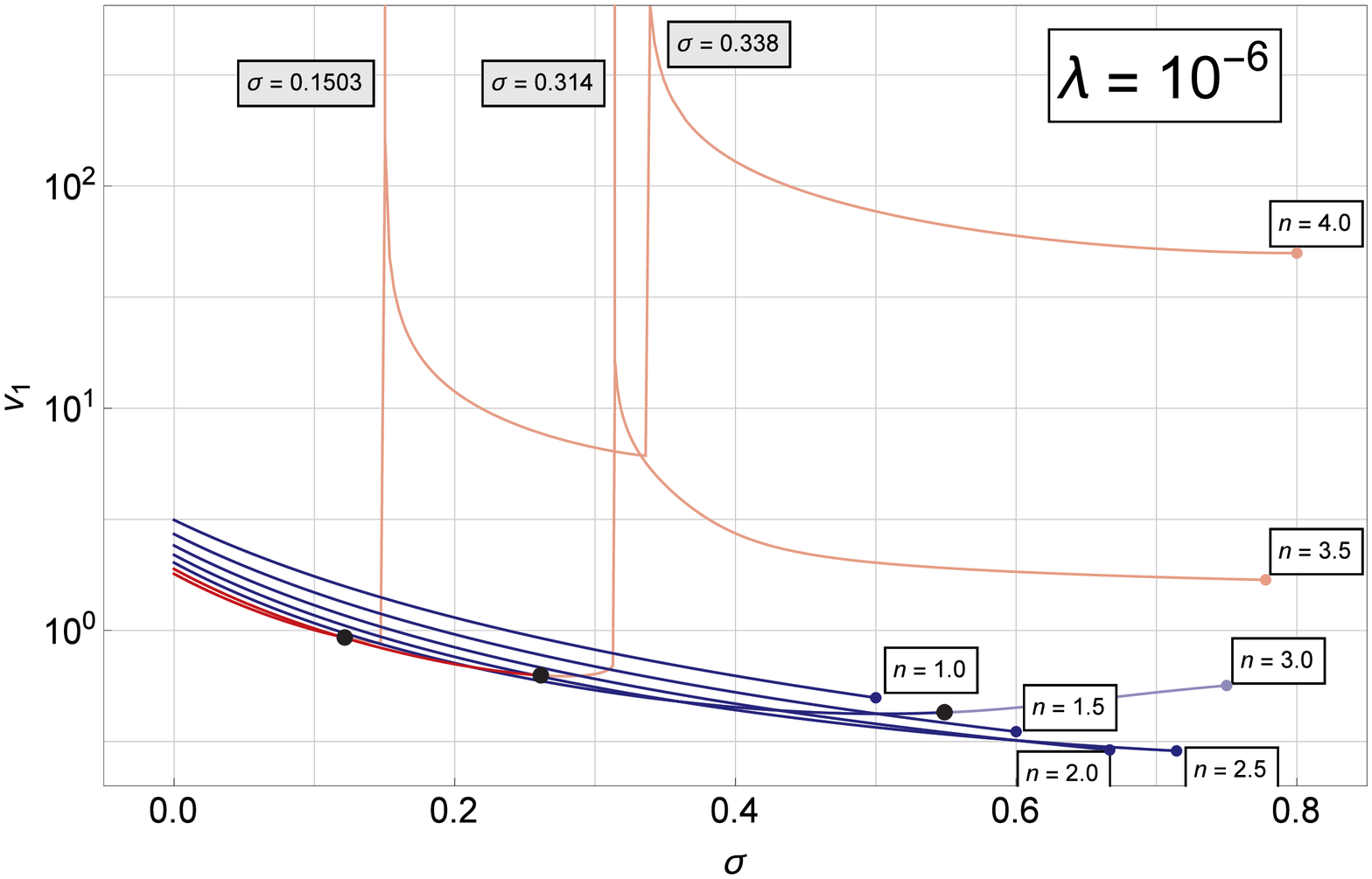}
\end{minipage}\hfill%
\begin{minipage}{0.48\linewidth}
\centering
\includegraphics[width=\linewidth]{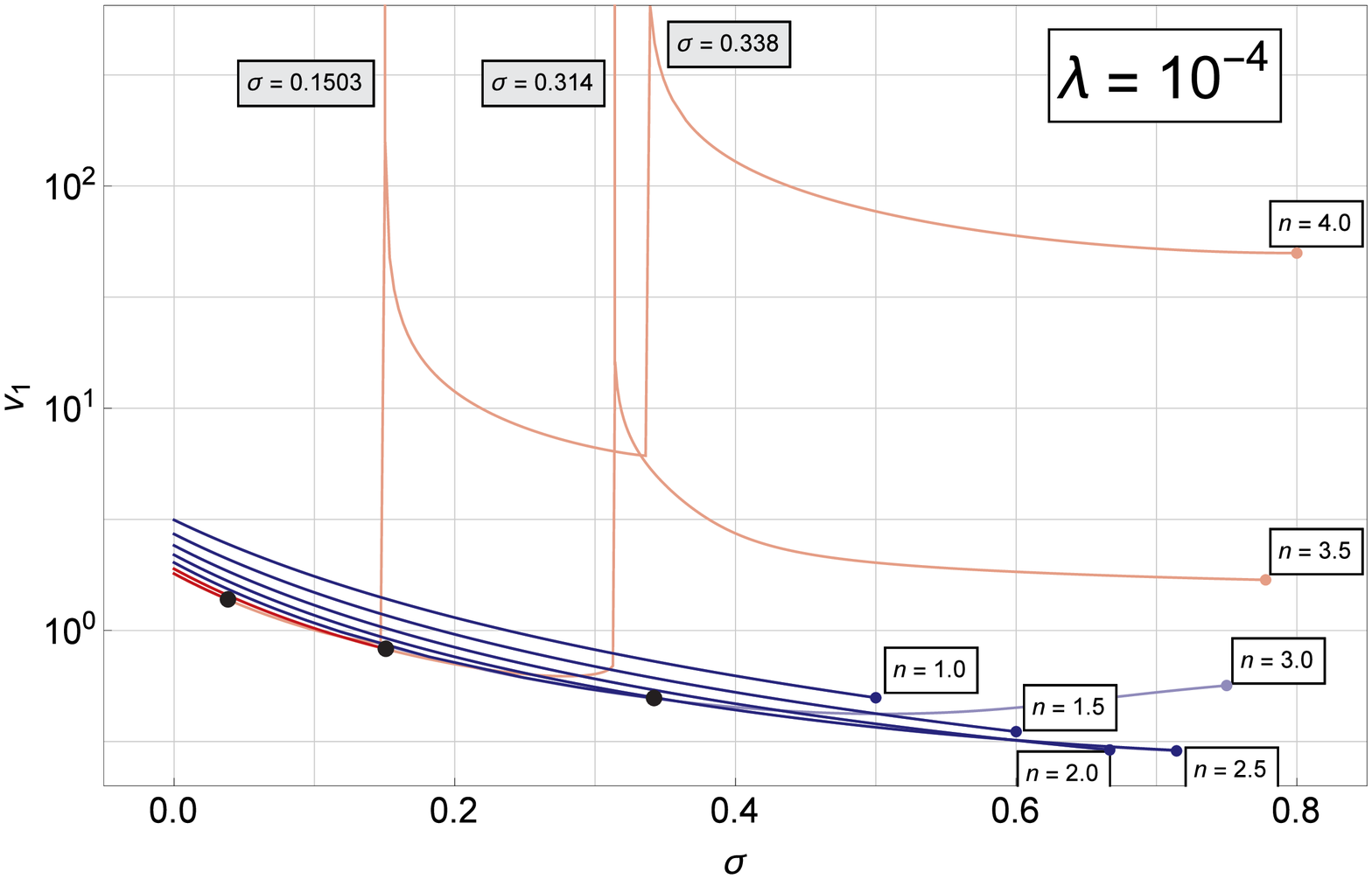}
\end{minipage}
\par\vspace{1.5\baselineskip}\par
\begin{minipage}{0.48\linewidth}
\centering
\includegraphics[width=\linewidth]{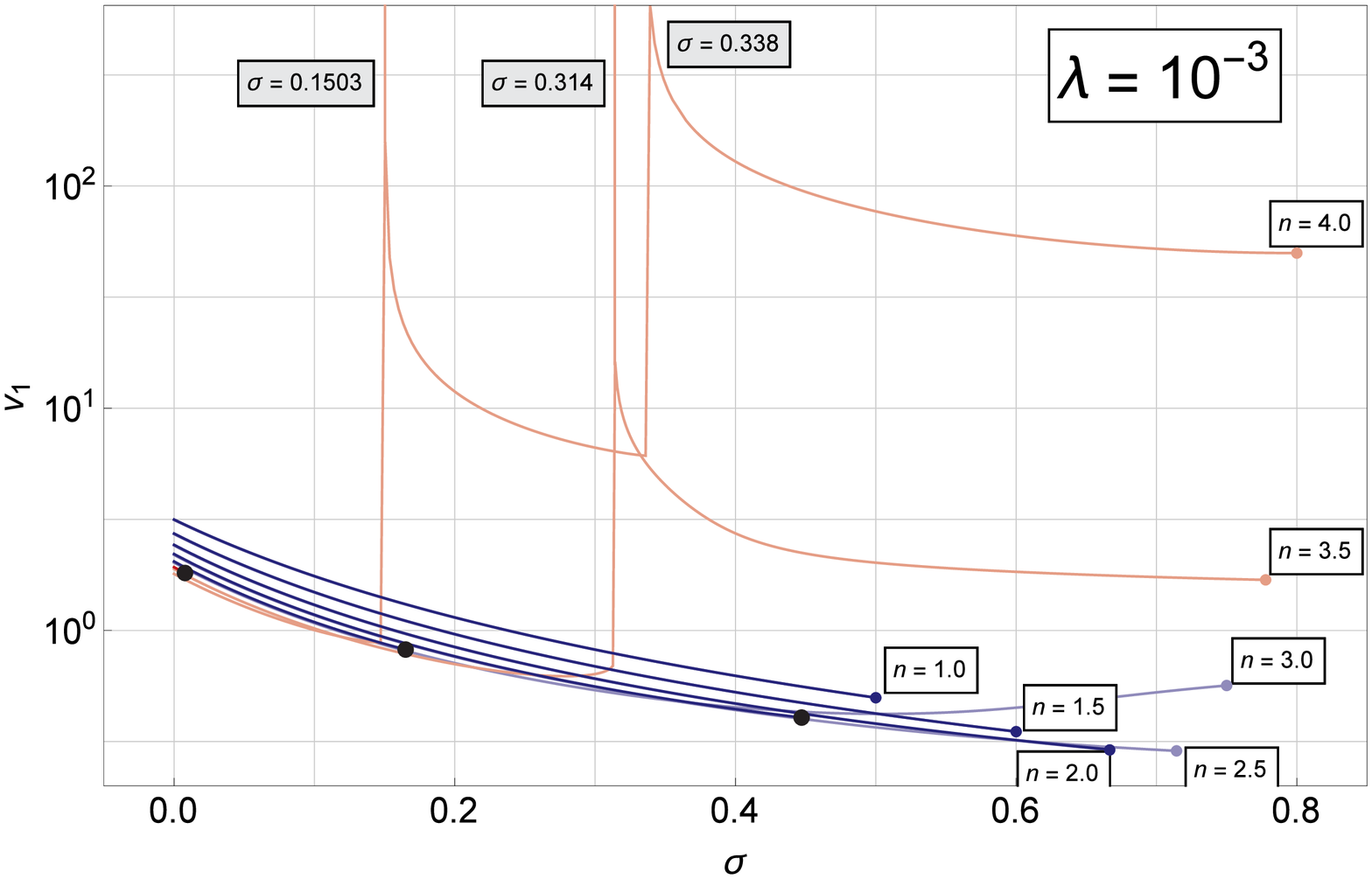}
\end{minipage}\hfill%
\begin{minipage}{0.48\linewidth}
\centering
\includegraphics[width=\linewidth]{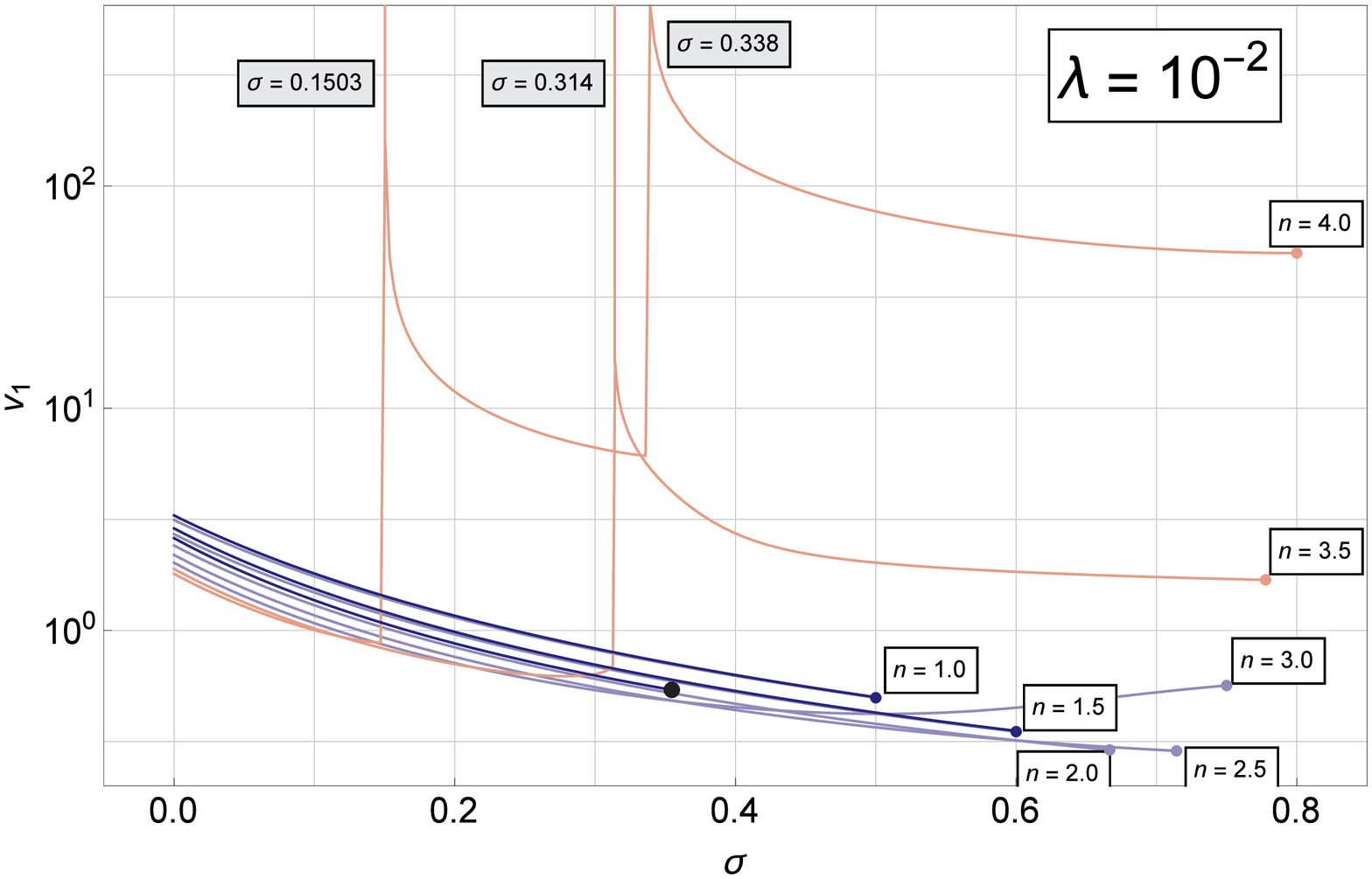}
\end{minipage}
\par\vspace{.8\baselineskip}\par
\caption{\label{SigV1}Dependences of the mass parameter
  $v_{1}\equiv v(\xi_{1})$ for the characteristic values of the polytropic
  index $n\in\{1.0, 1.5, 2, 2.5, 3, 3.5, 4\}$ with $\sigma$ varying up to the
  causal limit for $\lambda = 10^{-12}$, $\lambda = 10^{-9}$ (top),
  $\lambda = 10^{-6}$, $\lambda = 10^{-4}$ (middle) and $\lambda = 10^{-3}$,
  $\lambda = 10^{-2}$ (bottom).}
\end{figure*}

Then we illustrate dependences of the mass parameter $v_{1}=v(\xi_{1})$ for
the same characteristic values of the polytropic index
$n = 1, 1.5, 2, 2.5, 3, 3.5, 4$, with $\sigma$ varying up to the causal
limit. The dependence of the dimensionless mass parameter of the polytropes on
the cosmological constant parameter $\lambda$ is presented in
Fig.\,\ref{SigV1}, where we again vary the parameter for the characteristic
values of $\lambda = 10^{-12}, 10^{-9}, 10^{-6}, 10^{-4}, 10^{-3},
10^{-2}$. In Fig.\,\ref{SigV1}, the curves $v_{1}(\sigma,n,\lambda)$ are
compared to the curves $v_{1}(\sigma,n,\lambda=0)$. At the critical points of
$\sigma_{\mathrm{f}}$, the dimensionless parameter $v_{1}$ also diverges for
$\lambda=0$, and even faster than for the dimensionless radius, indicating
again that the critical polytrope has to be unlimited and is not well defined
for $\lambda=0$. The causal limit of validity of the curves
$v_{1}(\sigma,n,\lambda)$ is depicted by the shaded points.

\subsubsection{Extension parameter}

For $\lambda=0$, the polytropes with $\sigma\ll 1$ have the dimensionless
radius $\xi_{1} \sim 1$, and it slightly increases with increasing $n$. With
increasing $\sigma$, the radius $\xi_{1}$ slightly decreases for $n=3/2$, it
remains almost constant for $n=2$, and it has a minimum near $\sigma \sim 0.1$
and then increases up to values of $\xi_{1} = 10$ for $n=2.5$ and
$\xi_{1} \sim 200$ for $n=3$. For $n=3.5$, $\xi_{1}$ diverges at
$\sigma_{\mathrm{f}}=0.314$ and then it decreases to $\xi_{1} \sim 10^3$ at
the causal limit. For $n=4$, the radius $\xi_{1}$ diverges at the two critical
points, between the critical points there is $\xi_{1} > 10^4$, and at the
causal limit there is $\xi_{1} \sim 3 \times 10^7$.

The cosmological constant has crucial influence on the polytropic
configurations, as it removes the singular behavior of $\xi_{1}$---for any
value of $\lambda > 0$, the polytropes with $n=3.5$ or $n=4$ are forbidden
around the critical values of $\sigma_{\mathrm{f}}$.

For $\lambda = 10^{-12}$, the restriction implies $\xi_{1} < 10^{4}$ for all
values of polytrope index $n$. For $n=4$, the branch of $\xi_{1}(\sigma)$
above the first critical point is forbidden, while for $n=3.5$, the polytropes
can exist both above and below the critical values of
$\sigma_{\mathrm{f}}$. For $n \leq 3$, the influence of $\lambda>0$ in the
functions $\xi_{1}(\sigma;n,\lambda)$ is negligible---see the top left plot in
Fig.\,\ref{SigXi1}.

A similar situation occurs for $\lambda = 10^{-9}$, but for $n > 3$, the
polytropes exist only under the critical value of the first
$\sigma_{\mathrm{f}}$. The functions $\xi_{1}(\sigma;n,\lambda)$ are smaller
than $10^3$ and as in the previous case they follow closely those
corresponding to $\lambda=0$ (see top right plot in Fig.\,\ref{SigXi1}).

For $\lambda = 10^{-6}$, the functions $\xi_{1}(\sigma;n,\lambda)$ with
$n=3, 3.5, 4$ have their terminal point at the same value of
$\xi_{1} \sim 10^{2}$ with the terminal value of $\sigma$ increasing with
decreasing $n$; now the influence of $\lambda>0$ slightly increases the
$\xi_{1}(\sigma;n,\lambda)$ functions above their counterparts with
$\lambda=0$ (see middle left plot in Fig.\,\ref{SigXi1}).

A similar situation occurs for $\lambda = 10^{-4}$, but in this case
$\xi_{1}(\sigma,n) < 30$ (middle right plot in Fig.\,\ref{SigXi1}).

For $\lambda = 10^{-3}$, the $n=4$ polytropes are fully suppressed, the
$n=3.5$ polytropes are allowed at vicinity of $\sigma = 0$ only, while for
$n=3$ ($n=2.5$), the polytropes are limited at $\sigma \sim 0.15$
($\sigma \sim 0.45$), and the functions $\xi_{1}(\sigma,n,\lambda) < 10$
demonstrating slight influence of $\lambda > 0$ for $\sigma$ near the maximal
allowed values; for the polytropes with $n=2, 1.5, 1$, the influence of the
cosmological constant is negligible---see bottom left plot in
Fig.\,\ref{SigXi1}.

For $\lambda = 10^{-2}$, only the polytropes with $n \leq 2$ are allowed. The
limiting value of the relativistic parameter is shifted down to
$\sigma \sim 0.35$ in the case of $n=2$ polytropes, while it is not influenced
by the cosmological constant for polytropes $n=1.5, 1$; generally the
functions $\xi_{1}(\sigma,n,\lambda) < 8$, their $\sigma$-profile is
influenced by the cosmological constant only for $n=2$ ---see bottom right
plot in Fig.\,\ref{SigXi1}.

\subsubsection{Mass parameter}

For $\lambda=0$, polytropes with $\sigma\ll 1$ have the dimensionless mass
parameter $v_{1} = v(\xi_{1}) \!\sim 1$, decreasing from $v_{1} \sim 2.75$ for
$n=1.5$ down to $v_{1} \sim 2$ for $n=4$. For $n=1.5, 2. 2.5$, the function
$v_{1}(\sigma,n)$ decreases with increasing $\sigma$, down to values
$v_{1} \sim 0.25$ near the causal limit. For $n=3$, $v_{1}(\sigma,n)$ has a
minimum at $\sigma \sim 0.45$ and then slightly increases with $\sigma$
increasing to the causality limit. For $n=3.5, 4$, the mass parameter of the
polytropes demonstrates again divergence at the critical points
$\sigma_{\mathrm{f}}$. At the causal limits, the mass parameter takes the
value of $v_{1} \sim 2$ ($v_{1} \sim 30$) for $n=3.5$ ($n=4$).

Similarly to the case of the extension parameter $\xi_{1}$, the role of the
cosmological constant represented by the parameter $\lambda$ in the mass
parameter function $v_{1}(\sigma,n,\lambda)$ is given by the cuts off governed
by the existence limits on the GRPs determined in Fig.\,\ref{SigLamCrit}.  The
cuts off are illustrated in Fig.\,\ref{SigV1} for the same values of the
parameter $\lambda$ as in the case of the extension parameter $\xi_{1}$. Now
we can see that in the regions of validity, the functions
$v_{1}(\sigma,n,\lambda)$ almost coincide with the functions
$v_{1}(\sigma,n,\lambda=0)$. Notice that even for $\lambda = 10^{-12}$, the
mass parameter $v_{1}(\sigma,n)<3$ for all considered values of $n$ and in the
whole allowed region of $\sigma$. Large values of the mass parameter
$v_{1}(\sigma,n,\lambda)$, say $v_{1} > 10^{4}$, can be obtained for GRPs with
$n = 3.5, 4$ in vicinity of the critical values of $\sigma_f$, if
$\lambda < 10^{-17}$.



\subsubsection{Compactness}

Compactness is defined as the dimensionless ratio of the mass and extension of
the polytrope, i.e., it is governed by the ratio $v_{1}/\xi_{1}$. As we
consider here the global compactness parameter, related to the complete
polytropic configurations given by the parameters $\xi_{1}$ and $v_{1}$, we use
the notation $\mathcal{C}_{1} = \mathcal{C}(\xi_{1})$.  Later we study also
the radial profiles of the compactness $\mathcal{C}(\xi)$, defined for
$\xi \in (0,\xi_{1})$ with related $v(\xi)$.

\begin{figure*}[t]
\begin{minipage}{0.48\linewidth}
\centering
\includegraphics[width=\linewidth]{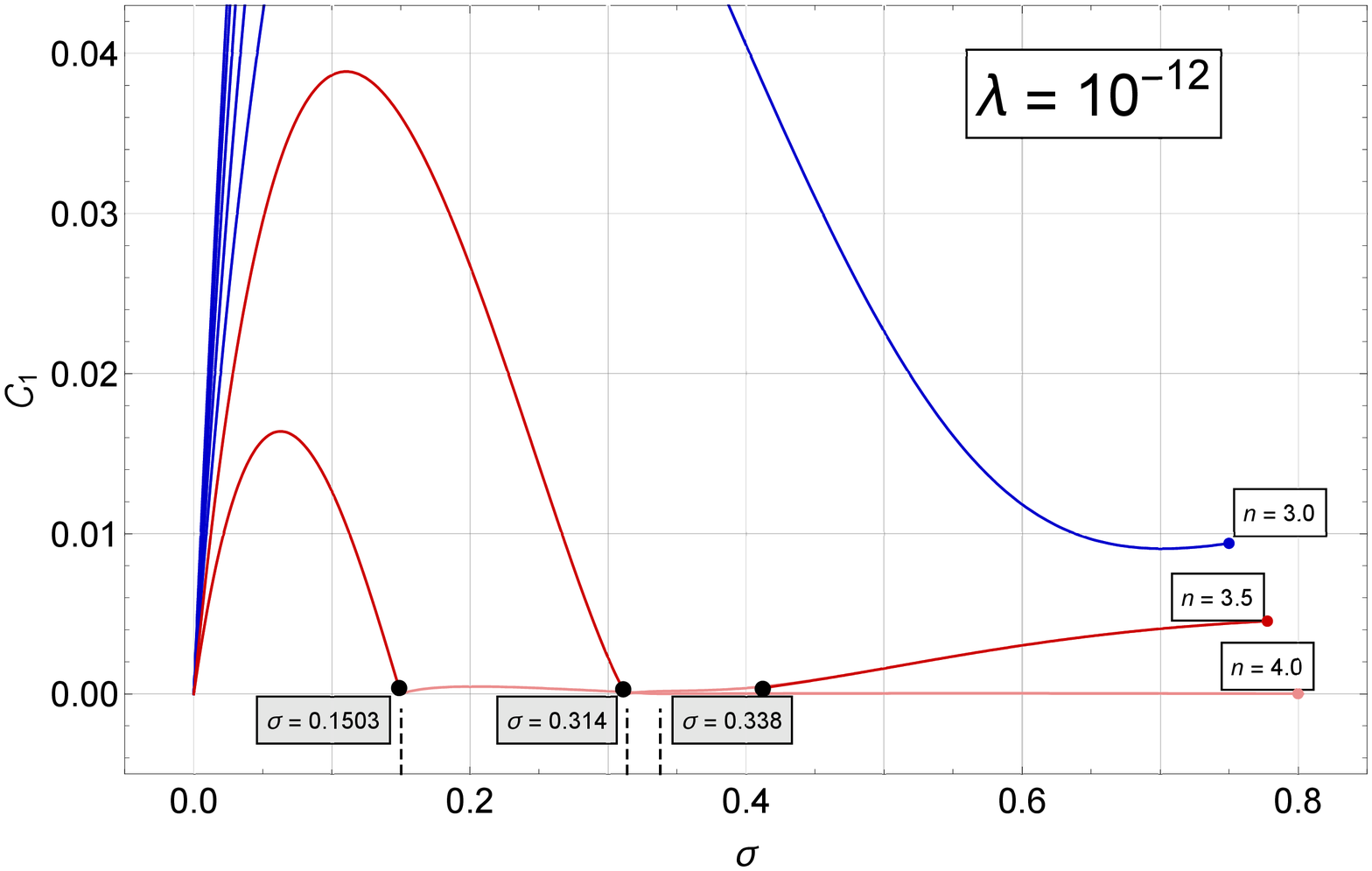}
\end{minipage}\hfill%
\begin{minipage}{0.48\linewidth}
\centering
\includegraphics[width=\linewidth]{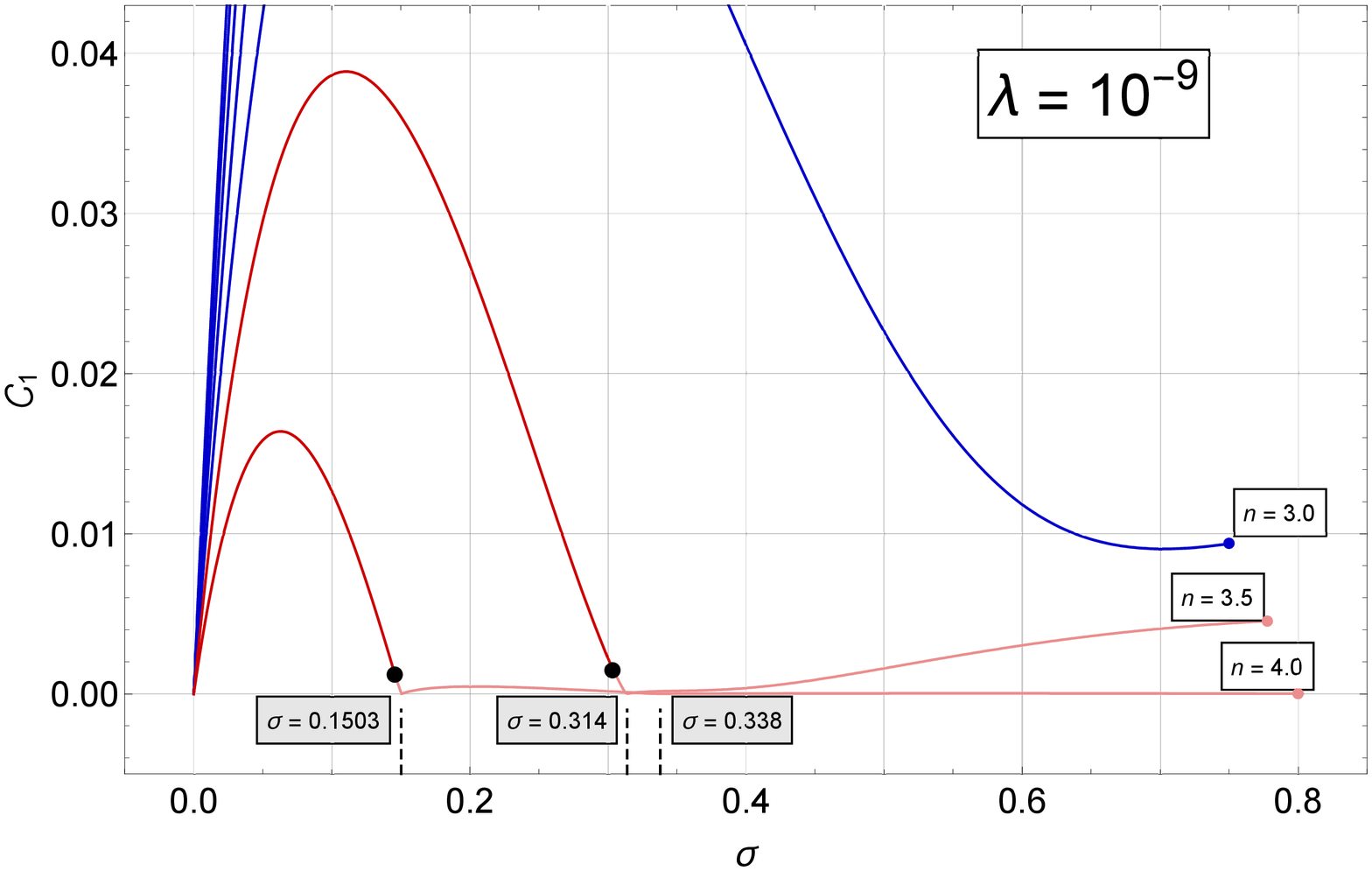}
\end{minipage}
\par\vspace{1.5\baselineskip}\par
\begin{minipage}{0.48\linewidth}
\centering
\includegraphics[width=\linewidth]{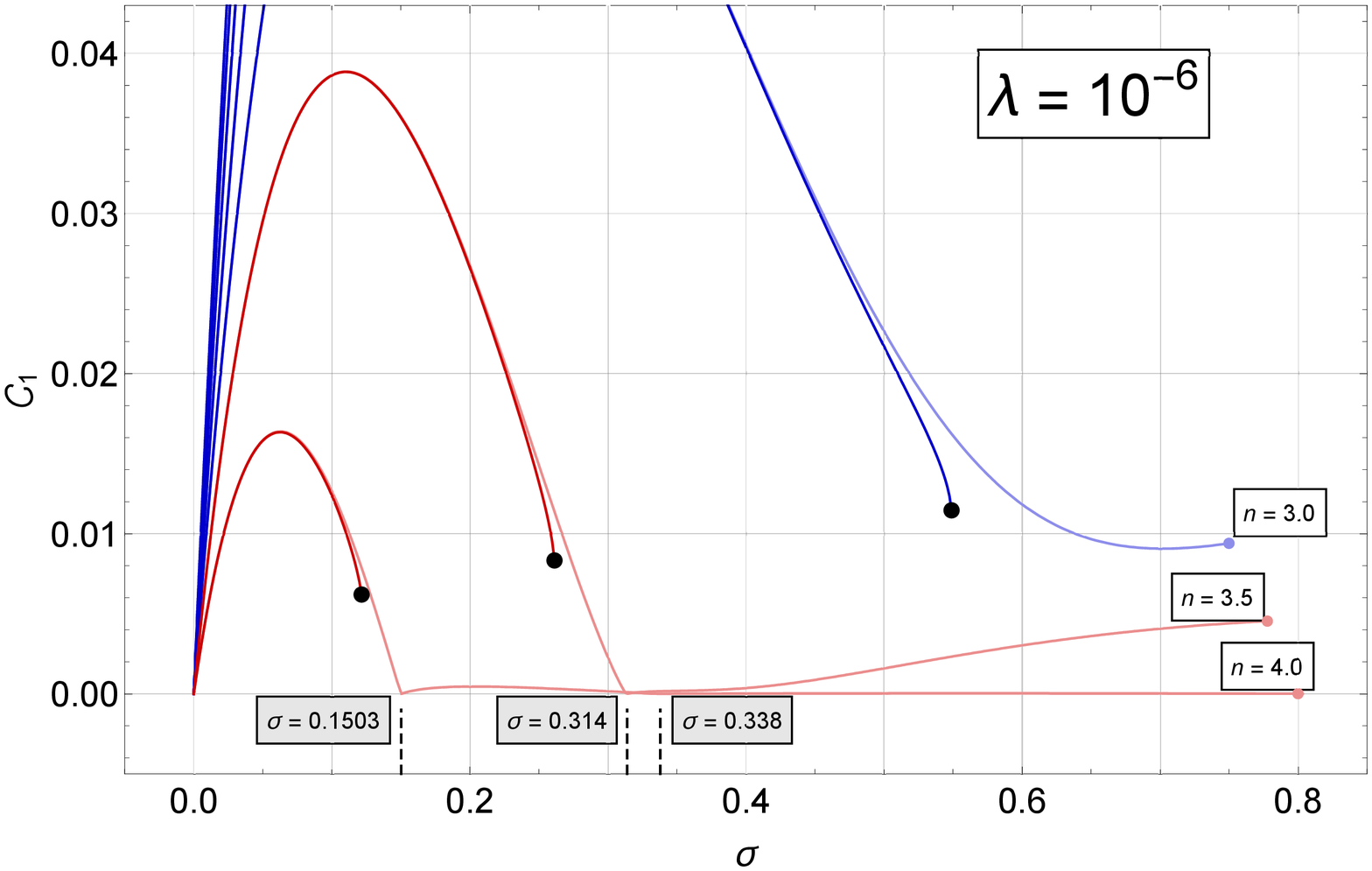}
\end{minipage}\hfill%
\begin{minipage}{0.48\linewidth}
\centering
\includegraphics[width=\linewidth]{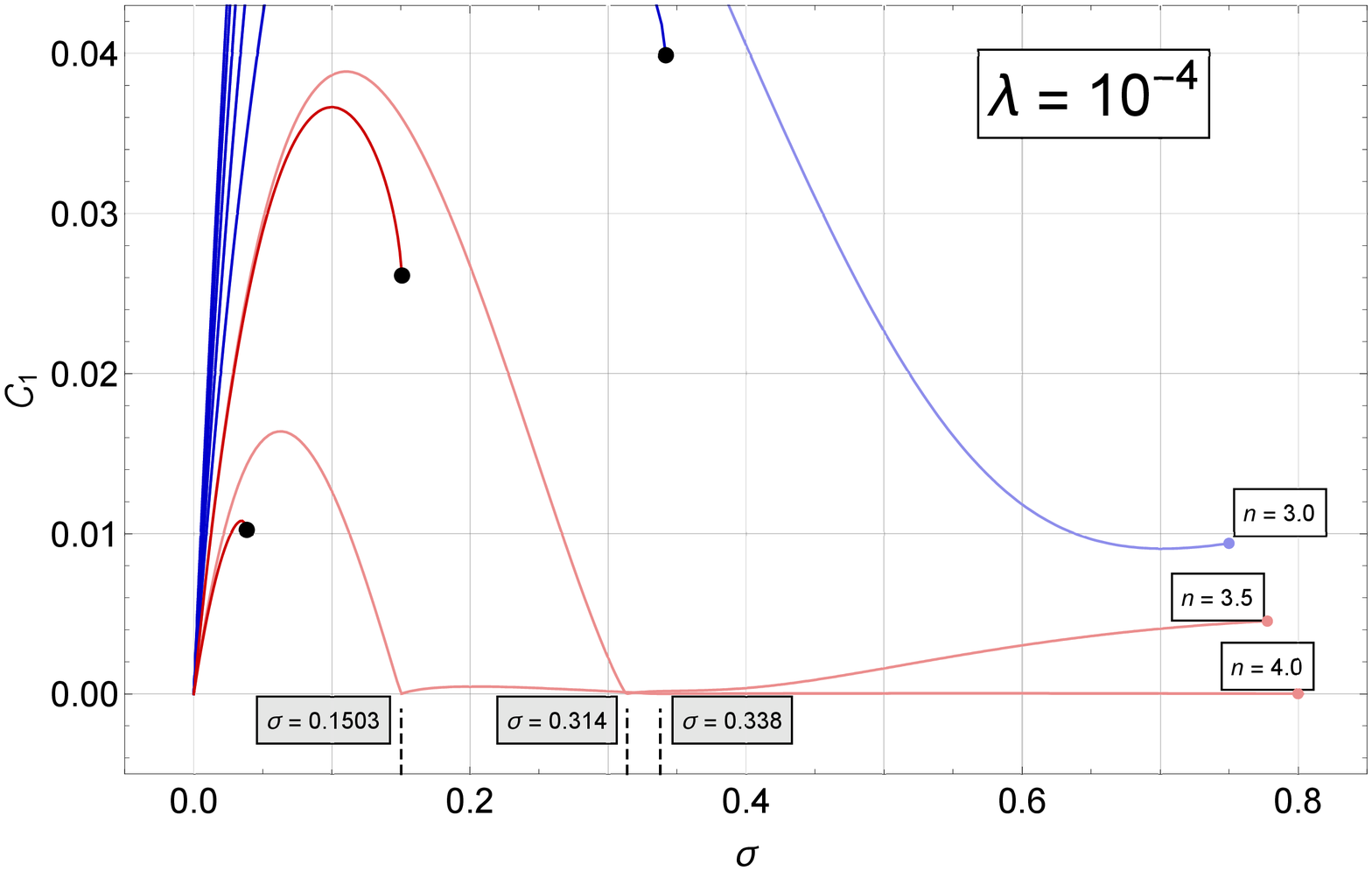}
\end{minipage}
\par\vspace{1.5\baselineskip}\par
\begin{minipage}{0.48\linewidth}
\centering
\includegraphics[width=\linewidth]{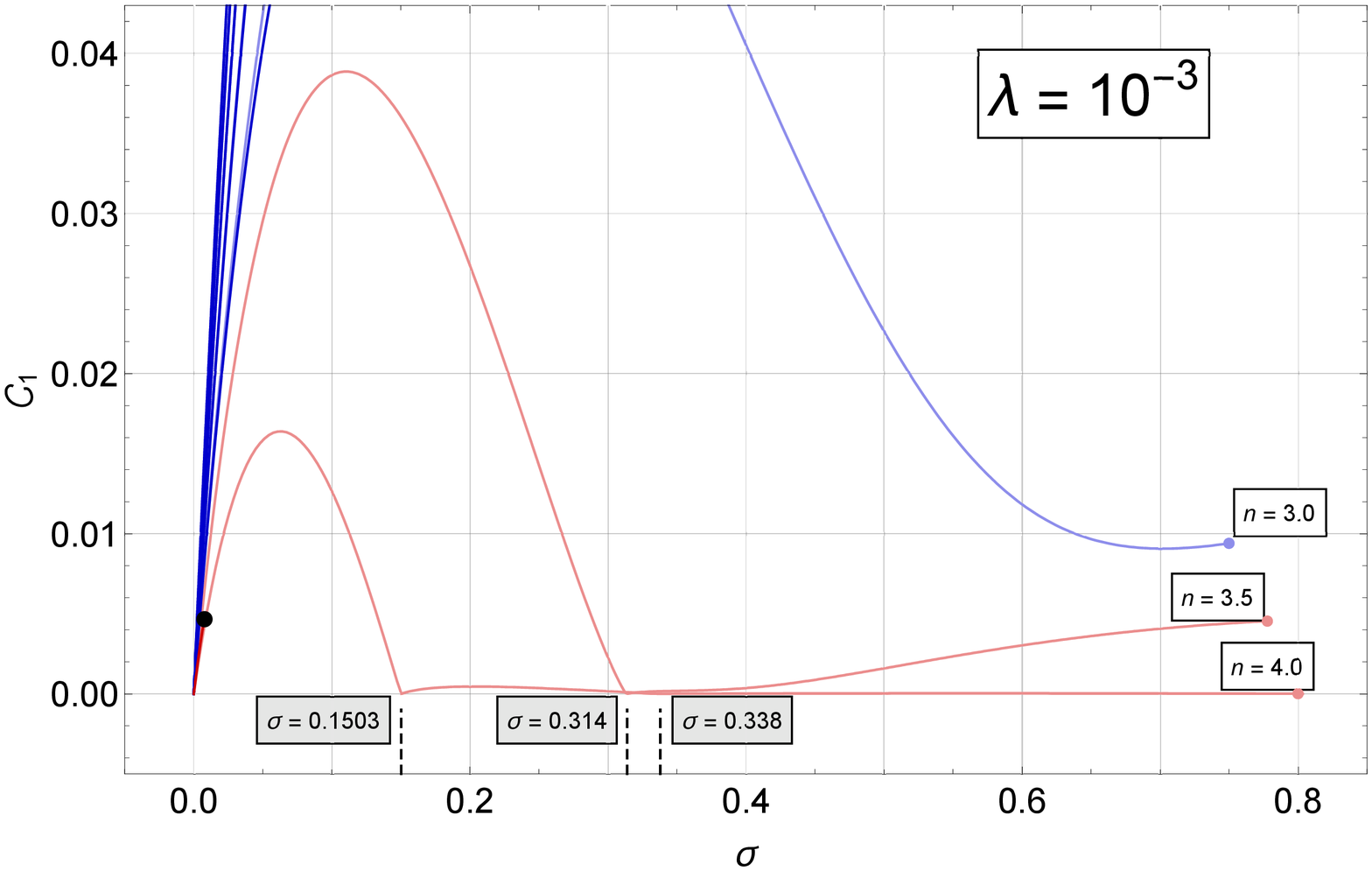}
\end{minipage}\hfill%
\begin{minipage}{0.48\linewidth}
\centering
\includegraphics[width=\linewidth]{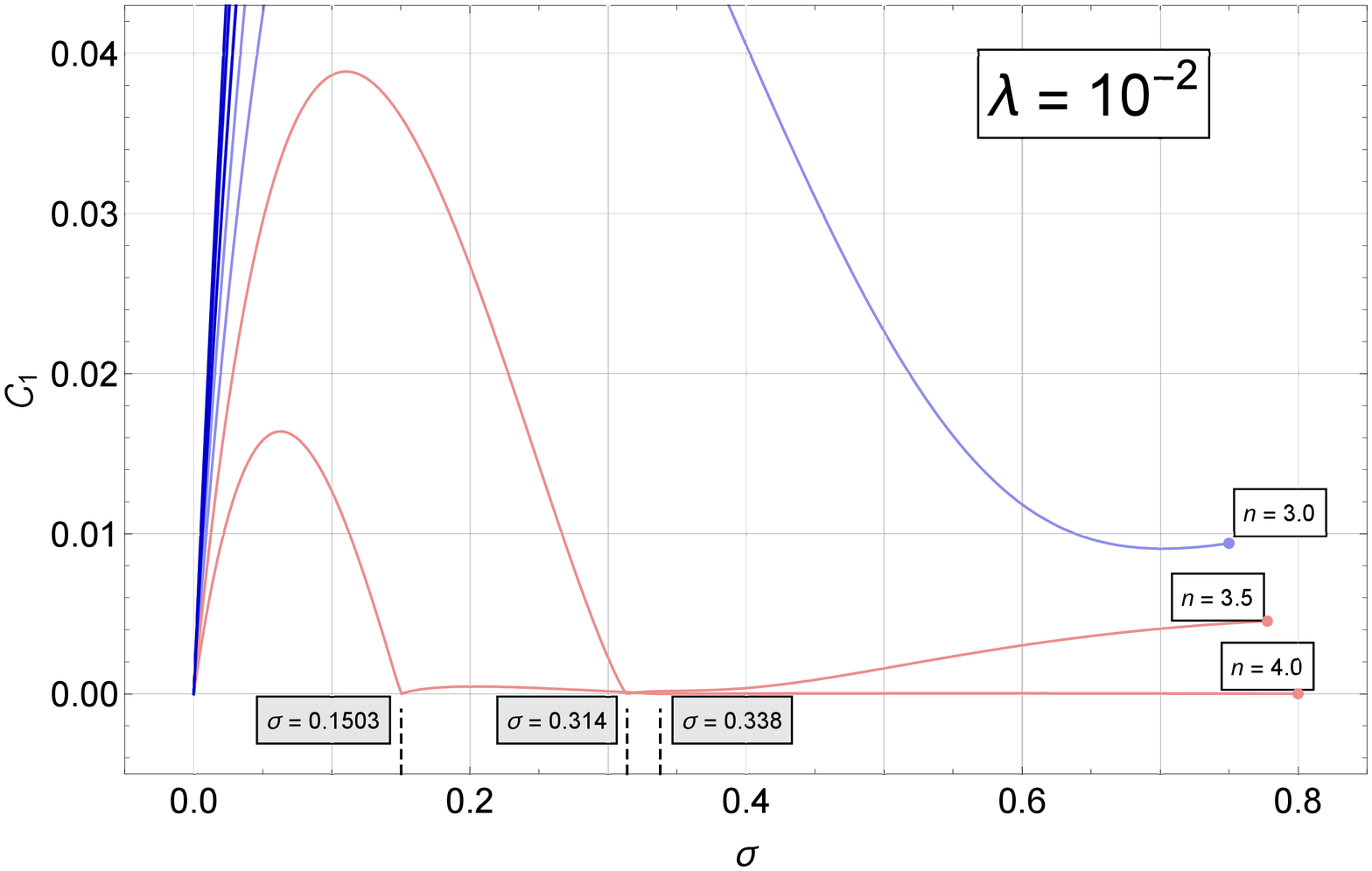}
\end{minipage}
\par\vspace{.8\baselineskip}\par
\caption{\label{SigComp}Dependences of the compactness $\mathcal{C}_{1}$ for the
  characteristic values of the polytropic index
  $n\in\{1.0, 1.5, 2, 2.5, 3, 3.5, 4\}$ with $\sigma$ varying up to the causal
  limit for $\lambda = 10^{-12}$, $\lambda = 10^{-9}$ (top),
  $\lambda = 10^{-6}$, $\lambda = 10^{-4}$ (middle) and $\lambda = 10^{-3}$,
  $\lambda = 10^{-2}$ (bottom).}
\end{figure*}

\begin{figure*}[t]
\begin{minipage}{0.48\linewidth}
\centering
\includegraphics[width=\linewidth]{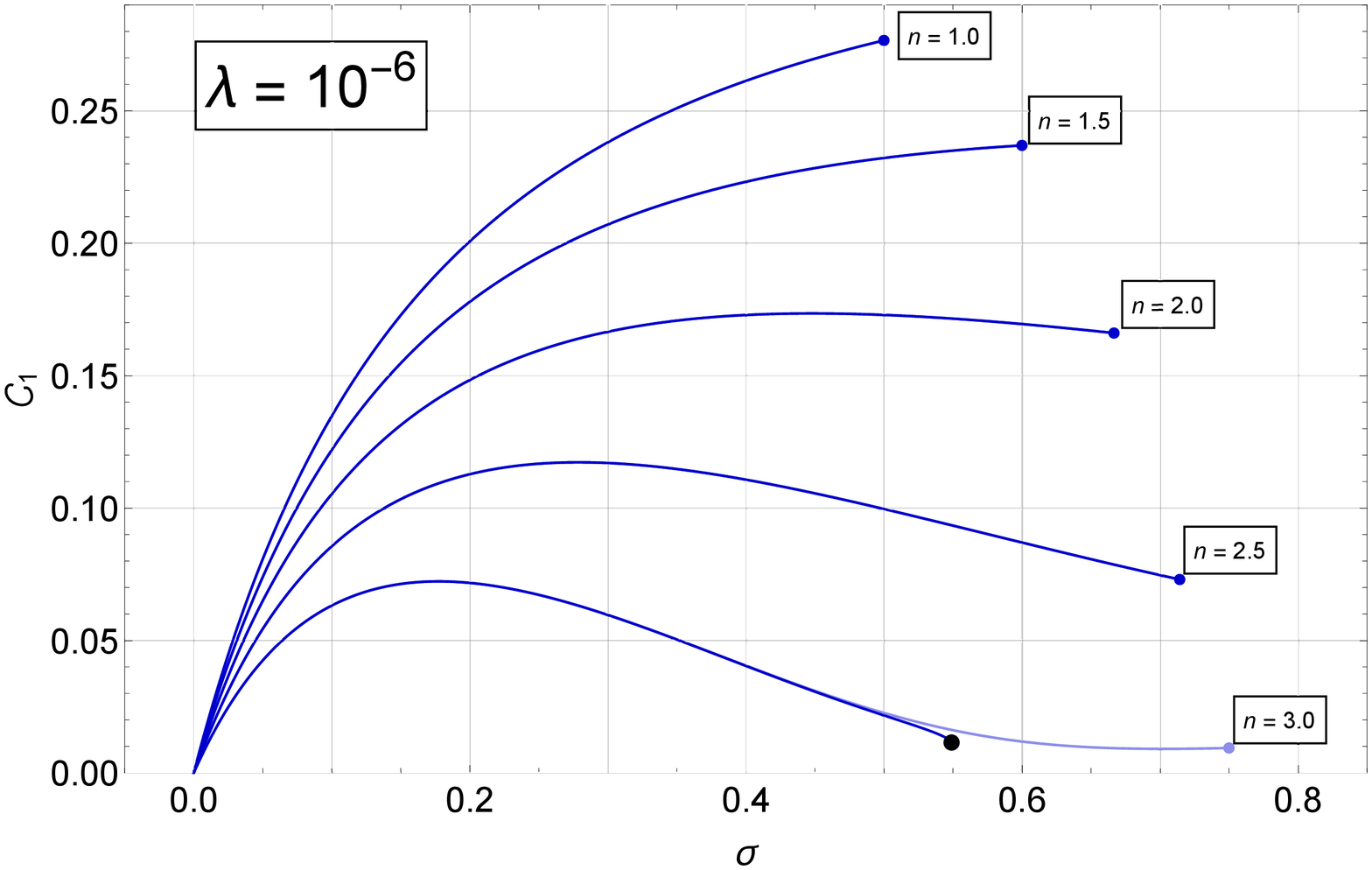}
\end{minipage}\hfill%
\begin{minipage}{0.48\linewidth}
\centering
\includegraphics[width=\linewidth]{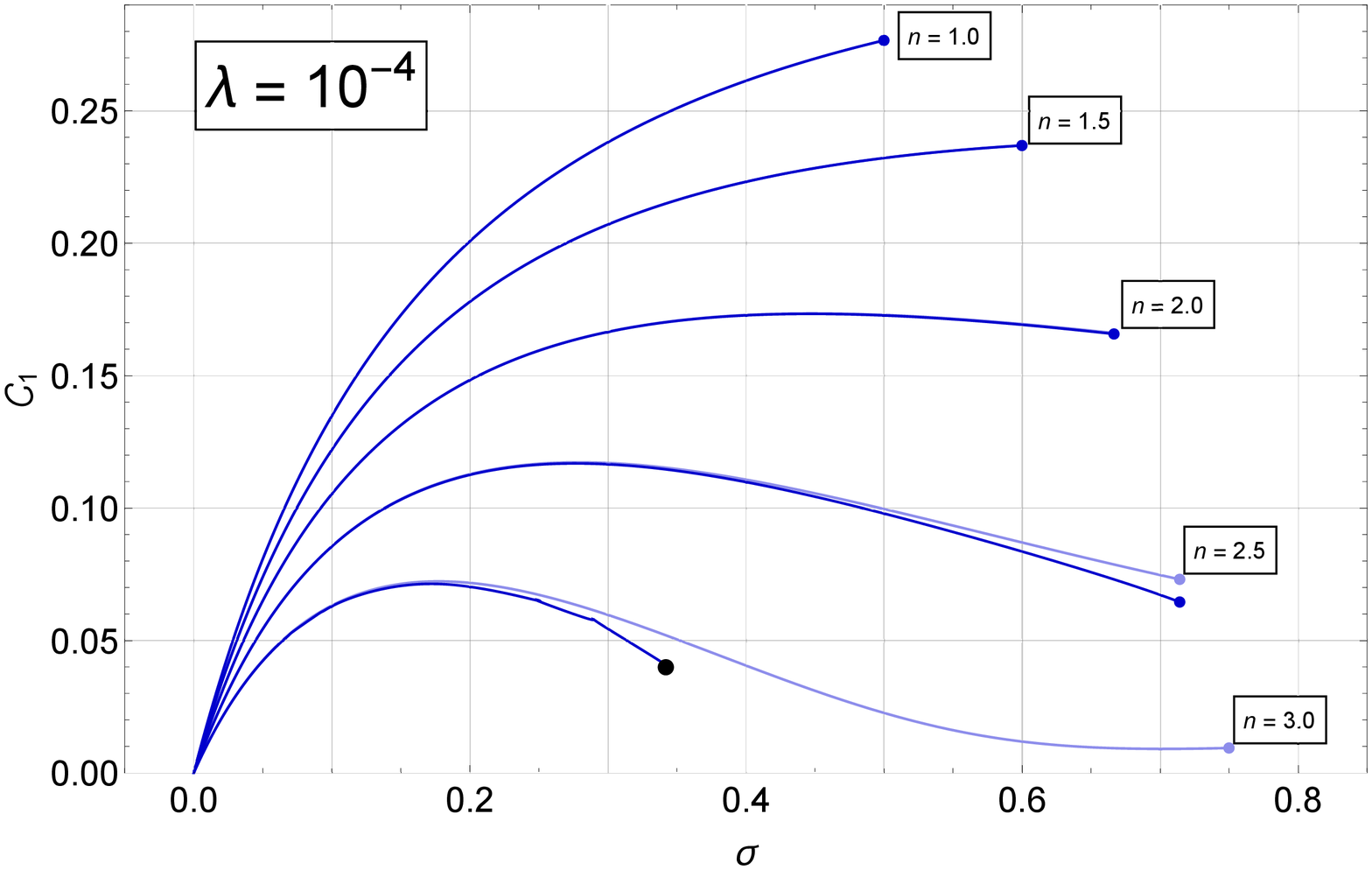}
\end{minipage}
\par\vspace{1.5\baselineskip}\par
\begin{minipage}{0.48\linewidth}
\centering
\includegraphics[width=\linewidth]{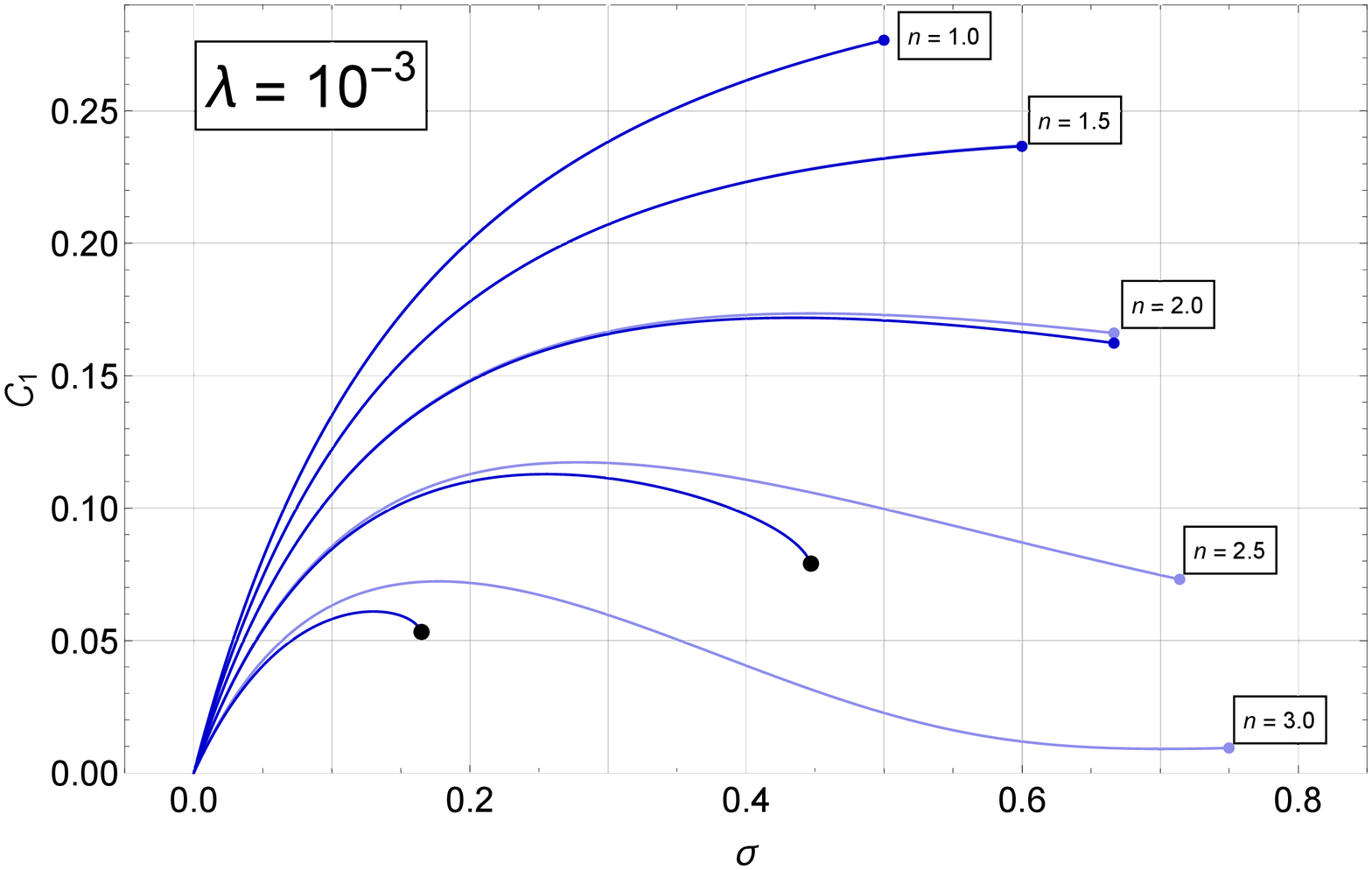}
\end{minipage}\hfill%
\begin{minipage}{0.48\linewidth}
\centering
\includegraphics[width=\linewidth]{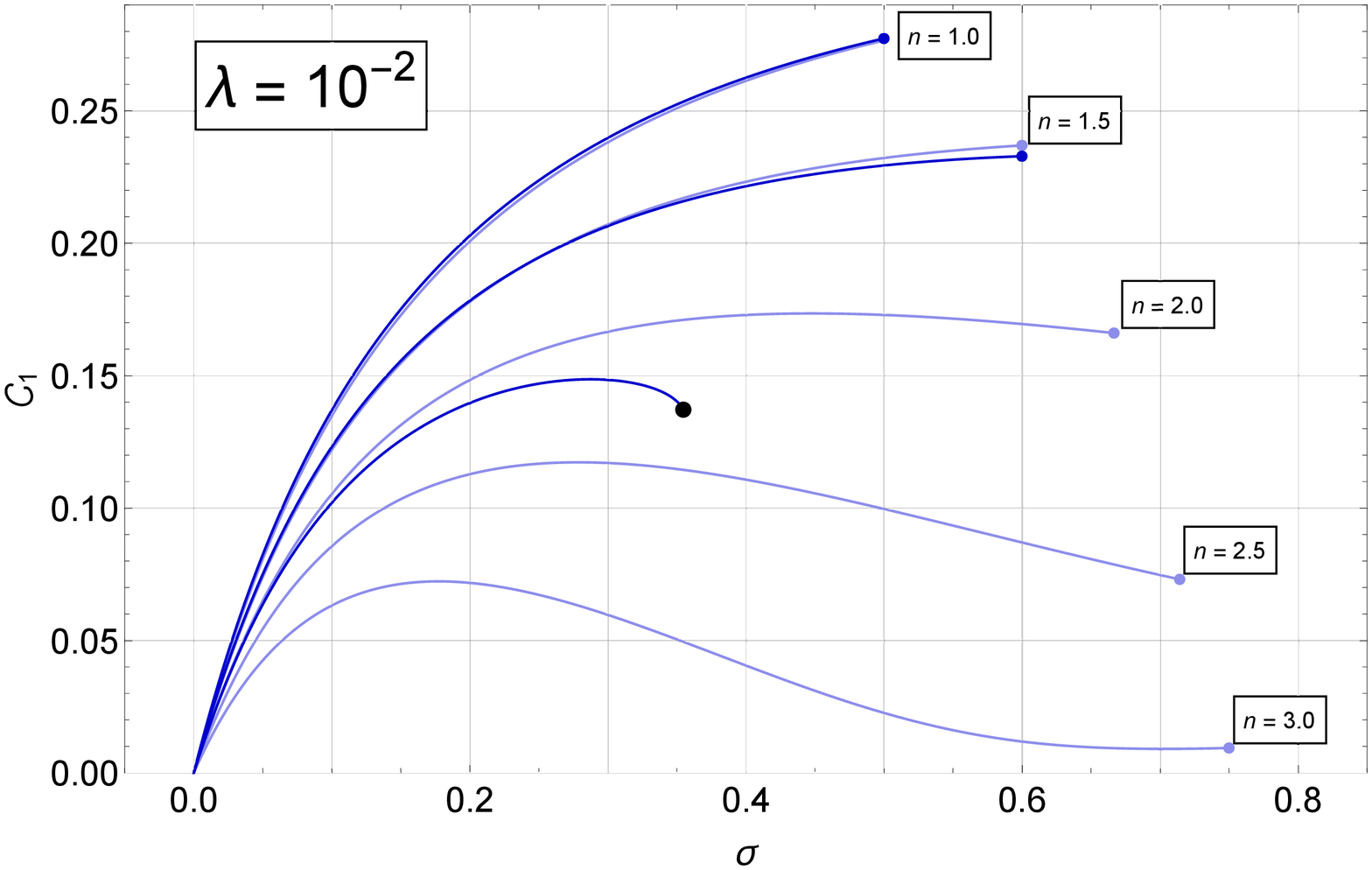}
\end{minipage}
\par\vspace{.8\baselineskip}\par
\caption{\label{SigCompEx}Extended dependences of the compactness
  $\mathcal{C}_{1}$ for the characteristic values of the polytropic index
  $n\in\{1.0, 1.5, 2, 2.5, 3, 3.5, 4\}$ with $\sigma$ varying up to the causal
  limit for $\lambda = 10^{-6}$, $\lambda = 10^{-4}$ (top) and
  $\lambda = 10^{-3}$, $\lambda = 10^{-2}$ (bottom). For $n \leq 3$, the
  compactness $\mathcal{C}_{1}$ $\sigma$-profiles are illustrated for the
  whole existence ranges, extending thus the detailed picture given in
  Fig.\,\ref{SigComp}.}
\end{figure*}

The influence of the cosmological constant on the compactness function
$\mathcal{C}_{1}(n;\sigma,\lambda)$ of the GRPs is represented in
Fig.\,\ref{SigComp} and Fig.\,\ref{SigCompEx} for the same values of the
polytrope index $n$ and the dimensionless parameter $\lambda$ as in the case
of $\xi_{1}$ and $v_{1}$.

All the functions are compared to the compactness function
$\mathcal{C}_{1}(n;\sigma,\lambda=0)$ for the same values of the polytropic
index $n$. There is $\mathcal{C}(n;\sigma=0,\lambda)=0$ for all values of the
cosmological parameter $\lambda$ allowing existence of the static
polytropes. In the case of $\lambda=0$, we can see that for $n=1.5$ the
compactness parameter increases with increasing relativistic parameter
$\sigma$ reaching the largest value of $\mathcal{C}_{1} \sim 0.235$ at the
causality limit. For $n=2, 2.5, 3$, the compactness parameter
$\mathcal{C}_{1}$ reaches a maximal value in the middle of the interval of
allowed values of $\sigma$ and then decreases to a minimum at the causality
limit of $\sigma$. The $\sigma$-profile of the compactness parameter strongly
decreases with increasing polytropic index $n$ - for $n=2$ there is
$\mathcal{C}_{1}<0.175$, while for $n=3$, there is $\mathcal{C}_{1}<0.1$. For
the polytropes with $n=3.5,4$, demonstrating the divergent behavior of the
extension parameter $\xi_{1}$ at the critical values of $\sigma_{\mathrm{f}}$,
the $\sigma$-profile of the compactness parameter contains zero points at the
critical points $\sigma_{\mathrm{f}}$, reaching a maximum between $\sigma=0$
and $\sigma_{\mathrm{f}}$ points. The compactness parameter $\mathcal{C}_{1}$
significantly decreases with increasing $n$. For $n=3.5$, there is
$\mathcal{C}_{1}<0.04$, while $\mathcal{C}_{1}<0.17$ for $n=4$
polytropes. Notice that in the case of the $n=4$ polytropes there is
$\mathcal{C}_{1}<10^{-4}$ for $\sigma > \sigma_{\mathrm{f}2}$---polytropes
with such extremely low compactness parameter $\mathcal{C}_{1}$ occur, as
their extension parameter $\xi_{1}$ has to be extremely high.

In the case of the compactness parameter function
$\mathcal{C}_{1}(n;\sigma,\lambda)$, the role of the repulsive cosmological
constant is again concentrated in the cut-off of the polytropes allowed for
fixed parameters $n$ and $\sigma$, when strong restrictions appear with $n$
increasing. Further, we can observe in Fig.\,\ref{SigComp} and
Fig.\,\ref{SigCompEx} significant modifications of the $\sigma$-profile in
addition to the limits implied by the restriction on the existence of the
polytropic equilibrium configurations. The modifications of the
$\mathcal{C}_{1}(n;\sigma,\lambda)$ $\sigma$-profiles become substantial for
$\lambda \geq 10^{-6}$ and the influence of the cosmological constant always
decreases compactness of the polytrope while the other parameters are kept
fixed.

\subsection{Energy of polytropes}

We can appropriately describe the GRPs by global characteristics reflecting
the result of interplay of the gravitational forces and the forces governing
properties of matter constituting the polytrope. We consider now the
representative global characteristics, gravitational energy and binding energy
of the complete equilibrium polytropic configurations characterized by
parameters $\xi_{1}$ and $v_{1}$. We thus denote them as
$\mathcal{G}_{1} = \mathcal{G}(\xi_{1})$ for the gravitational energy and
$\mathcal{B}_{1} = \mathcal{B}(\xi_{1})$ for the binding energy. Later we
shall consider also their radial profiles $\mathcal{G}(\xi)$ and
$\mathcal{B}(\xi)$.

\subsubsection{Gravitational energy}

\begin{figure*}[t]
\begin{minipage}{0.48\linewidth}
\centering
\includegraphics[width=\linewidth]{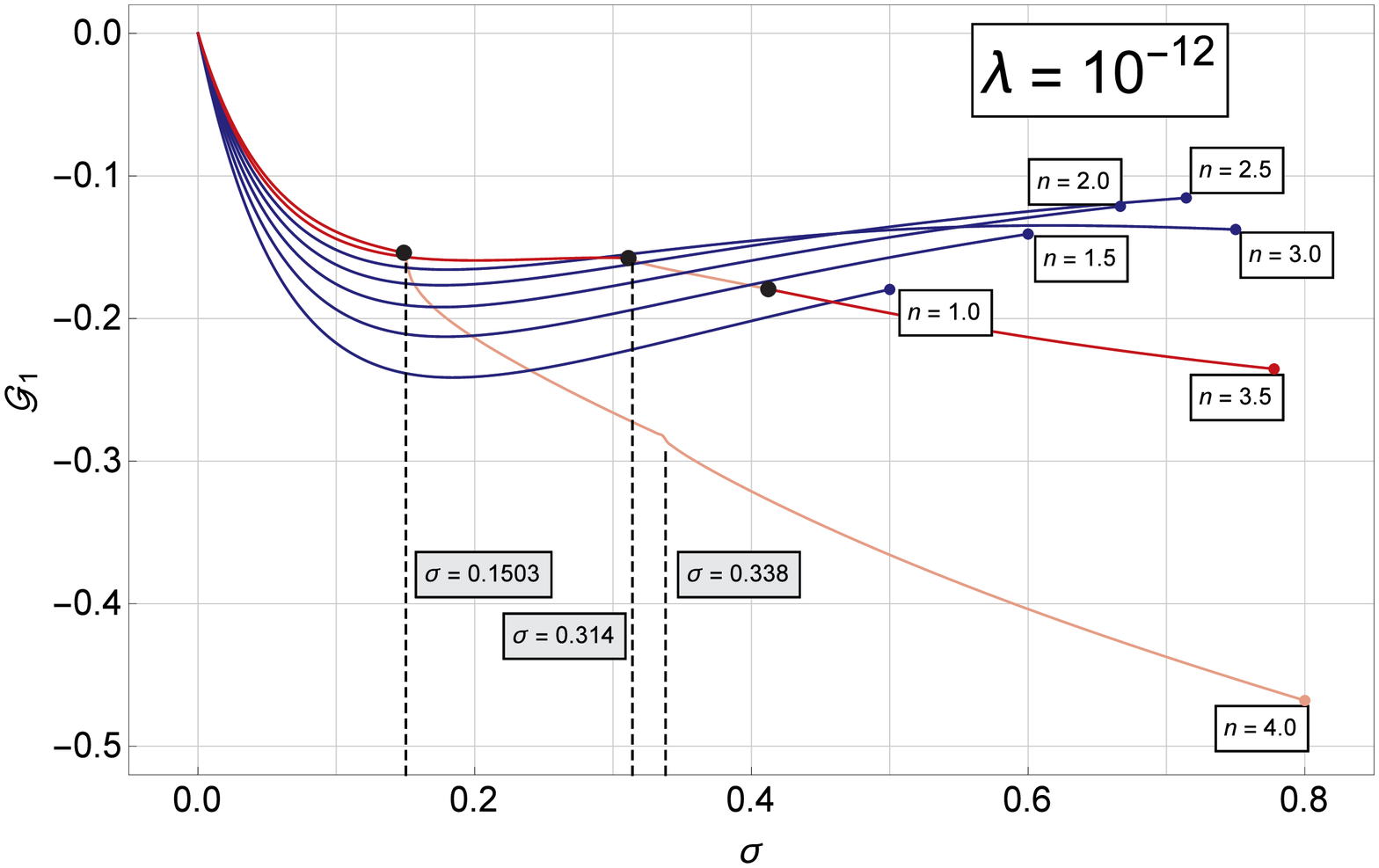}
\end{minipage}\hfill%
\begin{minipage}{0.48\linewidth}
\centering
\includegraphics[width=\linewidth]{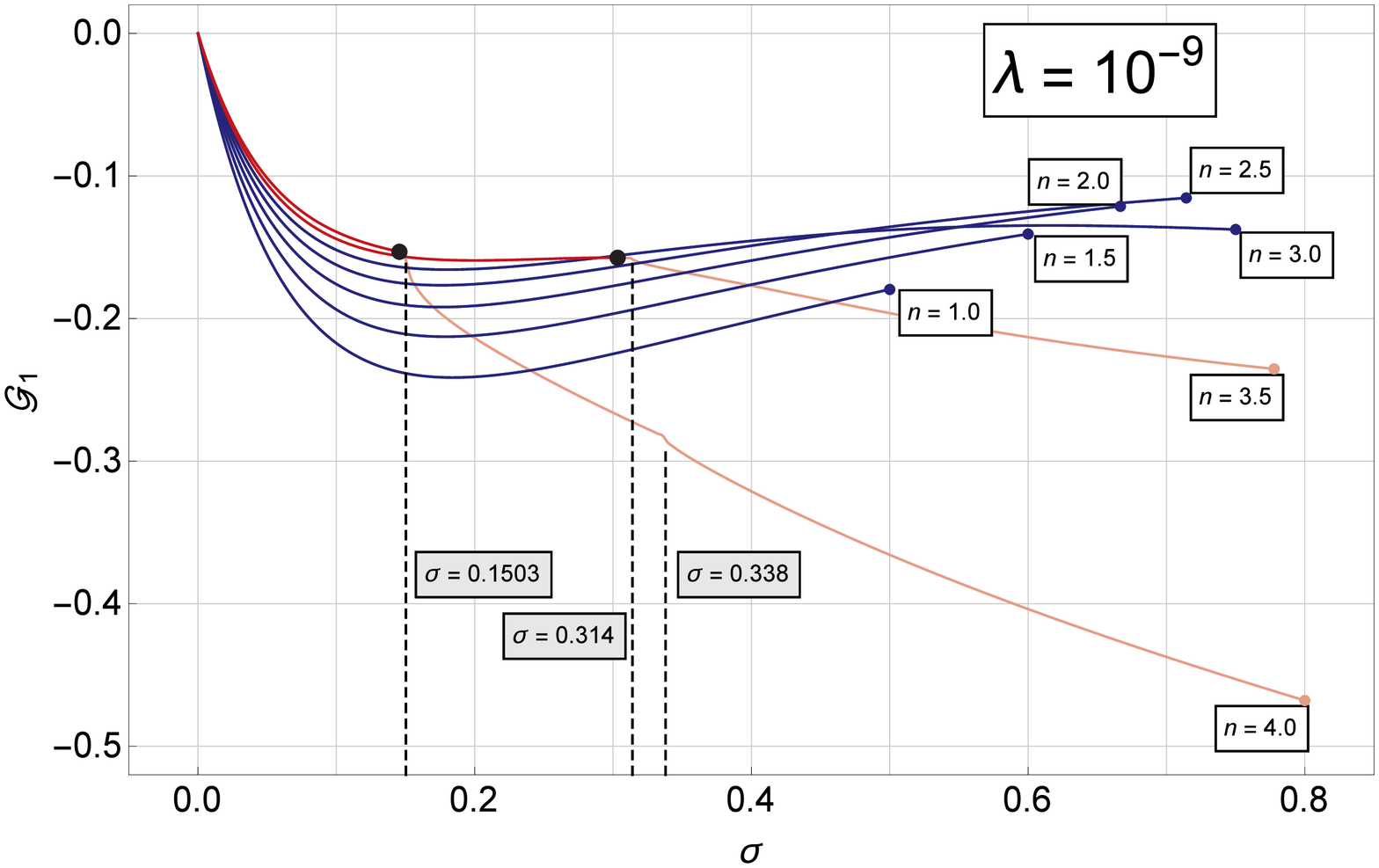}
\end{minipage}
\par\vspace{1.5\baselineskip}\par
\begin{minipage}{0.48\linewidth}
\centering
\includegraphics[width=\linewidth]{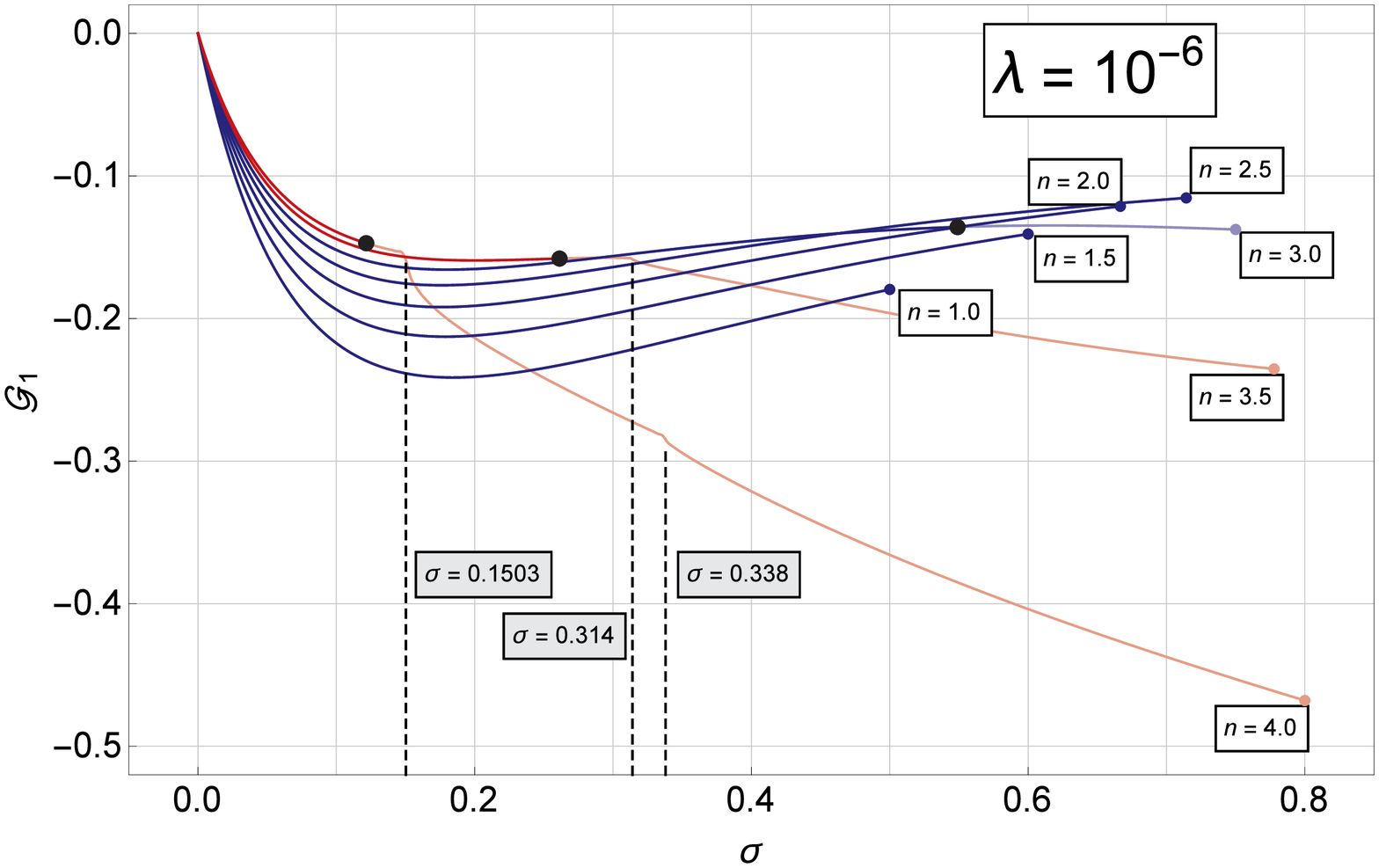}
\end{minipage}\hfill%
\begin{minipage}{0.48\linewidth}
\centering
\includegraphics[width=\linewidth]{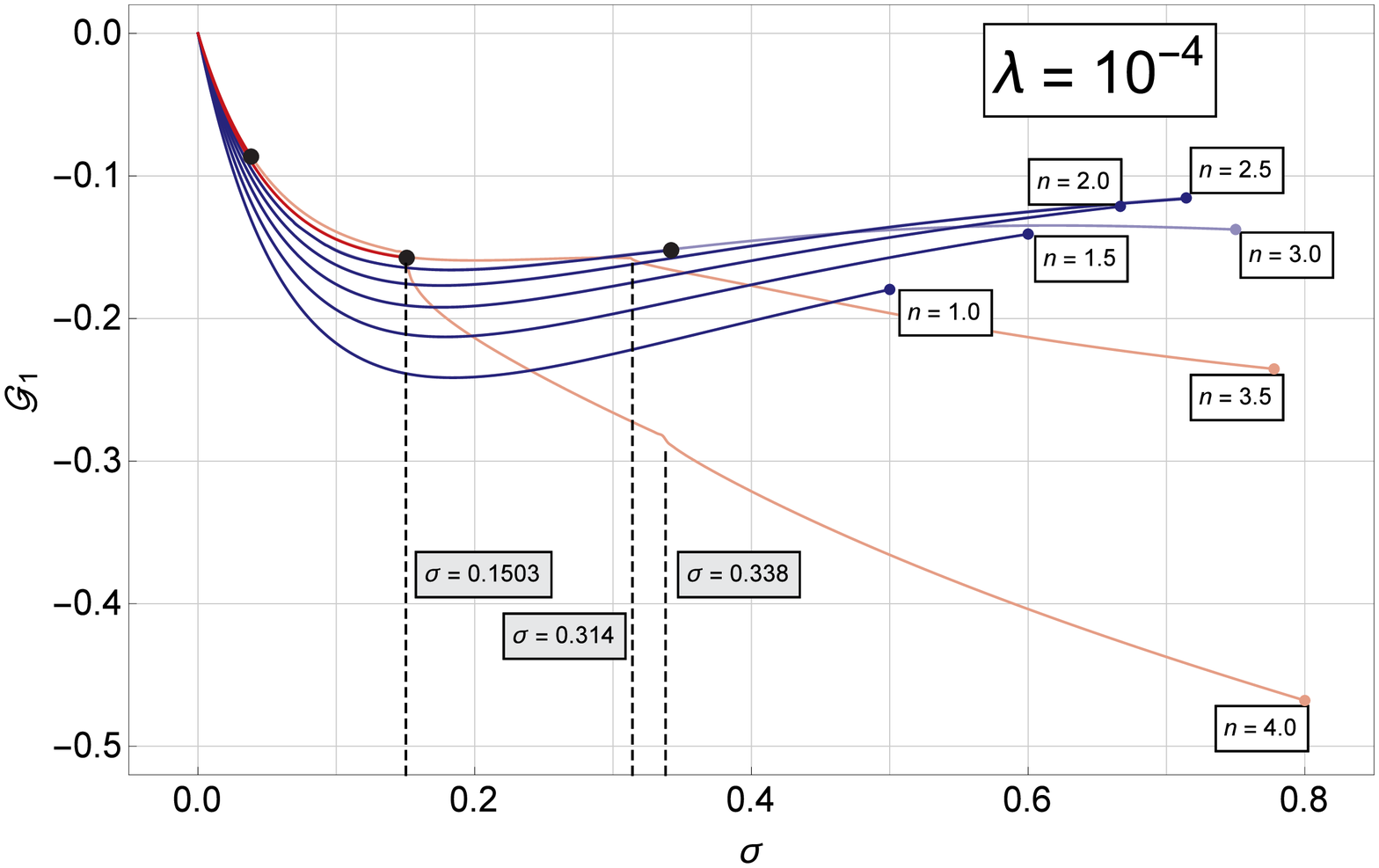}
\end{minipage}
\par\vspace{1.5\baselineskip}\par
\begin{minipage}{0.48\linewidth}
\centering
\includegraphics[width=\linewidth]{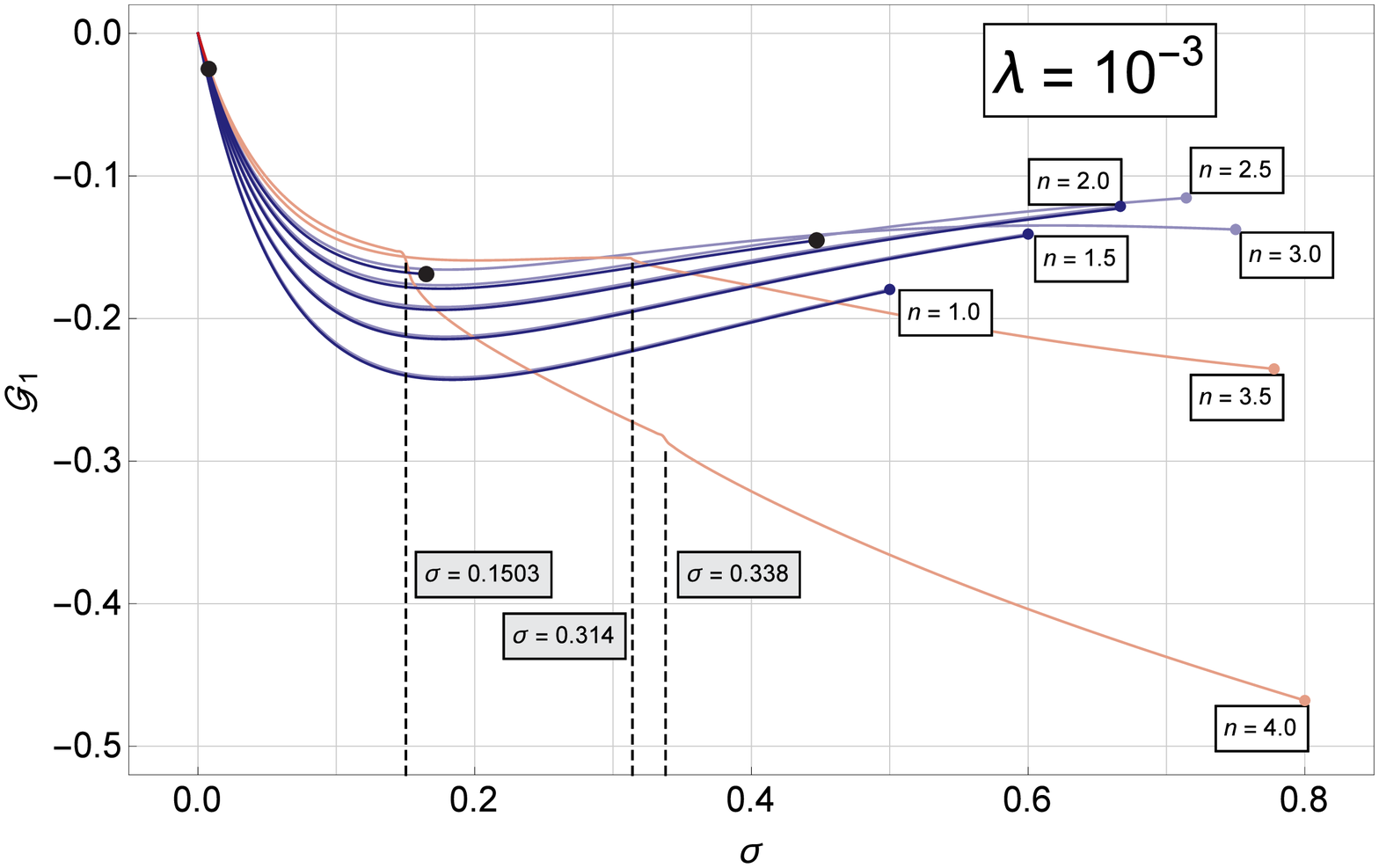}
\end{minipage}\hfill%
\begin{minipage}{0.48\linewidth}
\centering
\includegraphics[width=\linewidth]{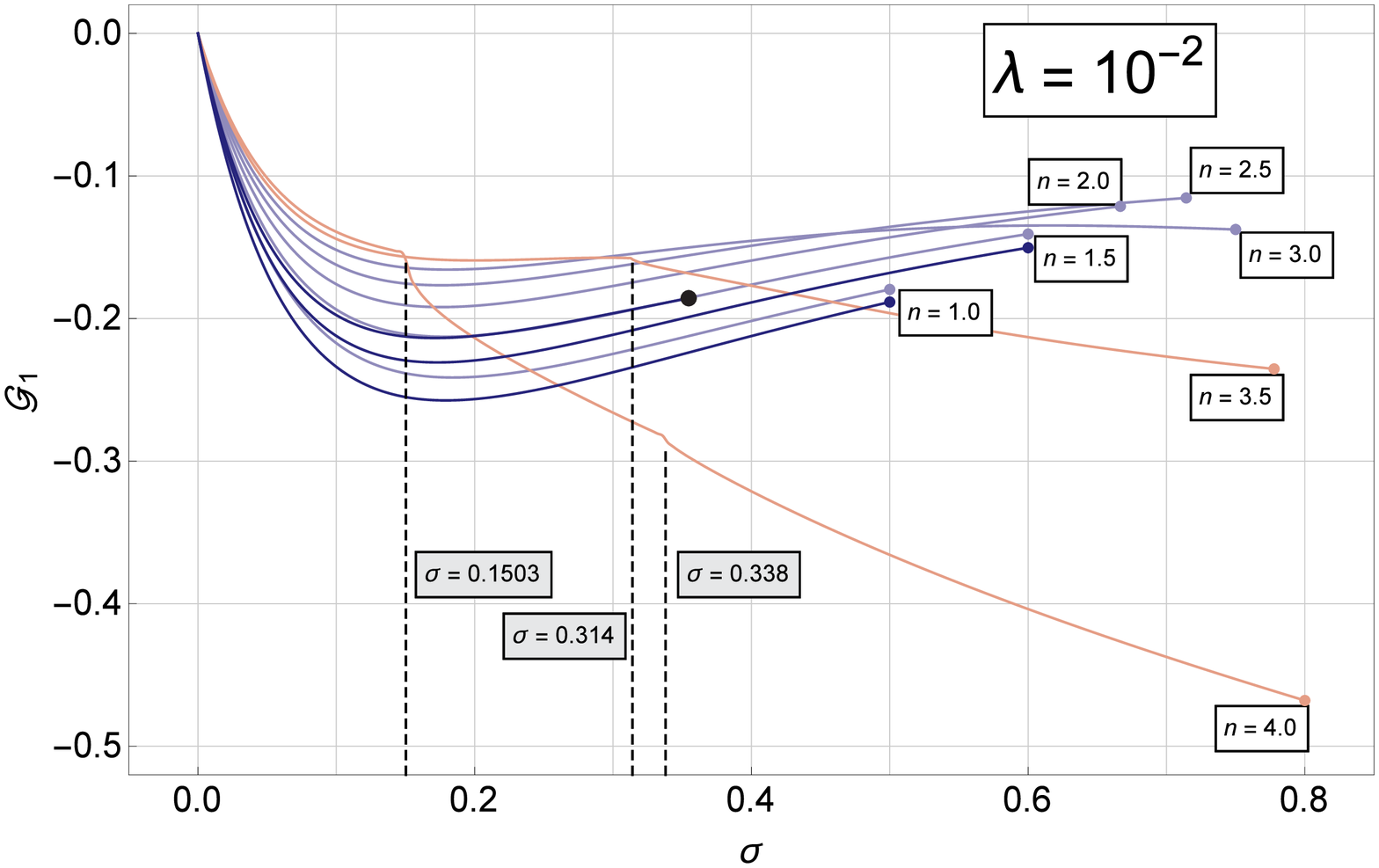}
\end{minipage}
\par\vspace{.8\baselineskip}\par
\caption{\label{SigGrPoEn}Dependences of the gravitational energy
  $\mathcal{G}_{1}$ for the characteristic values of the polytropic index
  $n\in\{1.0, 1.5, 2, 2.5, 3, 3.5, 4\}$ with $\sigma$ varying up to the causal
  limit for $\lambda = 10^{-12}$, $\lambda = 10^{-9}$ (top),
  $\lambda = 10^{-6}$, $\lambda = 10^{-4}$ (middle) and $\lambda = 10^{-3}$,
  $\lambda = 10^{-2}$ (bottom).}
\end{figure*}

The dimensionless gravitational energy $\mathcal{G}_{1}$ represents a global
characteristic of binding effects of gravity in equilibrium and has to be
negative for any polytrope. The role of the cosmological constant in the
behavior of the gravitational energy of the polytropes is represented in
Fig.\,\ref{SigGrPoEn} for the same values of the polytrope index $n$ and the
dimensionless parameter $\lambda$ as in the case of quantities $\xi_{1}$ and
$v_{1}$. The gravitational energy function $\mathcal{G}_{1}(n;\sigma,\lambda)$
is always compared to the function $\mathcal{G}_{1}(n;\sigma,\lambda=0)$. In
the case of $\lambda=0$ there is $\mathcal{G}_{1}(n;\sigma=0)=0$, and the
gravitational energy of polytropes with $n=1.5, 2, 2.5$ reaches a minimal
value in the middle of the interval of allowed values of the relativistic
parameter $\sigma$ and then increases to a maximum at the causality limit on
the value of $\sigma$. In the case of $n=3$ polytropes, the $\sigma$-profile
of the gravitational energy has a maximum following the minimum.  At the
minimum of the $\sigma$-profile, the gravitational energy significantly
decreases with decreasing polytropic index $n$ (gravitational binding
increases), demonstrating shift from the value of $\mathcal{G}_{1} \sim -0.17$
in the case $n=3$ to the value of $\mathcal{G}_{1} \sim -0.21$ in the case
$n=1.5$. For the polytropes with $n=3.5, 4$, having the divergence of
$\xi_{1}$ at the critical values of $\sigma_{\mathrm{f}}$, the
$\sigma$-profile of the gravitational energy is continuous at the critical
points, but its derivative has a jump there. In the region of large values of
$\sigma$, the gravitational energy demonstrates a strong decrease, and in the
case of the $n=4$ polytropes, $\mathcal{G}_{1} \sim -0.47$ at the causality
limit, demonstrating thus strong gravitational binding.

For $\mathcal{G}_{1}(n;\sigma,\lambda)$, Fig.\,\ref{SigGrPoEn} demonstrates
that the role of the cosmological constant parameter is again reflected mainly
by the cut-off in allowed values of the parameter $\sigma$ for polytropes with
fixed parameter $n$---strong restrictions occur with $n$ increasing, in
similarity with the previously considered cases. We can also observe a slight
modification of the $\sigma$-profile in addition to the limits implied by the
restriction on the existence of the polytropic equilibrium
configurations. However, modifications of the
$\mathcal{G}_{1}(n;\sigma,\lambda)$ $\sigma$-profiles large enough to be
recognized occur only for $\lambda \geq 10^{-3}$---increasing of $\lambda$
always decreases the gravitational energy of the polytrope while other
parameters are fixed.

\begin{figure}[t]
\begin{center}
\includegraphics[width=\linewidth]{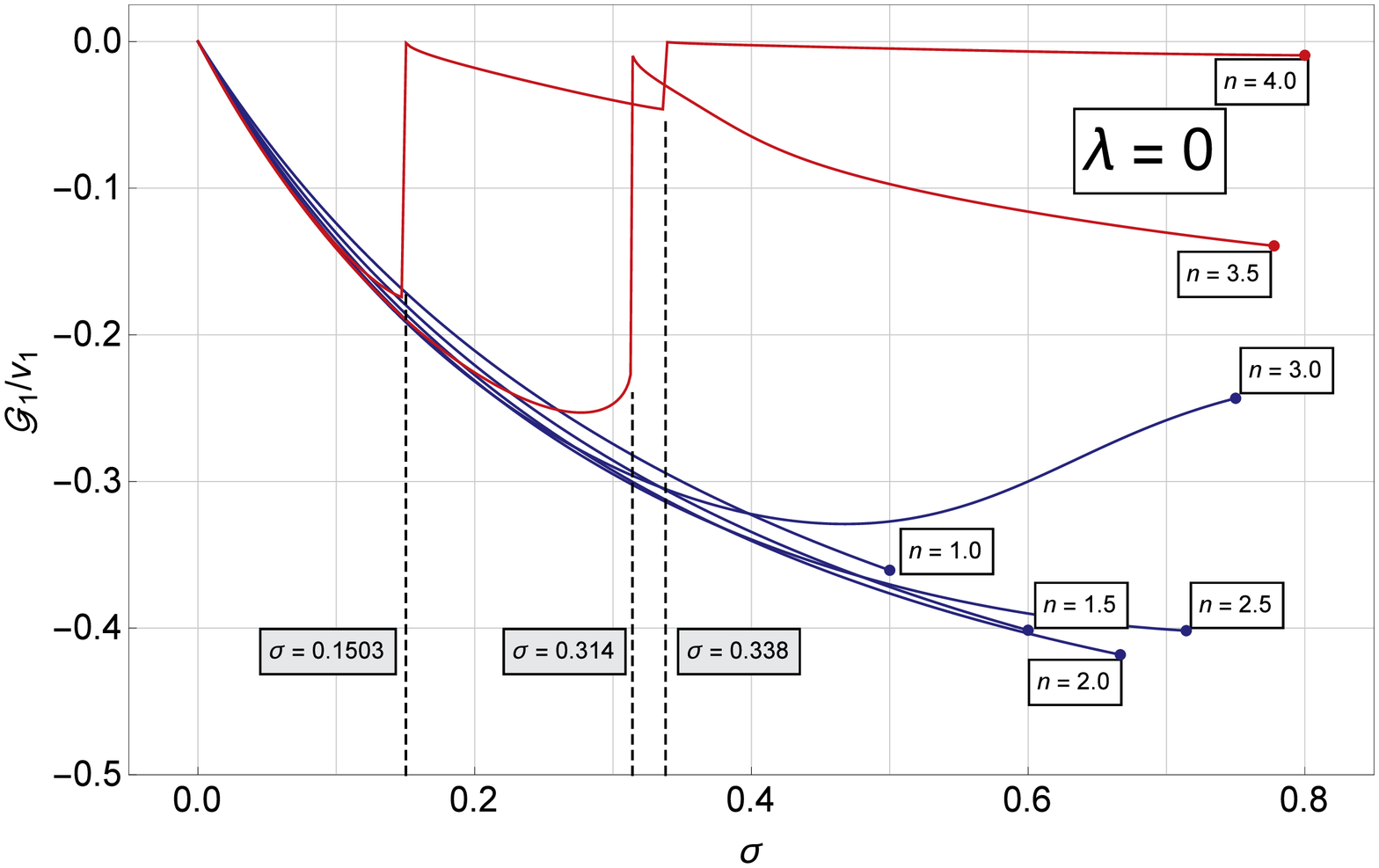}
\end{center}
\caption{\label{SigGrPoEnReLa0}Dependences of the relative gravitational
  energy $\mathcal{G}_{1}/v_{1}$ for the characteristic values of the polytropic
  index $n\in\{1.0, 1.5, 2, 2.5, 3, 3.5, 4\}$ with $\sigma$ varying up to the
  causal limit for $\lambda = 0$.}
\end{figure}

For completeness, we give also the $\sigma$-profiles for the relative
gravitational energy defined as $\mathcal{G}_{1}/v_{1}(n;\sigma,\lambda)$,
i.e., the gravitational energy is related to the dimensionless gravitational
mass of the polytrope. There is $\mathcal{G}/v_{1}(n;\sigma=0)=0$. For
$\lambda = 0$, the relative gravitational energy
$\mathcal{G}_{1}/v_{1}(n;\sigma,\lambda=0)$ is illustrated in
Fig.\,\ref{SigGrPoEnReLa0}. We can see that the character of the
$\sigma$-profiles for polytropes with $n=1, 1.5, 2, 2.5, 3$ is the same as for
the gravitational energy, but its magnitude is larger than for the
gravitational energy. On the other hand, a substantial change occurs in the
$\mathcal{G}_{1}/v_{1}$ $\sigma$-profiles of the $n=3.5,4$ polytropes, as a
jump to a zero point has to occur at the critical points of relativistic
parameter $\sigma_{\mathrm{f}}$ due to the behavior of $v_{1}$, and the
profiles of the $n=3.5,4$ polytropes are located above the profiles of
polytropes with lower $n$ at the region of large values of $\sigma$, contrary
to the case of gravitational energy.

\begin{figure*}[t]
\begin{minipage}{0.48\linewidth}
\centering
\includegraphics[width=\linewidth]{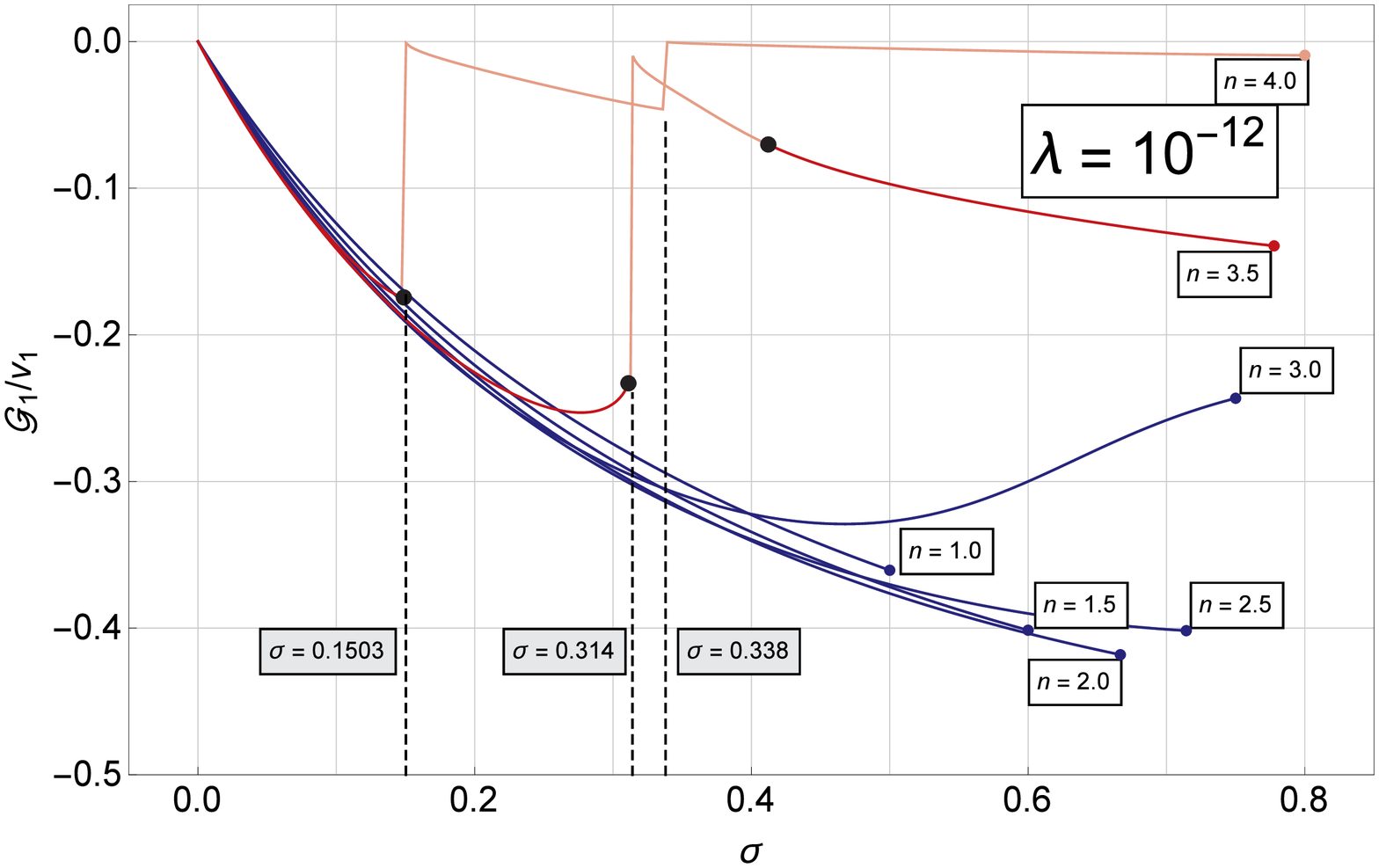}
\end{minipage}\hfill%
\begin{minipage}{0.48\linewidth}
\centering
\includegraphics[width=\linewidth]{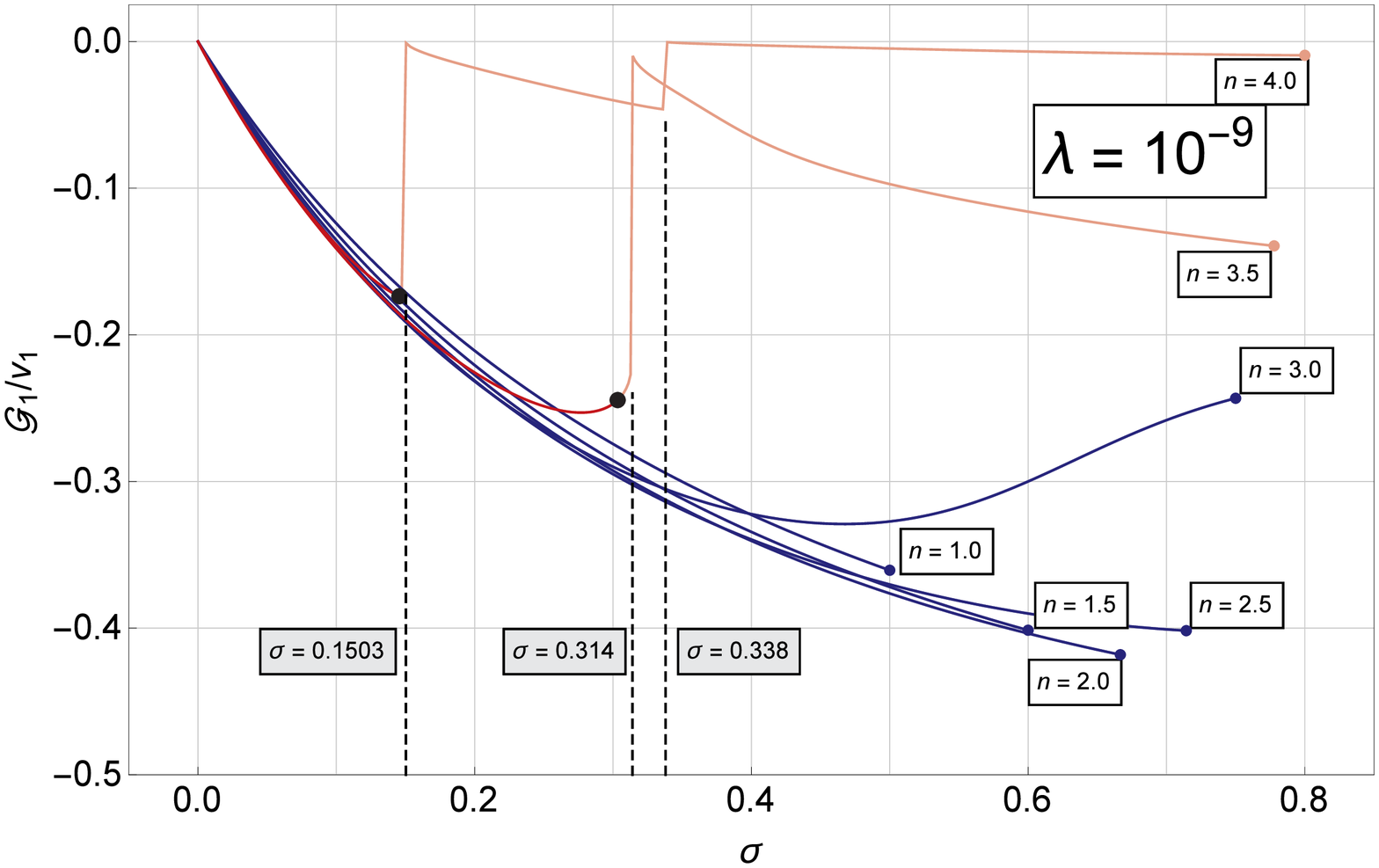}
\end{minipage}
\par\vspace{1.5\baselineskip}\par
\begin{minipage}{0.48\linewidth}
\centering
\includegraphics[width=\linewidth]{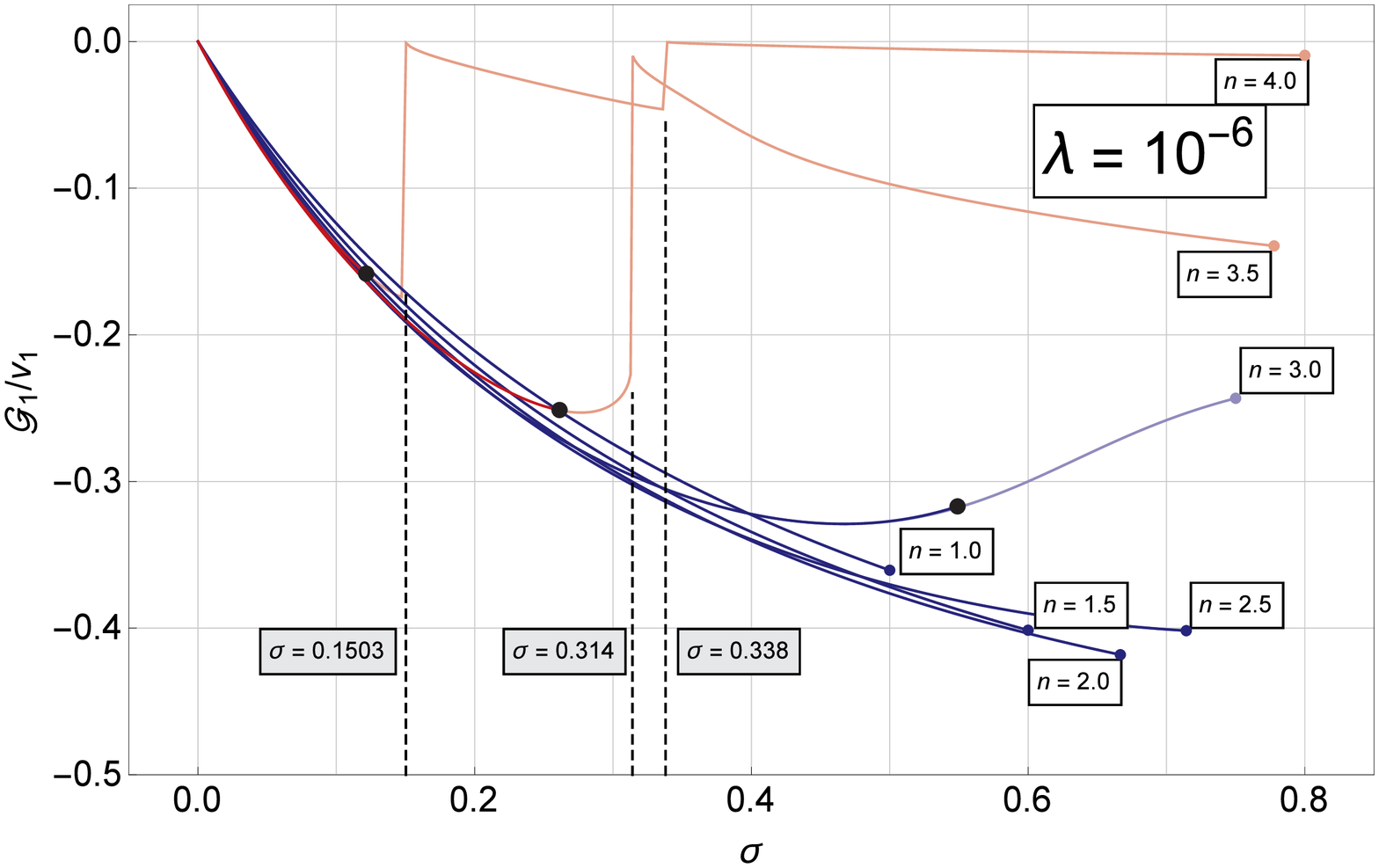}
\end{minipage}\hfill%
\begin{minipage}{0.48\linewidth}
\centering
\includegraphics[width=\linewidth]{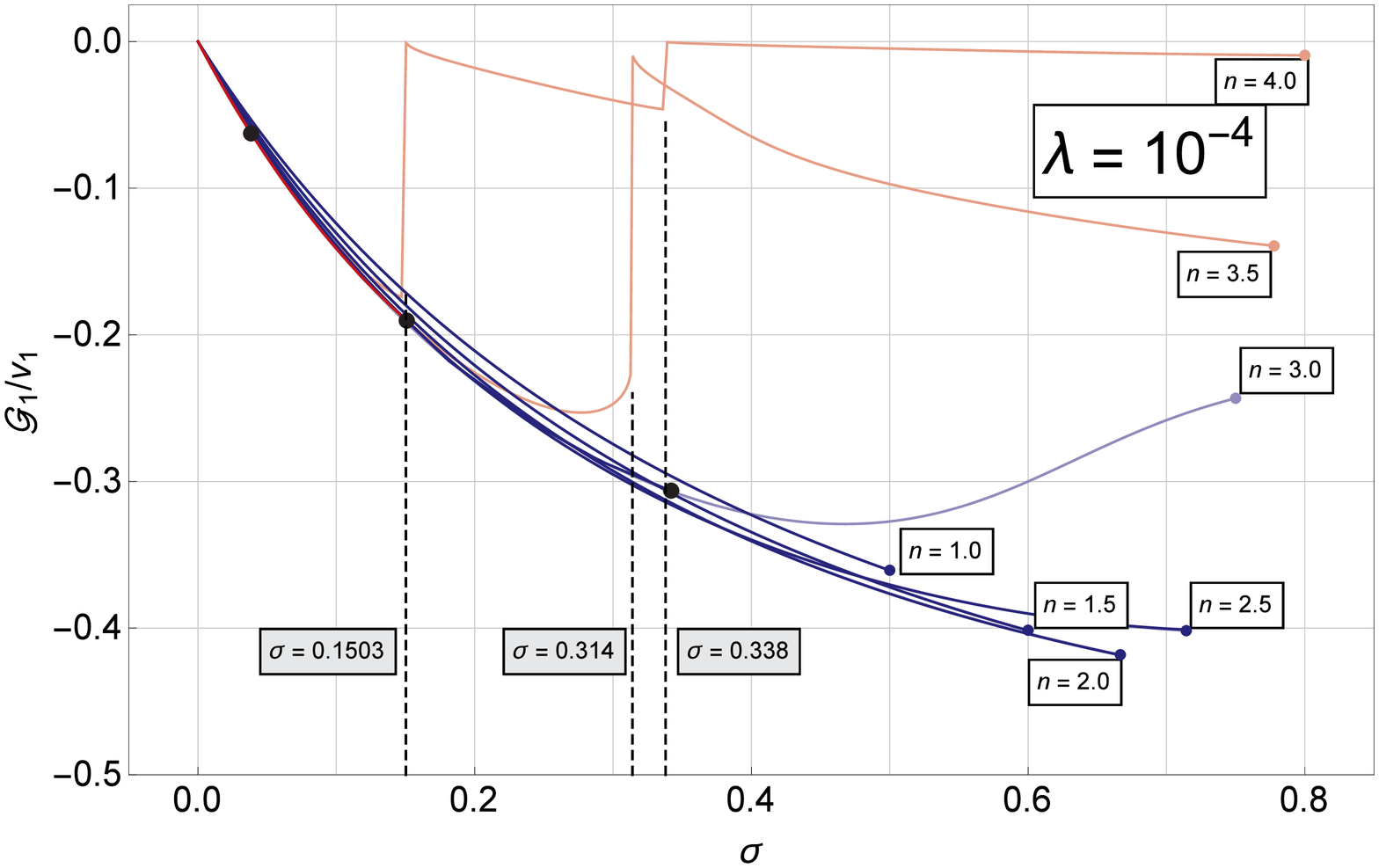}
\end{minipage}
\par\vspace{1.5\baselineskip}\par
\begin{minipage}{0.48\linewidth}
\centering
\includegraphics[width=\linewidth]{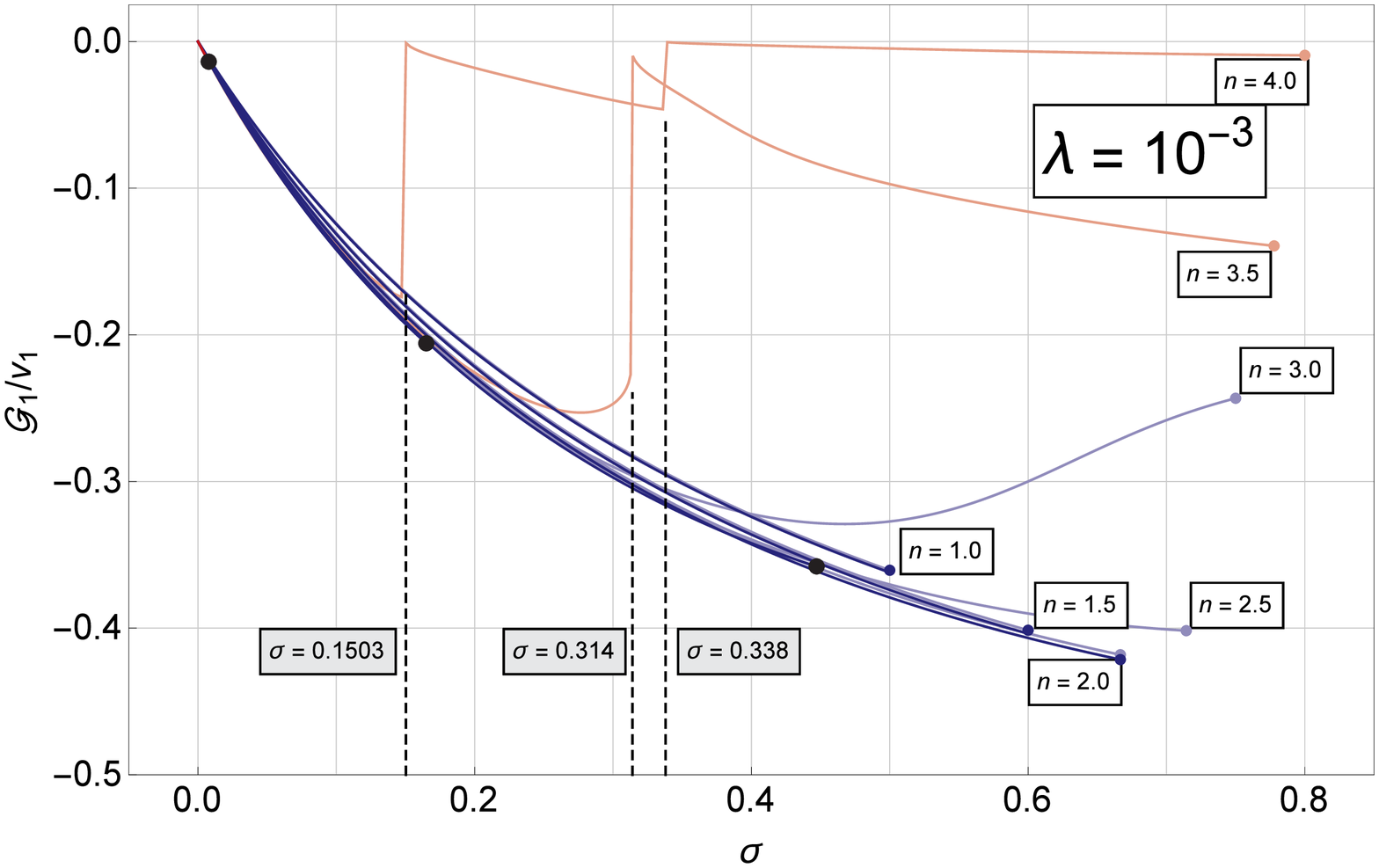}
\end{minipage}\hfill%
\begin{minipage}{0.48\linewidth}
\centering
\includegraphics[width=\linewidth]{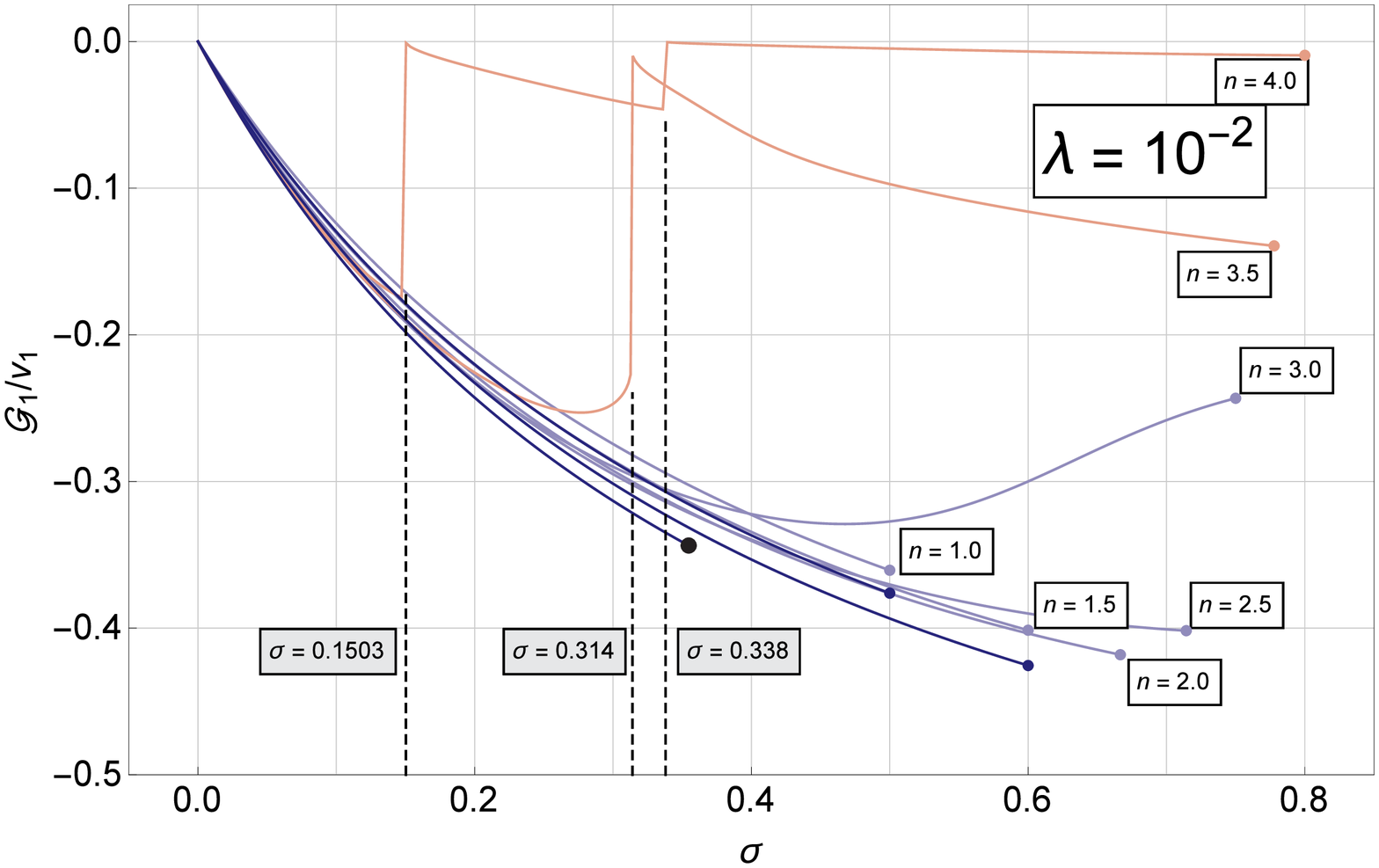}
\end{minipage}
\par\vspace{.8\baselineskip}\par
\caption{\label{SigGrPoEnRe}Dependences of the relative gravitational energy
  $\mathcal{G}_{1}/v_{1}$ for the characteristic values of the polytropic index
  $n\in\{1.0, 1.5, 2, 2.5, 3, 3.5, 4\}$ with $\sigma$ varying up to the causal
  limit for $\lambda = 10^{-12}$, $\lambda = 10^{-9}$ (top),
  $\lambda = 10^{-6}$, $\lambda = 10^{-4}$ (middle) and $\lambda = 10^{-3}$,
  $\lambda = 10^{-2}$ (bottom).}
\end{figure*}

The influence of the cosmological constant on the $\sigma$-profiles of the
relative gravitational energy $\mathcal{G}_{1}/v_{1}(n;\sigma,\lambda)$ is
illustrated in Fig.\,\ref{SigGrPoEnRe}. We can see that for polytropes with
fixed parameter $n$, the influence is represented mainly by the cut-off at the
allowed values of the parameter $\sigma$, while the modifications of the
profiles due to non-zero $\lambda$-term are very small and they decrease the
global parameter $\mathcal{G}_{1}/v_{1}$ of the GRPs.

\subsubsection{Binding energy}

The dimensionless binding energy $\mathcal{B}_{1}$ represents combination of the
binding effects of gravity and the internal energy of the polytrope. It can be
thus positive or negative, according to domination of negative gravitational
or positive internal energy.

\begin{figure*}[t]
\begin{minipage}{0.48\linewidth}
\centering
\includegraphics[width=\linewidth]{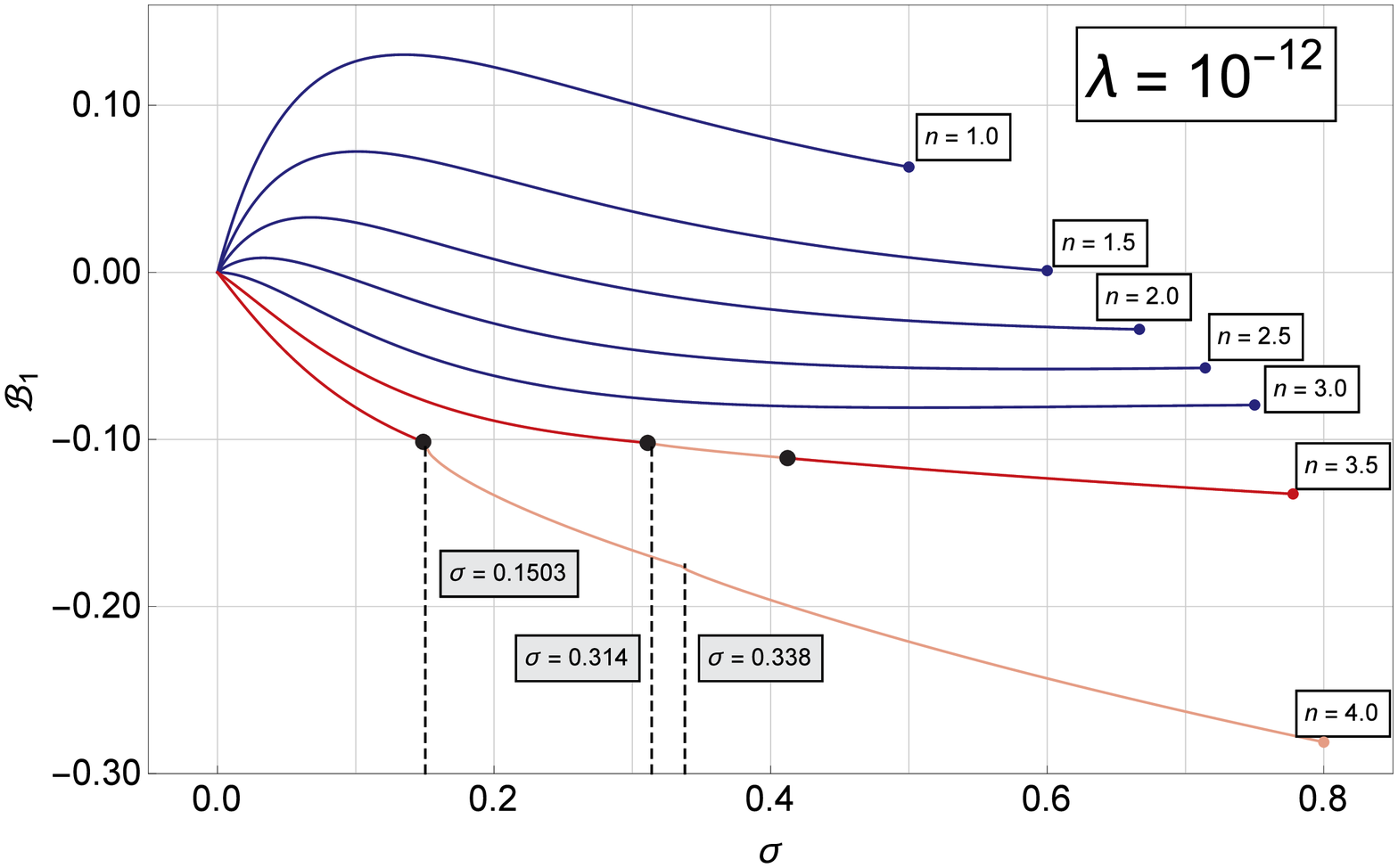}
\end{minipage}\hfill%
\begin{minipage}{0.48\linewidth}
\centering
\includegraphics[width=\linewidth]{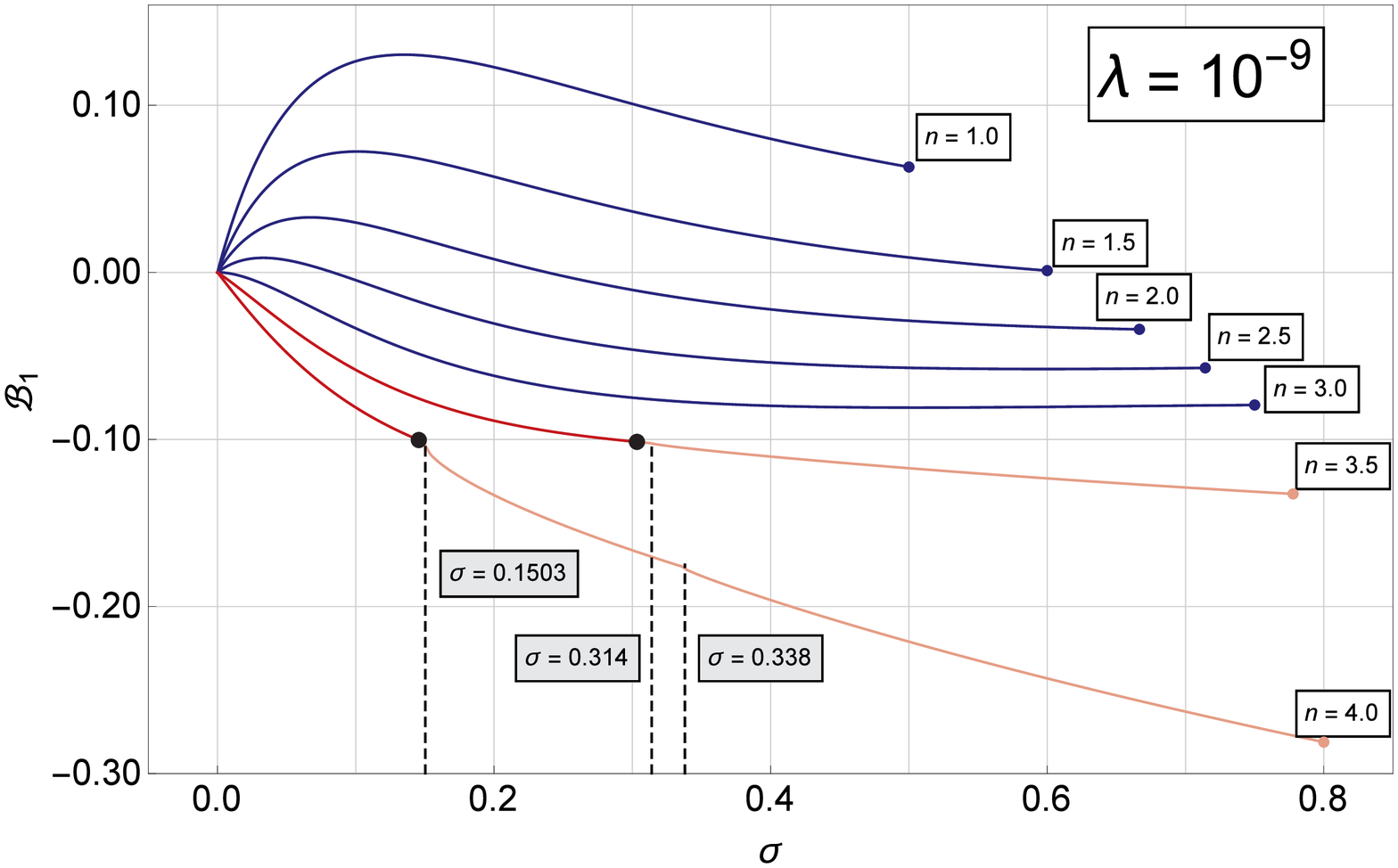}
\end{minipage}
\par\vspace{1.5\baselineskip}\par
\begin{minipage}{0.48\linewidth}
\centering
\includegraphics[width=\linewidth]{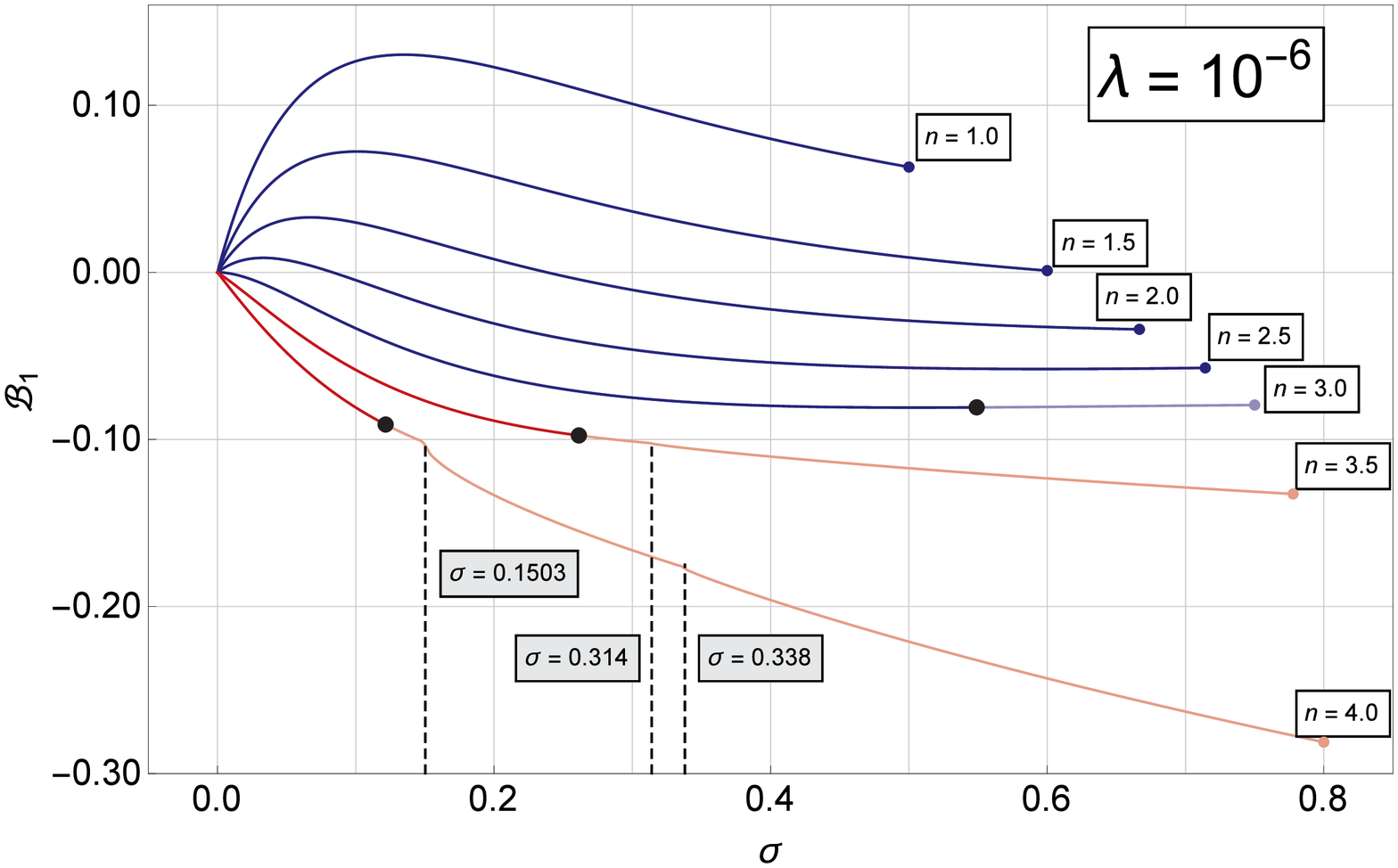}
\end{minipage}\hfill%
\begin{minipage}{0.48\linewidth}
\centering
\includegraphics[width=\linewidth]{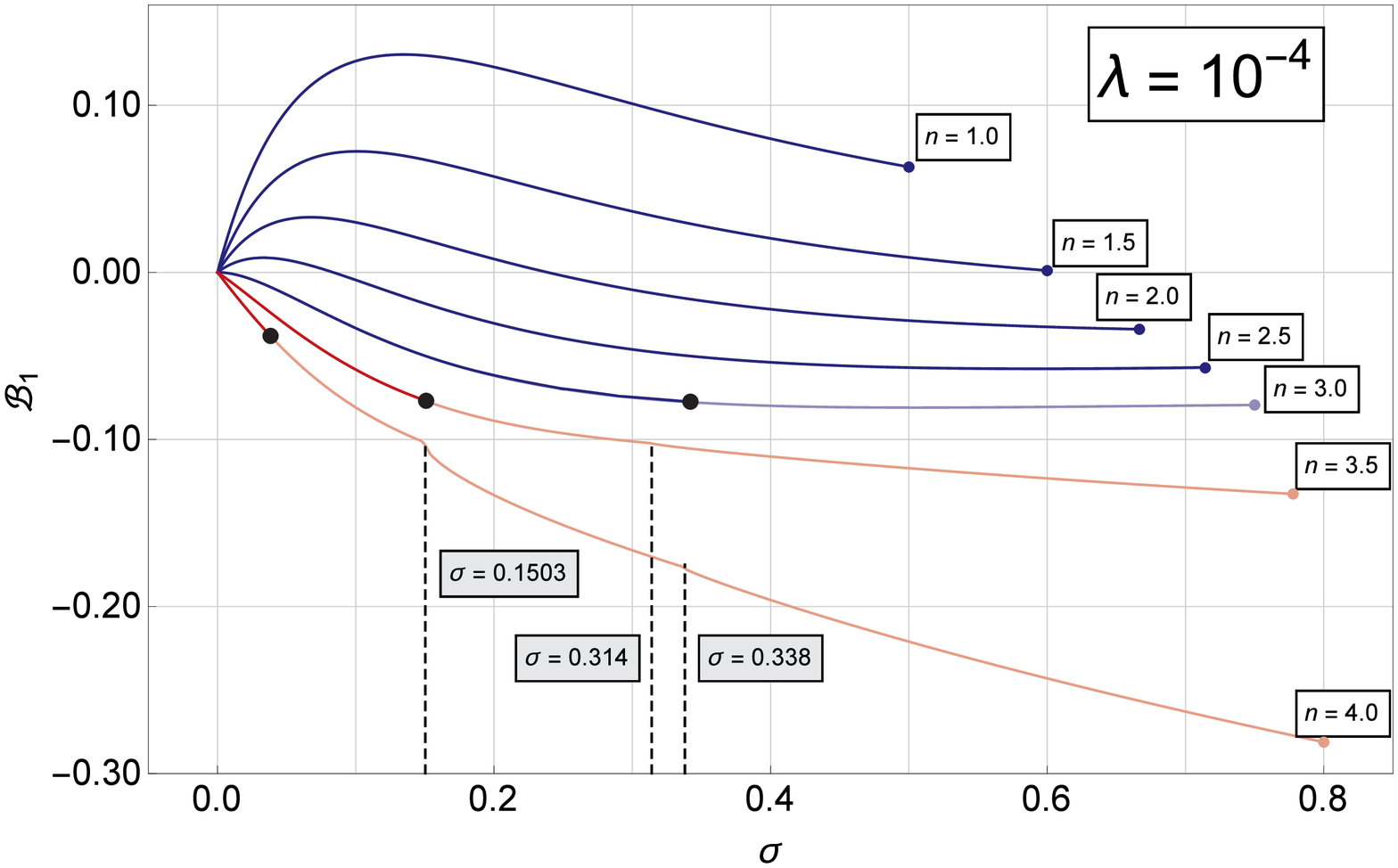}
\end{minipage}
\par\vspace{1.5\baselineskip}\par
\begin{minipage}{0.48\linewidth}
\centering
\includegraphics[width=\linewidth]{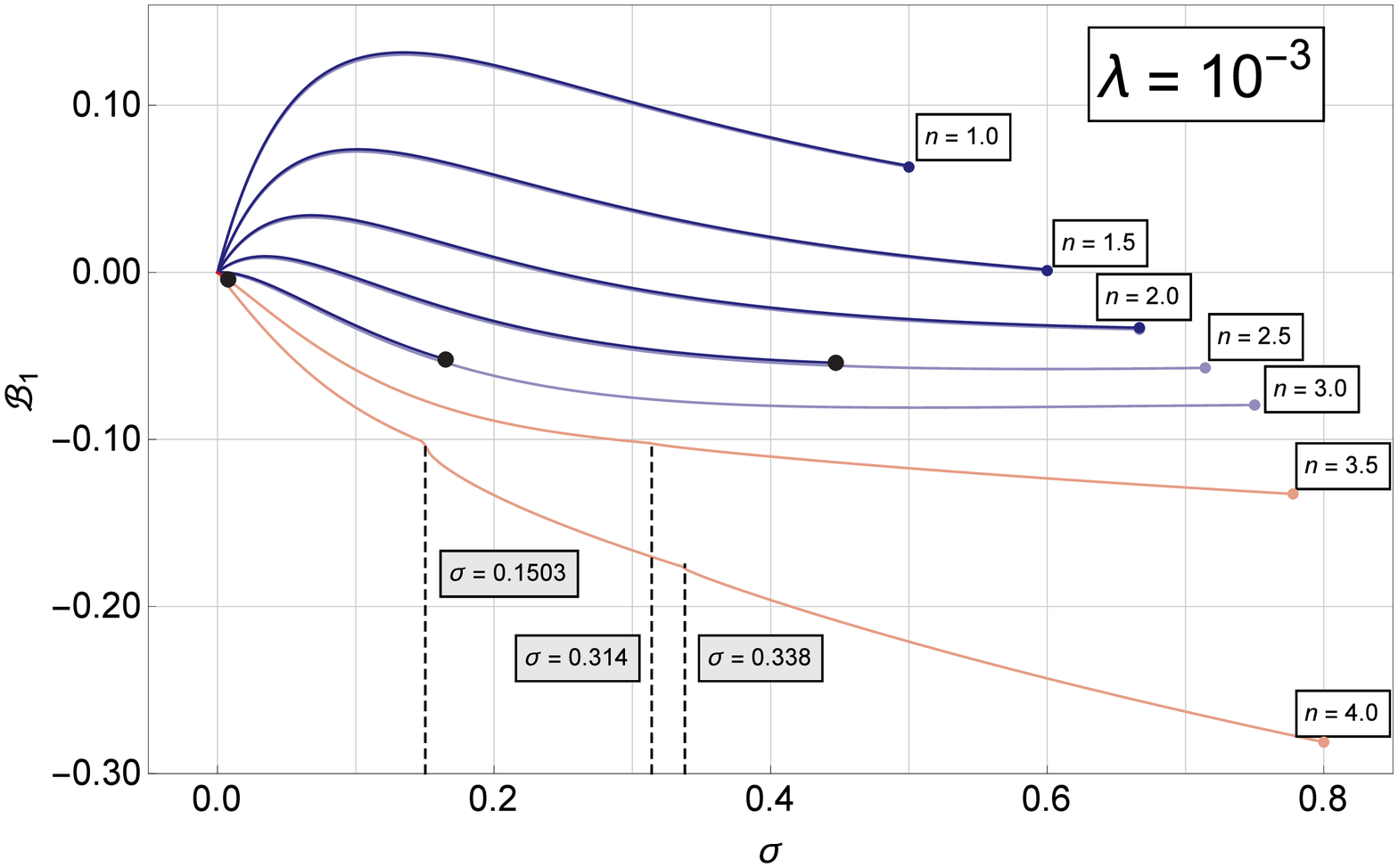}
\end{minipage}\hfill%
\begin{minipage}{0.48\linewidth}
\centering
\includegraphics[width=\linewidth]{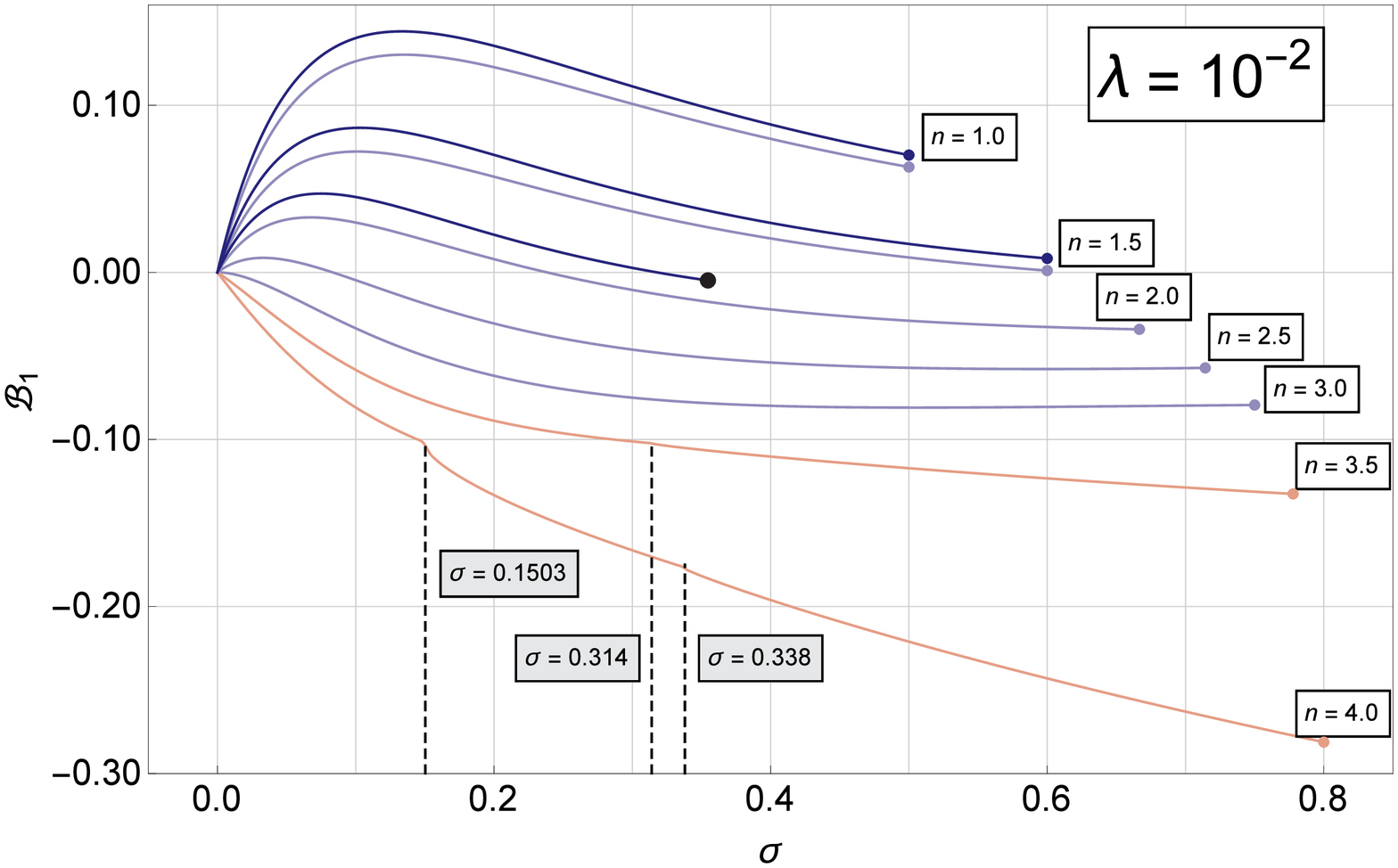}
\end{minipage}
\par\vspace{.8\baselineskip}\par
\caption{\label{SigBindEn}Dependences of the binding energy $\mathcal{B}_{1}$
  for the characteristic values of the polytropic index
  $n\in\{1.0, 1.5, 2, 2.5, 3, 3.5, 4\}$ with $\sigma$ varying up to the causal
  limit for $\lambda = 10^{-12}$, $\lambda = 10^{-9}$ (top),
  $\lambda = 10^{-6}$, $\lambda = 10^{-4}$ (middle) and $\lambda = 10^{-3}$,
  $\lambda = 10^{-2}$ (bottom).}
\end{figure*}

The role of the cosmological constant in the behavior of the binding energy of
the polytropes is represented in Fig.\,\ref{SigBindEn} for the same values of
the polytrope index $n$ and the dimensionless parameter $\lambda$ as in the
case of quantities $\xi_{1}$ and $v_{1}$. The binding energy function
$\mathcal{B}_{1}(n;\sigma,\lambda)$ is always compared to the function
$\mathcal{B}_{1}(n;\sigma,\lambda=0)$. In the case of $\lambda=0$, there is
$\mathcal{B}_{1}(n;\sigma=0)=0$. For each of the $n=1, 1.5, 2, 2.5$
polytropes, the binding energy $\sigma$-profile reaches a maximum at some
$\sigma < 0.1$, and then decreases to a minimum at the causality limit of the
relativistic parameter $\sigma$. In the case of $n=1, 1.5$ polytropes, the
$\sigma$-profile gives positive dimensionless binding energy at whole allowed
range of $\sigma$. For the $n=2, 2.5$ polytropes, there is a zero point of the
binding energy $\sigma$-profile, and behind this point the binding energy is
negative. On the other hand, for the polytropes with $n=3, 3.5, 4$, the
binding energy is negative for all the allowed values of $\sigma$, and the
gravitational energy prevails the internal energy of the configuration. For
the polytropes with $n=3.5, 4$, having divergence of $\xi_{1}$ at the critical
values of $\sigma_{\mathrm{f}}$, the $\sigma$-profiles of the binding energy
are continuous at the critical points, but their derivative has a jump
there. In the region of large values of $\sigma$, the binding energy decreases
in the case of the $n=4$ polytropes to the value $\mathcal{B}_{1} \sim -0.27$
at the causality limit.

Influence of the cosmological constant parameter $\lambda$ on the binding energy 
function $\mathcal{B}_{1}(n;\sigma,\lambda)$, demonstrated
in Fig.\,\ref{SigBindEn}, is concentrated to the cut-off in allowed values of
the parameter $\sigma$ for polytropes with fixed parameter $n$. We can also
observe slight modifications of the $\sigma$-profile of the binding energy
function $\mathcal{B}_{1}(n;\sigma,\lambda)$, but modifications of the
$\sigma$-profiles that are large enough to be recognizable occur only for
$\lambda \geq 10^{-3}$---increasing of the parameter $\lambda$ always
increases the binding energy of the polytrope, if its other parameters are
fixed.

\begin{figure*}[t]
\begin{minipage}{0.48\linewidth}
\centering
\includegraphics[width=\linewidth]{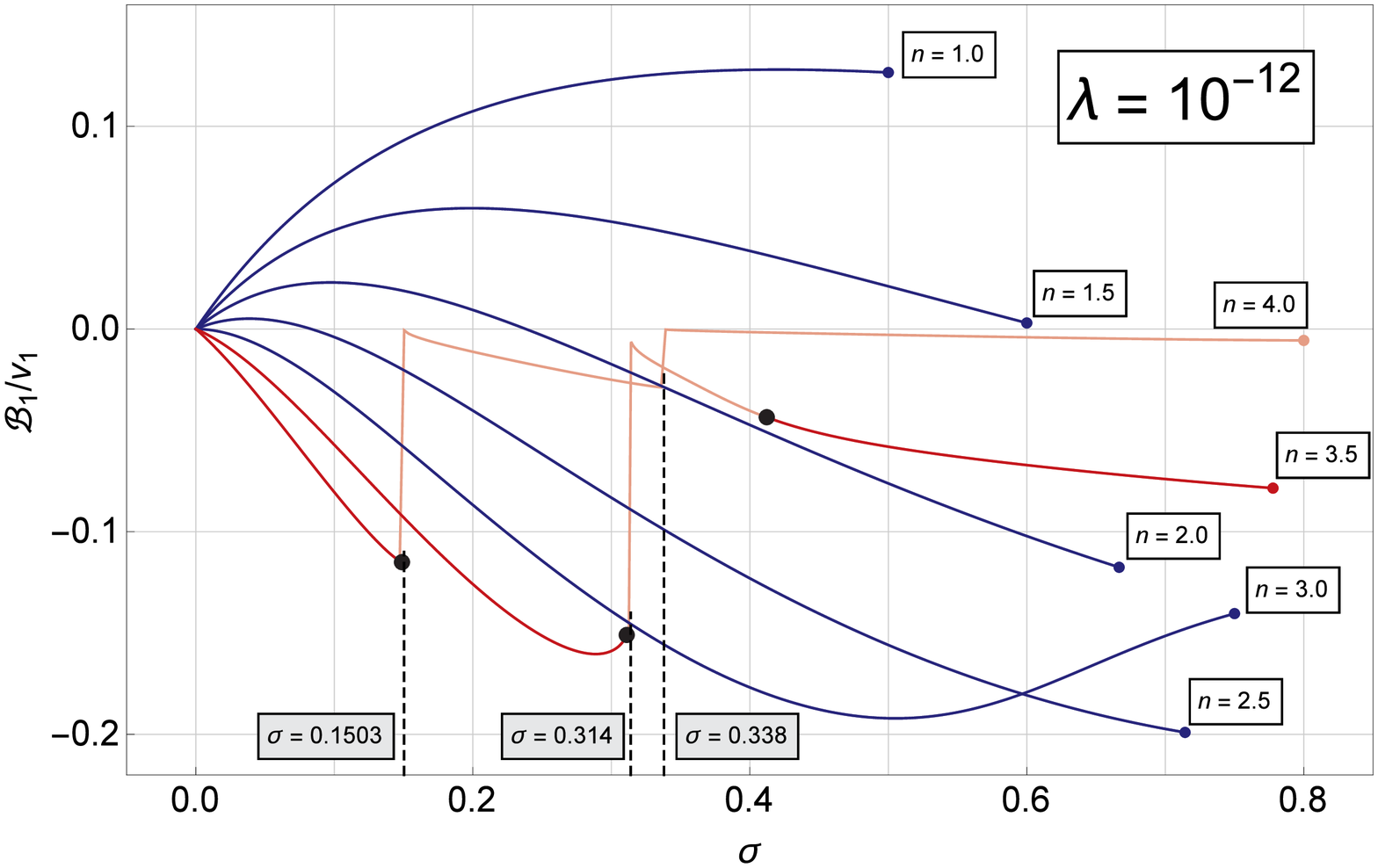}
\end{minipage}\hfill%
\begin{minipage}{0.48\linewidth}
\centering
\includegraphics[width=\linewidth]{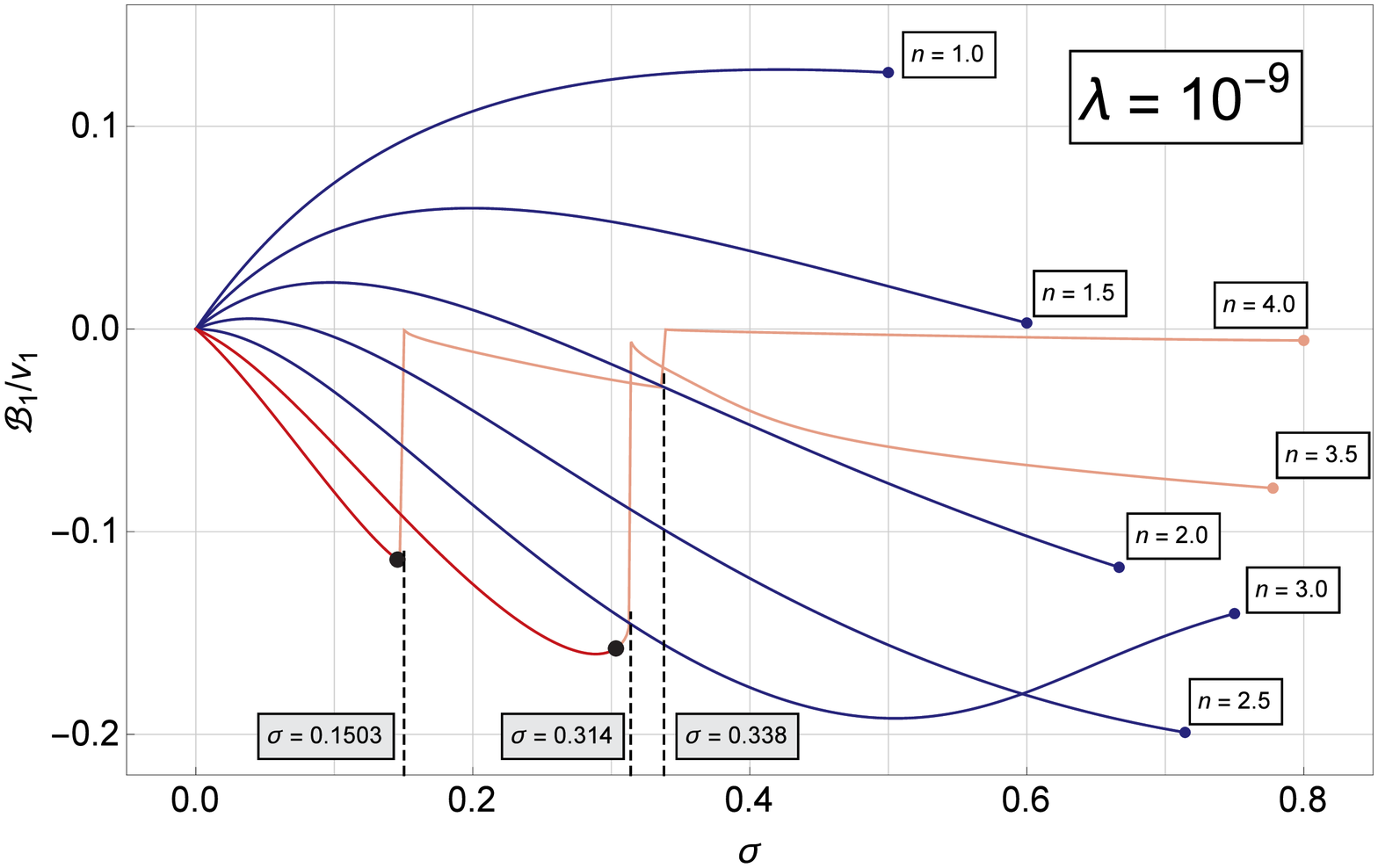}
\end{minipage}
\par\vspace{1.5\baselineskip}\par
\begin{minipage}{0.48\linewidth}
\centering
\includegraphics[width=\linewidth]{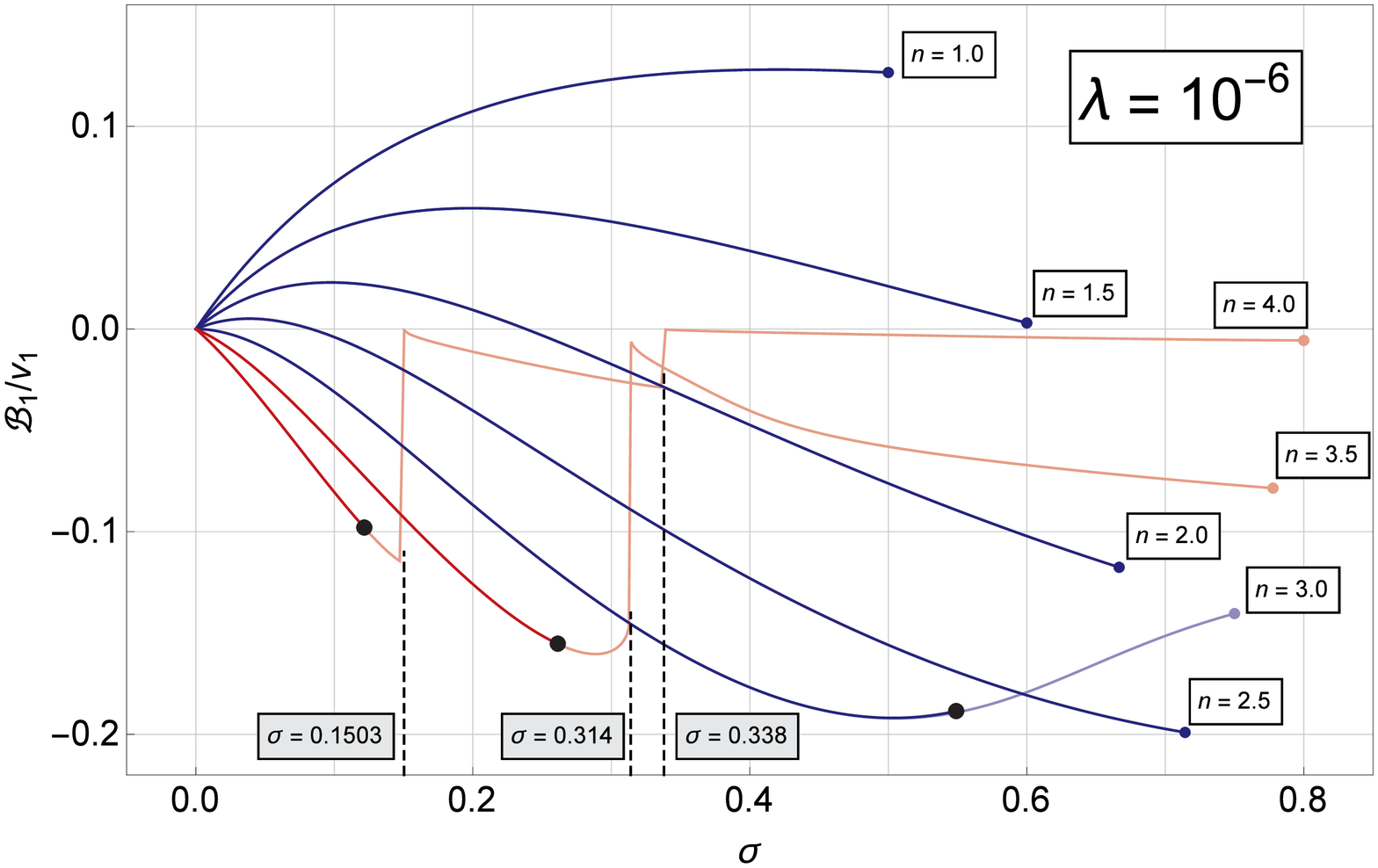}
\end{minipage}\hfill%
\begin{minipage}{0.48\linewidth}
\centering
\includegraphics[width=\linewidth]{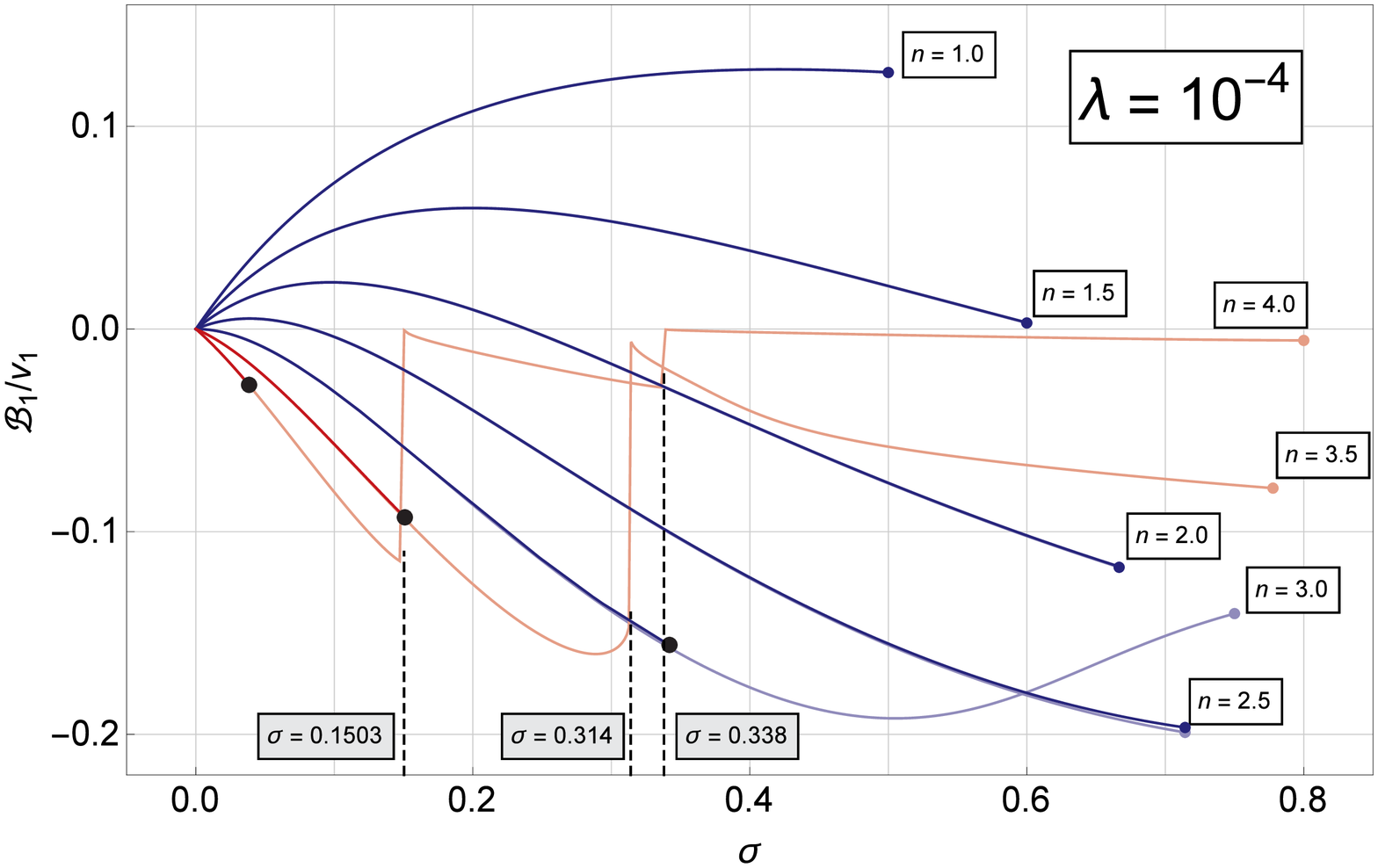}
\end{minipage}
\par\vspace{1.5\baselineskip}\par
\begin{minipage}{0.48\linewidth}
\centering
\includegraphics[width=\linewidth]{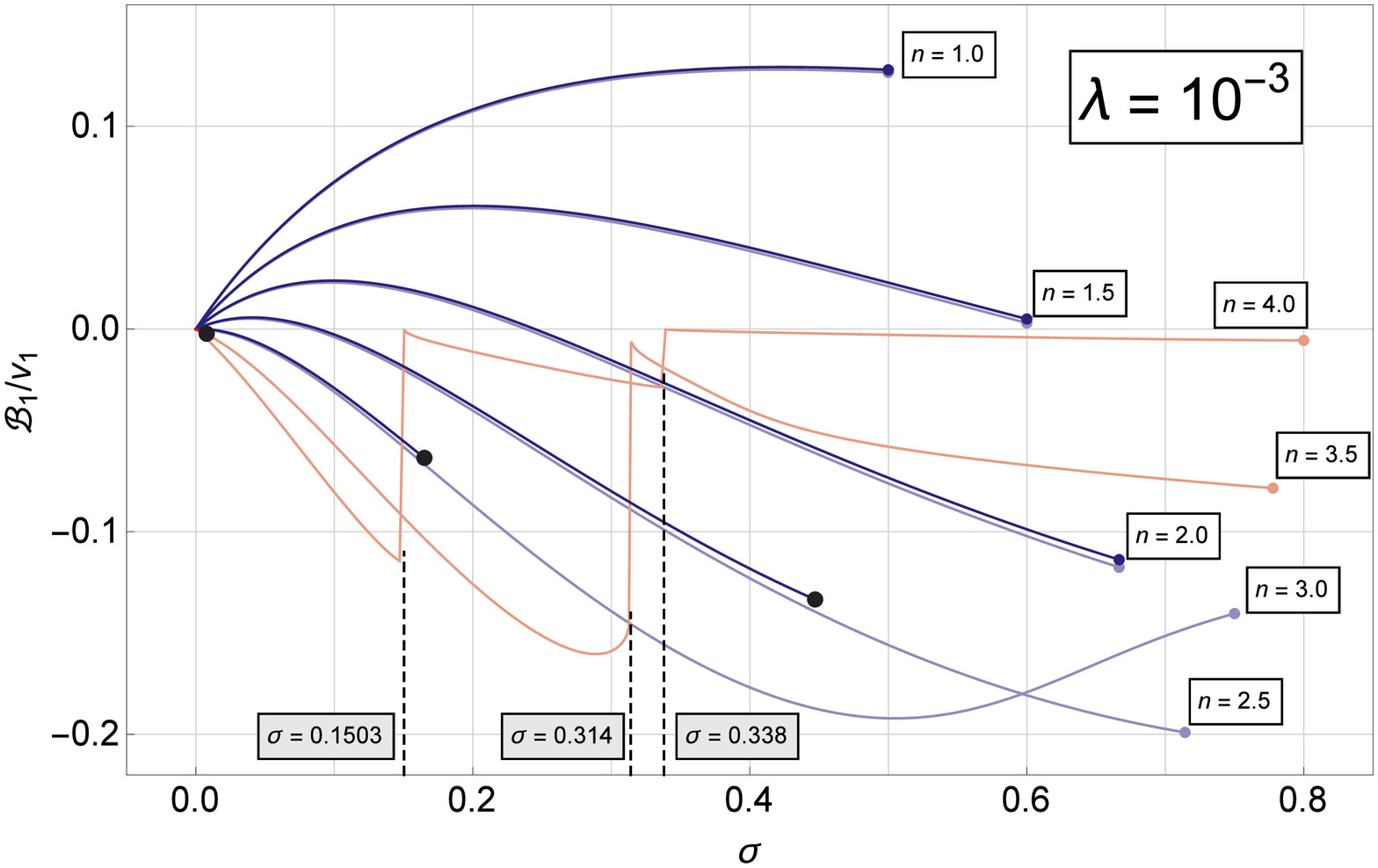}
\end{minipage}\hfill%
\begin{minipage}{0.48\linewidth}
\centering
\includegraphics[width=\linewidth]{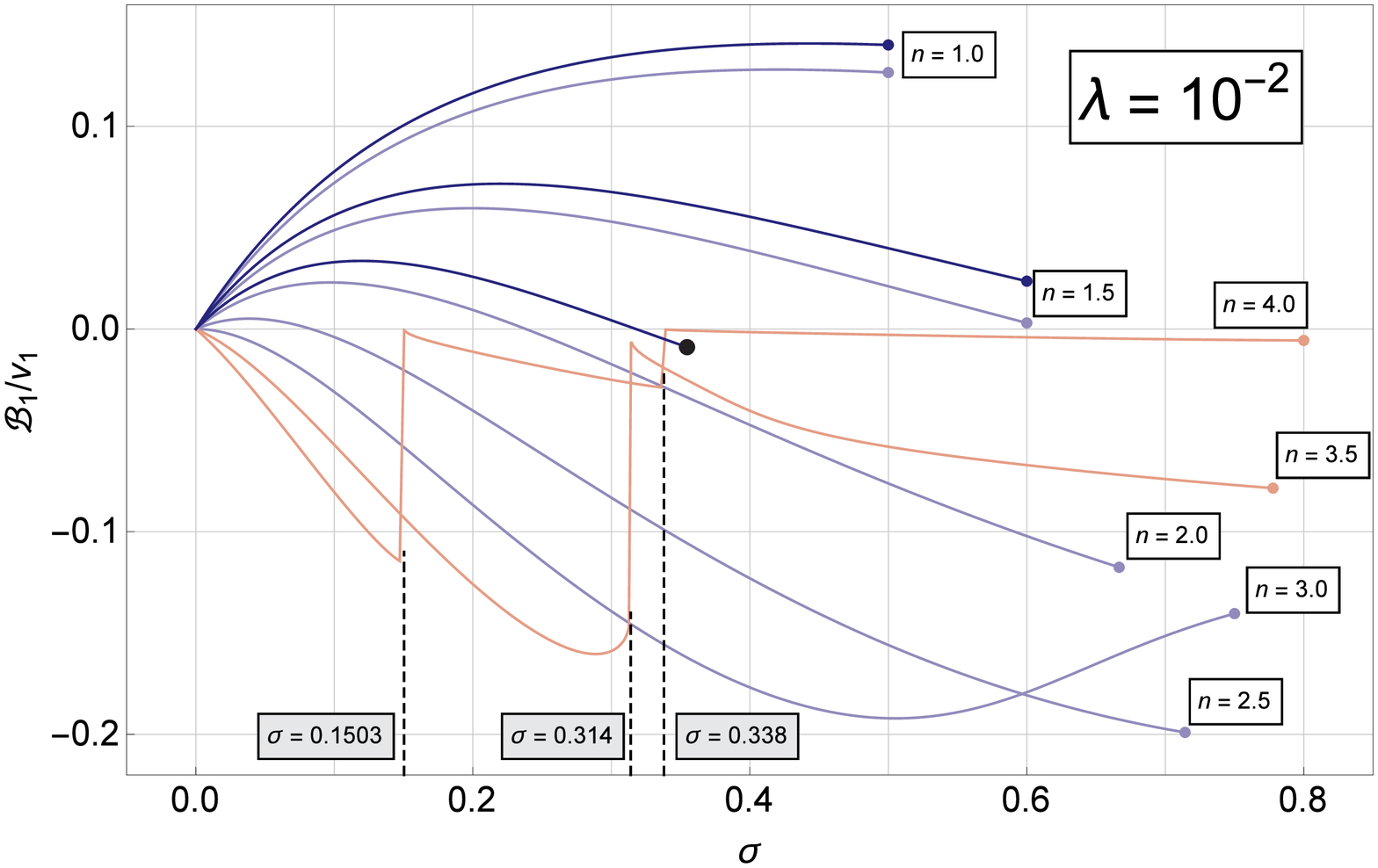}
\end{minipage}
\par\vspace{.8\baselineskip}\par
\caption{\label{SigBindEnRe}Dependences of the relative binding energy
  $\mathcal{B}_{1}/v_{1}$ for the characteristic values of the polytropic index
  $n\in\{1.0, 1.5, 2, 2.5, 3, 3.5, 4\}$ with $\sigma$ varying up to the causal
  limit for $\lambda = 10^{-12}$, $\lambda = 10^{-9}$ (top),
  $\lambda = 10^{-6}$, $\lambda = 10^{-4}$ (middle) and $\lambda = 10^{-3}$,
  $\lambda = 10^{-2}$ (bottom).}
\end{figure*}

For completeness, we present in Fig.\,\ref{SigBindEnRe} also the
$\sigma$-profiles of the relative binding energy related to the whole
polytrope that is defined by $\mathcal{B}_{1}/v_{1}(n;\sigma,\lambda)$; again,
the binding energy is related to the dimensionless gravitational mass of the
polytrope. For $\lambda = 0$, there is
$\mathcal{B}_{1}/v_{1}(n;\sigma=0)=0$. We can see that the character of the
$\sigma$-profiles of $\mathcal{B}_{1}/v_{1}$ for the polytropes with
$n=1, 1.5, 2, 2.5$ is the same as for the binding energy, but their magnitude
is larger since the gravitational mass parameter $v_{1} < 1$. For the $n=3$
polytropes, the $\sigma$-profiles of the relative binding energy function
$\mathcal{B}_{1}/v_{1}$ have a clear minimum, while for the $n=3.5,4$
polytropes they demonstrate jump to the zero point at the critical points of
$\sigma_{\mathrm{f}}$ due to the behavior of mass parameter $v_{1}$. For the
$n=4$ polytropes, the relative binding energy is extremely small for
$\sigma > \sigma_{\mathrm{f}2}$.

The influence of the cosmological constant parameter $\lambda$ on the
$\sigma$-profiles of the relative binding energy function
$\mathcal{B}_{1}/v_{1}(n;\sigma,\lambda)$ is illustrated in
Fig.\,\ref{SigBindEnRe}. Again, there is the cut-off in allowed values of the
parameter $\sigma$ for polytropes with fixed parameter $n$, and very small
modifications of the $\sigma$-profiles relative to those with $\lambda=0$
occur. They slightly grow with increasing $\lambda$, leading to small increase
of the relative binding energy $\mathcal{B}_{1}/v_{1}$.

\section{\label{rprof}Radial profiles of the GRPs}

Full understanding of the GRPs can be obtained by studying in detail the
character of their internal spacetime structure represented by the metric
coefficients, and the distribution of the physical quantities in their
interior, namely the energy density, pressure, gravitational mass, and
profiles of the compactness, gravitational, binding, and kinetic energy. The
gravitational phenomena are properly characterized also by the embedding
diagrams of the ordinary and the optical geometry of the central planes of the
polytropes. In the spherically symmetric spacetimes, we have thus to find the
radial profiles of the metric coefficients and the physical quantities
mentioned above.

\subsection{Construction of the profiles}

We illustrate the detailed behavior of the polytropic spheres in dependence on
the parameters $n$, $\sigma$ and $\lambda$, demonstrating in appropriately
selected cases the radial profiles of the energy density, pressure,
gravitational mass and metric coefficients. These are completed by the
embedding diagrams of the ordinary and optical geometry, and by the radial
profiles of the gravitational energy, binding energy and the kinetic energy,
and by the corresponding radial profiles of these energies related to the
dimensionless mass parameter $v_{1}$ of the polytropes, and the radial
profiles of the compactness parameter. The results of the numerical
calculations of the radial profiles are presented in series of figures. For
fixed values of the parameters $n$ and $\sigma$ the radial profiles are
constructed for $\lambda=0$ that are compared to radial profiles constructed
for appropriately chosen value of the cosmological parameter $\lambda$
enabling clear demonstration of the role of the cosmological
constant---naturally, for given values of the polytropic index $n$ and the
relativistic parameter $\sigma$, we chose the value of the cosmological
parameter $\lambda$ close to the critical cosmological parameter limiting the
polytropes, $\lambda_{\mathrm{crit}}(n,\sigma)$, guaranteeing clear
illustration of the influence of the cosmological constant.

We construct the radial profiles in the case of four characteristic values of
the polytropic index $n$, restricting thus the wide selection of the
polytropic indexes used in constructing the global characteristics of the
polytropes. We choose the most relevant polytropic indexes $n=1.5,3$ and the
indexes $n=0.5,3.5$, enabling to give a clear illustration of all the possible
cases of the behavior of the GRPs. Note that the case of the $n=1$ polytropes
is similar to $n=0.5$ polytropes, the polytropes $n=2,2.5$ represent
transition of the $n=1.5$ polytropes to the relativistic $n=3$ polytropes. The
$n=4$ polytropes are similar to the case of $n=3.5$ polytropes, but they are
more extreme in the vicinity of the critical points of the relativistic
parameter, $\sigma_{\mathrm{f}}$. The values of the parameter $\sigma$ are
selected from the whole allowed interval, up to the critical value determined
by the causality limit. For all of the considered polytropic indexes $n$, we
choose a very small relativistic parameter representing the non-relativistic
limit of the polytropes, $\sigma=10^{-3}$, and the largest one representing
the causal limit. We also select some intermediate value of $\sigma$ in order
to represent the characteristic intermediate polytropic configurations. We use
one such $\sigma$ for $n<3$, but more such intermediate values of $\sigma$ for
$n=3$ and $n=3.5$ polytropes. For each value of the polytropic index $n$, we
construct four sequences of radial profiles related to: (a)~the metric
coefficients $-g_{tt}$, $g_{rr}$, energy density $\rho/\rhocent$, pressure
$p/\pcent$, and mass parameter $v$, (b)~gravitational energy $e_{\mathrm{g}}$,
binding energy $e_{\mathrm{b}}$, kinetic energy $e_{\mathrm{k}}$, (c)~relative
gravitational energy $e_{\mathrm{g}}/v$, relative binding energy
$e_{\mathrm{b}}/v$, relative kinetic energy $e_{\mathrm{k}}/v$ and compactness
$\mathcal{C}$, (d)~embedding diagrams of the ordinary space $z_{\mathrm{ord}}$
and the optical space $z_{\mathrm{opt}}$.

\begin{figure*}[t]
\begin{minipage}{0.32\linewidth}
\centering
\includegraphics[width=\linewidth]{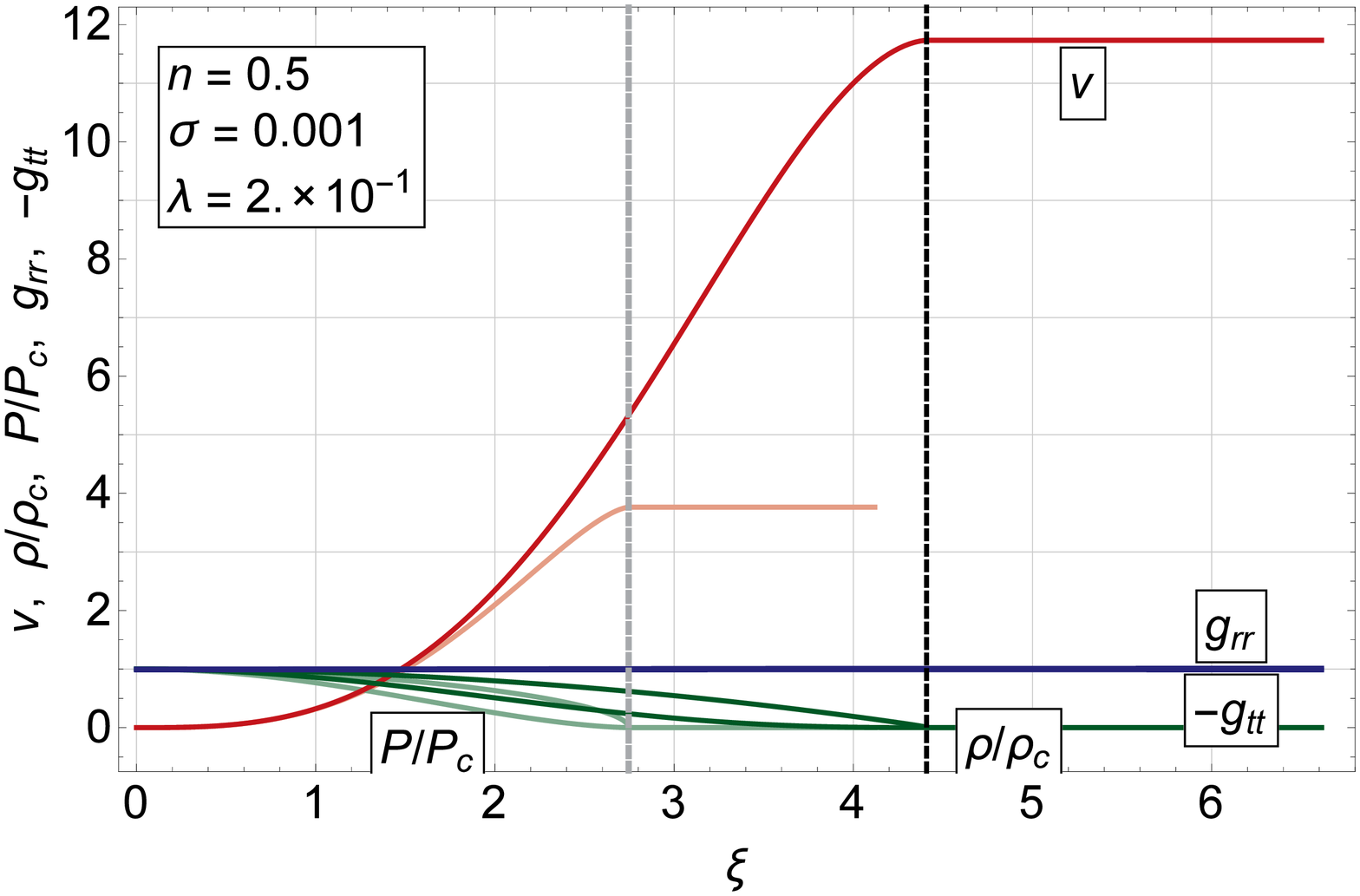}
\end{minipage}\hfill%
\begin{minipage}{0.32\linewidth}
\centering
\includegraphics[width=\linewidth]{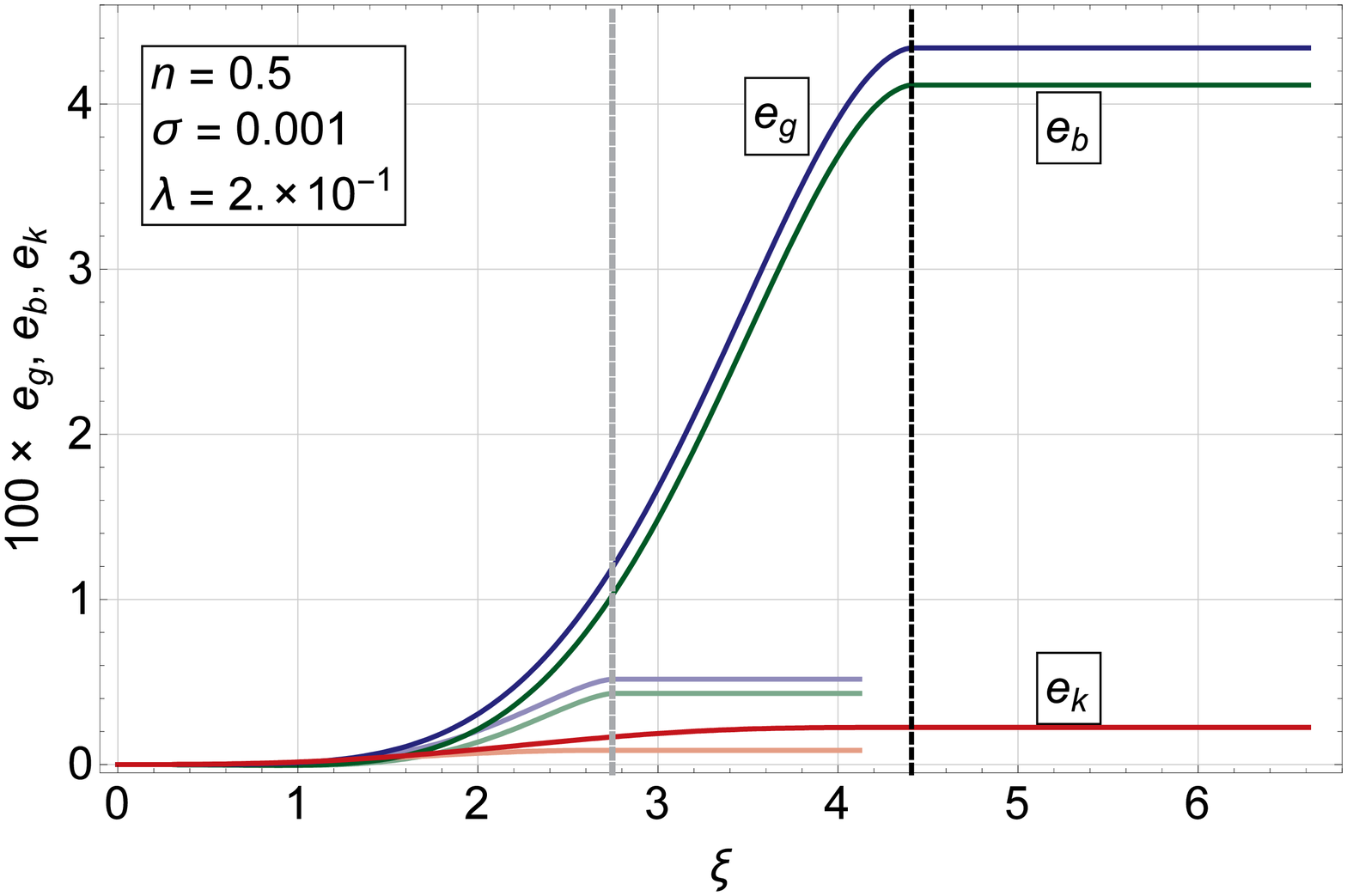}
\end{minipage}\hfill%
\begin{minipage}{0.32\linewidth}
\centering
\includegraphics[width=\linewidth]{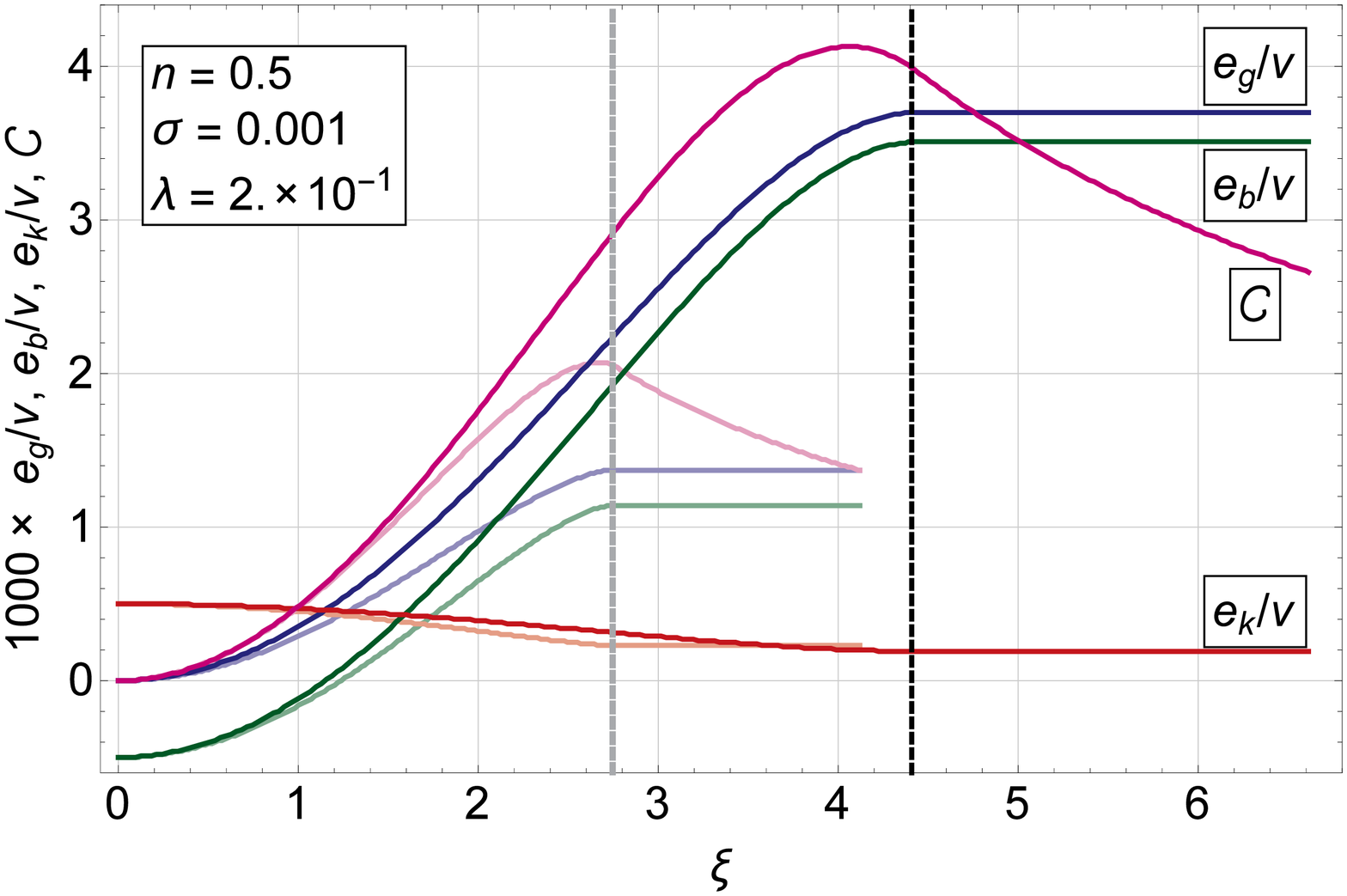}
\end{minipage}
\par\vspace{1.5\baselineskip}\par
\begin{minipage}{0.32\linewidth}
\centering
\includegraphics[width=\linewidth]{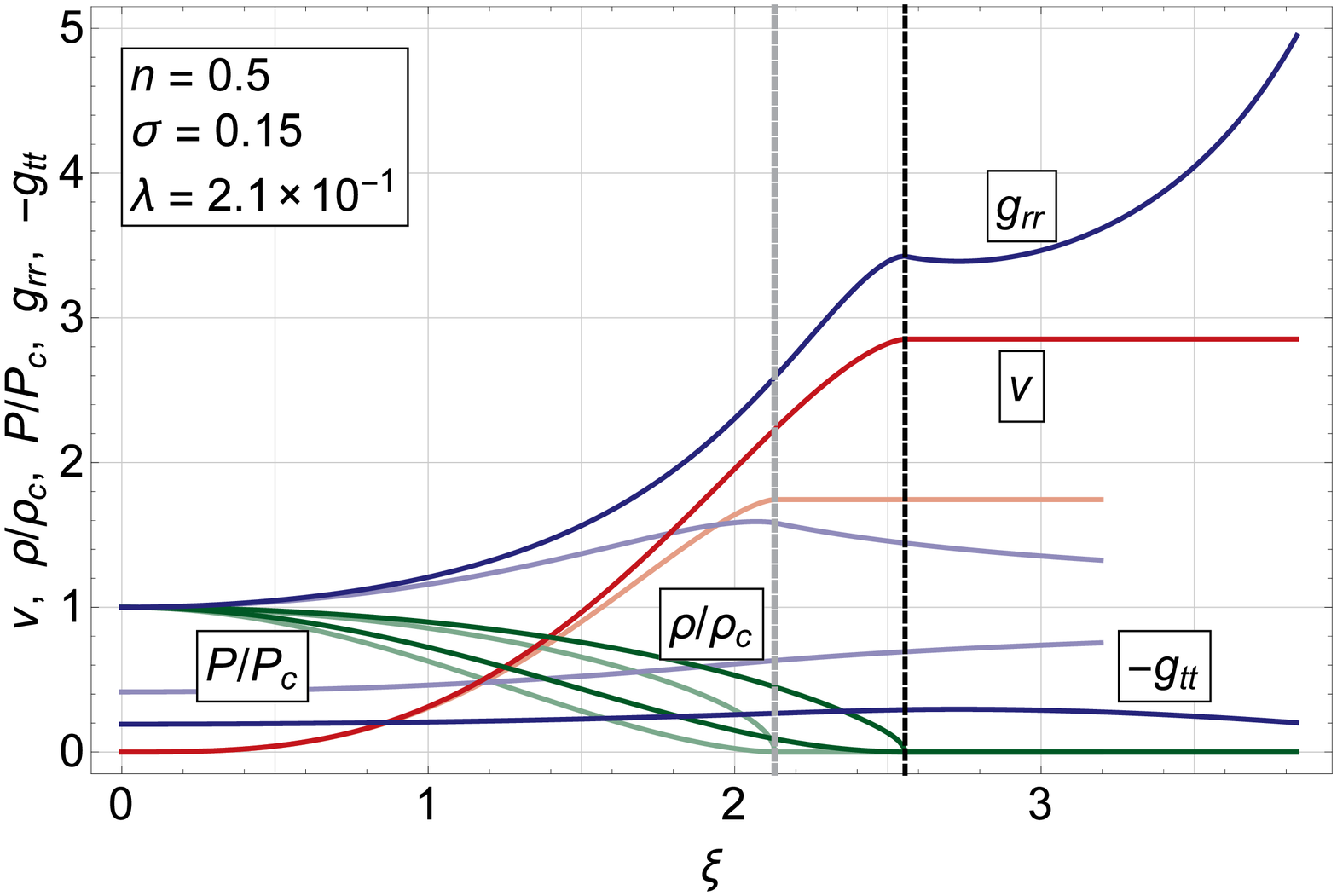}
\end{minipage}\hfill%
\begin{minipage}{0.32\linewidth}
\centering
\includegraphics[width=\linewidth]{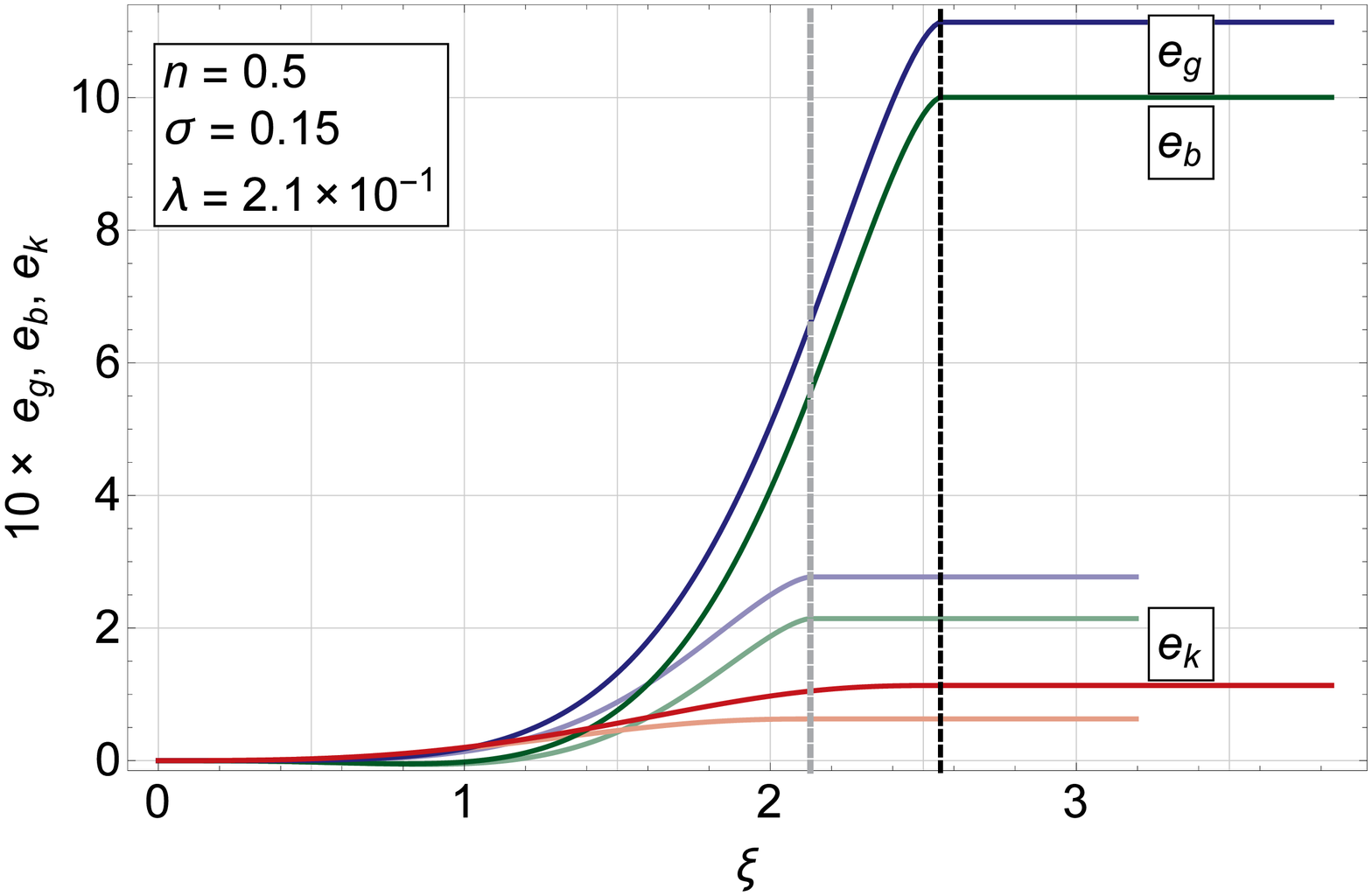}
\end{minipage}\hfill%
\begin{minipage}{0.32\linewidth}
\centering
\includegraphics[width=\linewidth]{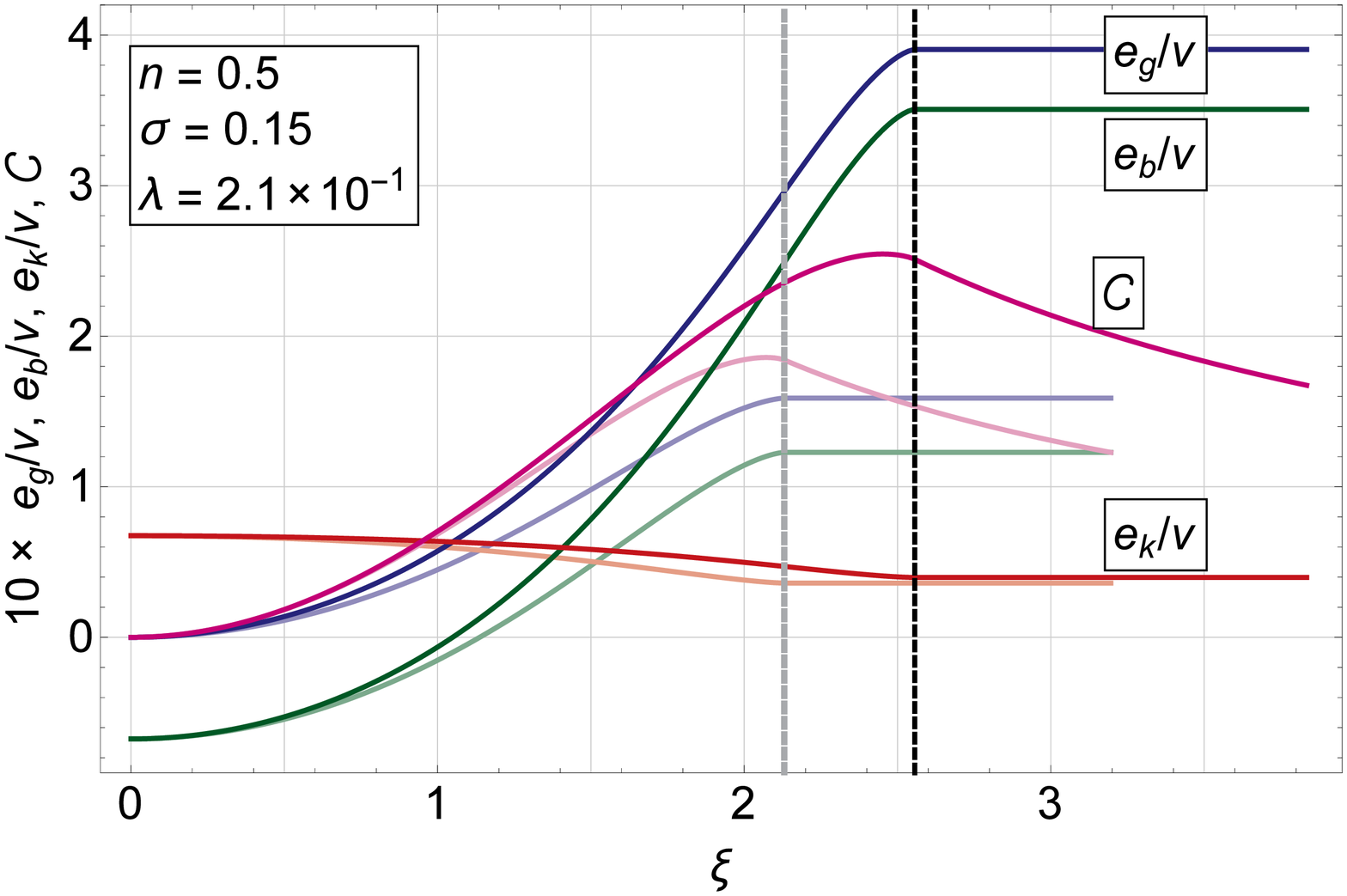}
\end{minipage}
\par\vspace{1.5\baselineskip}\par
\begin{minipage}{0.32\linewidth}
\centering
\includegraphics[width=\linewidth]{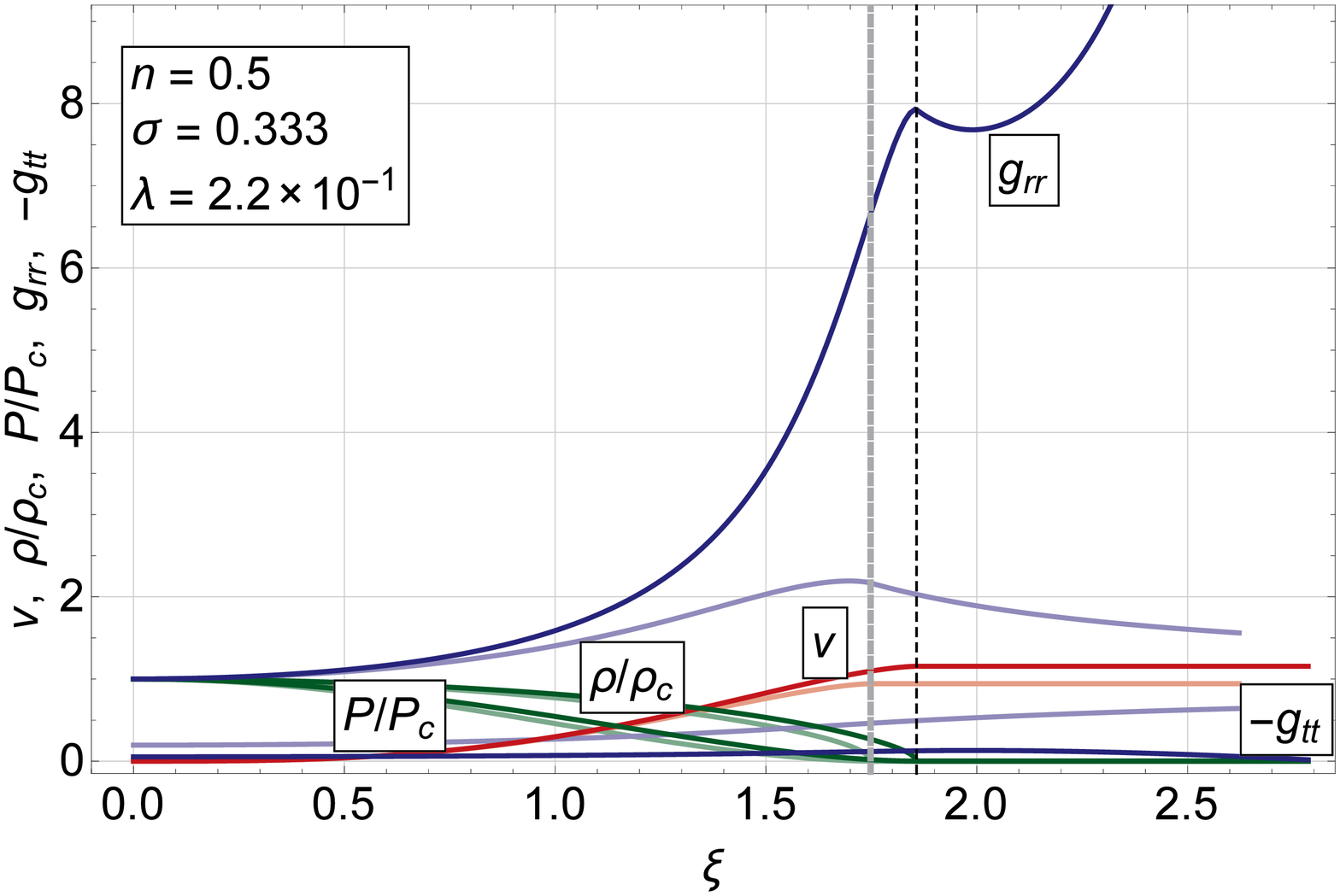}
\end{minipage}\hfill%
\begin{minipage}{0.32\linewidth}
\centering
\includegraphics[width=\linewidth]{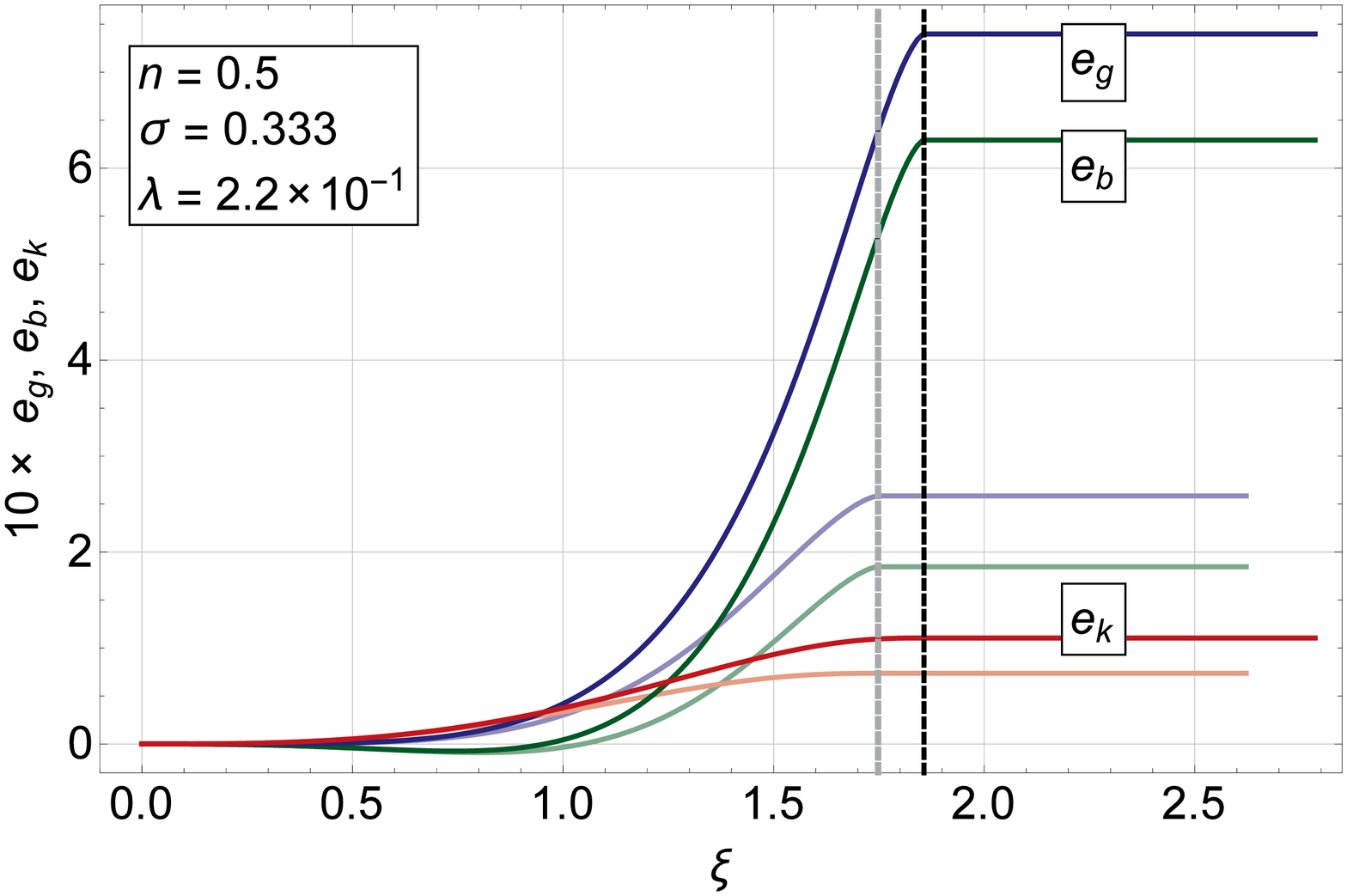}
\end{minipage}\hfill%
\begin{minipage}{0.32\linewidth}
\centering
\includegraphics[width=\linewidth]{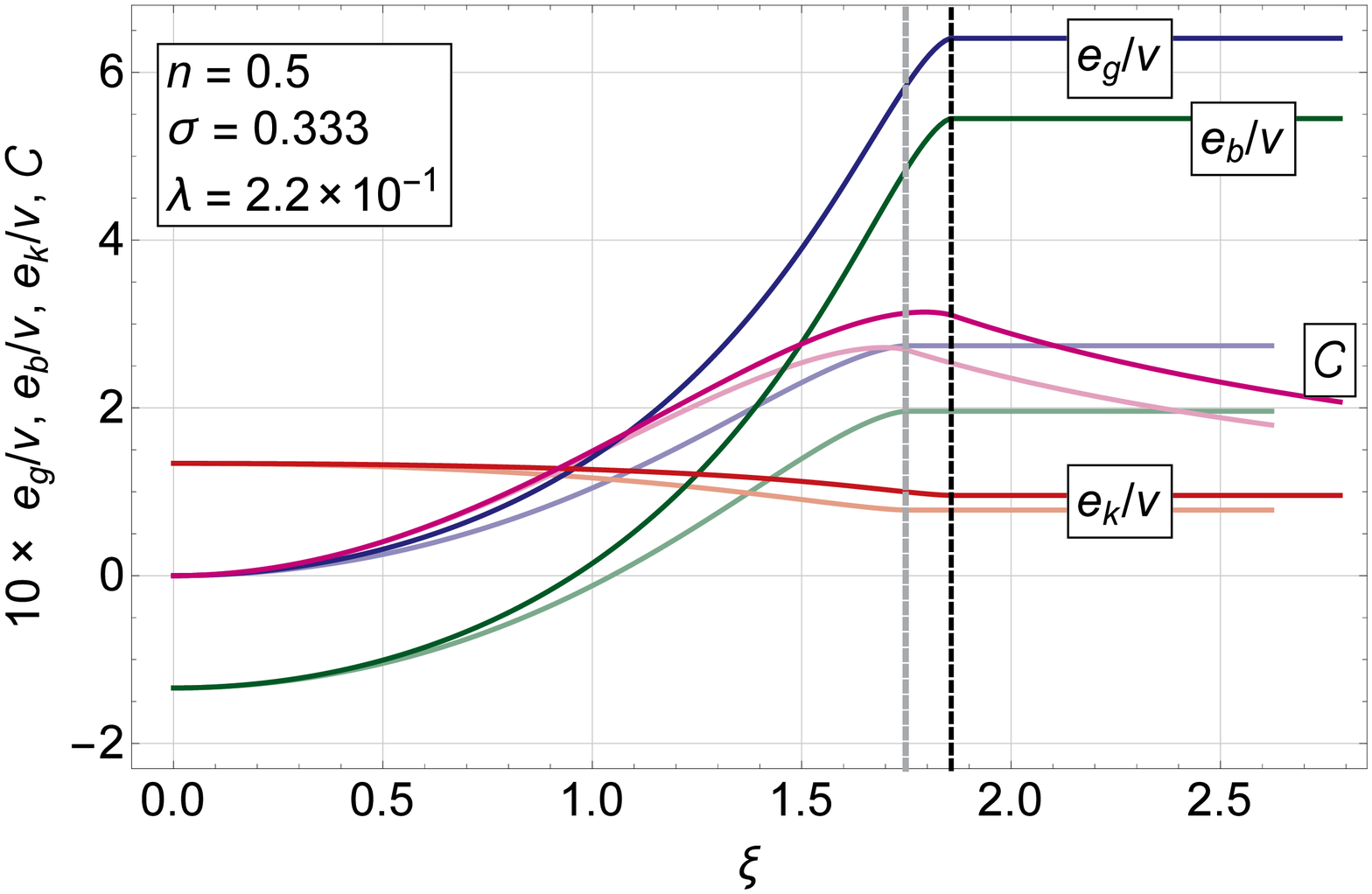}
\end{minipage}
\par\vspace{.8\baselineskip}\par
\caption{\label{ProN05}Profile plots for polytropic index
  $n=0.5$. \textit{Left column:} Mass, density, pressure, and metric
  coefficients. \textit{Middle column:} Gravitational, binding and kinetic
  energy. \textit{Right column:} Relative gravitational, binding, kinetic
  energy and compactness.}
\end{figure*}

\begin{figure}[t]
\includegraphics[width=\linewidth]{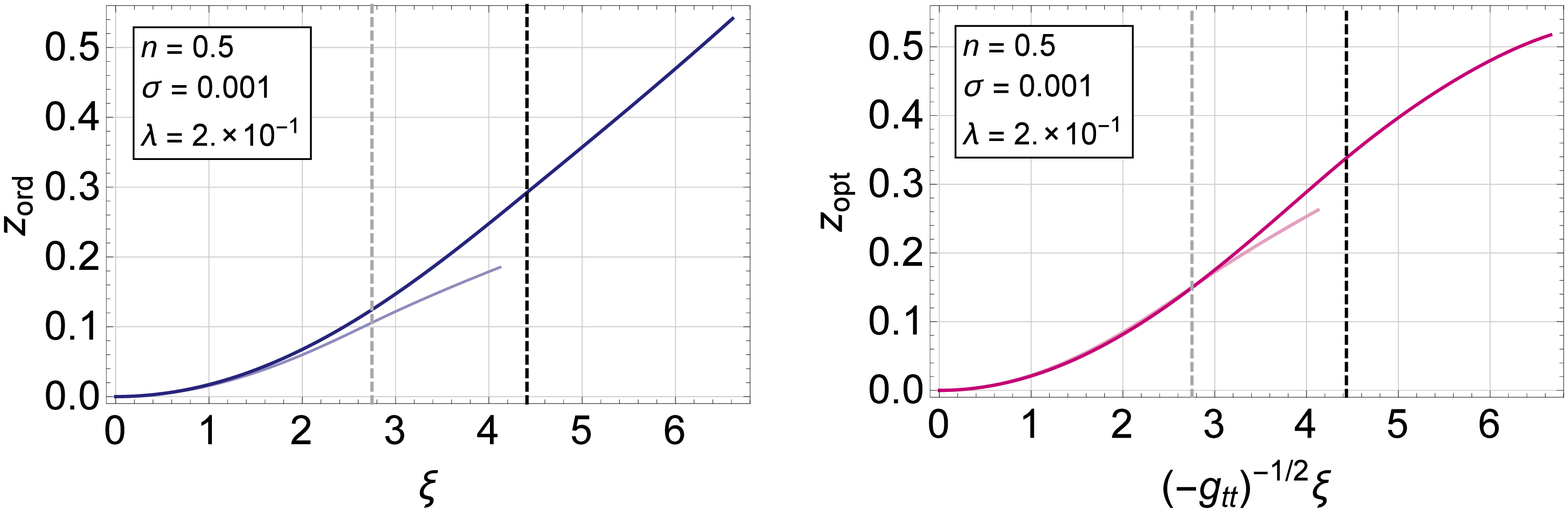}
\par\vspace{0.8\baselineskip}\par
\includegraphics[width=\linewidth]{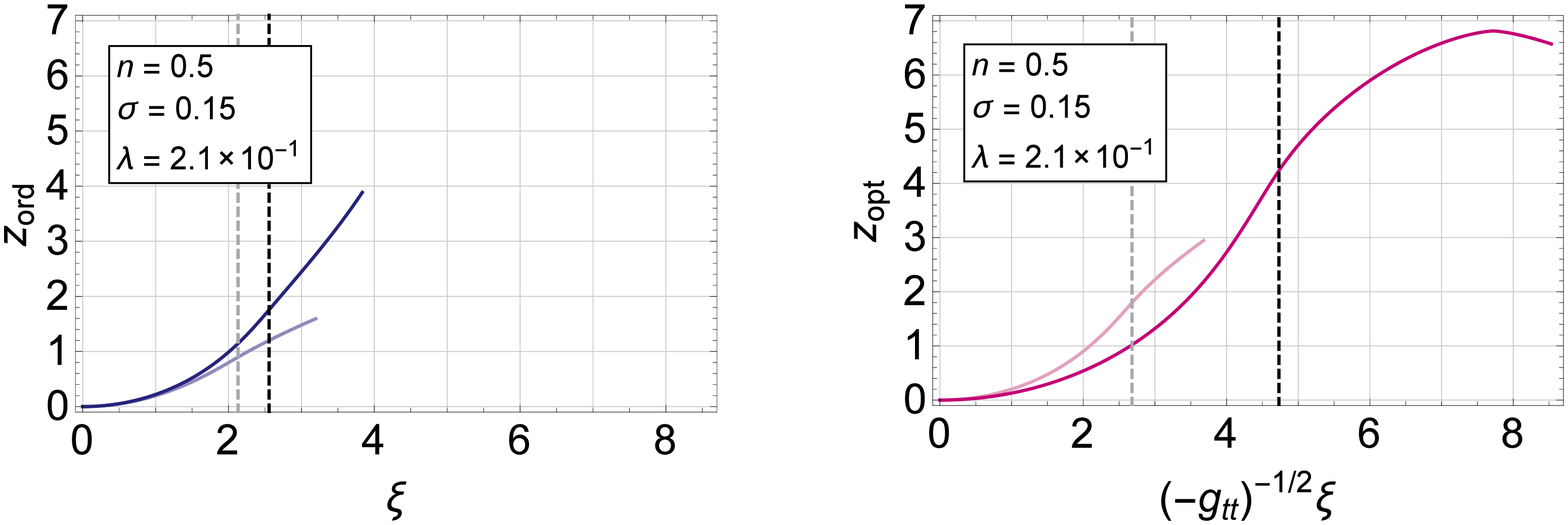}
\par\vspace{0.8\baselineskip}\par
\includegraphics[width=\linewidth]{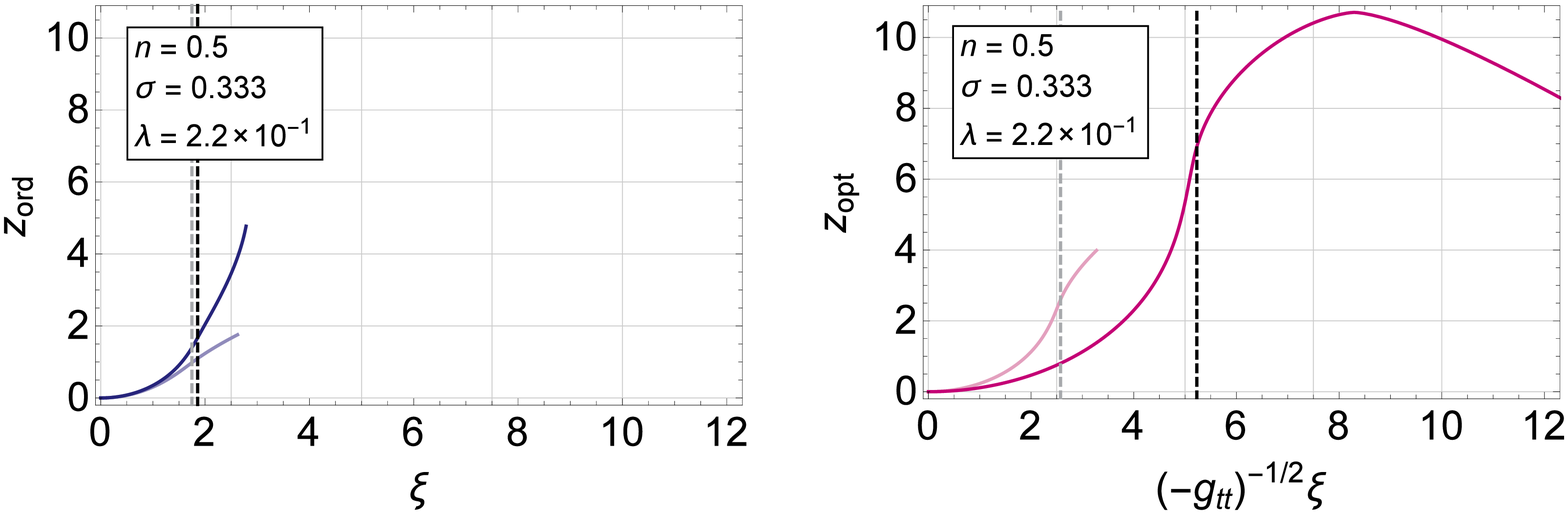}
\caption{\label{EmbN05}Embedding diagrams for polytropic index
  $n=0.5$.\textit{Left column:} Ordinary geometry. \textit{Right column:}
  Optical geometry.}
\end{figure}

\begin{figure}[t]
\includegraphics[width=\linewidth]{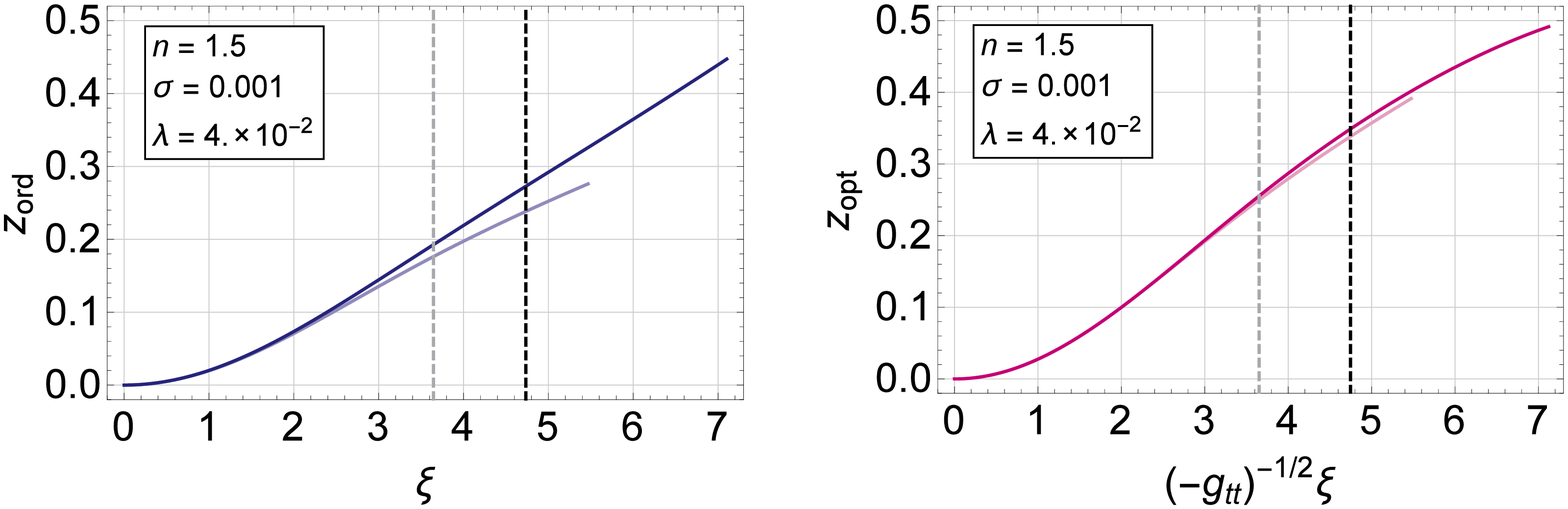}
\par\vspace{0.8\baselineskip}\par
\includegraphics[width=\linewidth]{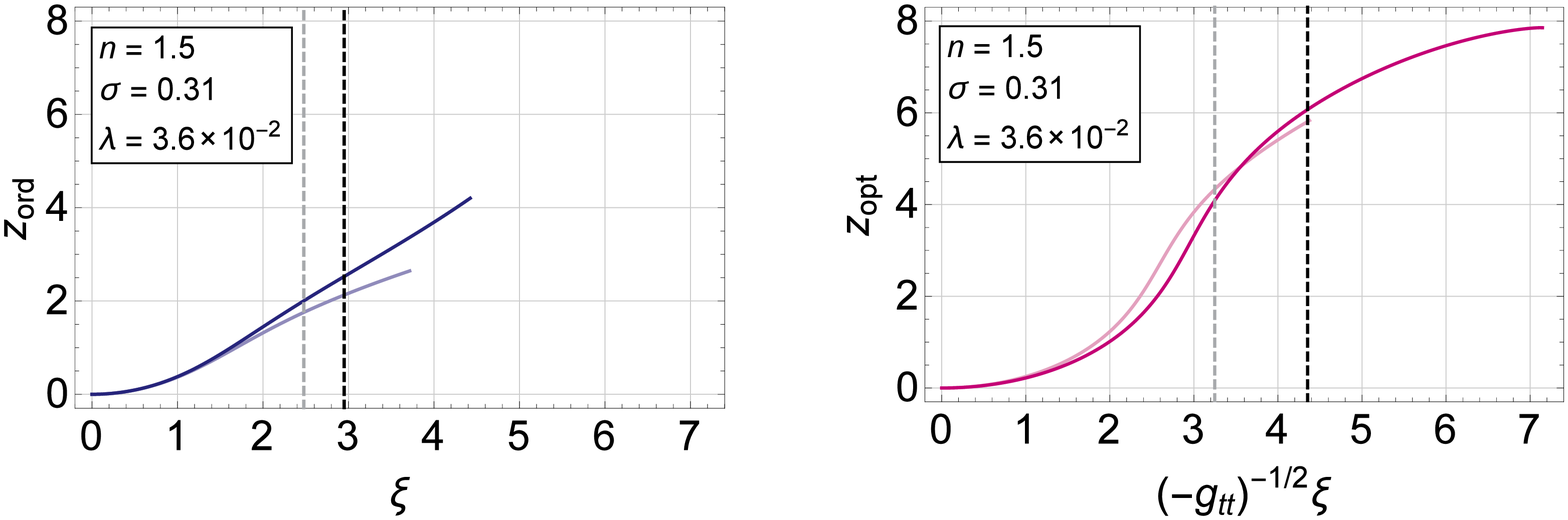}
\par\vspace{0.8\baselineskip}\par
\includegraphics[width=\linewidth]{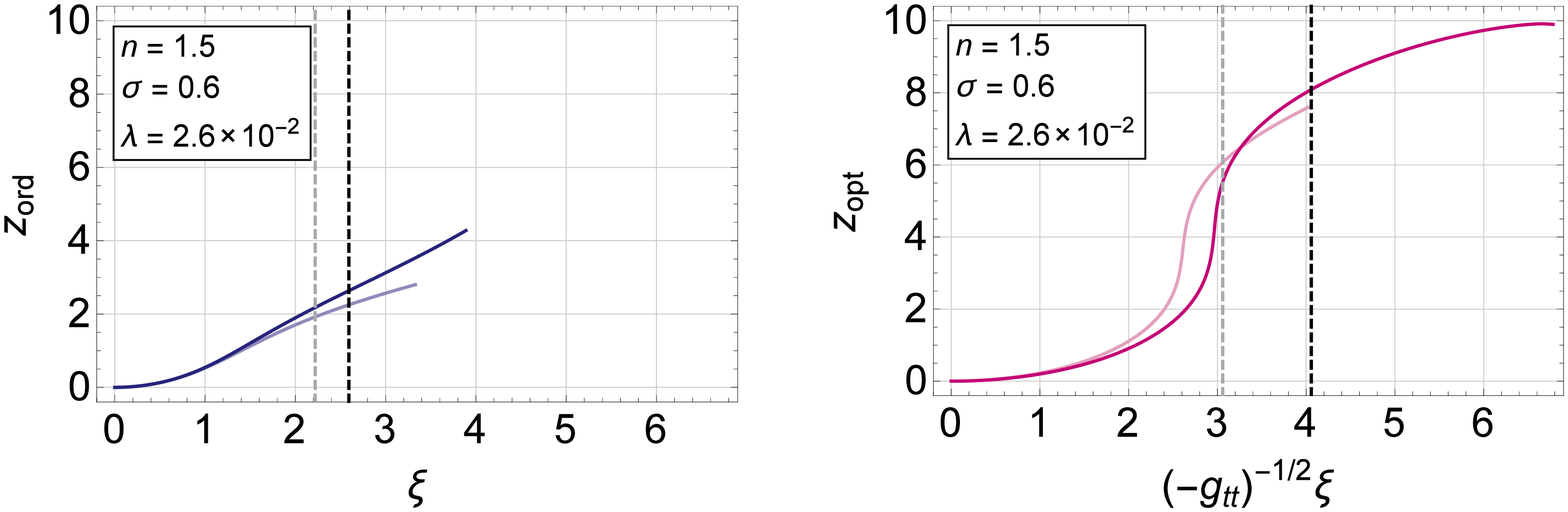}
\caption{\label{EmbN15}Embedding diagrams for polytropic index
  $n=1.5$.\textit{Left column:} Ordinary geometry. \textit{Right column:}
  Optical geometry.}
\end{figure}

\begin{figure*}[t]
\begin{minipage}{0.32\linewidth}
\centering
\includegraphics[width=\linewidth]{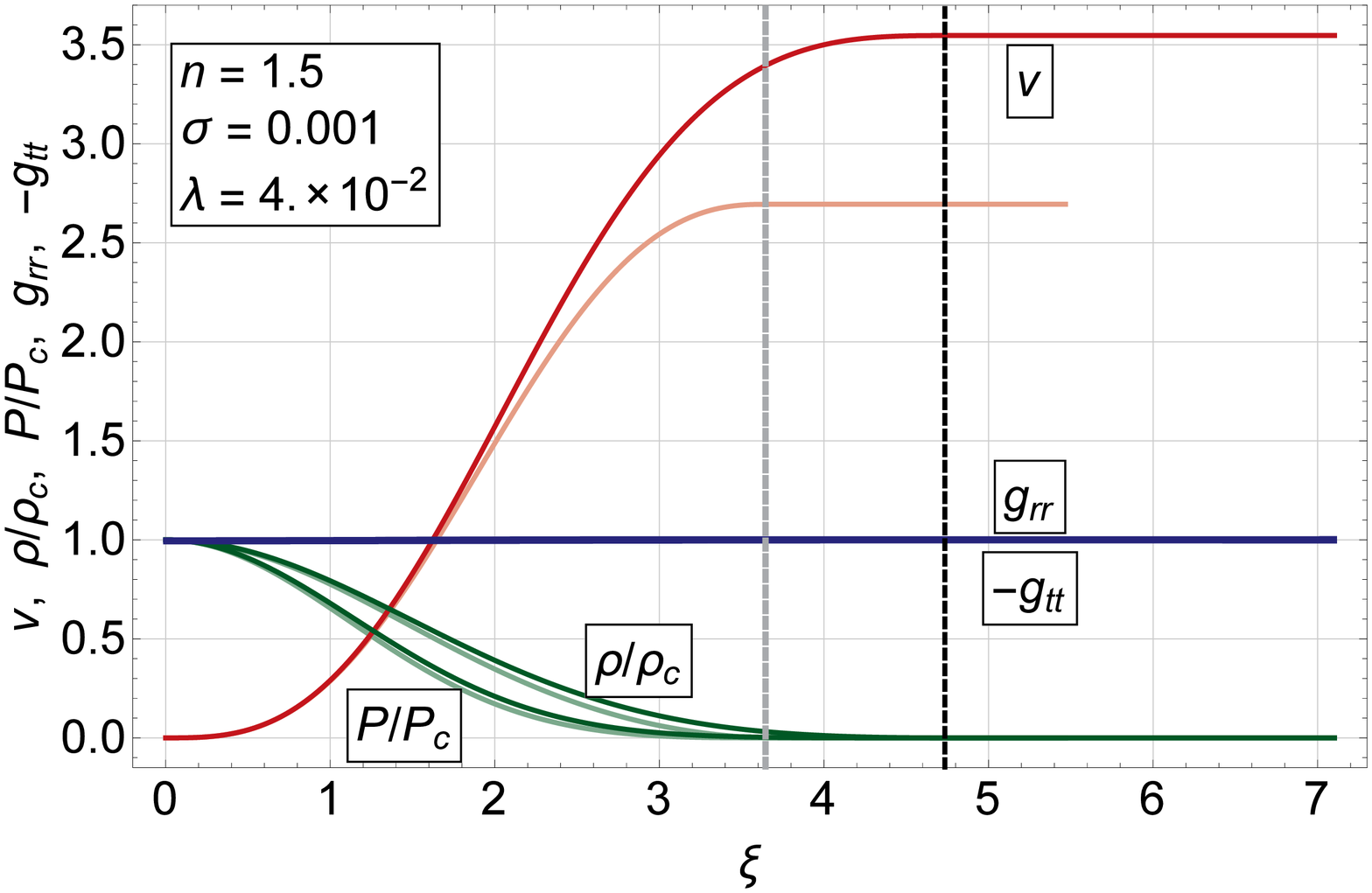}
\end{minipage}\hfill%
\begin{minipage}{0.32\linewidth}
\centering
\includegraphics[width=\linewidth]{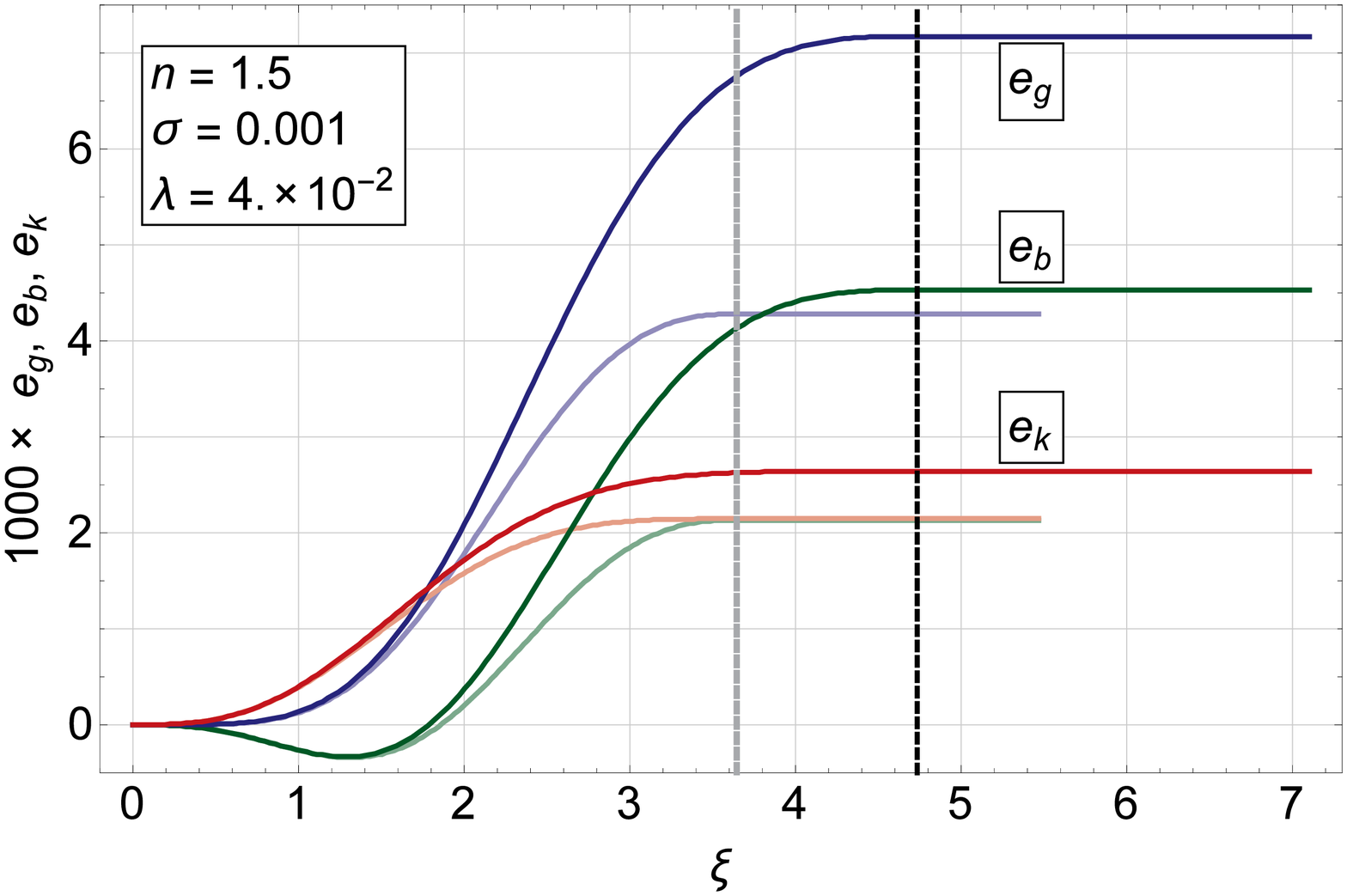}
\end{minipage}\hfill%
\begin{minipage}{0.32\linewidth}
\centering
\includegraphics[width=\linewidth]{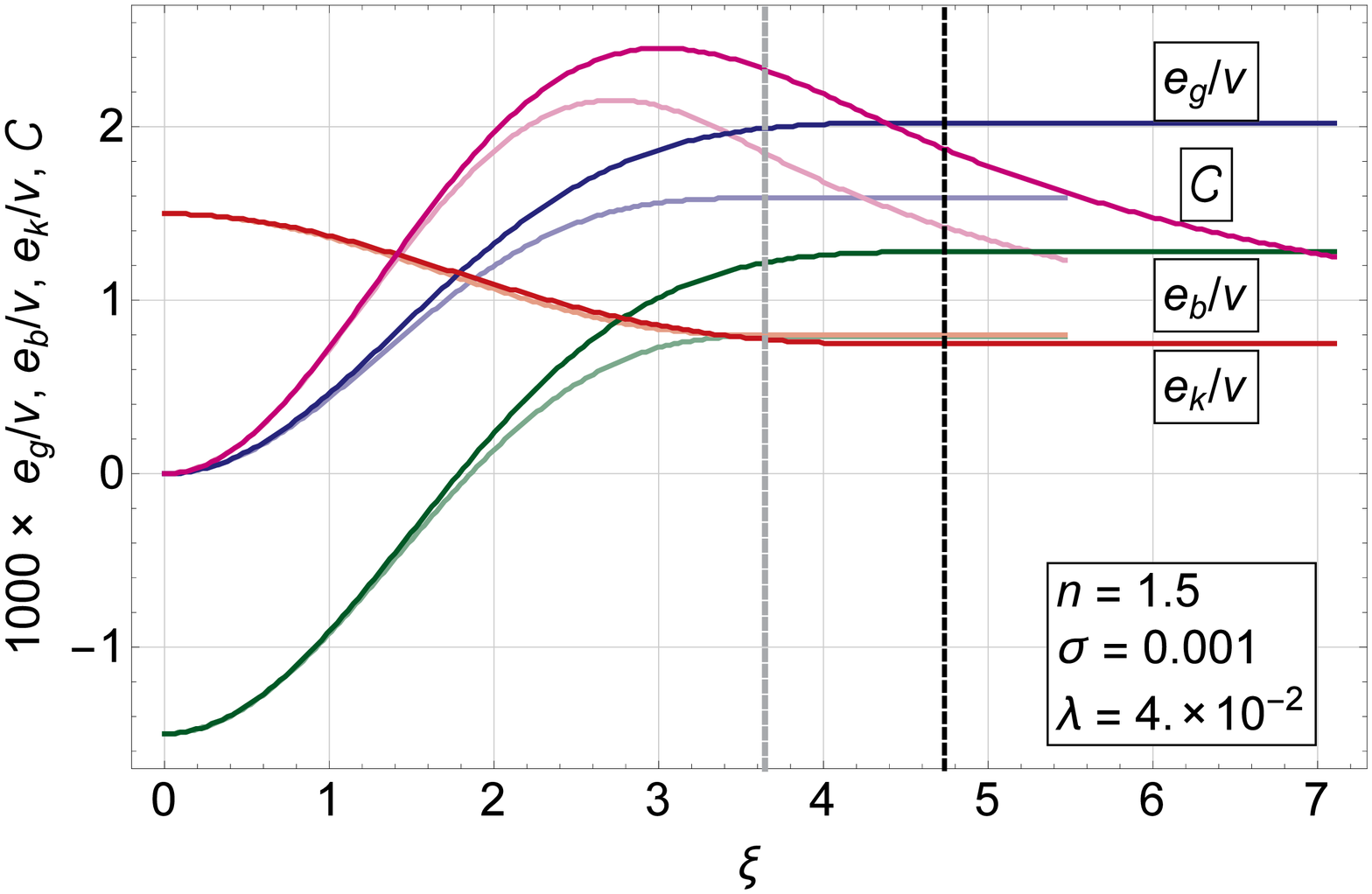}
\end{minipage}
\par\vspace{1.5\baselineskip}\par
\begin{minipage}{0.32\linewidth}
\centering
\includegraphics[width=\linewidth]{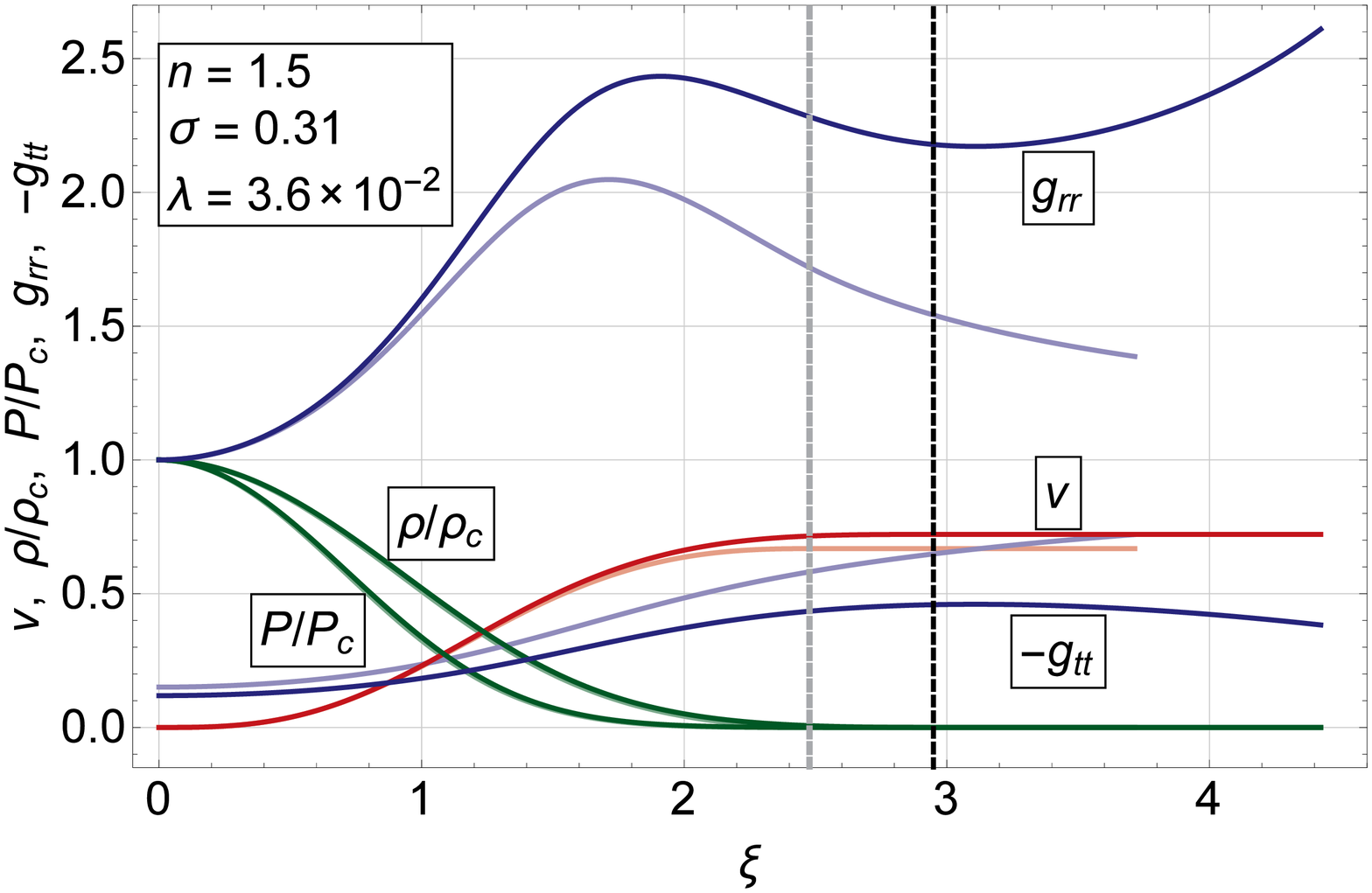}
\end{minipage}\hfill%
\begin{minipage}{0.32\linewidth}
\centering
\includegraphics[width=\linewidth]{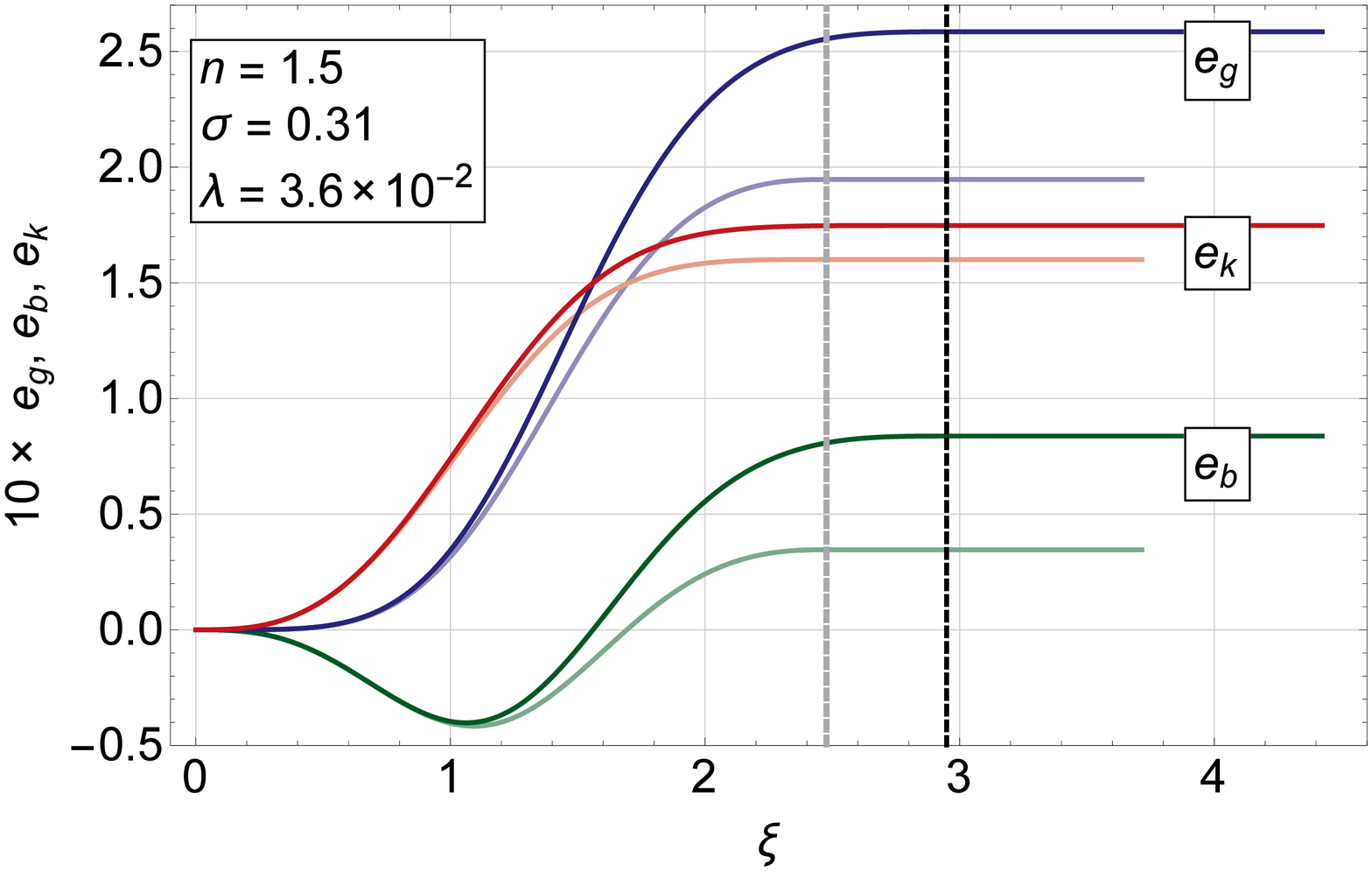}
\end{minipage}\hfill%
\begin{minipage}{0.32\linewidth}
\centering
\includegraphics[width=\linewidth]{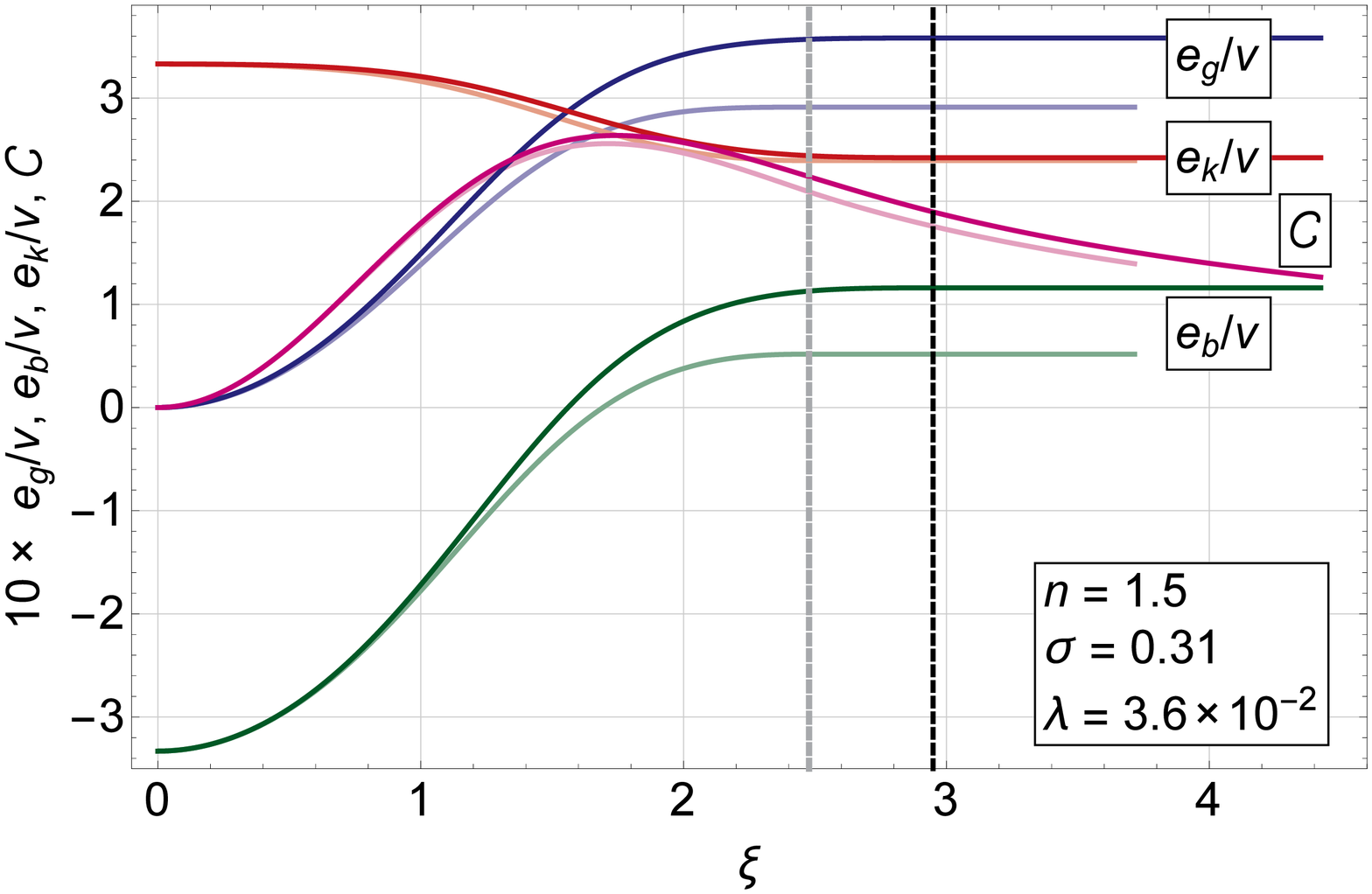}
\end{minipage}
\par\vspace{1.5\baselineskip}\par
\begin{minipage}{0.32\linewidth}
\centering
\includegraphics[width=\linewidth]{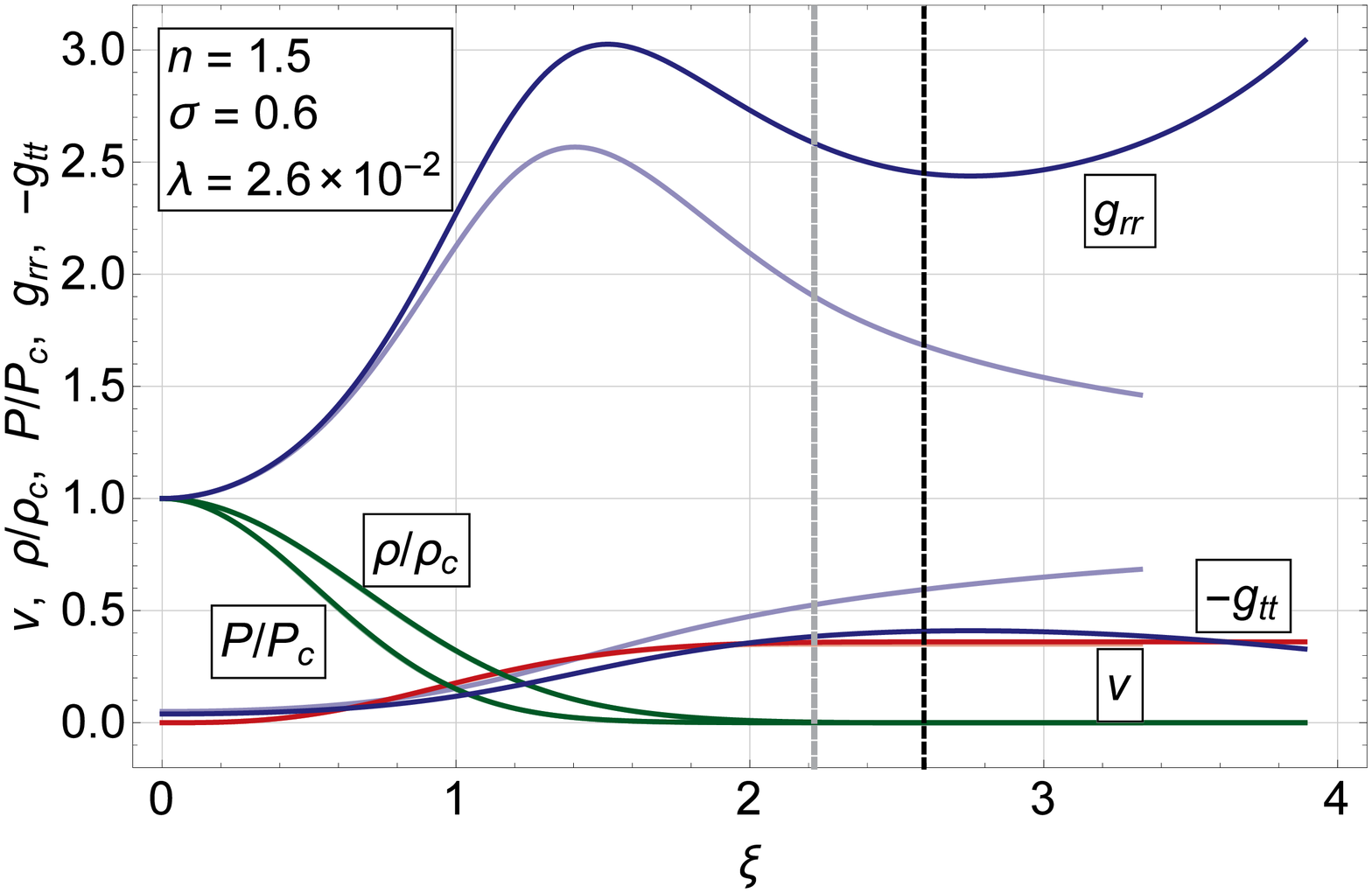}
\end{minipage}\hfill%
\begin{minipage}{0.32\linewidth}
\centering
\includegraphics[width=\linewidth]{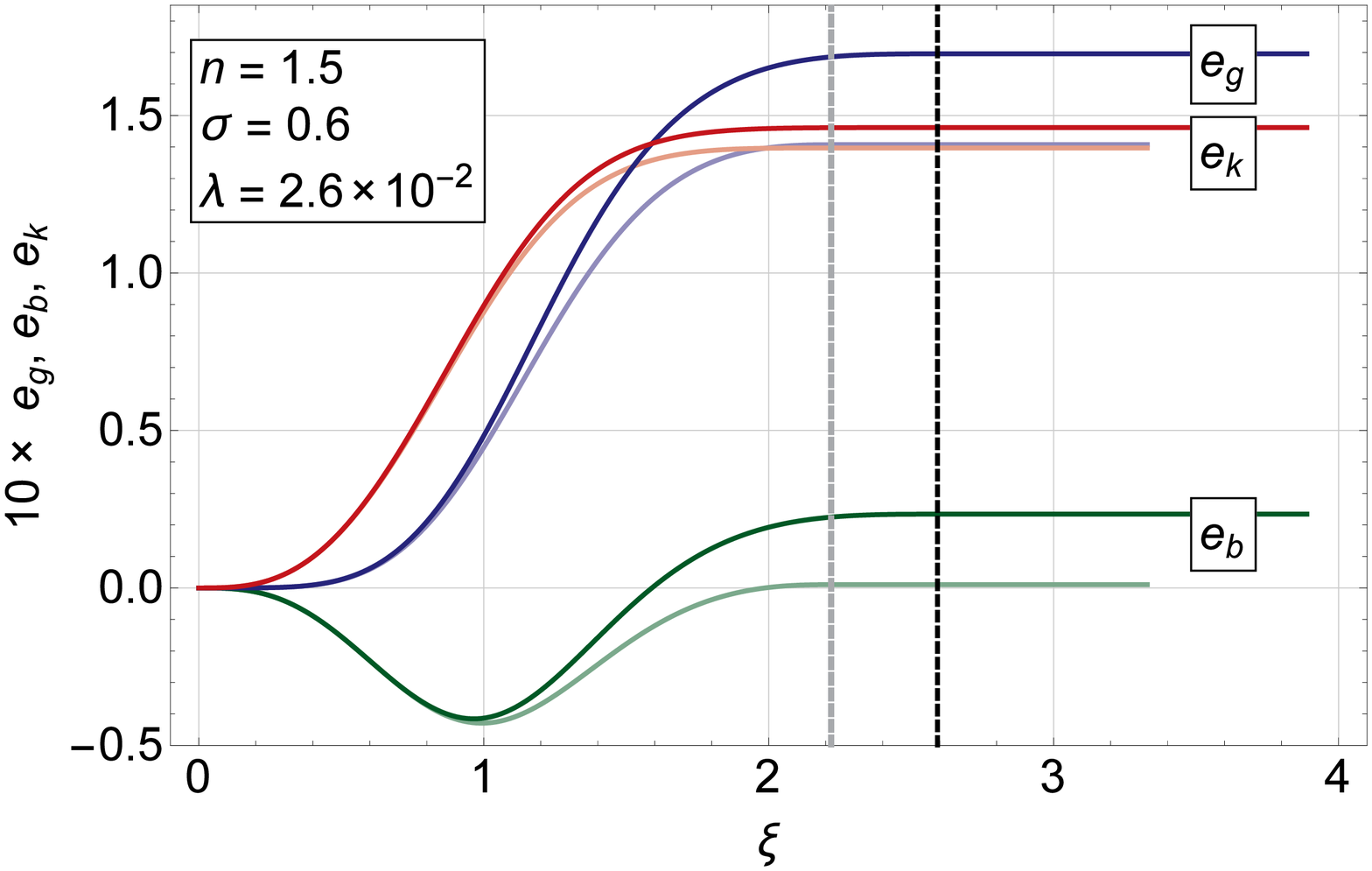}
\end{minipage}\hfill%
\begin{minipage}{0.32\linewidth}
\centering
\includegraphics[width=\linewidth]{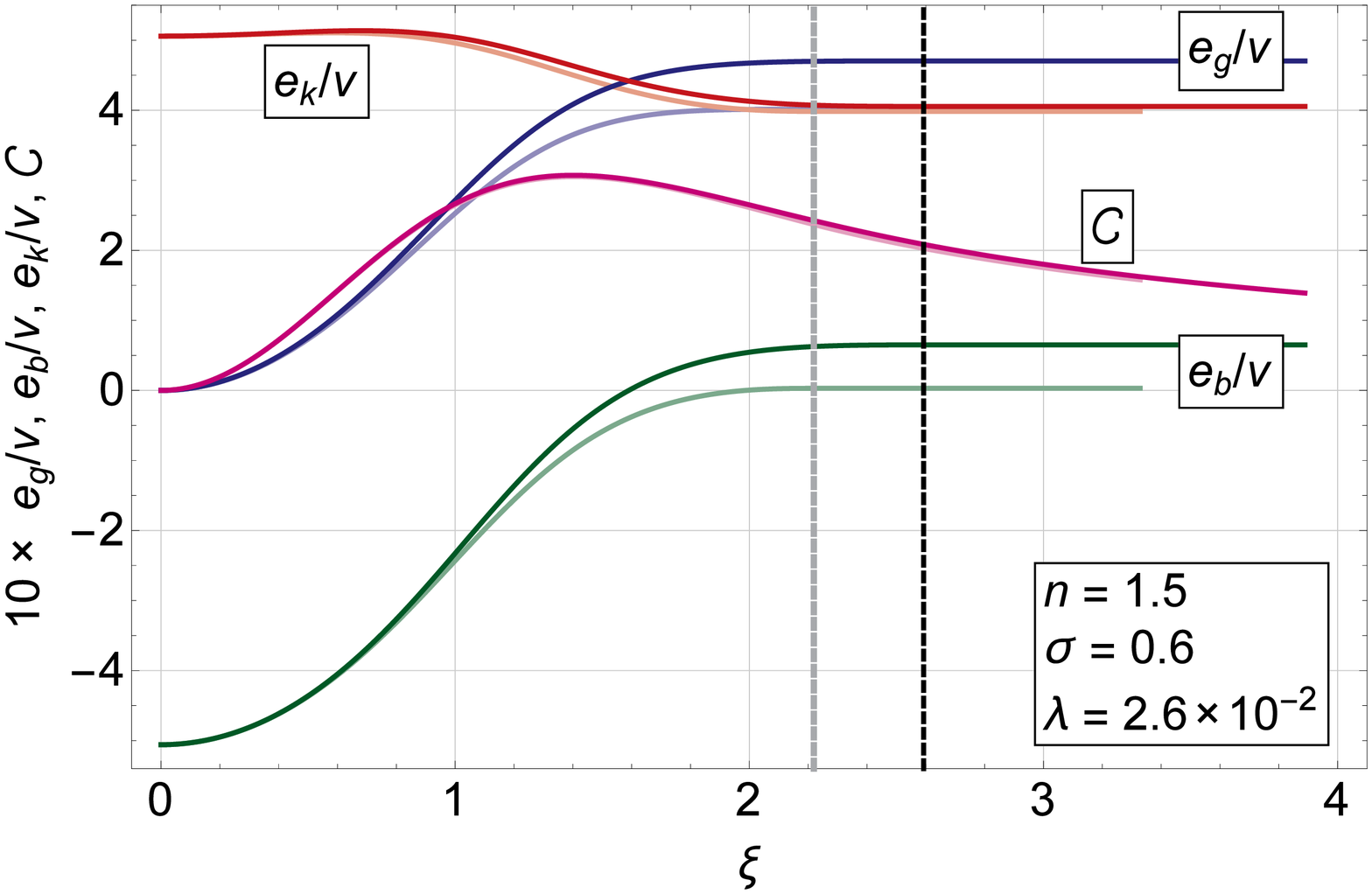}
\end{minipage}
\par\vspace{.8\baselineskip}\par
\caption{\label{ProN15}Profile plots for polytropic index
  $n=1.5$. \textit{Left column:} Mass, density, pressure, and metric
  coefficients. \textit{Middle column:} Gravitational, binding and kinetic
  energy. \textit{Right column:} Relative gravitational, binding, kinetic
  energy and compactness.}
\end{figure*}

\begin{figure}[t]
\includegraphics[width=\linewidth]{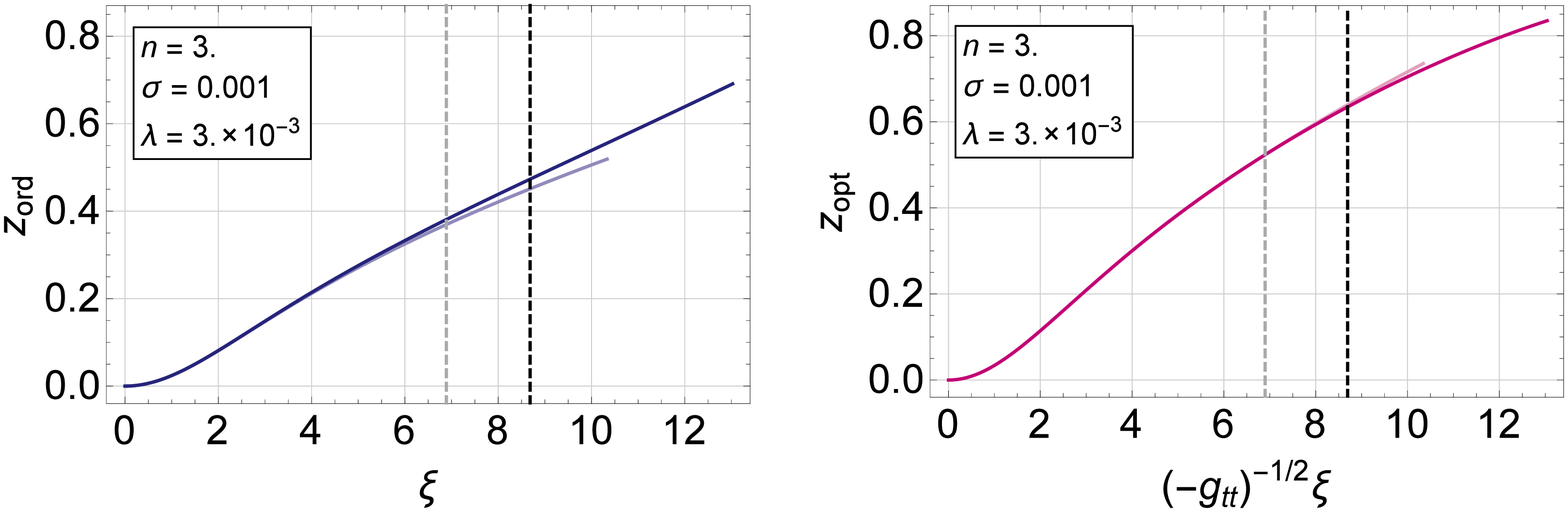}
\par\vspace{0.8\baselineskip}\par
\includegraphics[width=\linewidth]{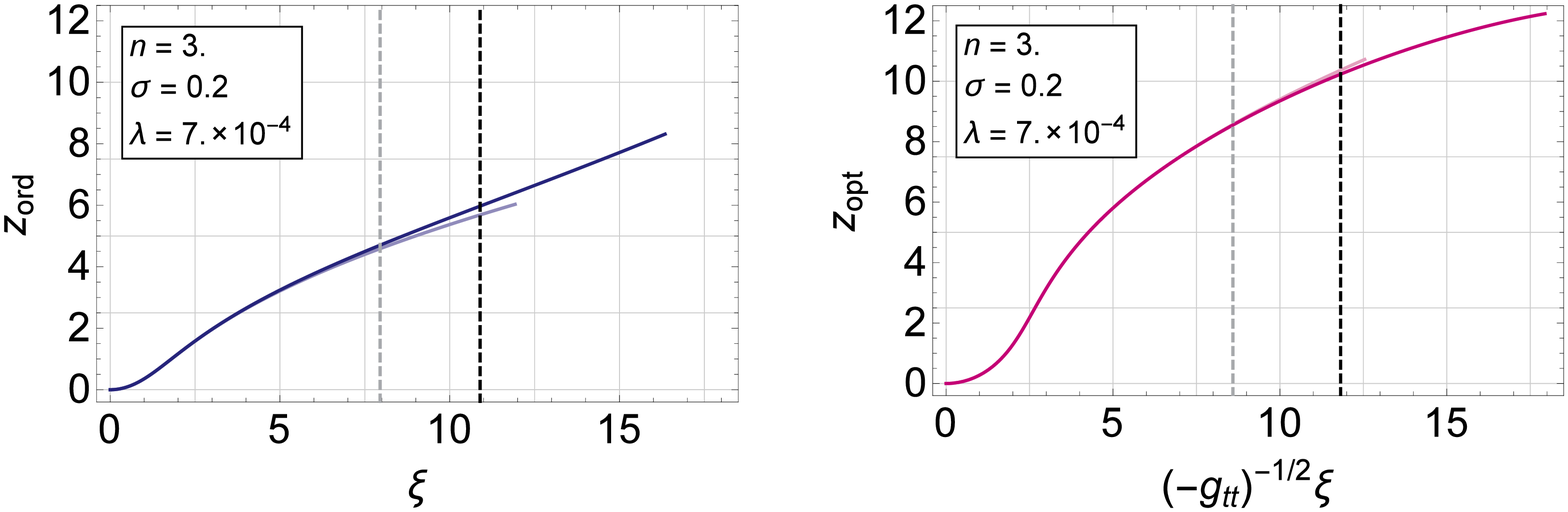}
\par\vspace{0.8\baselineskip}\par
\includegraphics[width=\linewidth]{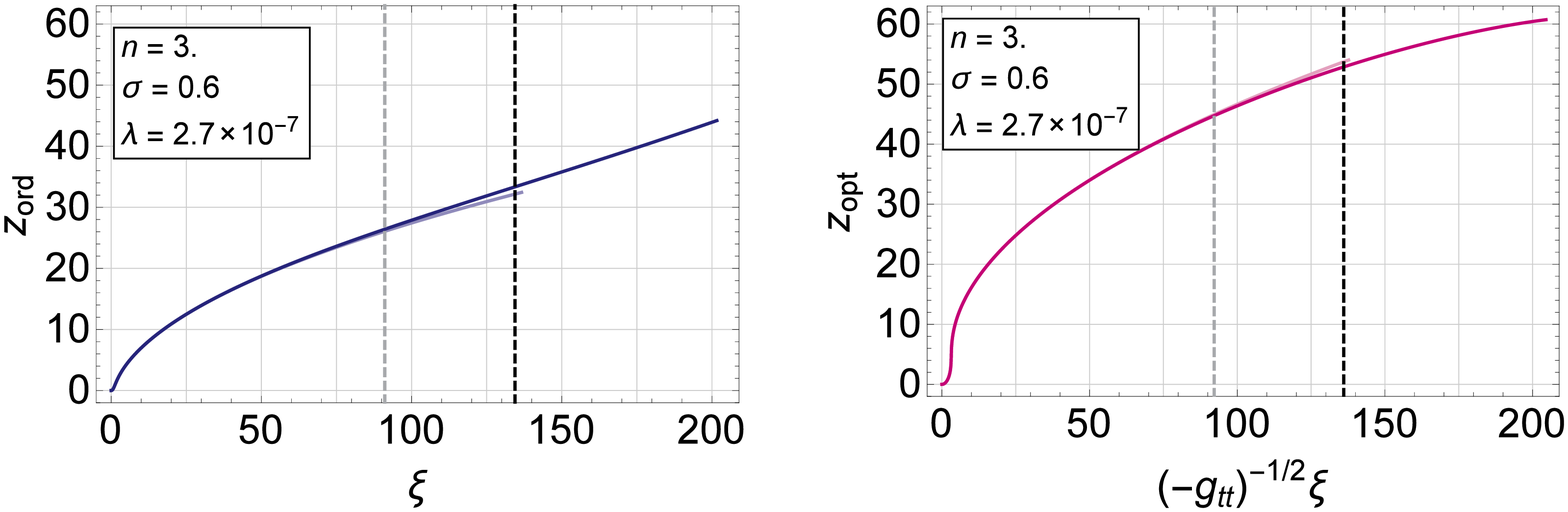}
\par\vspace{0.8\baselineskip}\par
\includegraphics[width=\linewidth]{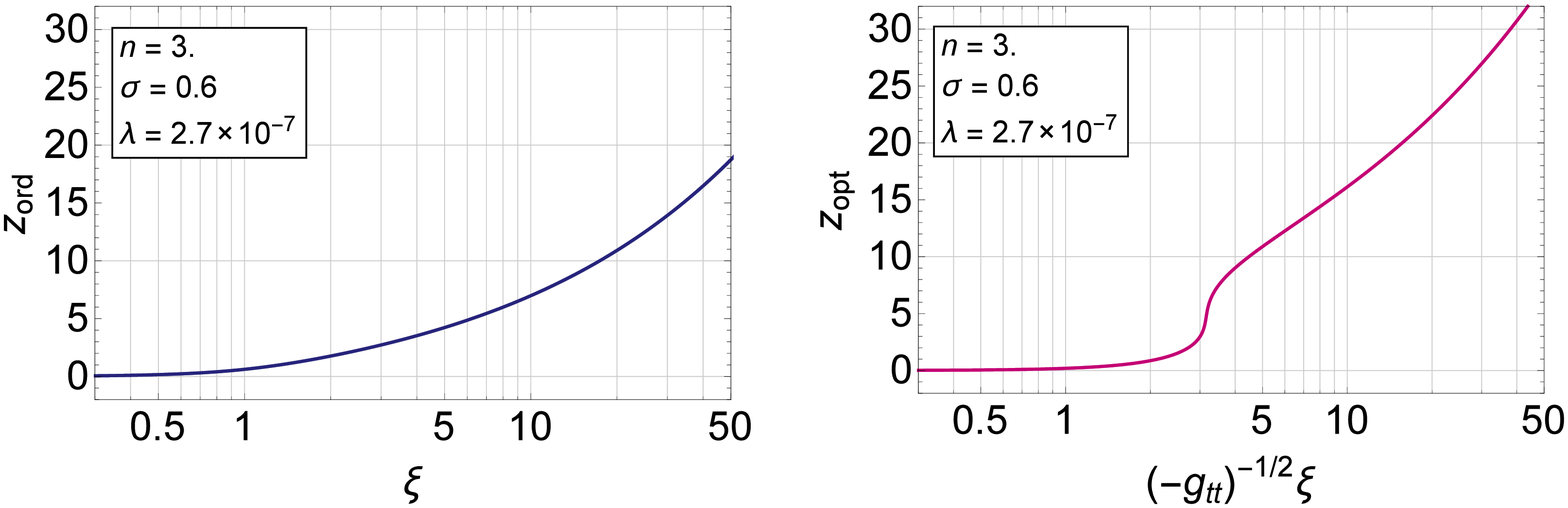}
\par\vspace{0.8\baselineskip}\par
\includegraphics[width=\linewidth]{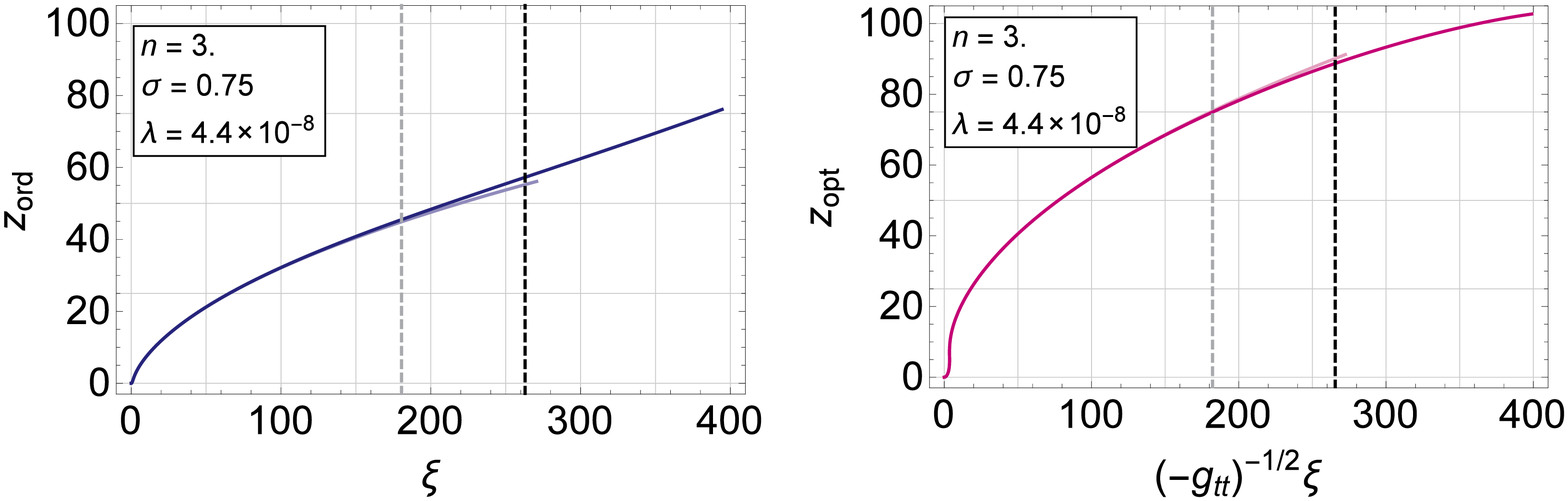}
\par\vspace{0.8\baselineskip}\par
\includegraphics[width=\linewidth]{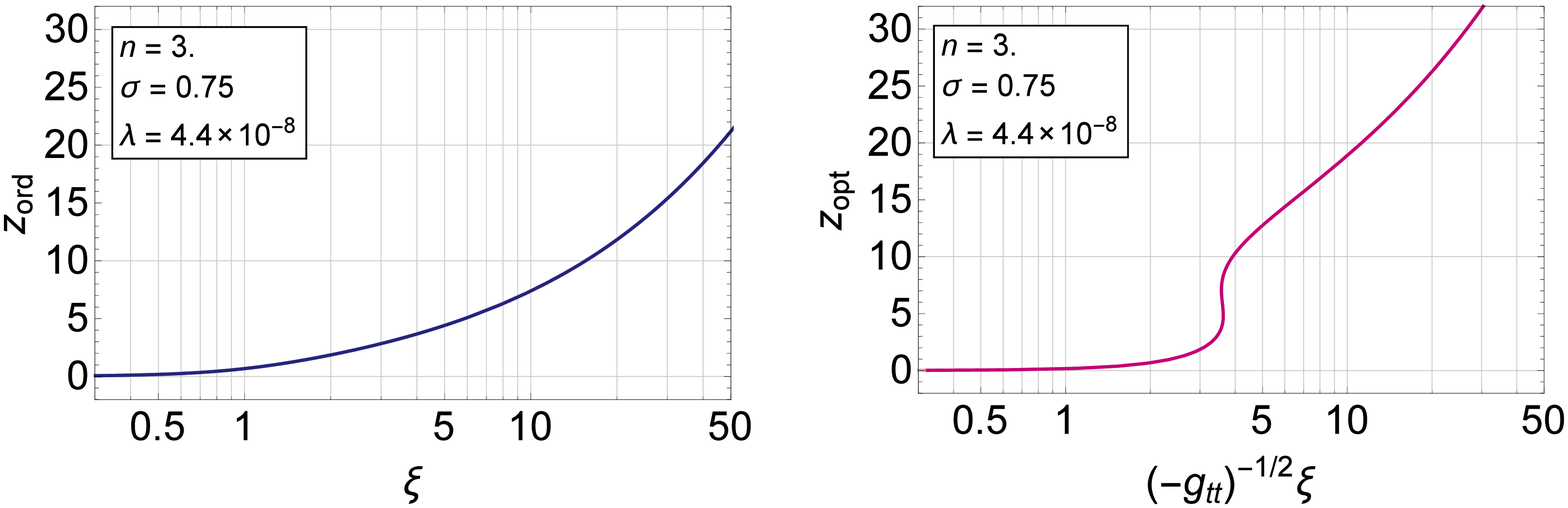}
\caption{\label{EmbN30}Embedding diagrams for polytropic index
  $n=3.0$.\textit{Left column:} Ordinary geometry. \textit{Right column:}
  Optical geometry. The S-shaped part in the third row plot is zoomed in the
  fourth row. The S-shaped part in the fifth row right plot is zoomed in the
  bottom row.}
\end{figure}

\begin{figure*}[t]
\begin{minipage}{0.32\linewidth}
\centering
\includegraphics[width=\linewidth]{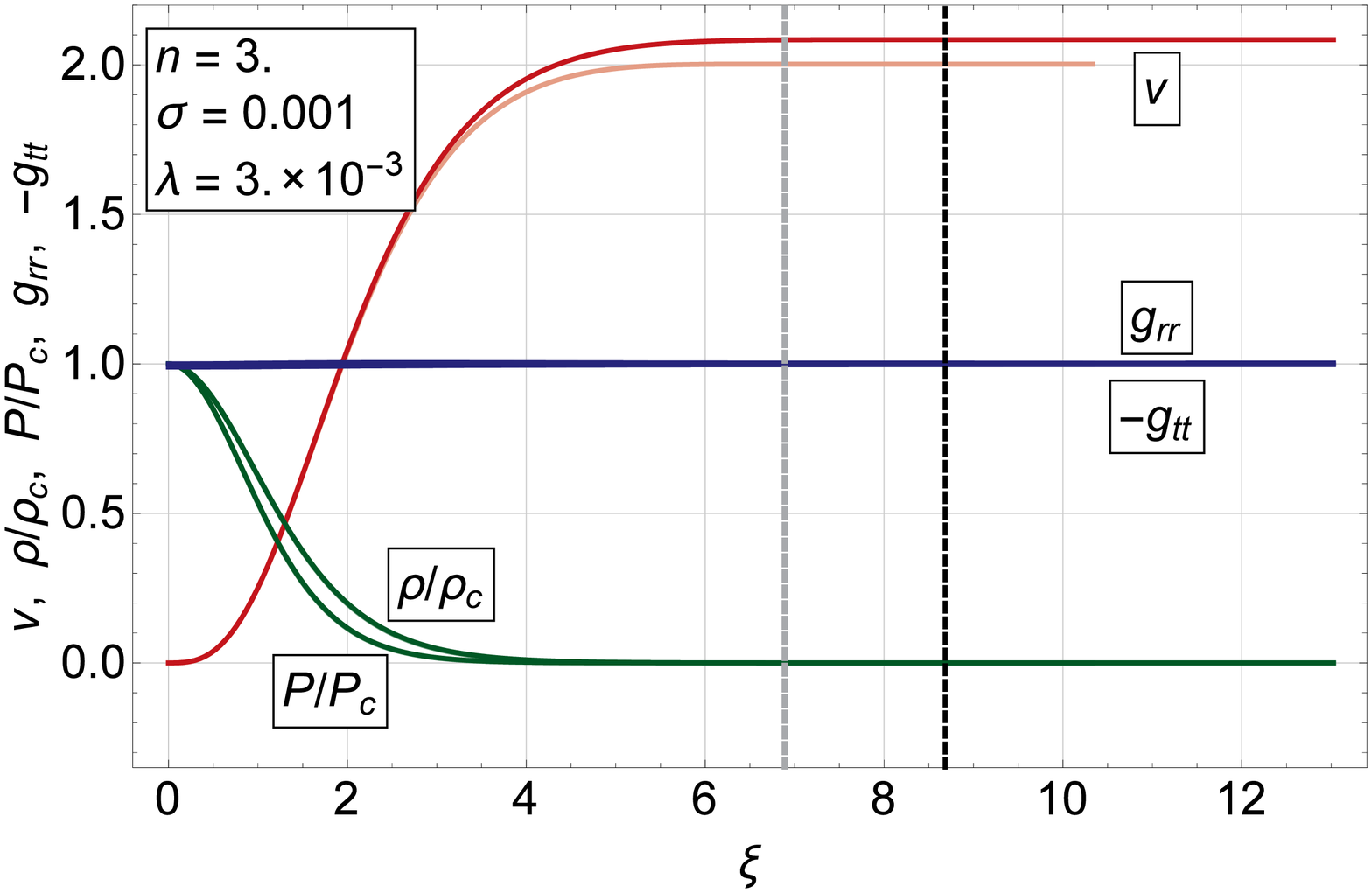}
\end{minipage}\hfill%
\begin{minipage}{0.32\linewidth}
\centering
\includegraphics[width=\linewidth]{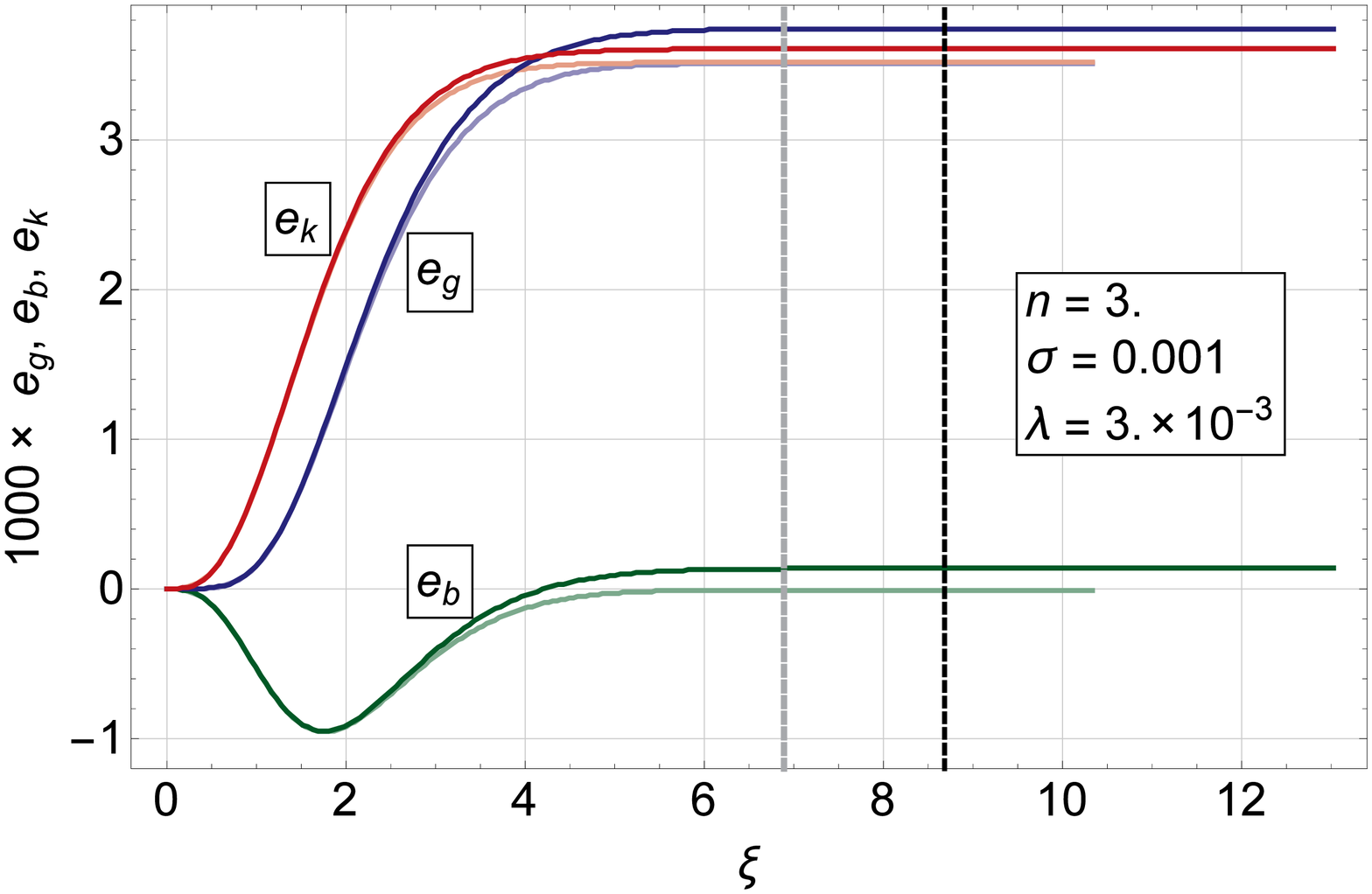}
\end{minipage}\hfill%
\begin{minipage}{0.32\linewidth}
\centering
\includegraphics[width=\linewidth]{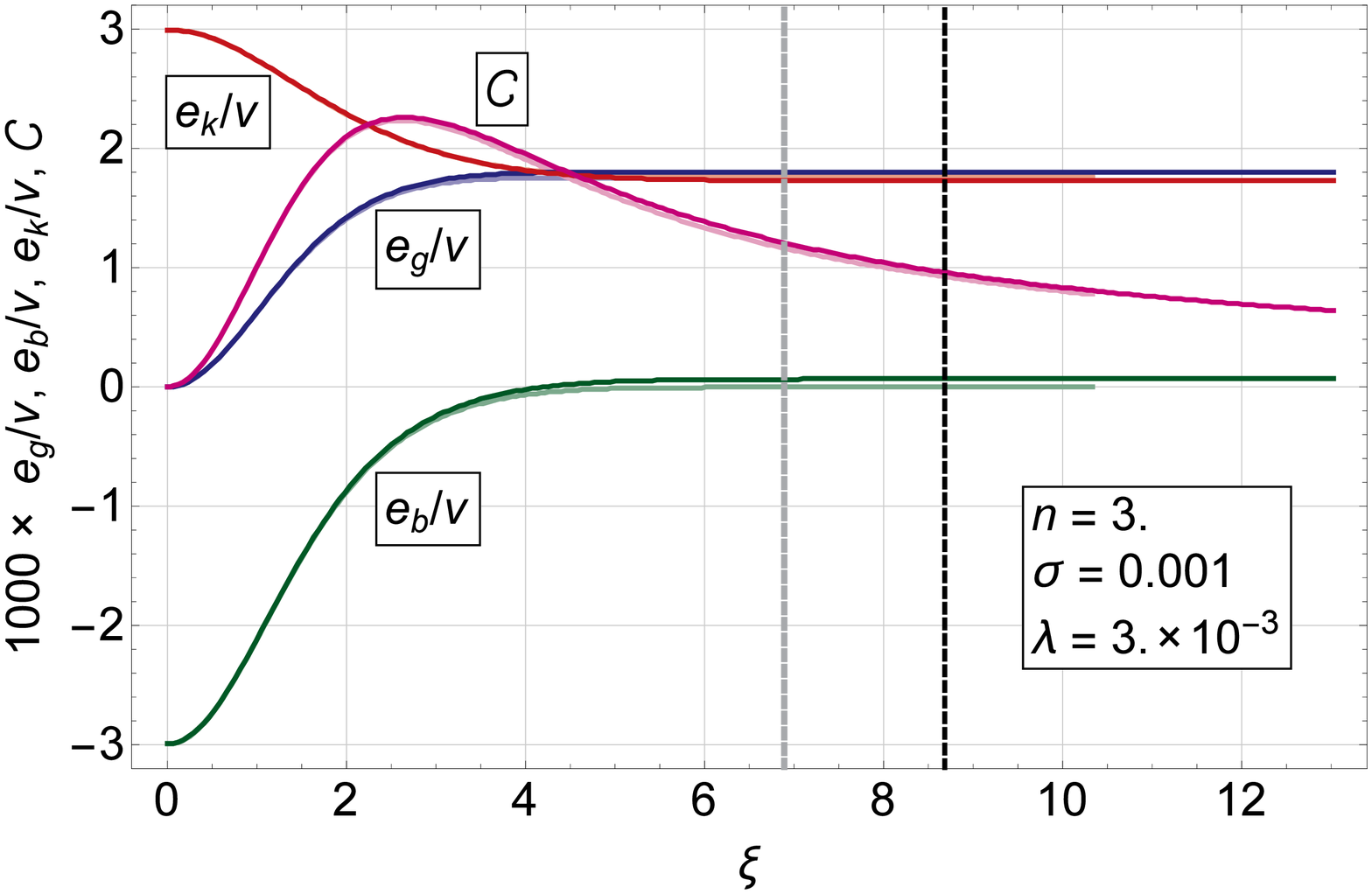}
\end{minipage}
\par\vspace{1.5\baselineskip}\par
\begin{minipage}{0.32\linewidth}
\centering
\includegraphics[width=\linewidth]{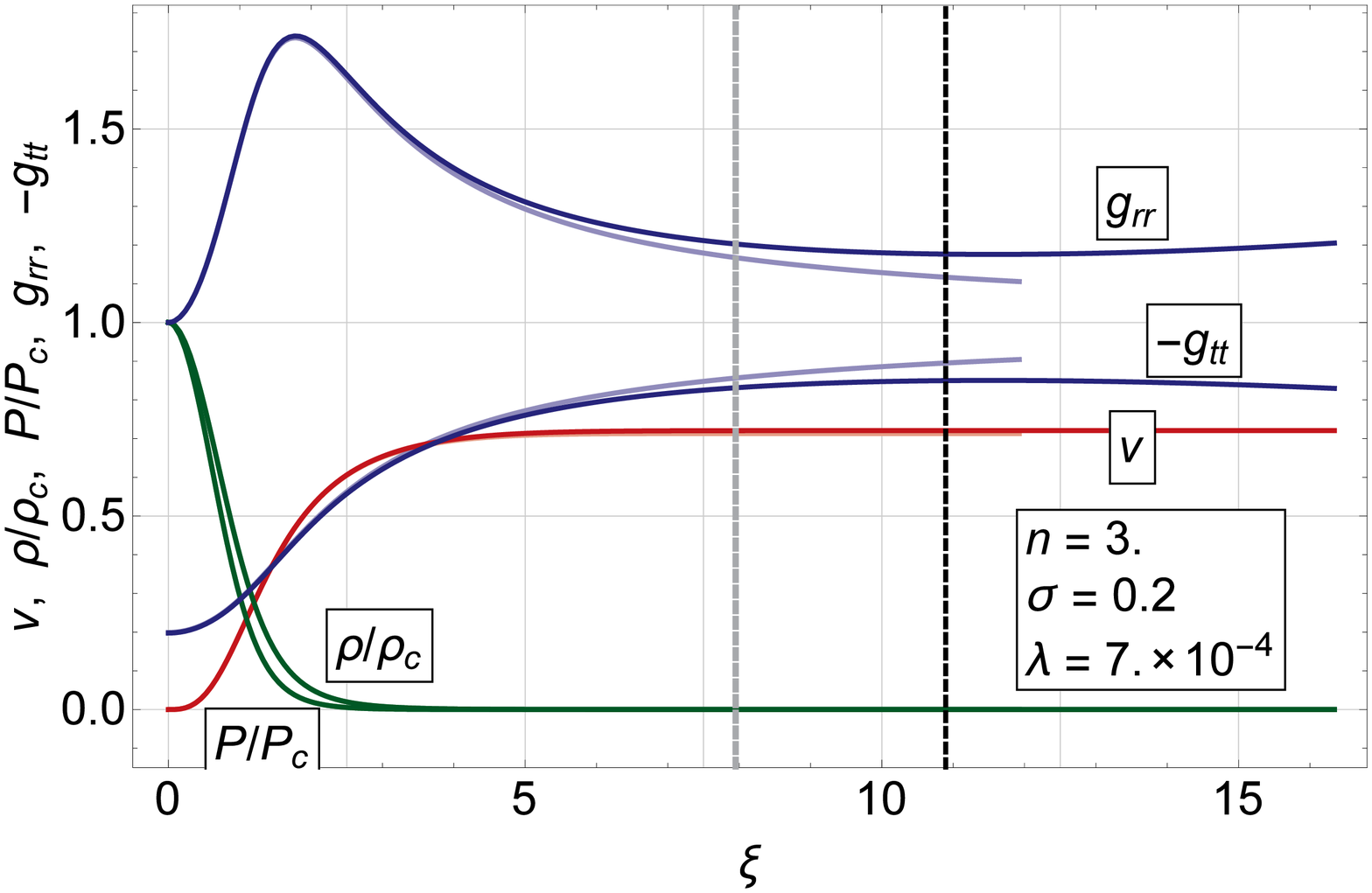}
\end{minipage}\hfill%
\begin{minipage}{0.32\linewidth}
\centering
\includegraphics[width=\linewidth]{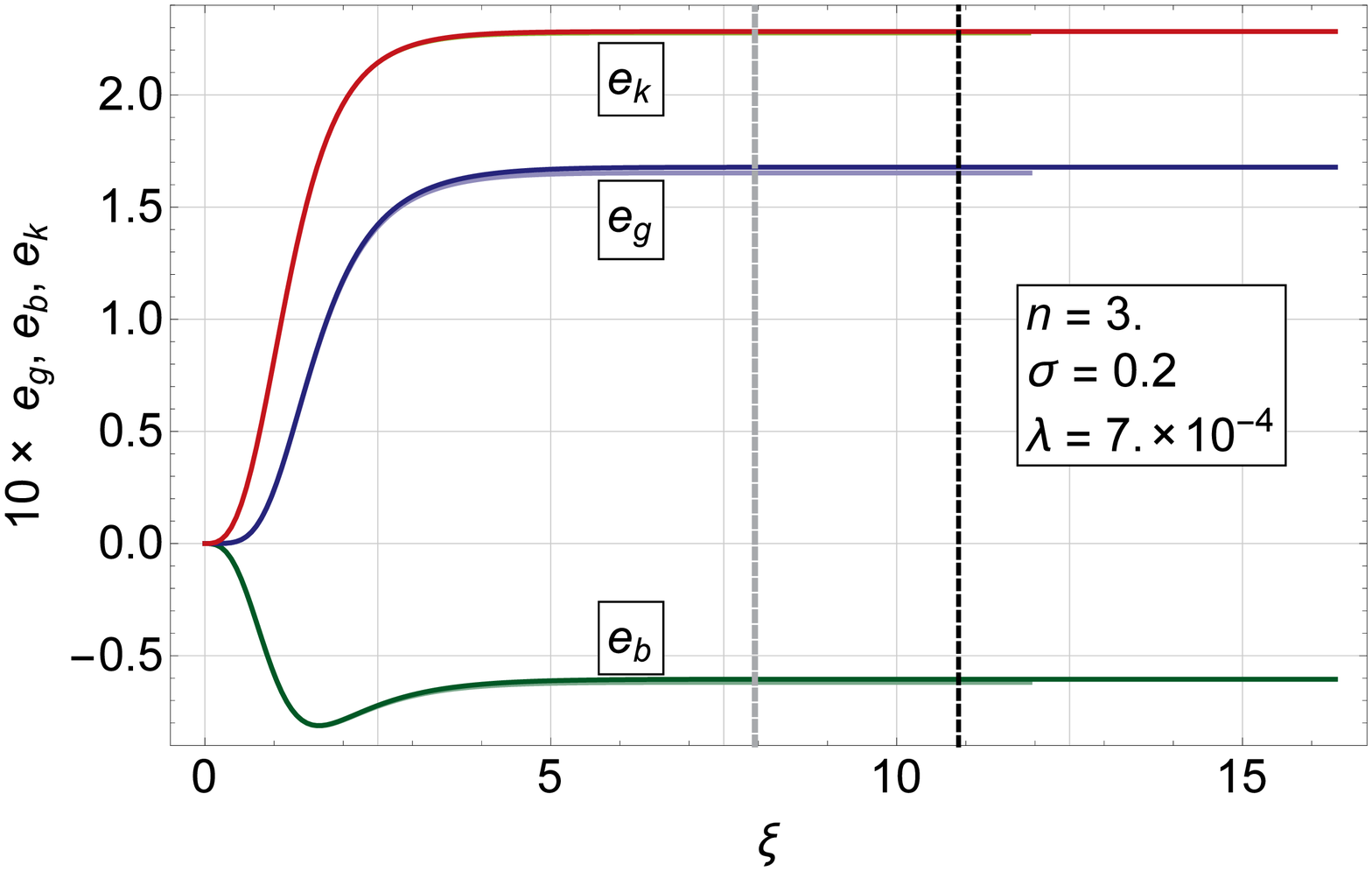}
\end{minipage}\hfill%
\begin{minipage}{0.32\linewidth}
\centering
\includegraphics[width=\linewidth]{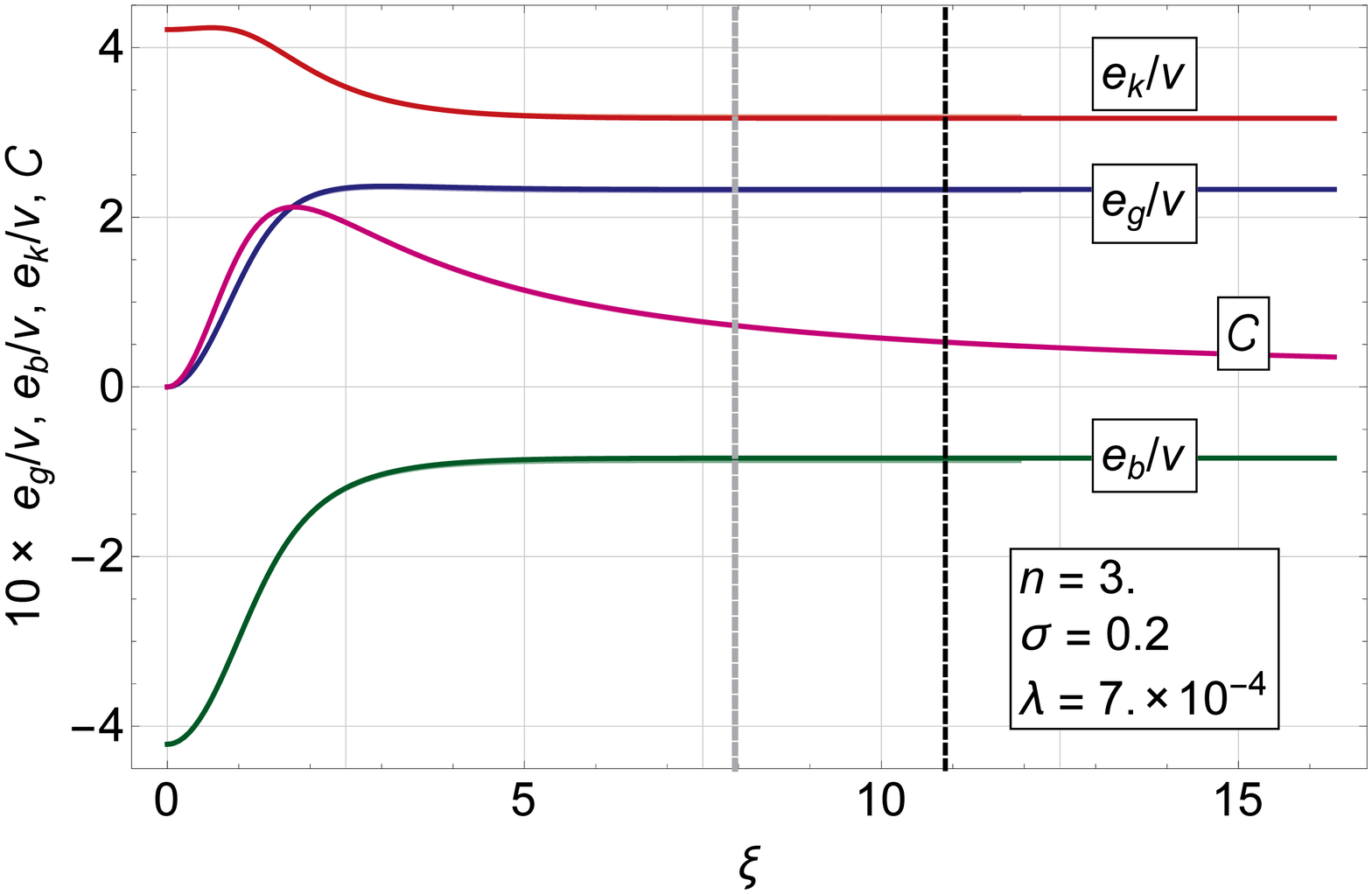}
\end{minipage}
\par\vspace{1.5\baselineskip}\par
\begin{minipage}{0.32\linewidth}
\centering
\includegraphics[width=\linewidth]{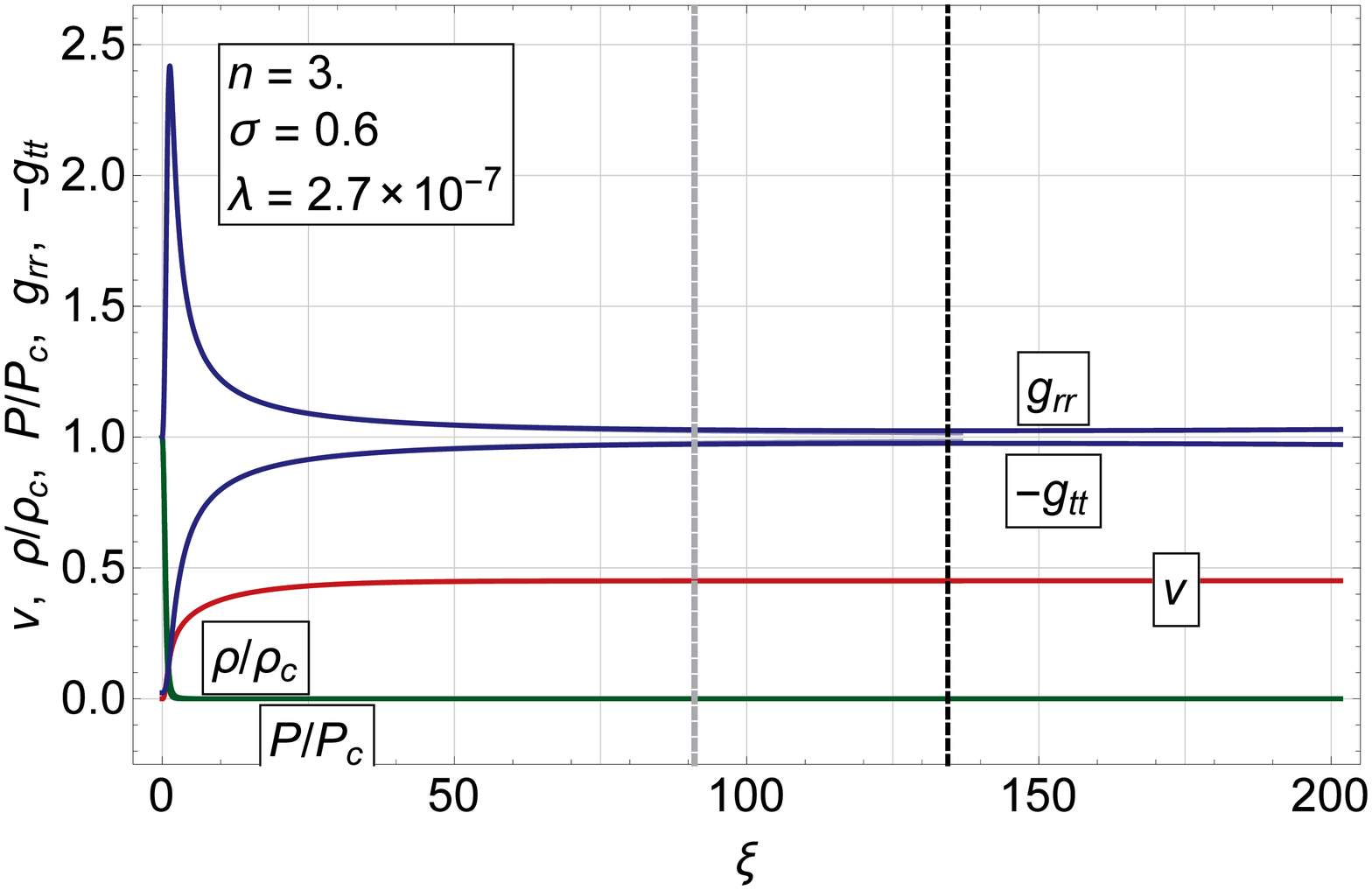}
\end{minipage}\hfill%
\begin{minipage}{0.32\linewidth}
\centering
\includegraphics[width=\linewidth]{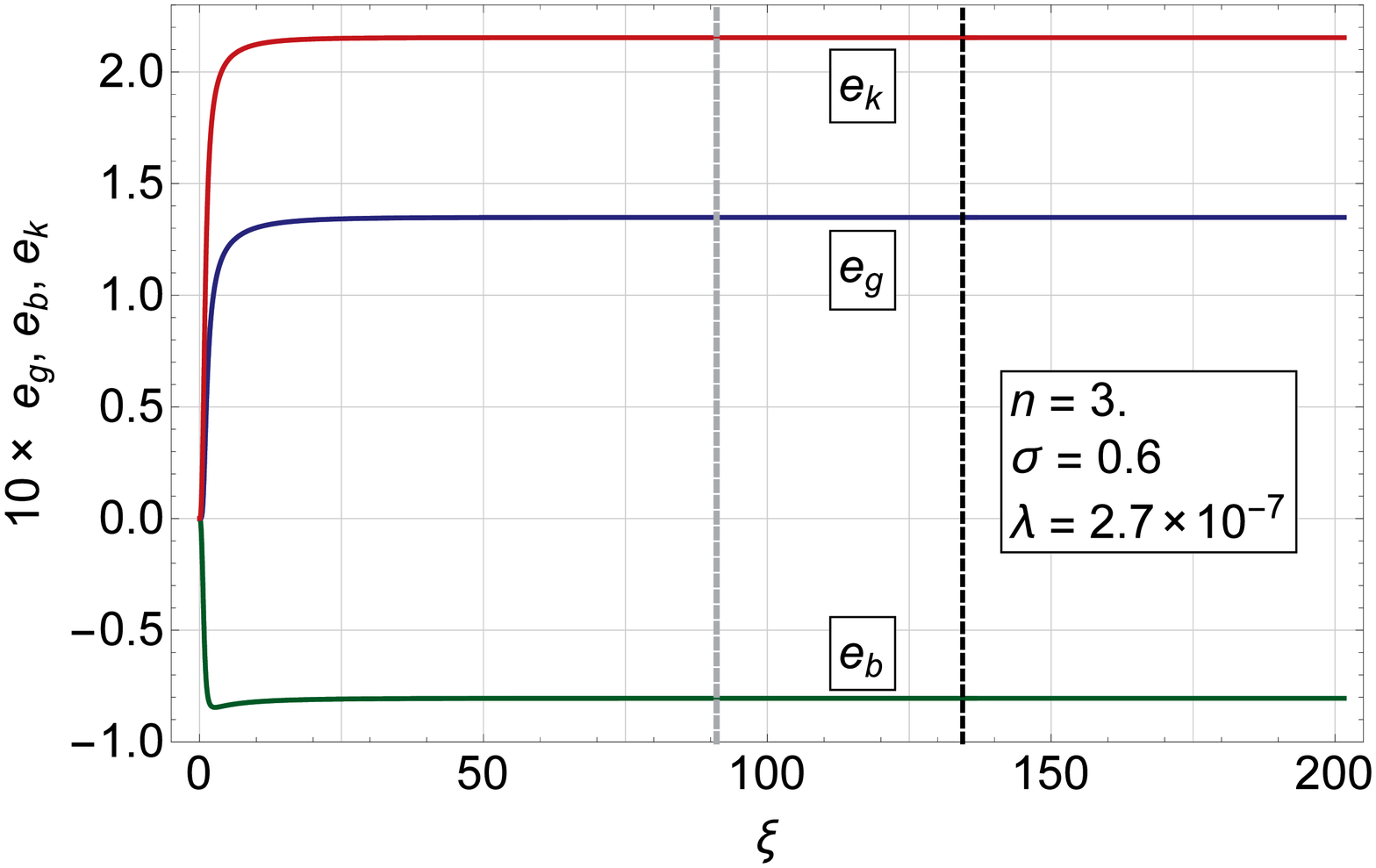}
\end{minipage}\hfill%
\begin{minipage}{0.32\linewidth}
\centering
\includegraphics[width=\linewidth]{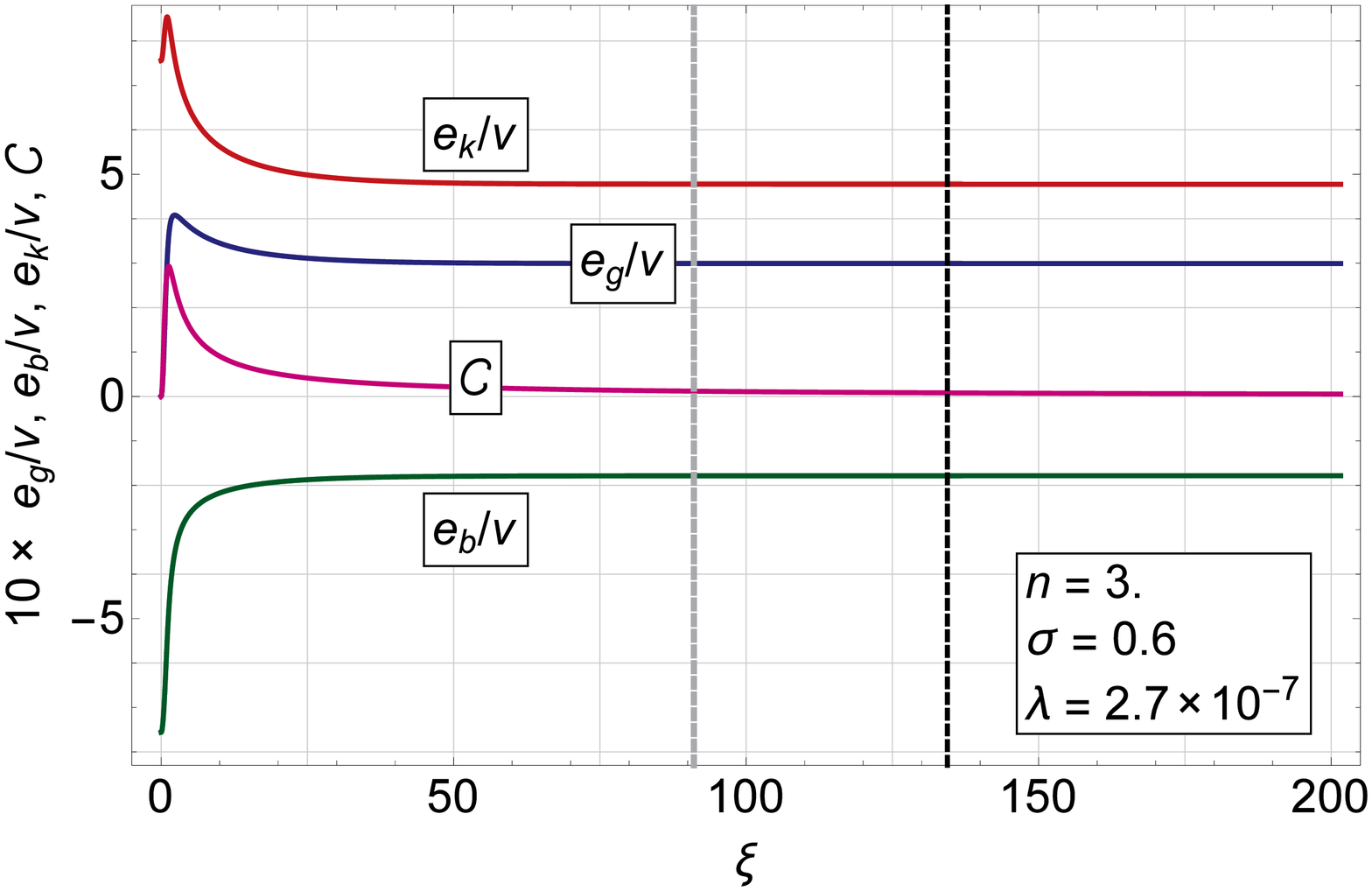}
\end{minipage}
\par\vspace{1.5\baselineskip}\par
\begin{minipage}{0.32\linewidth}
\centering
\includegraphics[width=\linewidth]{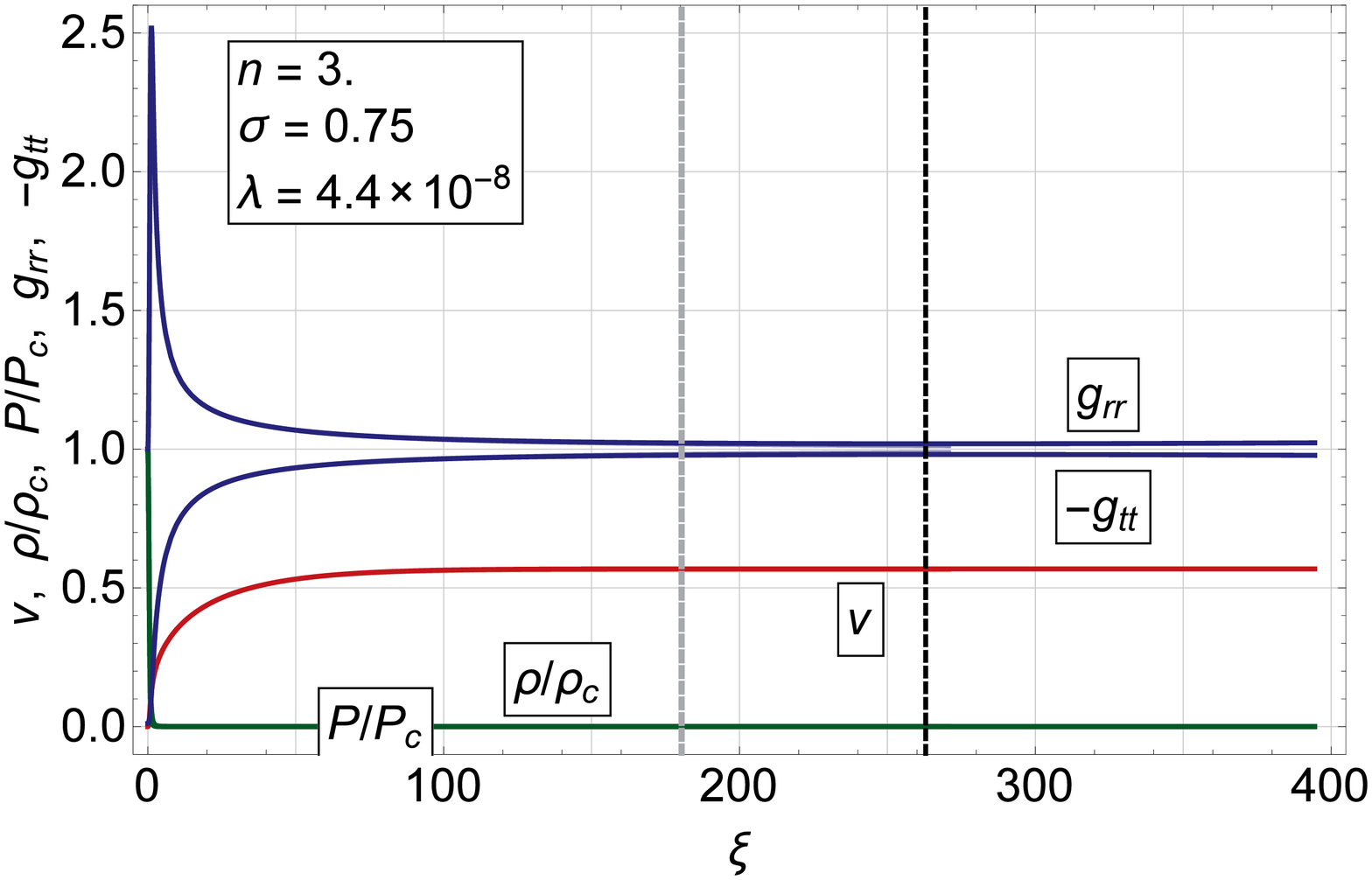}
\end{minipage}\hfill%
\begin{minipage}{0.32\linewidth}
\centering
\includegraphics[width=\linewidth]{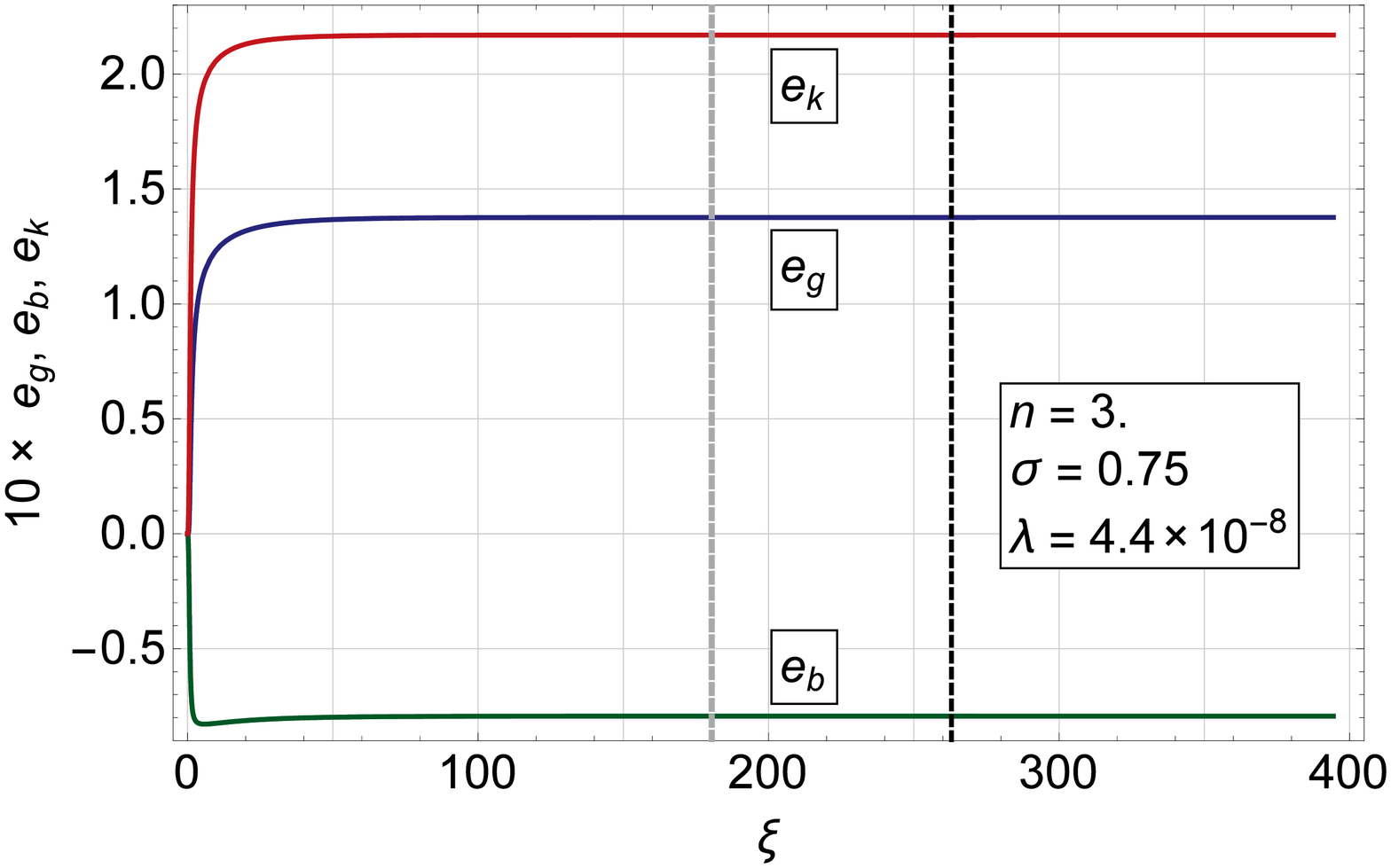}
\end{minipage}\hfill%
\begin{minipage}{0.32\linewidth}
\centering
\includegraphics[width=\linewidth]{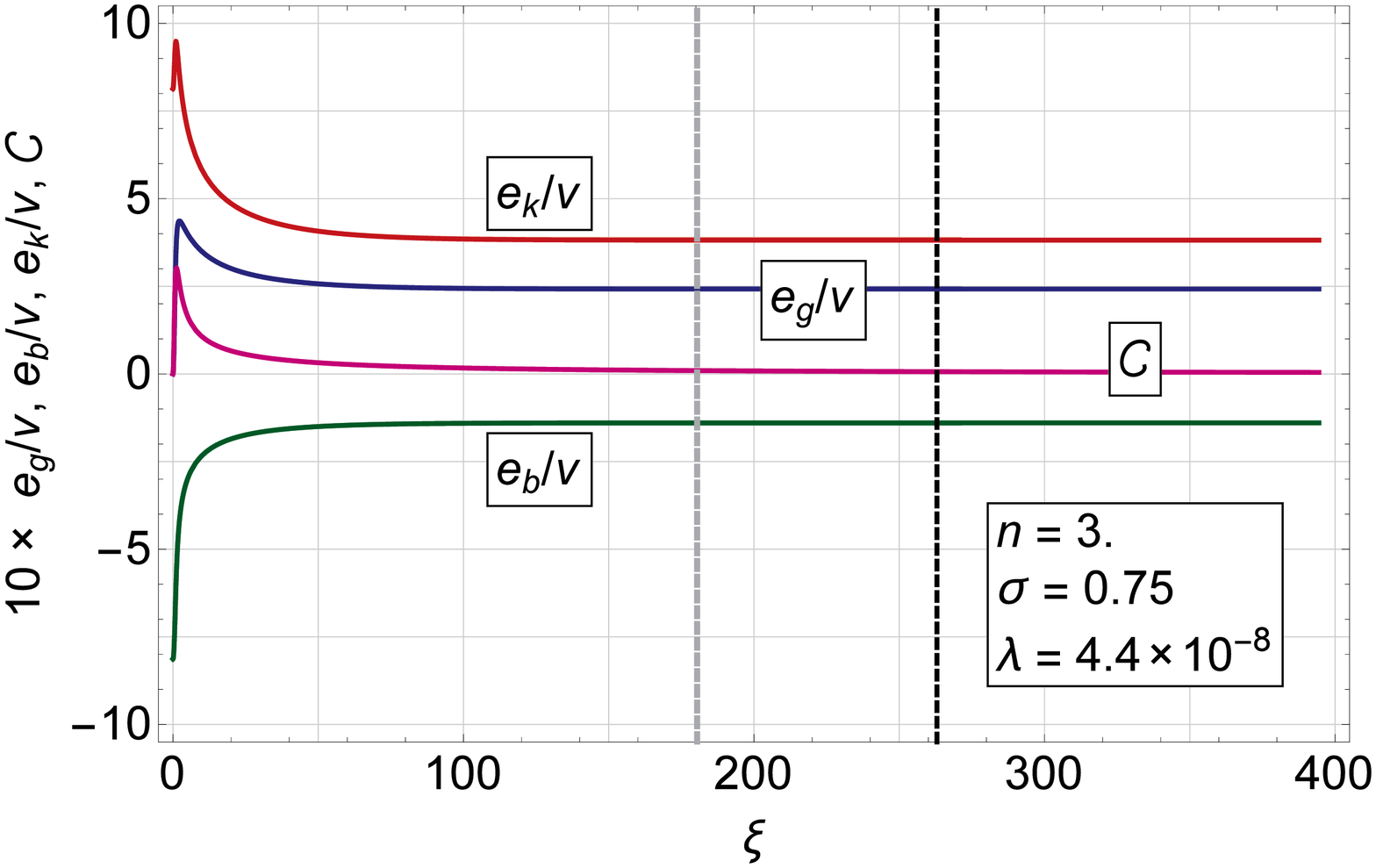}
\end{minipage}
\par\vspace{.8\baselineskip}\par
\caption{\label{ProN30}Profile plots for polytropic index
  $n=3.0$. \textit{Left column:} Mass, density, pressure, and metric
  coefficients. \textit{Middle column:} Gravitational, binding and kinetic
  energy. \textit{Right column:} Relative gravitational, binding, kinetic
  energy and compactness.}
\end{figure*}

\begin{figure*}[t]
\begin{minipage}{0.32\linewidth}
\centering
\includegraphics[width=\linewidth]{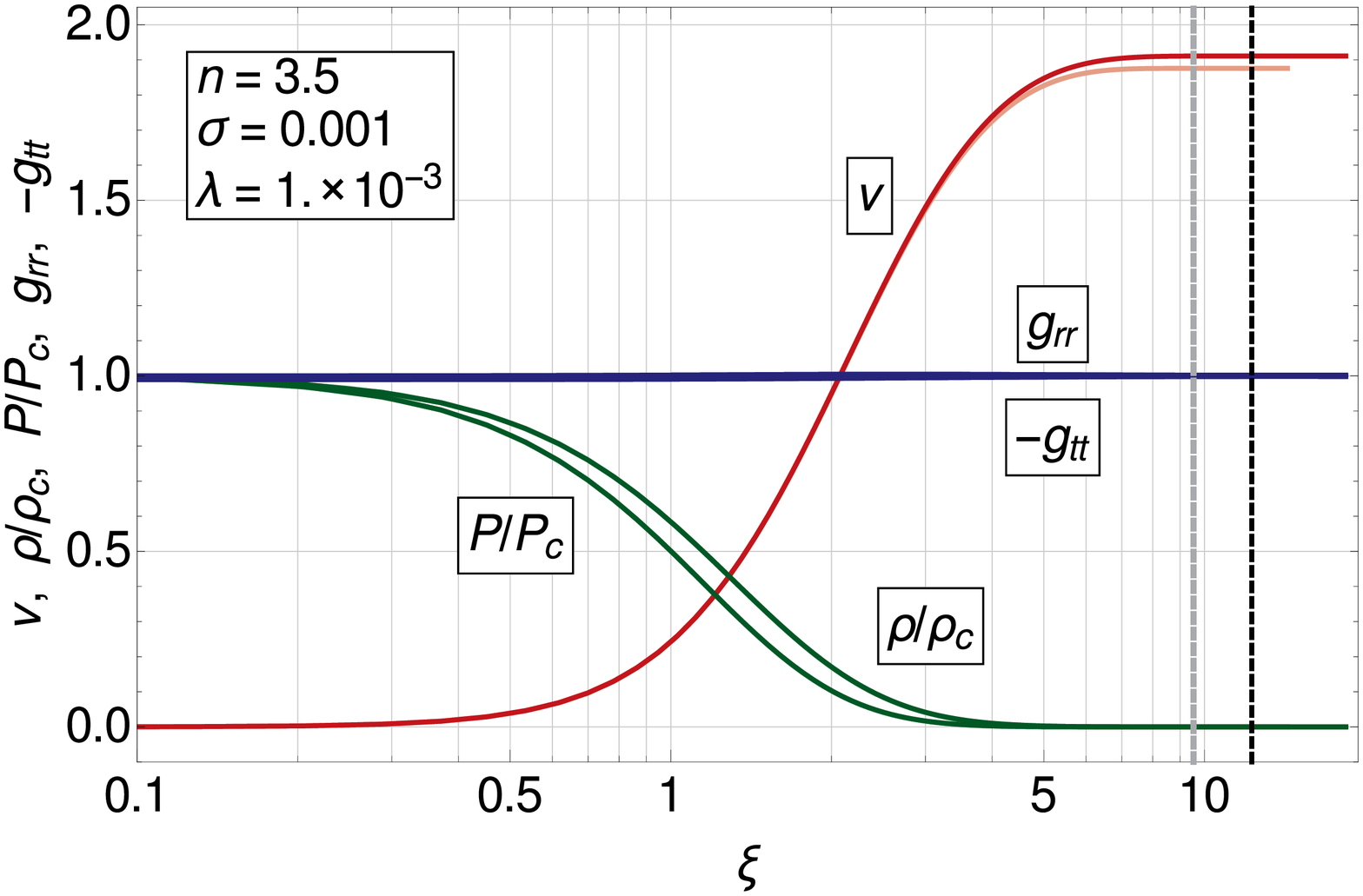}
\end{minipage}\hfill%
\begin{minipage}{0.32\linewidth}
\centering
\includegraphics[width=\linewidth]{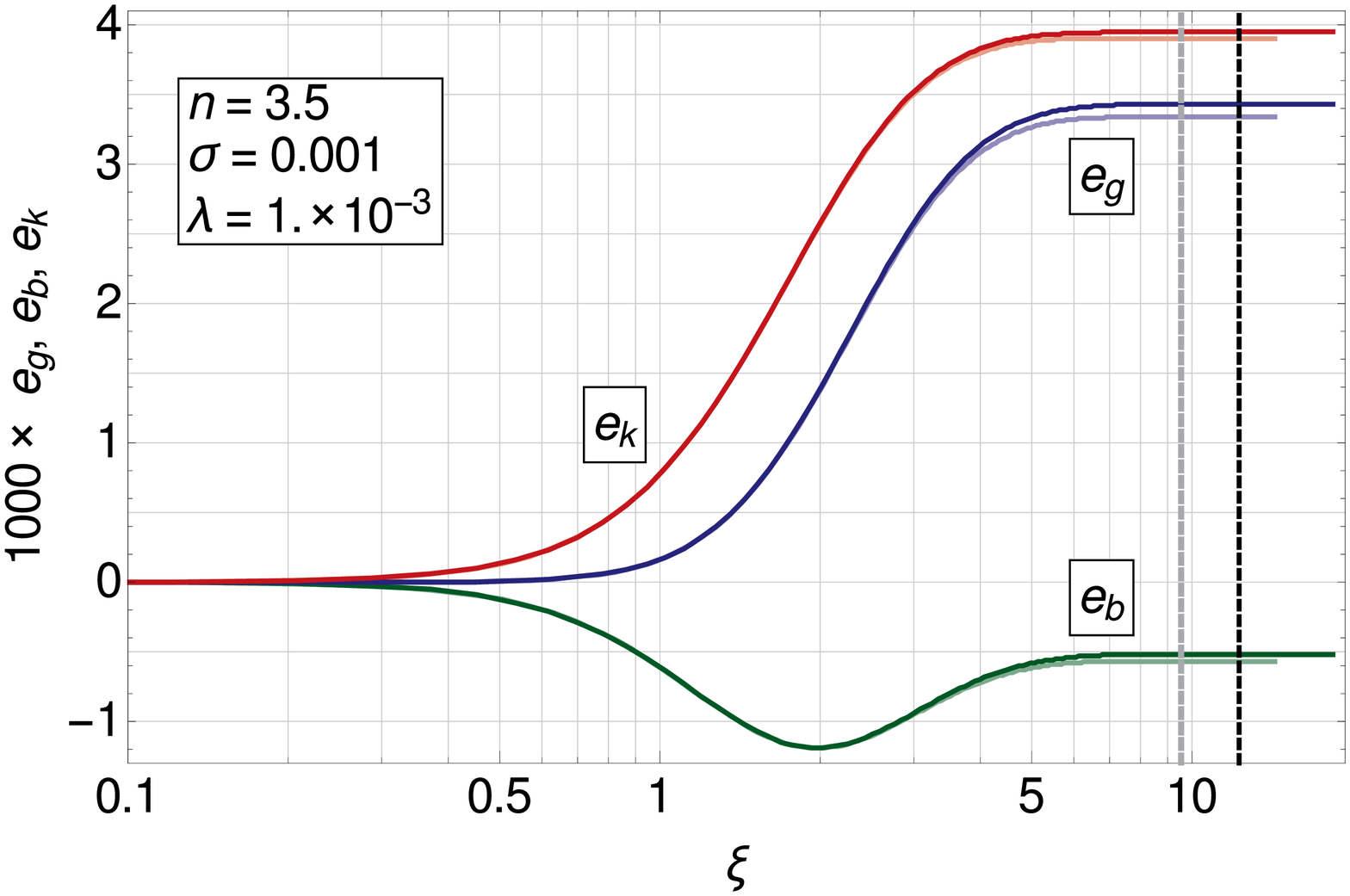}
\end{minipage}\hfill%
\begin{minipage}{0.32\linewidth}
\centering
\includegraphics[width=\linewidth]{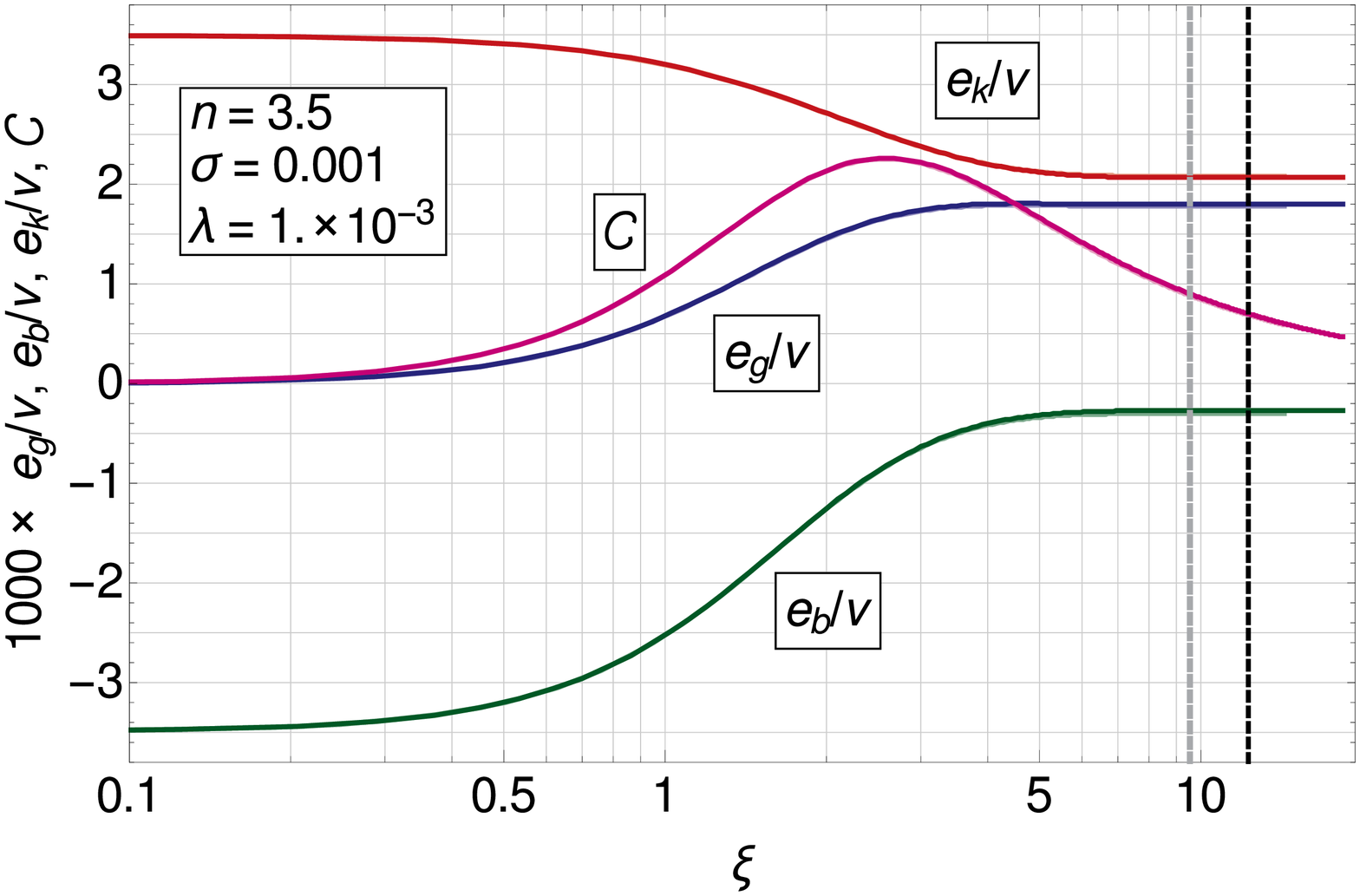}
\end{minipage}
\begin{minipage}{0.32\linewidth}
\centering
\includegraphics[width=\linewidth]{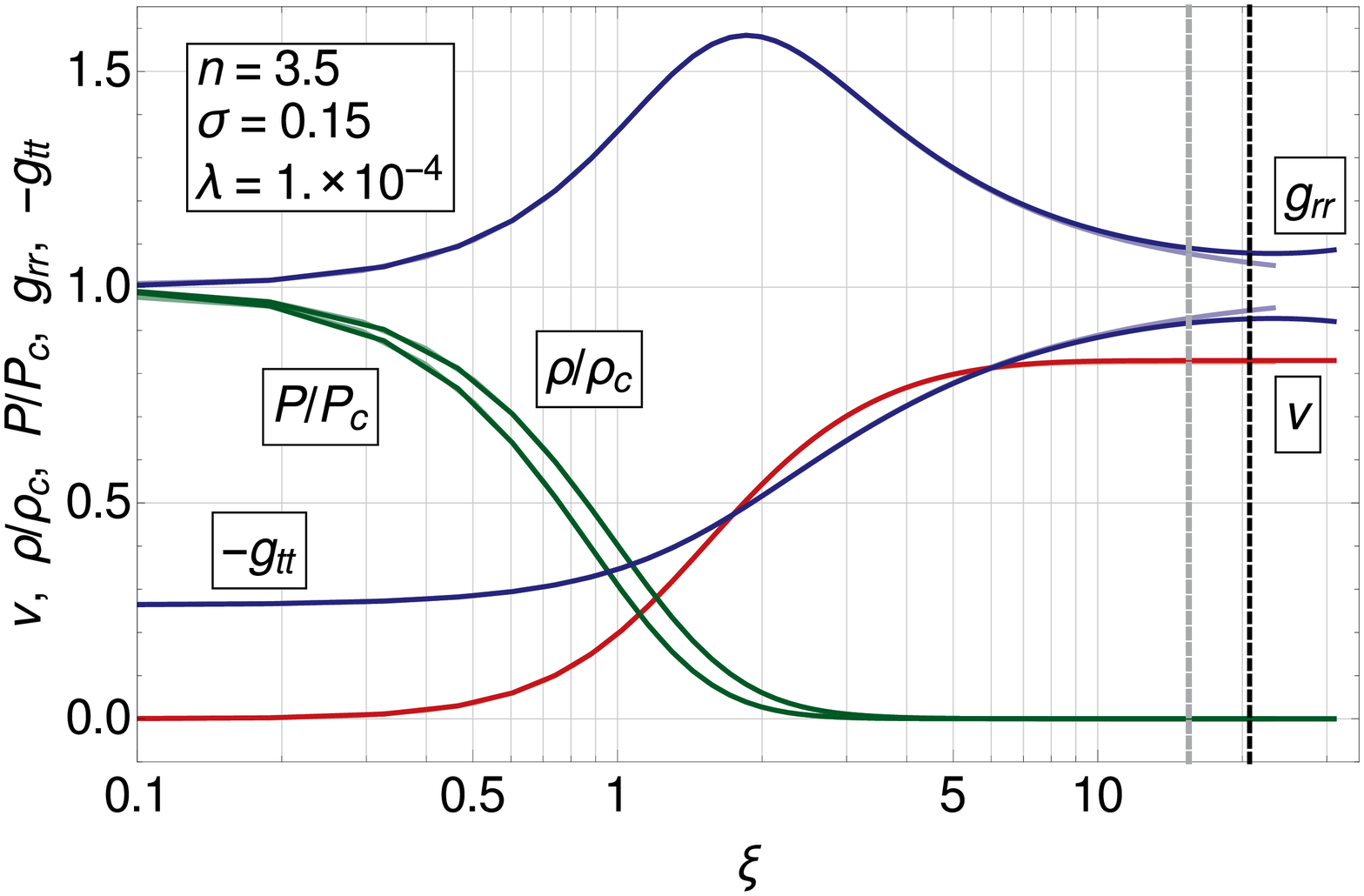}
\end{minipage}\hfill%
\begin{minipage}{0.32\linewidth}
\centering
\includegraphics[width=\linewidth]{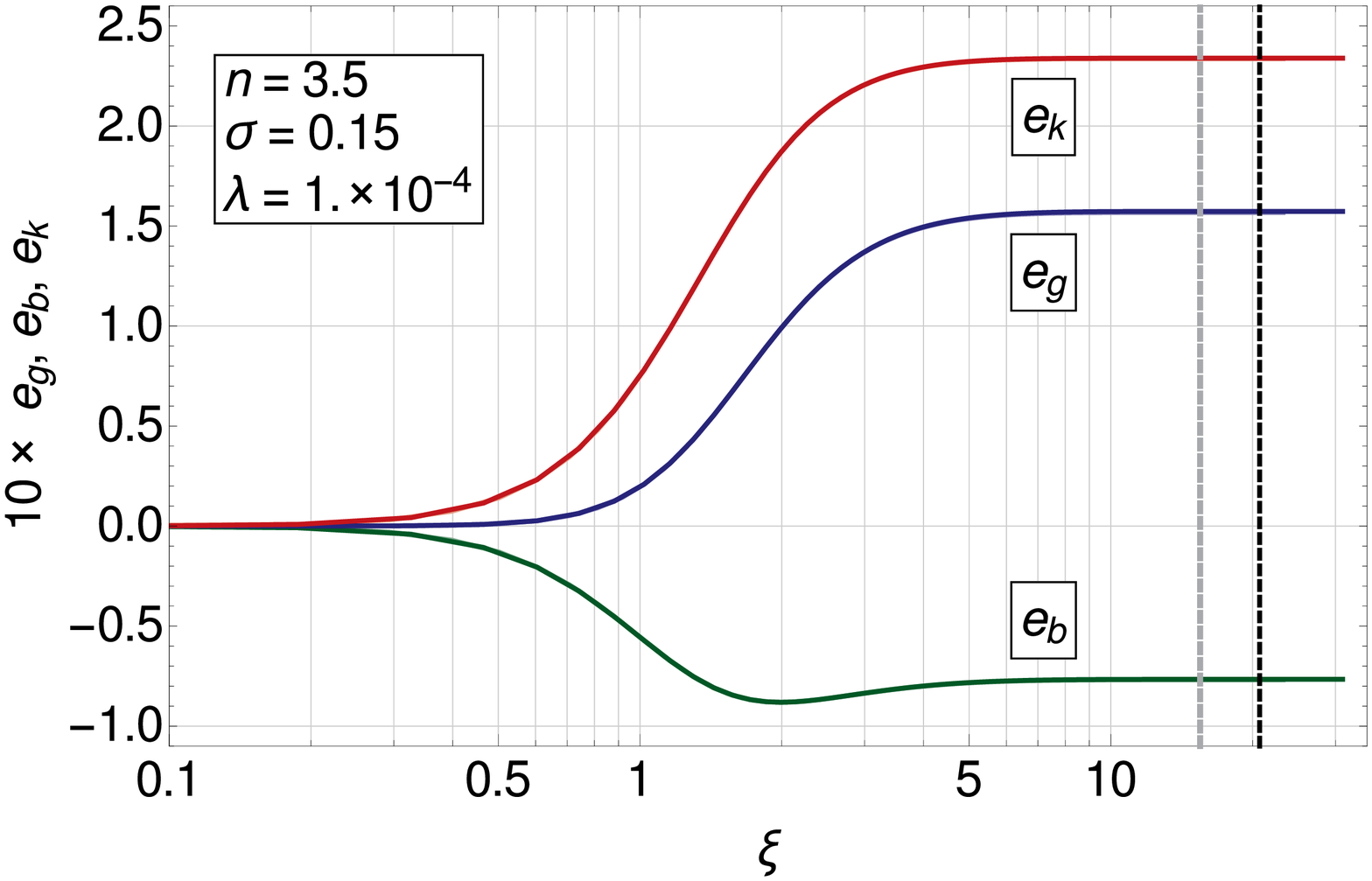}
\end{minipage}\hfill%
\begin{minipage}{0.32\linewidth}
\centering
\includegraphics[width=\linewidth]{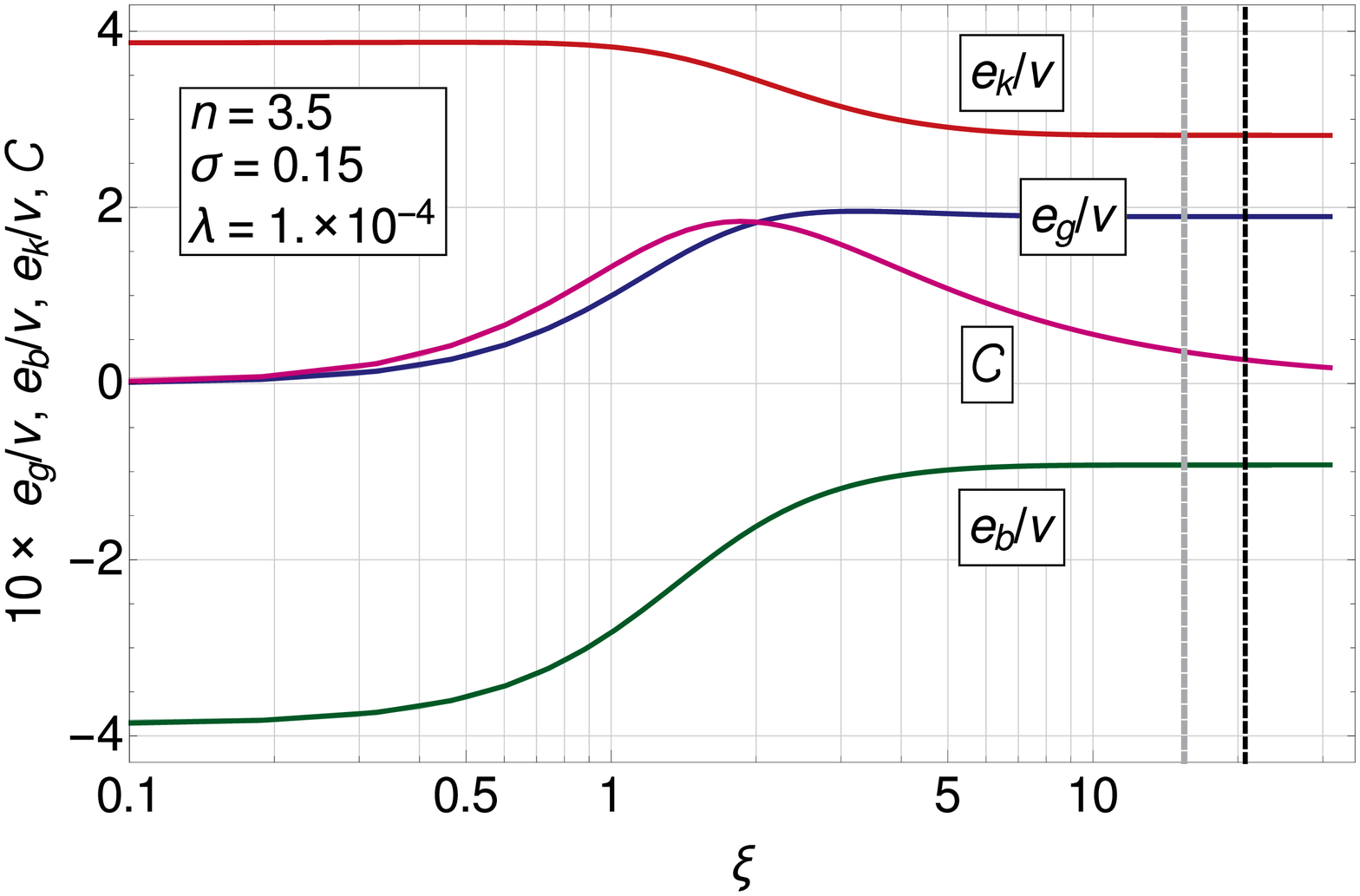}
\end{minipage}
\begin{minipage}{0.32\linewidth}
\centering
\includegraphics[width=\linewidth]{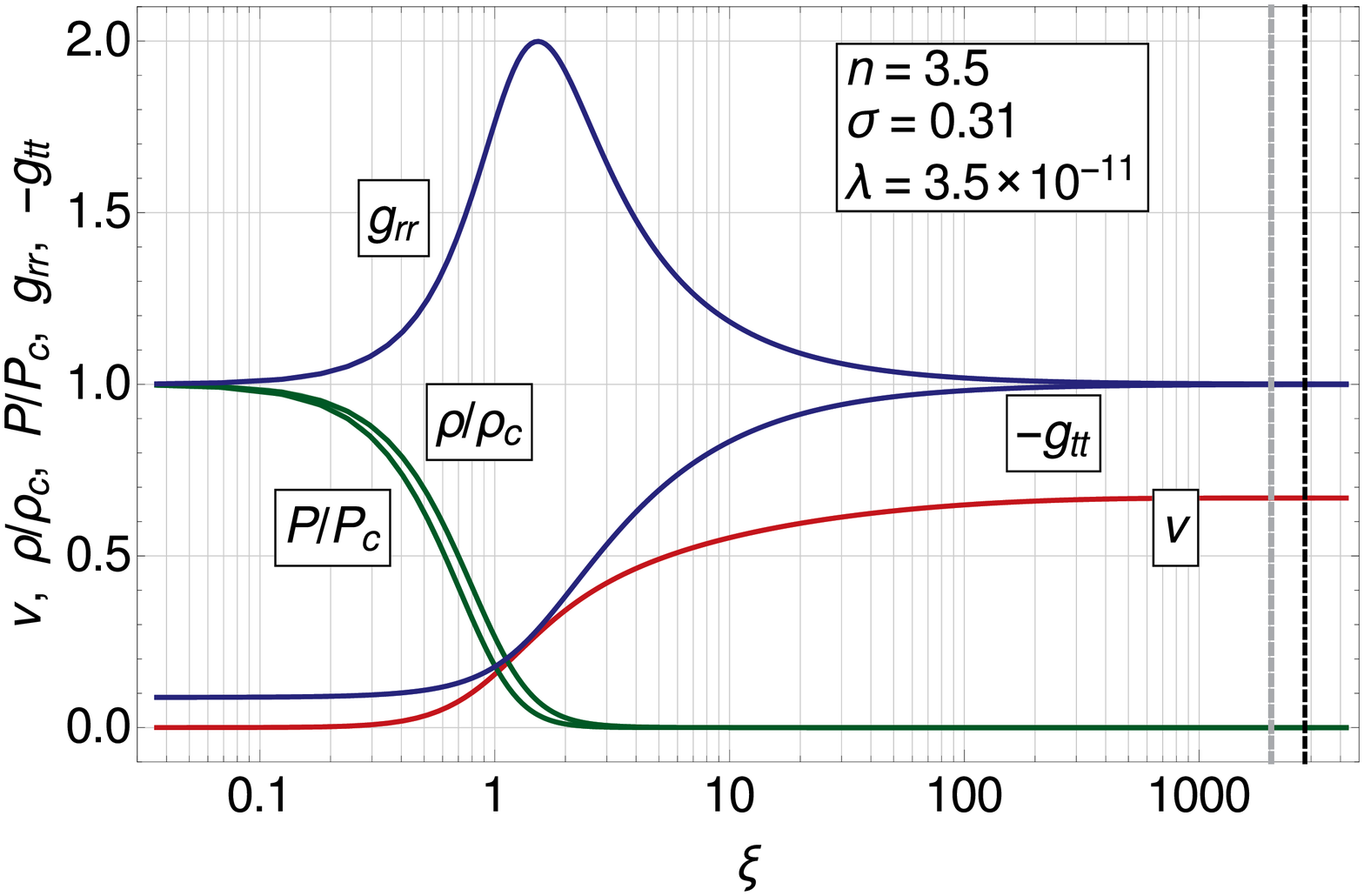}
\end{minipage}\hfill%
\begin{minipage}{0.32\linewidth}
\centering
\includegraphics[width=\linewidth]{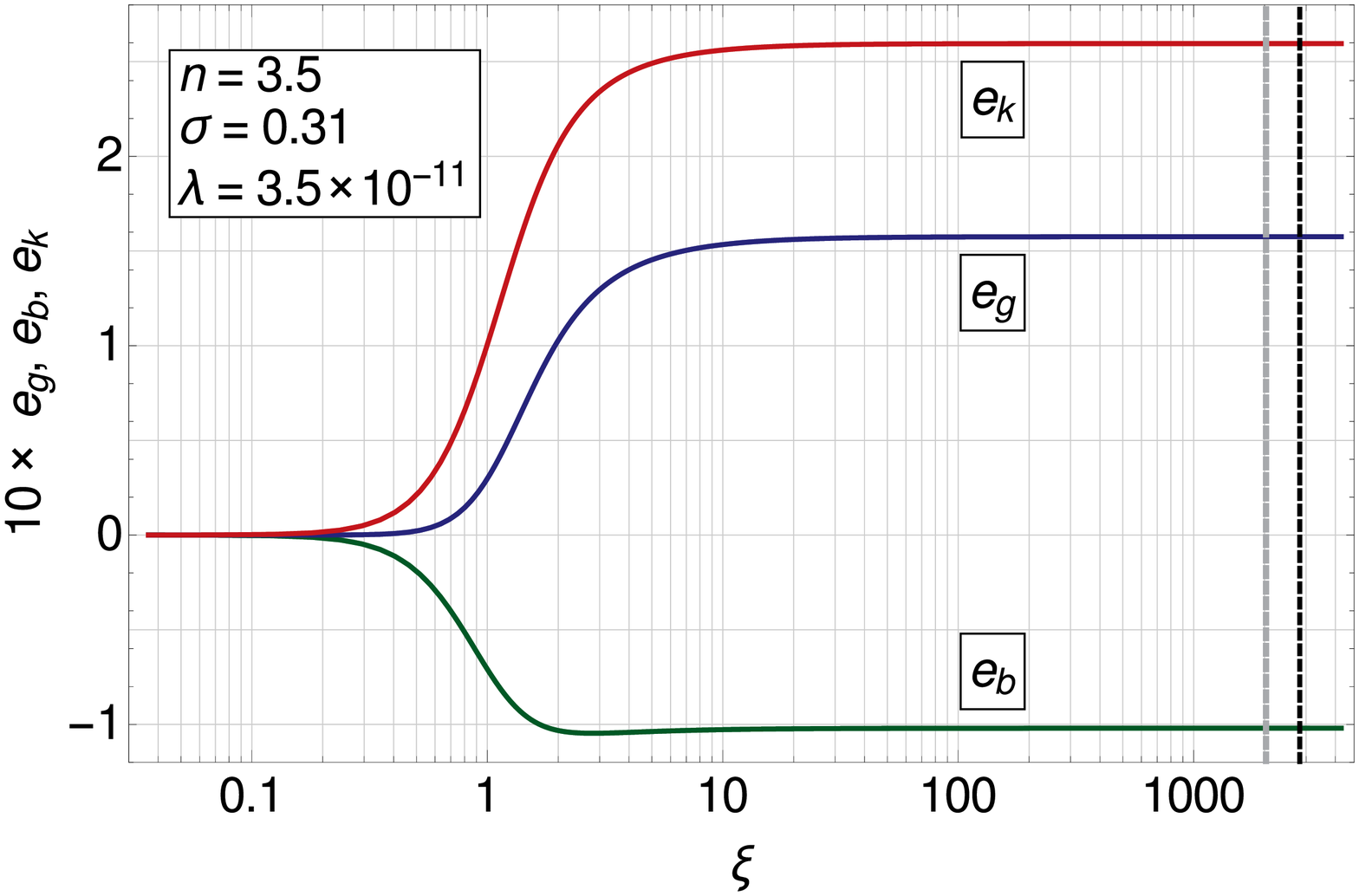}
\end{minipage}\hfill%
\begin{minipage}{0.32\linewidth}
\centering
\includegraphics[width=\linewidth]{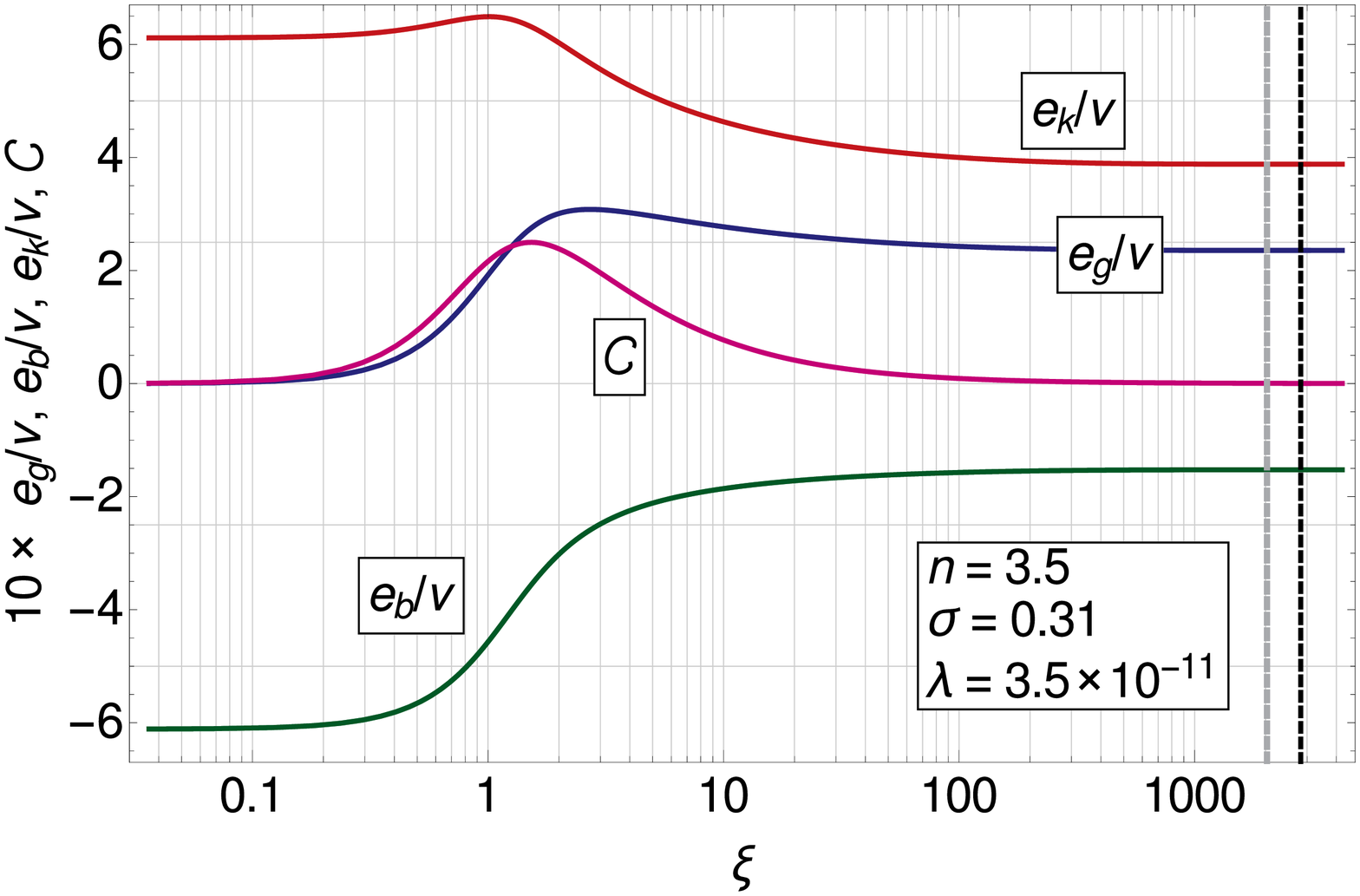}
\end{minipage}
\begin{minipage}{0.32\linewidth}
\centering
\includegraphics[width=\linewidth]{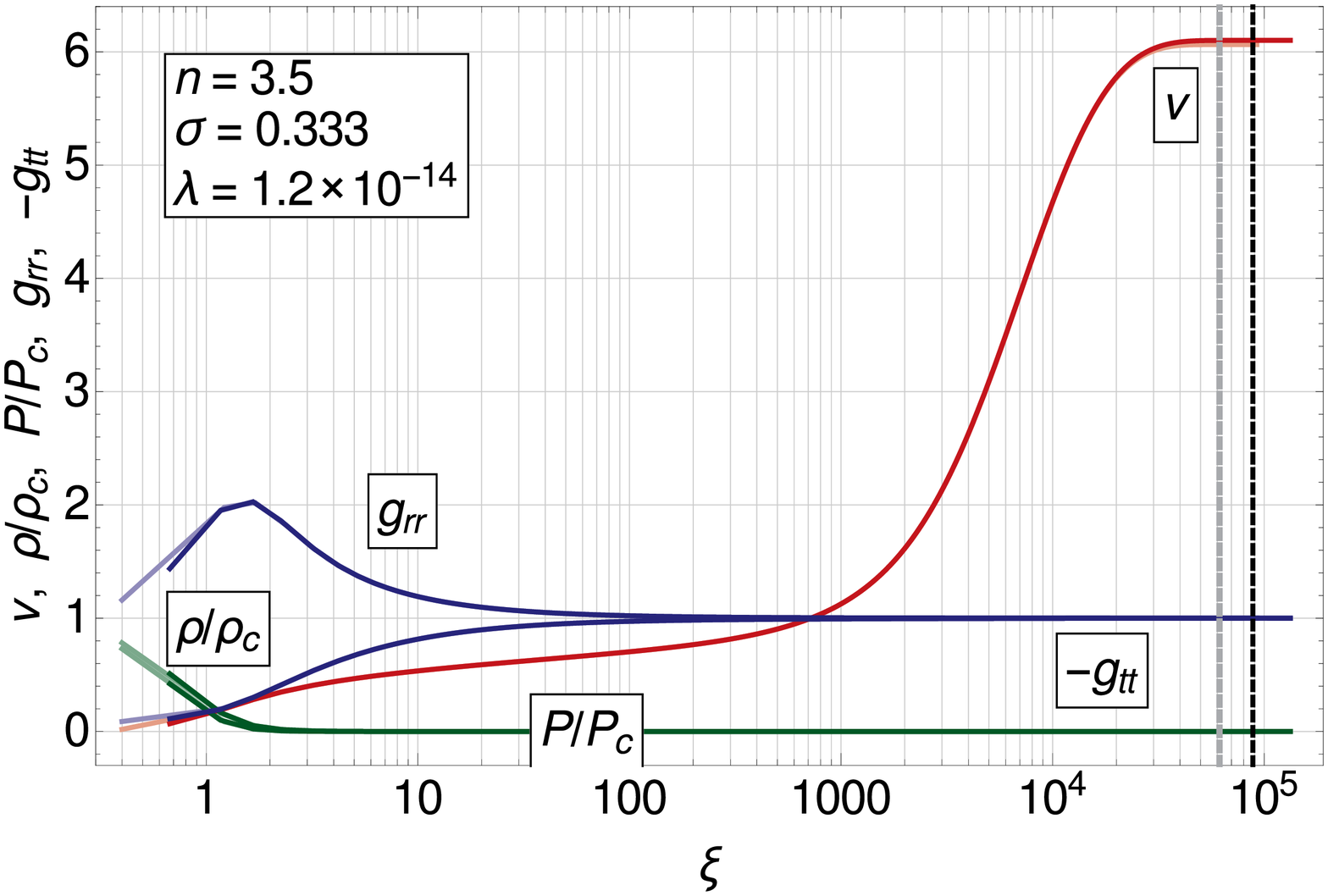}
\end{minipage}\hfill%
\begin{minipage}{0.32\linewidth}
\centering
\includegraphics[width=\linewidth]{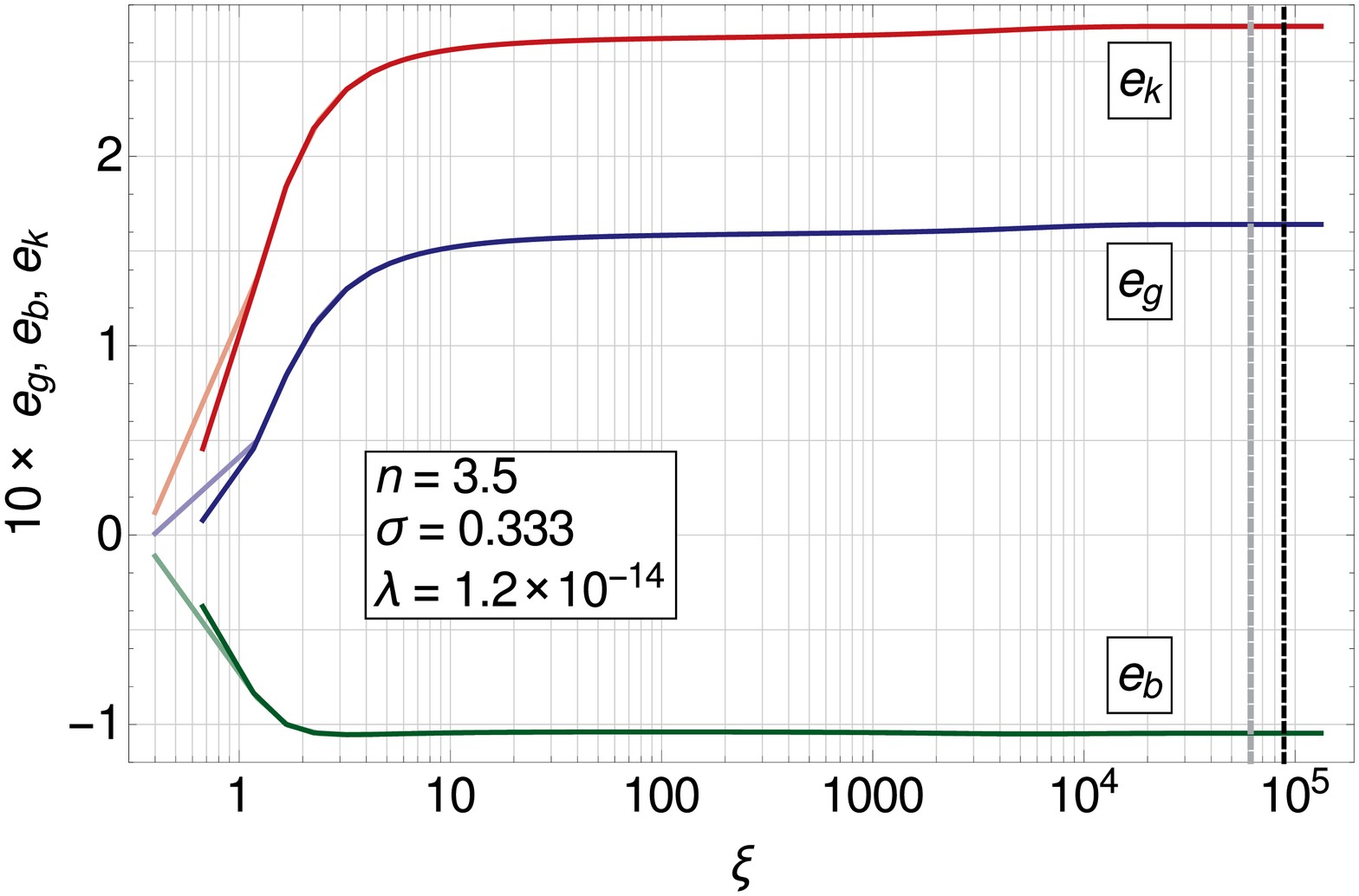}
\end{minipage}\hfill%
\begin{minipage}{0.32\linewidth}
\centering
\includegraphics[width=\linewidth]{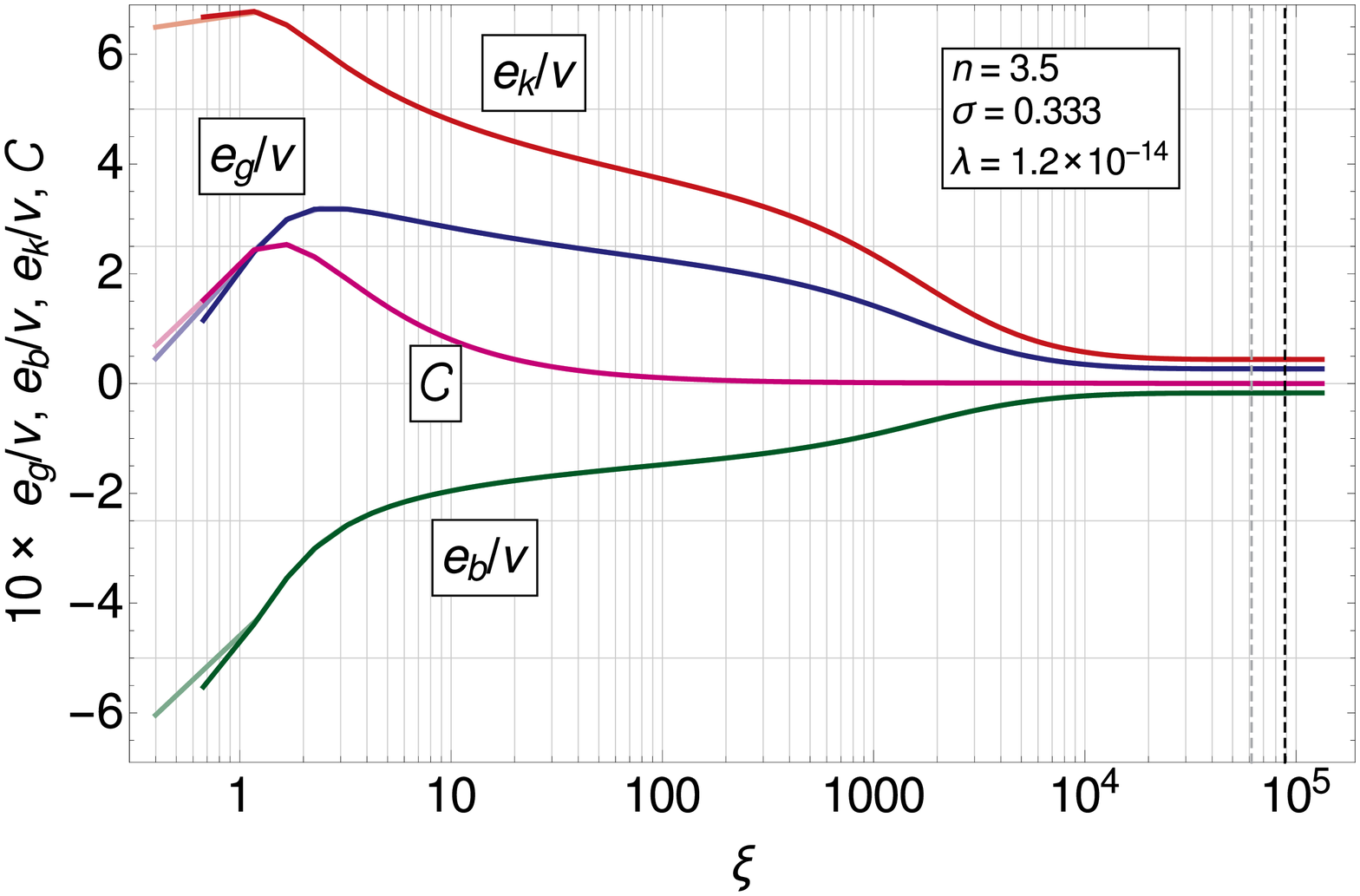}
\end{minipage}
\begin{minipage}{0.32\linewidth}
\centering
\includegraphics[width=\linewidth]{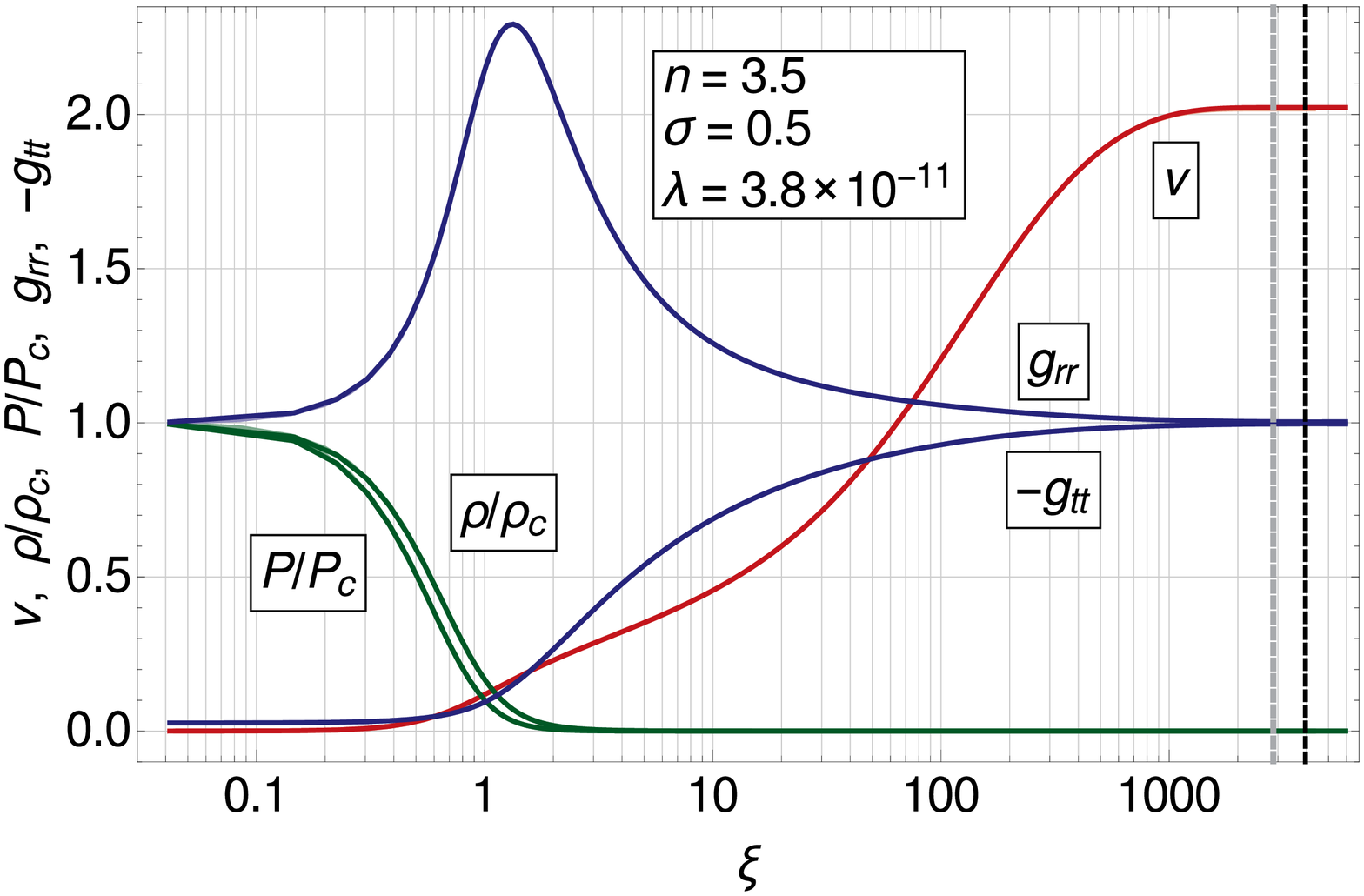}
\end{minipage}\hfill%
\begin{minipage}{0.32\linewidth}
\centering
\includegraphics[width=\linewidth]{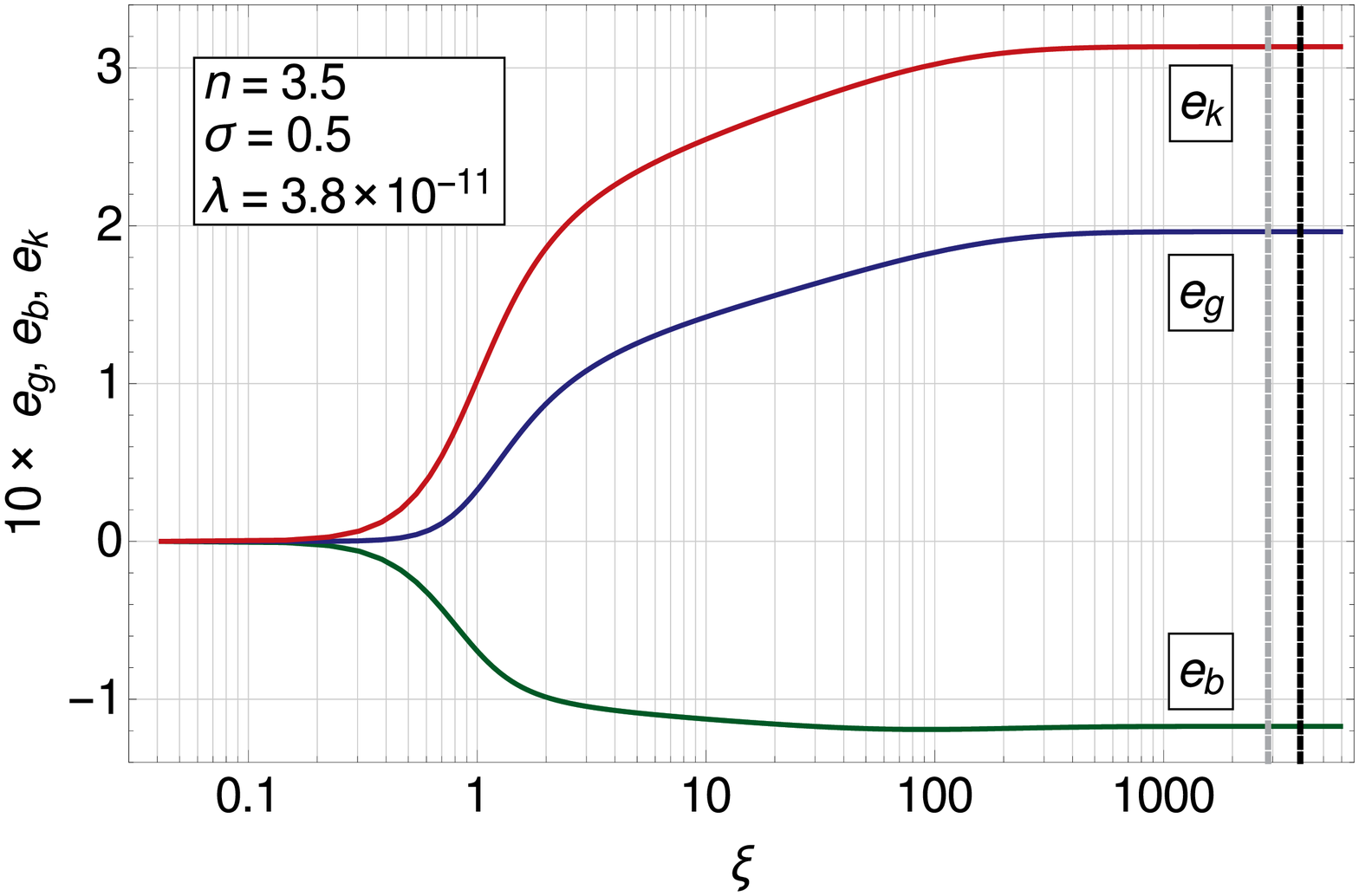}
\end{minipage}\hfill%
\begin{minipage}{0.32\linewidth}
\centering
\includegraphics[width=\linewidth]{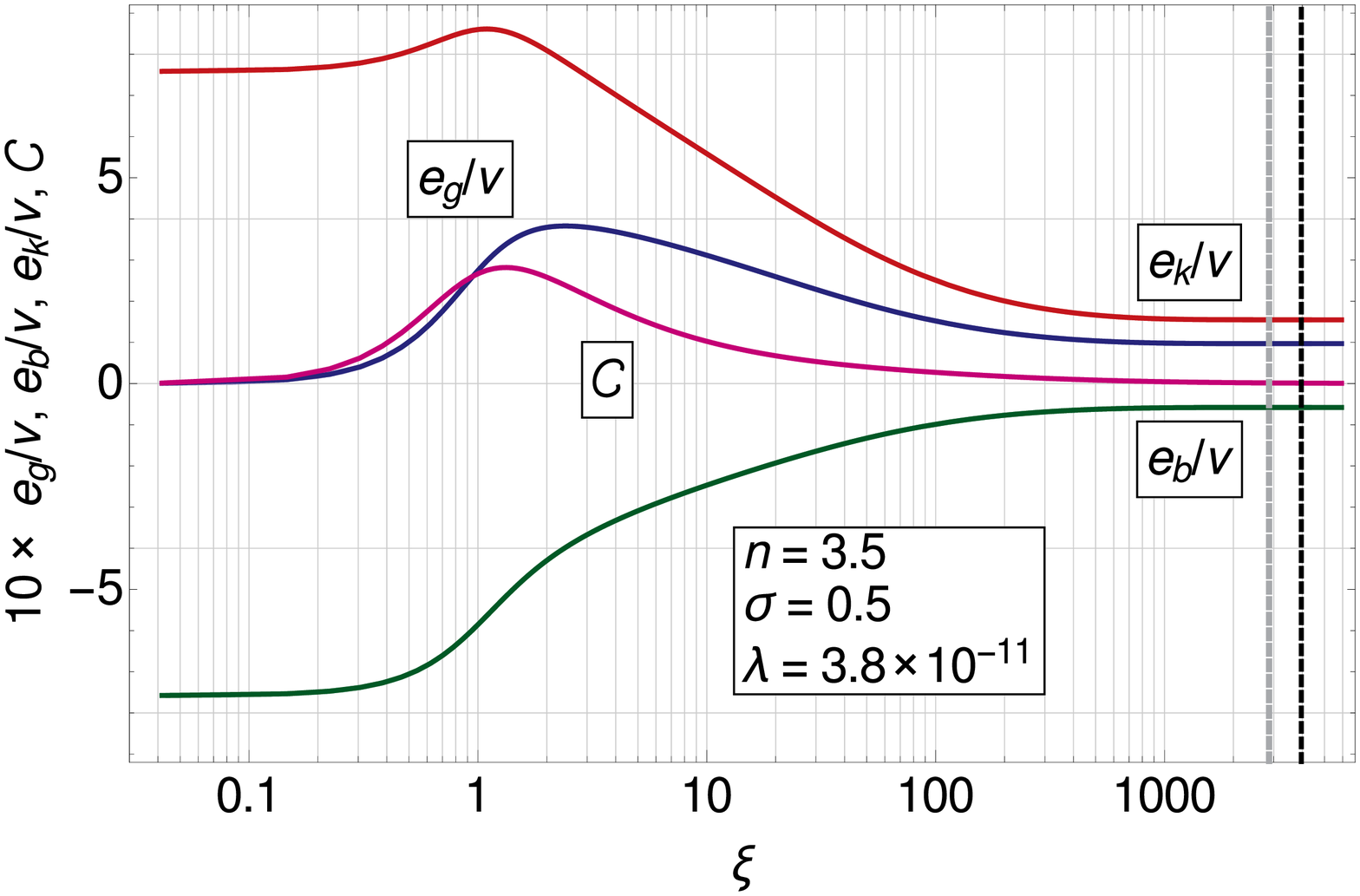}
\end{minipage}
\begin{minipage}{0.32\linewidth}
\centering
\includegraphics[width=\linewidth]{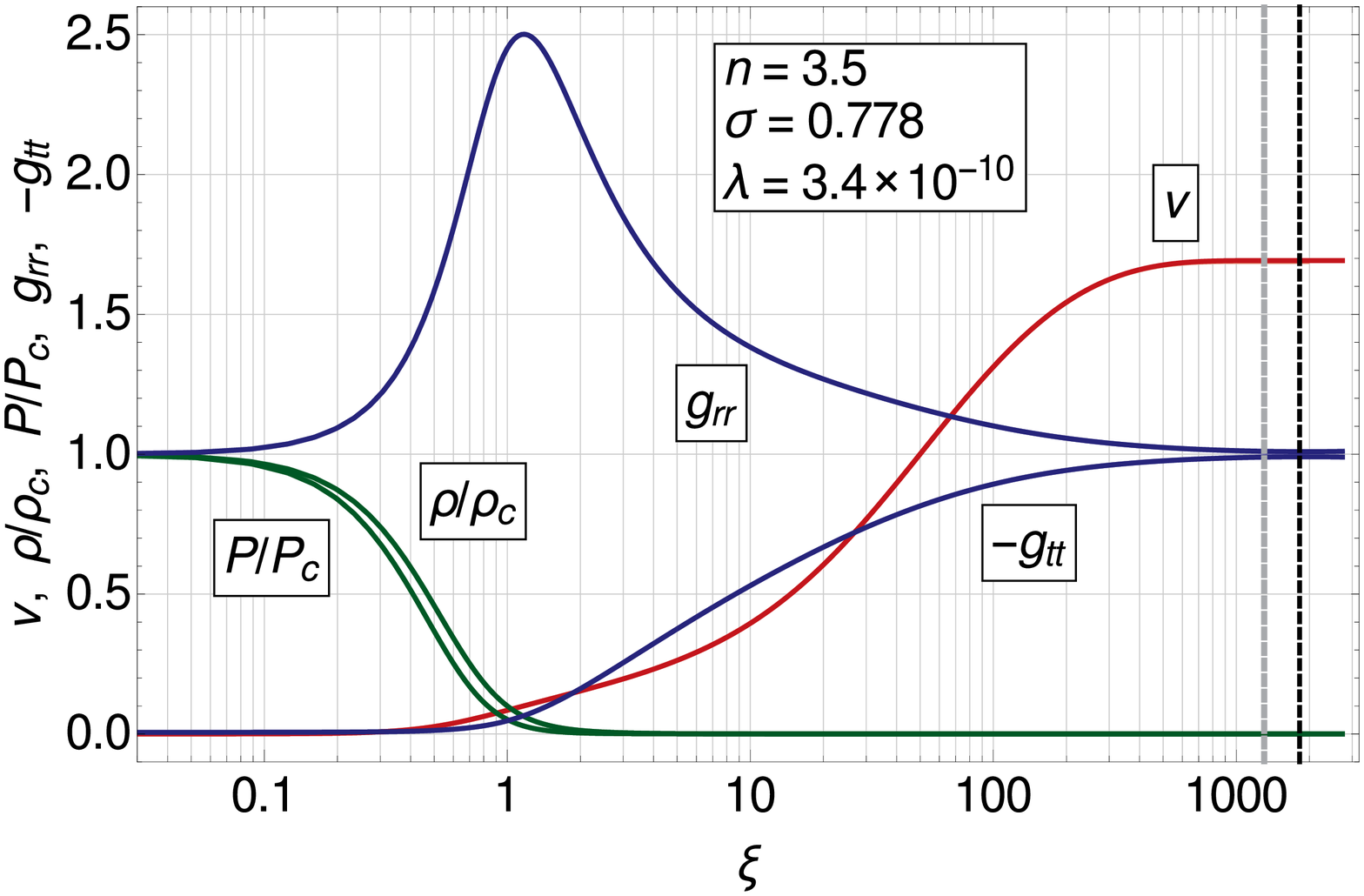}
\end{minipage}\hfill%
\begin{minipage}{0.32\linewidth}
\centering
\includegraphics[width=\linewidth]{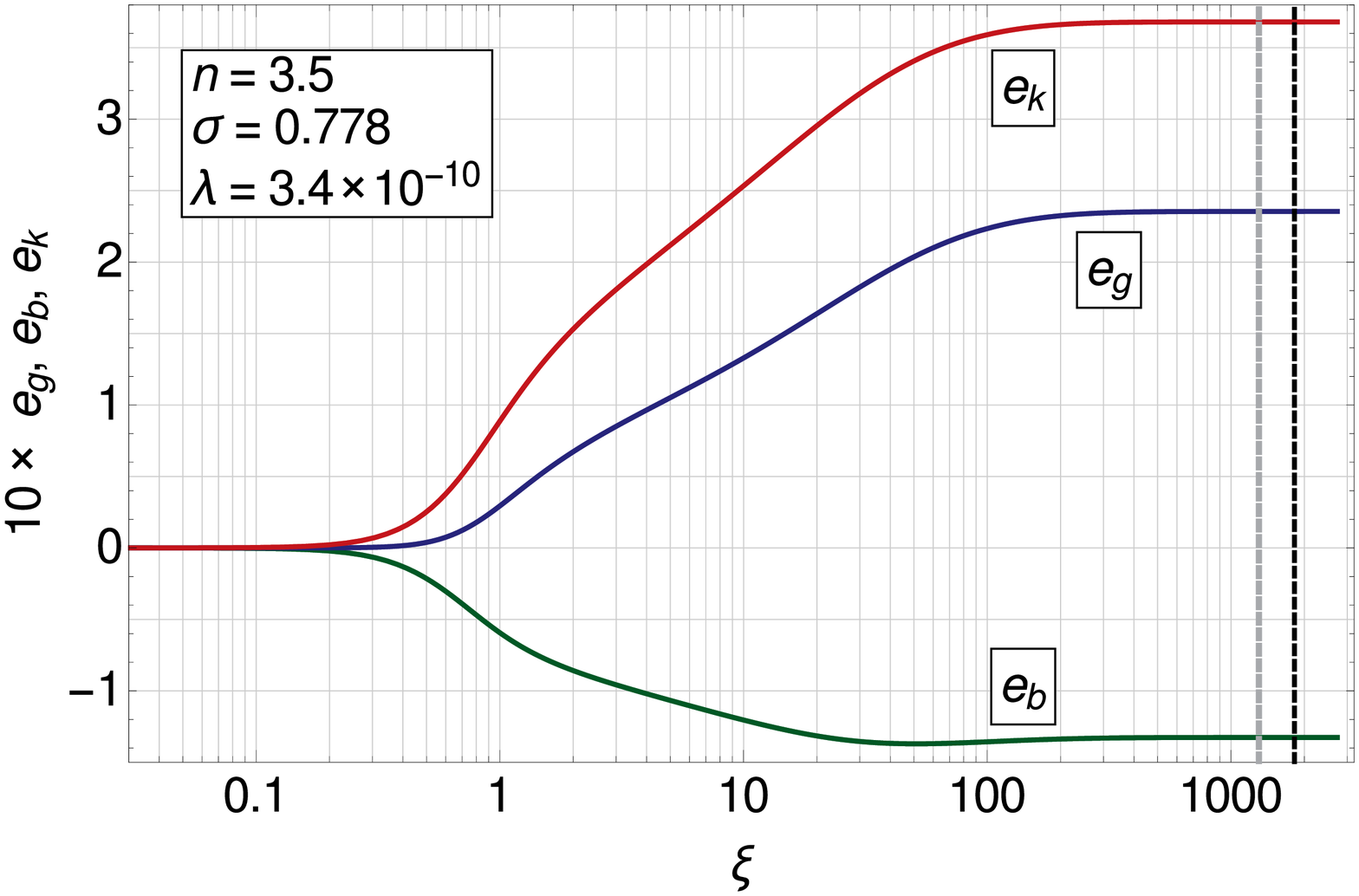}
\end{minipage}\hfill%
\begin{minipage}{0.32\linewidth}
\centering
\includegraphics[width=\linewidth]{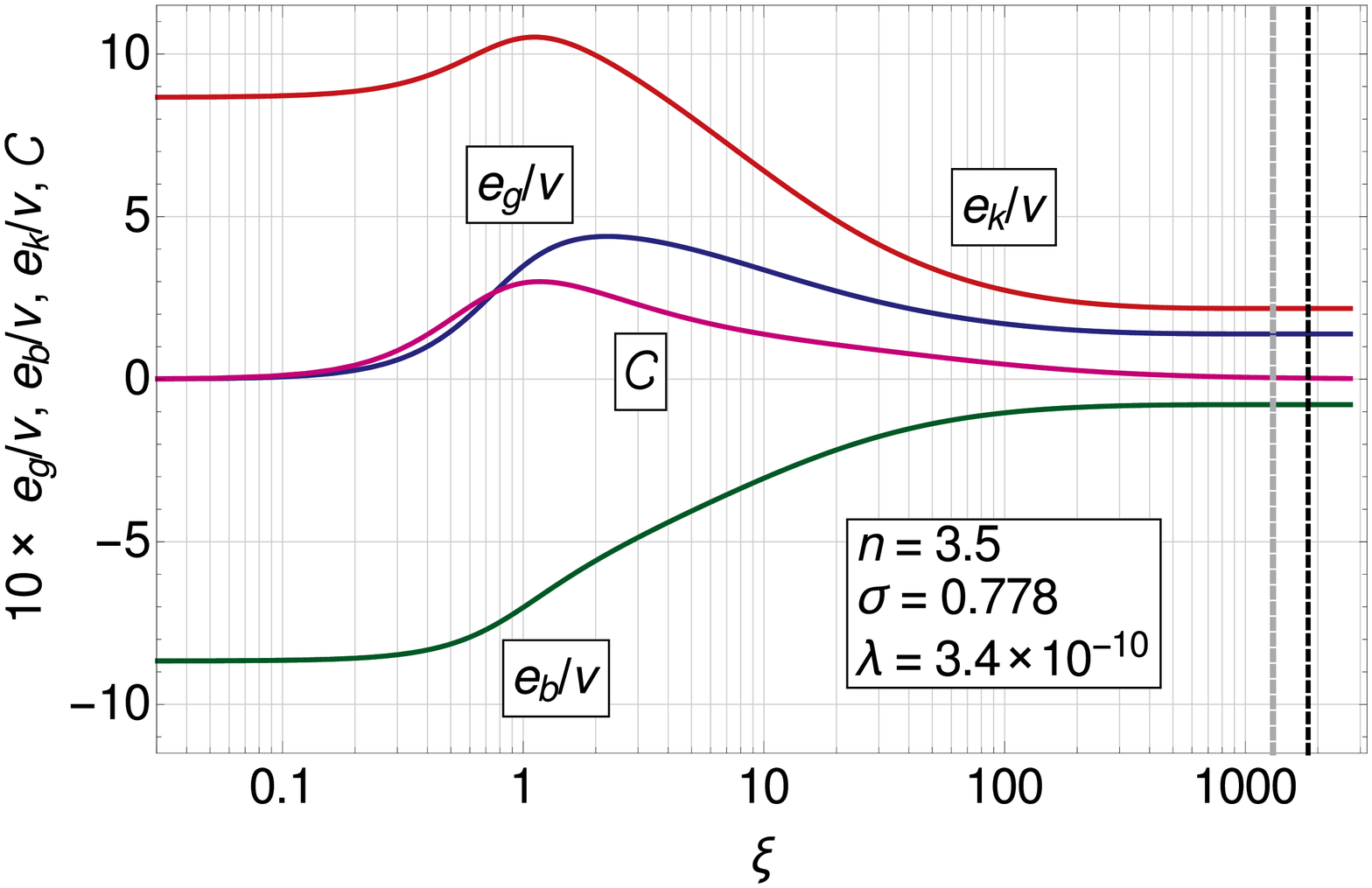}
\end{minipage}
\caption{\label{ProN35}Profile plots for polytropic index
  $n=3.5$. \textit{Left column:} Mass, density, pressure, and metric
  coefficients. \textit{Middle column:} Gravitational, binding and kinetic
  energy. \textit{Right column:} Relative gravitational, binding, kinetic
  energy and compactness.}
\end{figure*}

\begin{figure}[t]
\includegraphics[width=\linewidth]{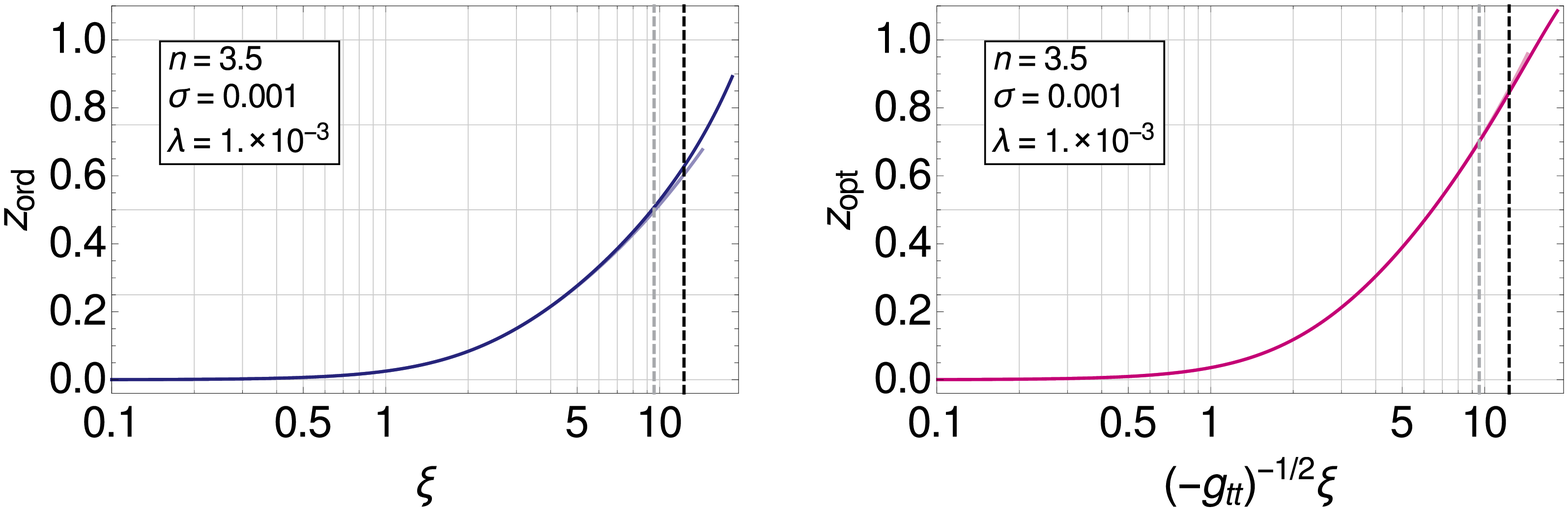}
\includegraphics[width=\linewidth]{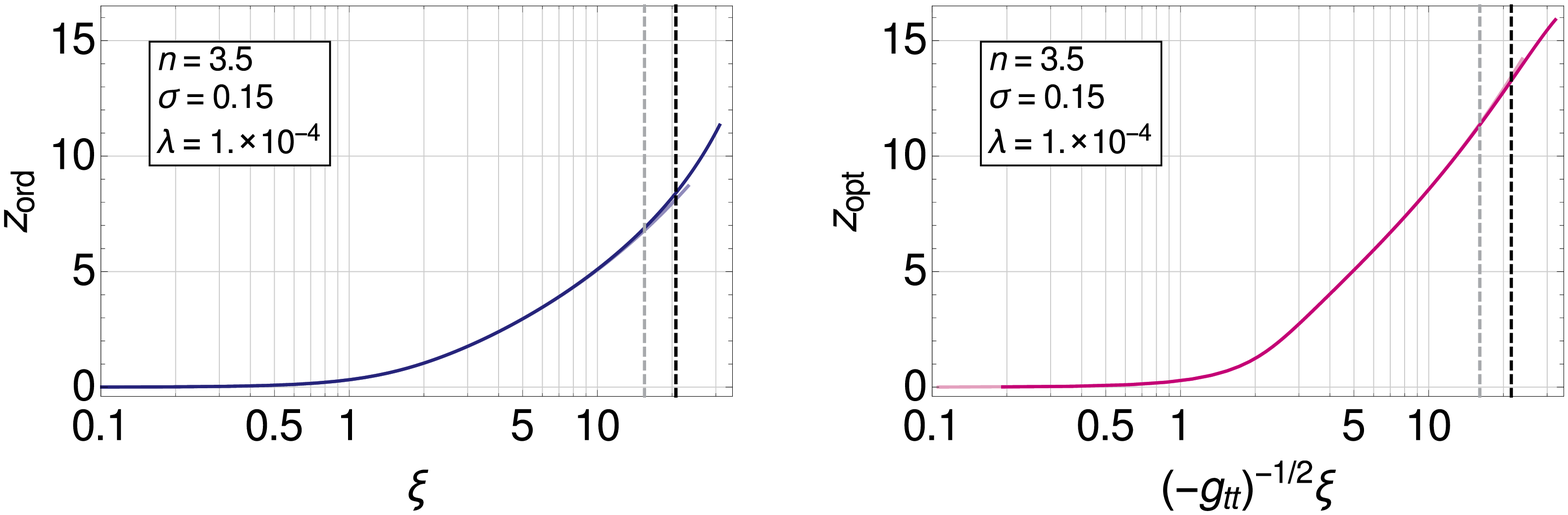}
\includegraphics[width=\linewidth]{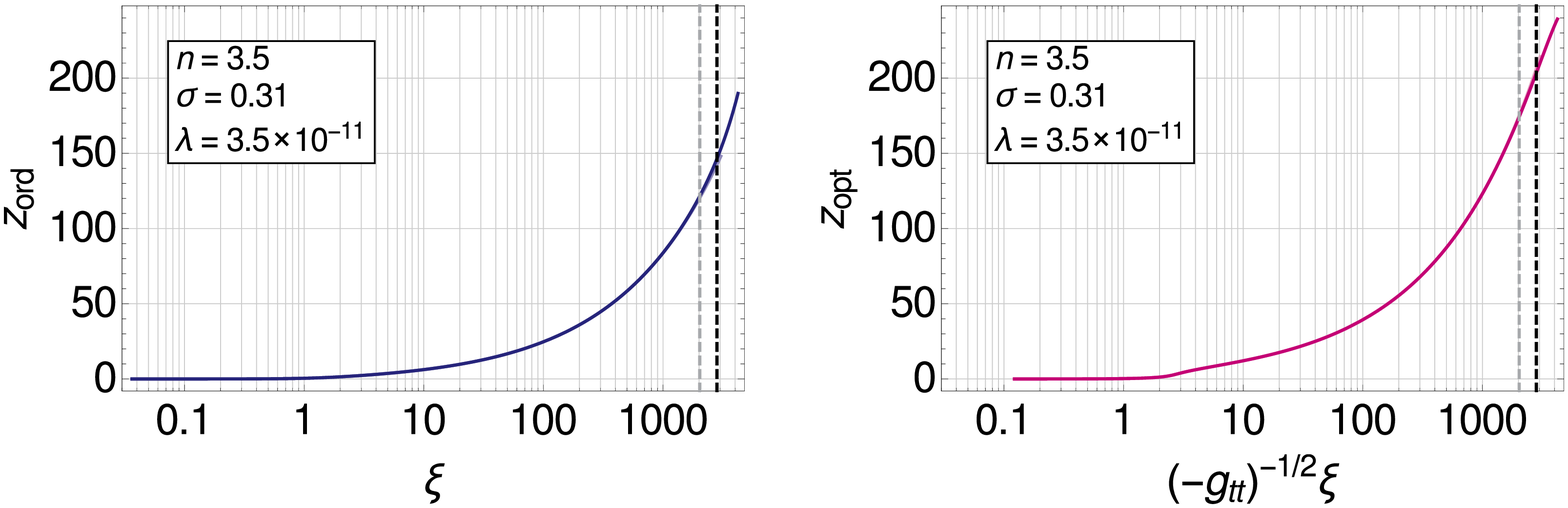}
\includegraphics[width=\linewidth]{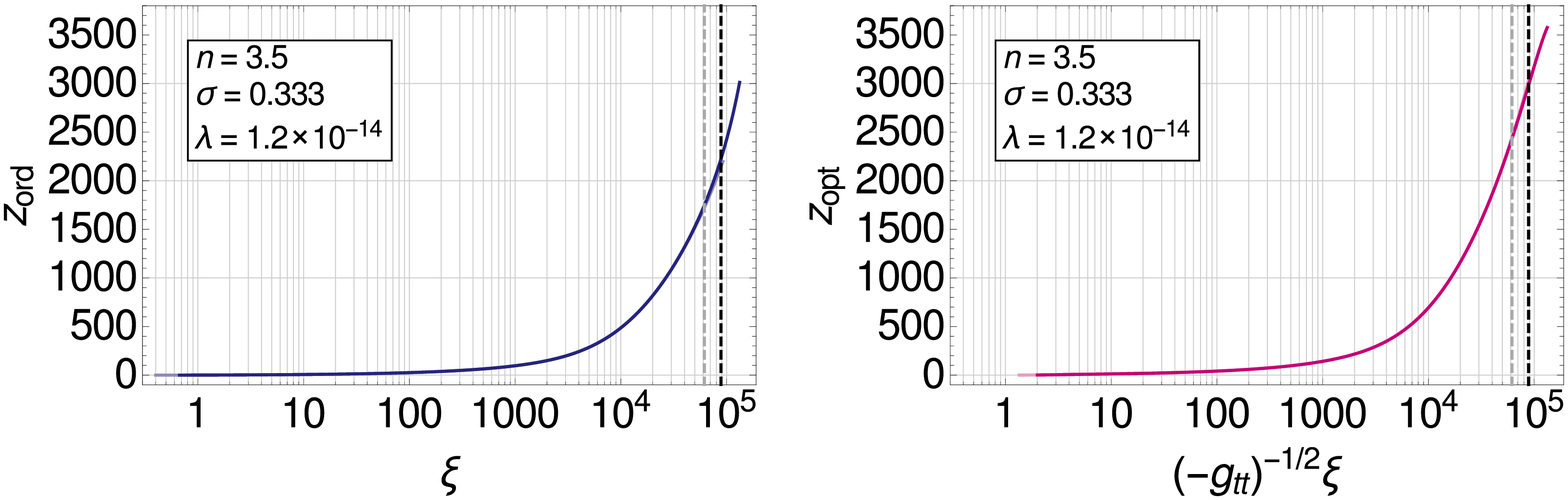}
\includegraphics[width=\linewidth]{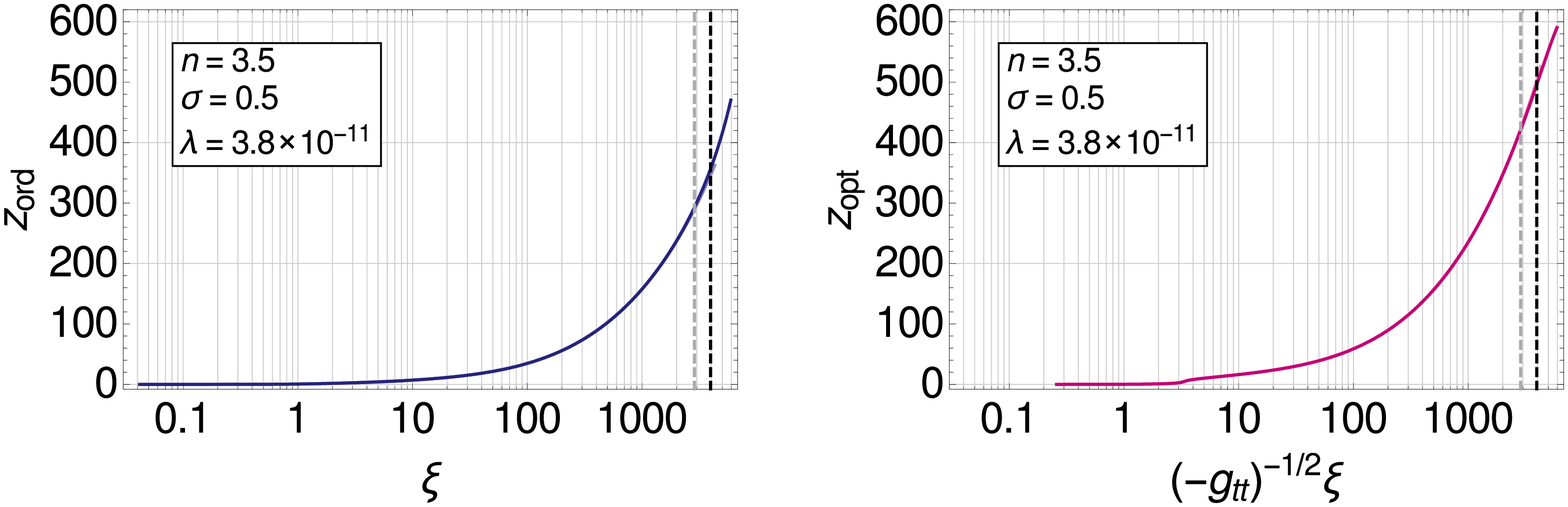}
\includegraphics[width=\linewidth]{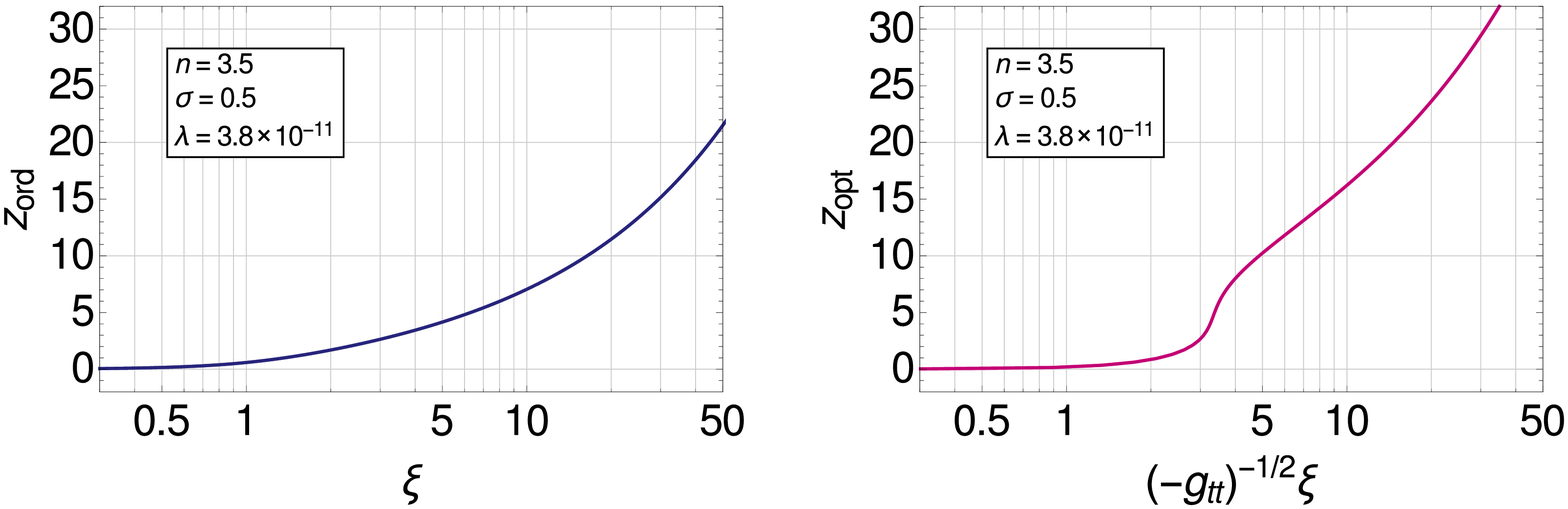}
\includegraphics[width=\linewidth]{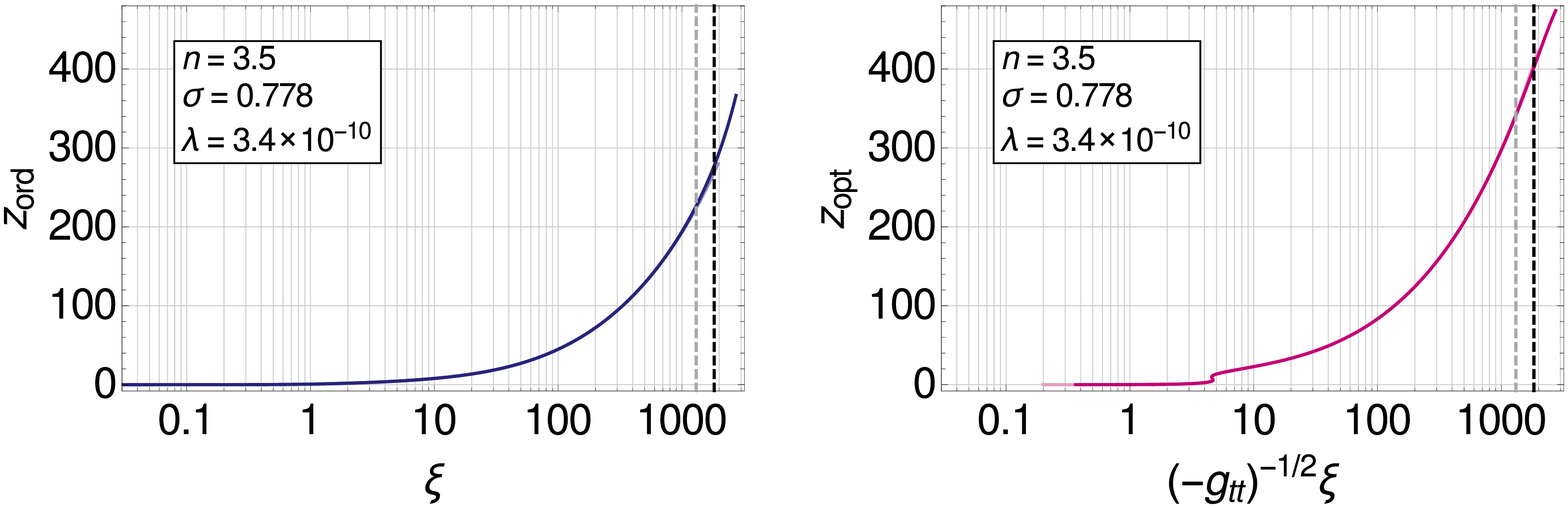}
\includegraphics[width=\linewidth]{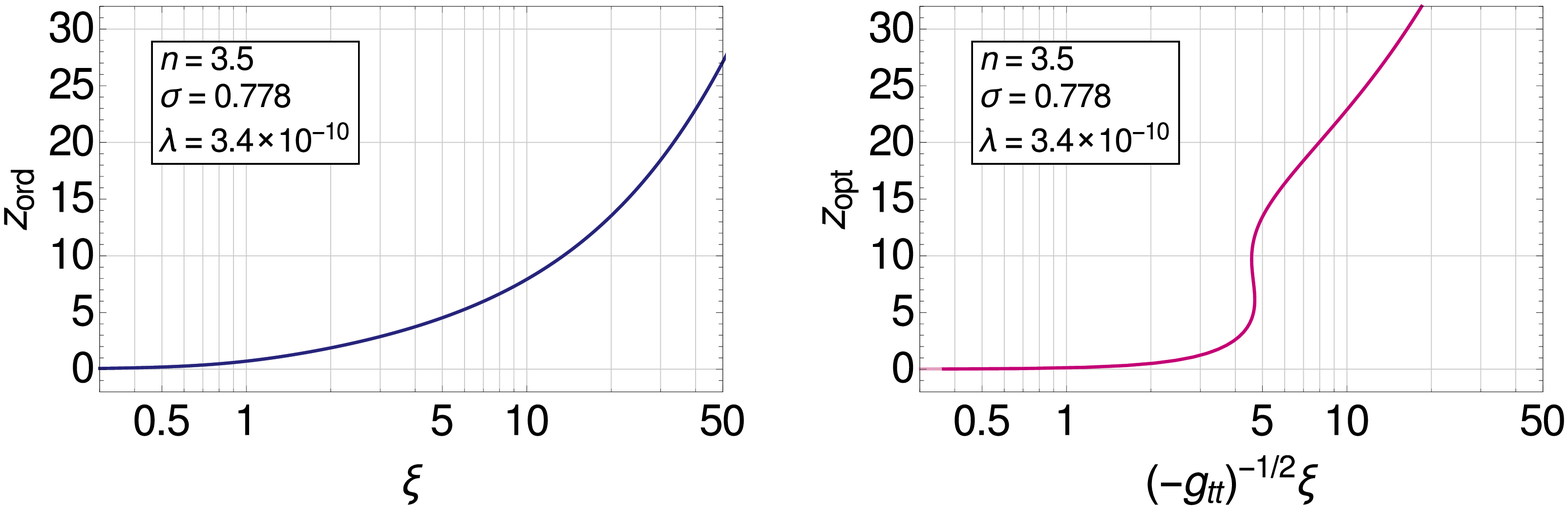}
\caption{\label{EmbN35}Embedding diagrams for polytropic index
  $n=3.5$.\textit{Left column:} Ordinary geometry. \textit{Right column:}
  Optical geometry. The S-shaped parts in fifth and seventh rows are repeated
  in zoomed form in sixth and eighth rows, respectively.}
\end{figure}

All the radial profiles and the embedding diagrams are given for the
polytropes with $n=0.5$ in Figs~\ref{ProN05} and~\ref{EmbN05}, respectively.
The polytrope $n=1.5$ case is reflected by Figs~\ref{ProN15} and~\ref{EmbN15},
respectively. The case of $n=3$ polytropes is illustrated in Figs~\ref{ProN30}
and~\ref{EmbN30}, respectively. The case of the $n=3.5$ polytropes is
represented in Figs~\ref{ProN35} and~\ref{EmbN35}, respectively, where the
profiles are given for the parameter $\sigma$ chosen on both sides of the
critical value of $\sigma_{\mathrm{f}}$.

\subsection{Properties of the profiles}

The dimensionless extension parameter $\xi_{1}$ increases with increasing
polytropic index $n$ and decreases with increasing relativistic parameter
$\sigma$, while the mass parameter $v_{1}=v(\xi_{1})$ decreases with
increasing polytropic index $n$, and increasing parameter $\sigma$. Notice
that in the polytropes with large values of $\xi_{1}$, the density (and
pressure) radial profiles are strongly decreasing near the origin $\xi=0$,
reaching values $\sim 0$ at $\xi \sim 2$ and decreasing exponentially slowly
while approaching $\xi_{1}$.  Similar behavior can be observed in such
polytropes also for the mass parameter $v(\xi)$ that is nearly equal to its
final value $v_{1}$ starting from $\xi \sim 2$.  We can see that for the
non-relativistic or slightly relativistic polytropic configurations (having
$\sigma < 0.2$), the dimensionless parameters determining the polytropic
sphere, $\xi_{1}$ and $v_{1}$, are of the order of $1$.  For $n \leq 3$
polytropes with the relativistic parameter increasing, the mass parameter
$v_{1}$ takes values smaller than $1$, while the extension parameter $\xi_{1}$
increases substantially, exceeding $1$, even by orders, for large values of
the relativistic parameter, comparable to the value of causality limit.  A
special behavior demonstrate polytropes having high values of the polytropic
index, e.g., $n=3.5$, especially for values of $\sigma$ close to the critical
values of $\sigma_{\mathrm{f}}$. Such configurations can have extremely large
extension parameter $\xi_{1}$, and mass parameter $v_{1} > 1$ even for large
values of the relativistic parameter, $\sigma > \sigma_{\mathrm{f}}$.  Recall
that at the critical values of the relativistic parameter,
$\sigma_{\mathrm{f}}$, static equilibrium polytropic configurations are not
well defined for $\lambda=0$, while any non-zero value of the cosmological
parameter cuts out the polytropic configurations with
$\sigma \sim \sigma_{\mathrm{f}}$---see Fig.\,\ref{SigLamCrit}.

As can be intuitively expected, the metric coefficients are nearly constant,
$g_{rr} \sim 1$ and $-g_{tt} \sim 1$, for very small values of the
relativistic parameter, $\sigma < 0.01$, being slightly dependent on the
polytropic index $n$; such spacetimes are nearly flat, demonstrating clearly
that the relativistic parameter $\sigma$ governs intensity of the general
relativistic effects in the polytropes. For $\sigma > 0.1$, the general
relativistic effects described by the metric coefficients $g_{rr}$ and
$g_{tt}$ become significant as the metric coefficients significantly vary
inside the polytrope. Outside the polytropes with
$\lambda \sim \lambda_{\mathrm{crit}}$, the gravitational field varies
strongly for small values of the polytropic index $n$ and large enough
parameter $\sigma > 0.1$, as the polytropes are compact enough while having
their surface located near the static radius. For such polytropes also the
radial profiles are strongly influenced by the cosmic repulsion (parameter
$\lambda$), as demonstrated in Figs~\ref{ProN05}--\ref{EmbN35}. With
increasing polytropic index, the polytropes with
$\lambda \sim \lambda_{\mathrm{crit}}$ demonstrate suppression of the role of
the cosmic repulsion in the character of the radial profiles. This suppression
is evident especially in the case of the $n=3.5$ polytropes with
$\sigma > \sigma_{\mathrm{f}}$ having large extension parameter $\xi_{1}$ and
$v_{1} >1$.

The magnitude of the gravitational binding energy is positive everywhere in
the polytrope, similarly to the kinetic energy of the polytrope. The binding
energy is negative in the central parts of the polytrope and becomes to be
positive in the outer region of the polytrope for whole the allowed interval
of $\sigma$, if $n=0.5, 1.5$. However, such a behavior occurs in the
polytropes with $n=3$ only for appropriately low values of $\sigma$---the
binding energy is negative at all radii of such polytropes for large enough
relativistic parameter. The critical value of $\sigma$ for altering the mixed
to fully negative radial profile of the binding energy strongly decreases with
increasing $n$, being $\sim 10^{-2}$ for $n=3$. Of course, the same properties
are valid for the gravitational, kinetic and binding energy related to the
dimensionless gravitational mass $v$ of the polytropes.

The embedding diagrams of the ordinary projected space give an illustration of
the curvature of the space inside the polytrope. We can clearly see that the
curvature of the ordinary space increases slightly with increasing polytropic
index $n$, but it increases very strongly with the relativistic parameter
$\sigma$ increasing while $n$ is fixed. The embedding diagrams of the optical
space can be extremely useful for understanding the properties of the
polytropes related to the possibility of existence of extremely curved regions
containing trapped null geodesics. Existence of such regions is indicated by
the radial profile of the optical space demonstrating two turning points that
could occur even deeply inside the polytrope, although no effect of this kind
has to be related to the external characteristics of the polytrope, determined
by its dimensionless radius $\xi_{1}$ and dimensionless mass $v_{1}$. Clearly
such extremely curved regions can occur only in the highly relativistic
spacetimes with sufficiently high values of the relativistic parameter
$\sigma$ related to the polytropes with high values of the polytropic index,
$n \geq 2$. For such GRPs with extremely curved regions, the global
compactness factor does not demonstrate the extremal compactness since such
GRPs have largely extended low density regions near their
surfaces. Technically this means that $\xi_{1}\gg 1$ and $v(\xi_{1})\sim 1$ so
that the global compactness drops down to $\mathcal{C} << 1/3$.

Concerning the effects of the cosmological constant, they can be clearly
important only in the extremely low-dense polytropic configurations, having
very small central density and high enough cosmological parameter
$\lambda$. We can state that quite generally, the influence of the
cosmological constant always increases the values of the extension and mass
parameters of the polytrope, its metric coefficients or the magnitude of the
gravitational energy. The influence of the cosmic repulsion on the structure
of polytropic configurations with the index $n \leq 3$ can be relevant for
relatively large values of $\lambda > 10^{-7}$ when some observable effects
could be expected, especially in the low density polytrope configurations; we
have convinced ourselves that the influence of the cosmological parameter is
negligible for $\lambda < 10^{-7}$. Such GRPs could be relevant and applicable
for very massive and very extended objects with very low density. The
influence of the cosmological constant on the extension, mass and the radial
profiles can be very large for the polytropes with low values of the
polytropic index, especially for $n=0.5$. On the other hand, for the
polytropes with $n=3,3.5$ the influence of the cosmological constant is strong
in the case of their extension, but it is small in the case of the mass
parameter and the radial profiles of all quantities for polytropes with very
small values of $\sigma$, and it is even negligible for $\sigma > 0.1$.

It is quite natural to consider the possibility to model dark matter halos as
polytropic spheres with $n=0.5$ or $n=1.5$, and test the role of the repulsive
cosmological constant in situations when
$\lambda \sim \lambda_{\mathrm{crit}}$. The large enough cosmological
parameter significantly restricts the polytropes in dependence on the
relativistic parameter $\sigma$. In the case of the polytropes with index
$n=3,3.5$, the situation is more complex, as these polytropes are influenced
in their extension by any $\lambda>0$ in the vicinity of the critical values
of the relativistic parameter $\sigma_{\mathrm{f}}$. Moreover, for values of
$\lambda$ large enough, existence of the polytropes is forbidden---the $n=4$
polytropes cannot exist for $\lambda > 10^{-3}$. Further, we can conclude that
the cosmological constant is irrelevant for very dense polytropes with high
central densities and extremely small cosmological parameter, except the
effect of restricting the extension of the polytropes with the relativistic
parameter $\sigma \sim \sigma_{\mathrm{f}}$.

\section{\label{polyrad}Polytrope radius modified by the cosmic repulsion}

\begin{figure*}[t]
\begin{minipage}{0.48\linewidth}
\centering
\includegraphics[width=\linewidth]{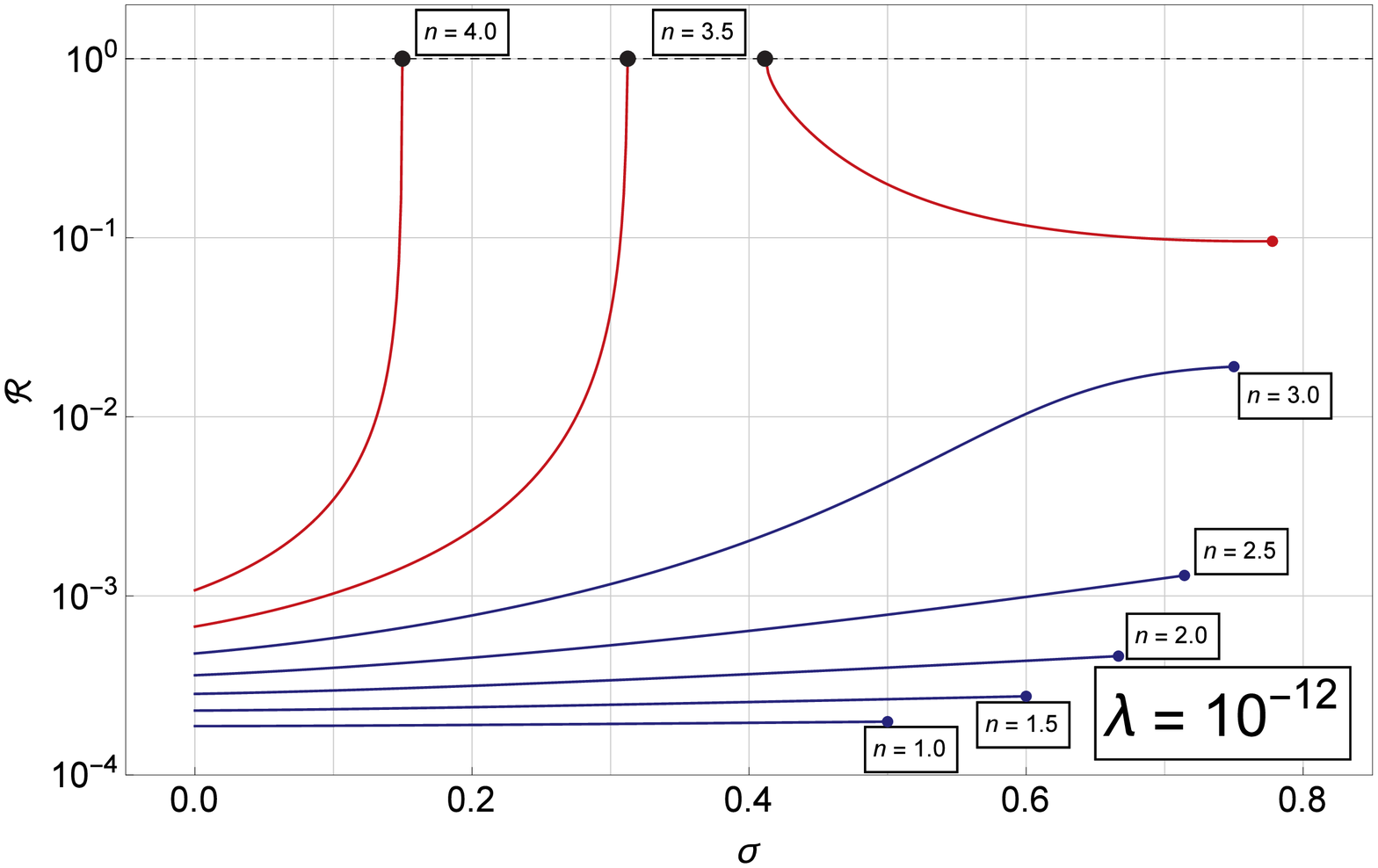}
\end{minipage}\hfill%
\begin{minipage}{0.48\linewidth}
\centering
\includegraphics[width=\linewidth]{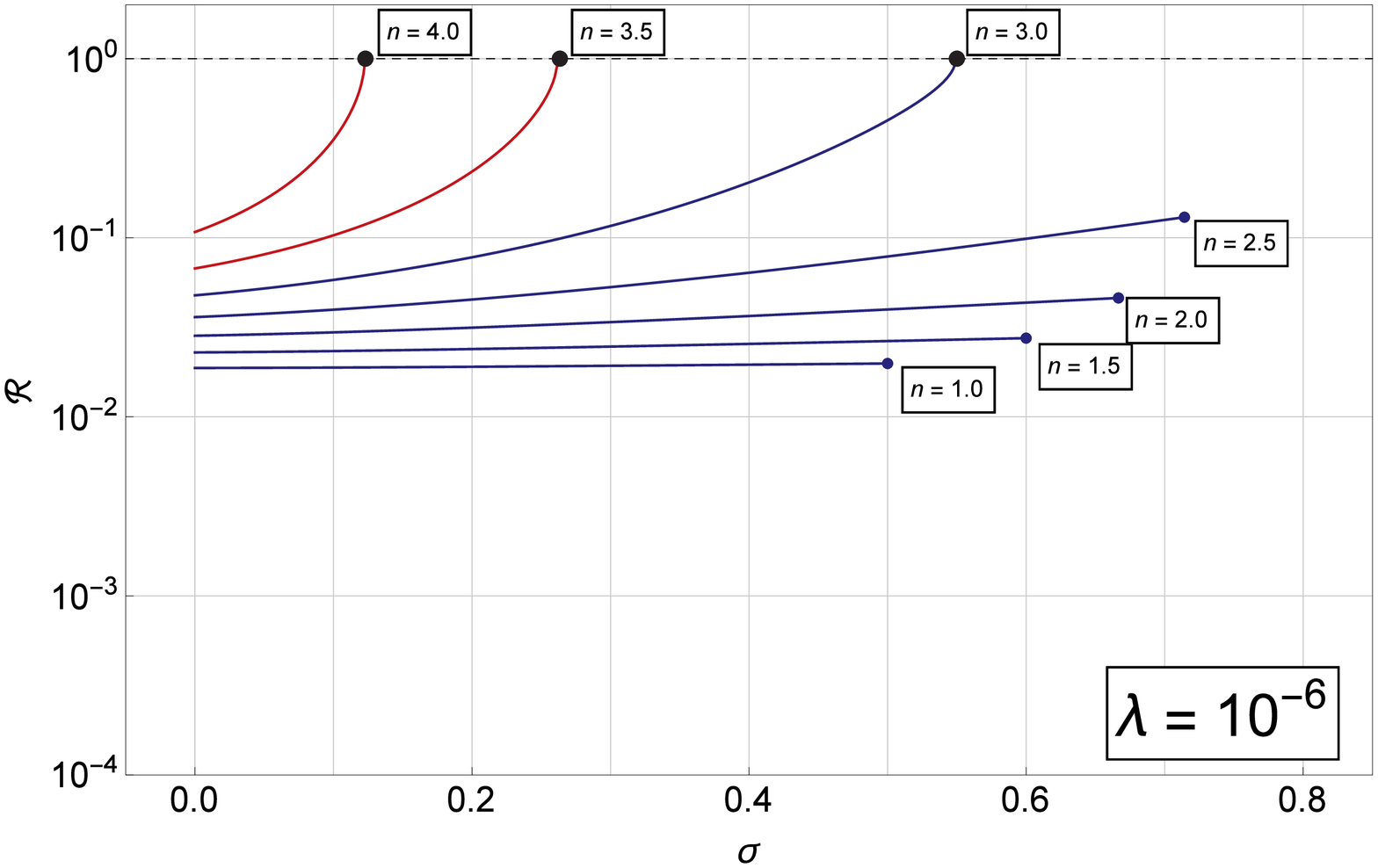}
\end{minipage}
\par\vspace{1.5\baselineskip}\par
\begin{minipage}{0.48\linewidth}
\centering
\includegraphics[width=\linewidth]{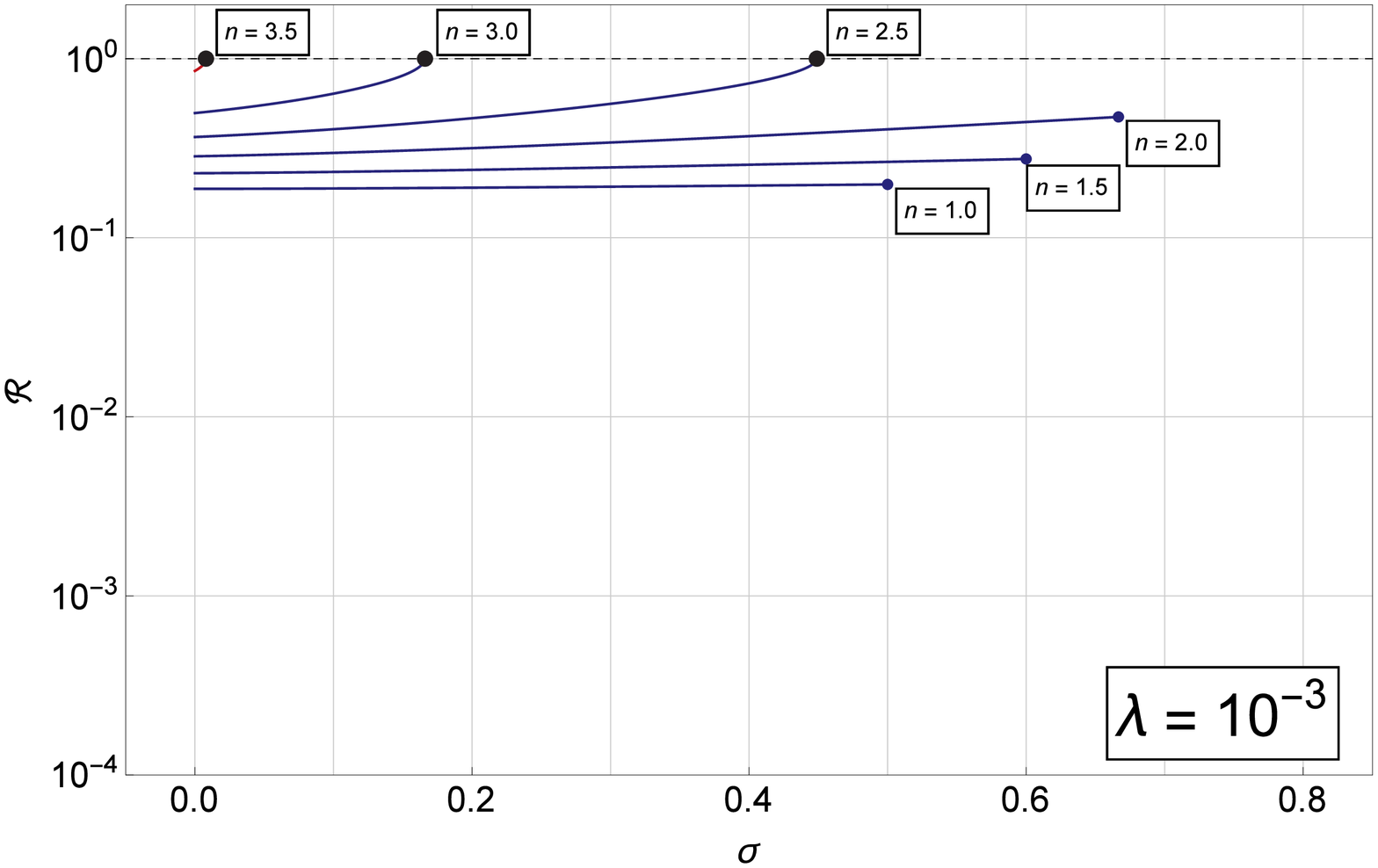}
\end{minipage}\hfill%
\begin{minipage}{0.48\linewidth}
  \caption{\label{SigR2rs}The results of numerical computations of the
    configuration-to-static radius ratio $\mathcal{R}$ for $\lambda=10^{-12}$
    (top left), $\lambda=10^{-6}$ (top right) and $\lambda=10^{-3}$ (left)
    with polytropic index $n$ ranging from 1.0 to 4.0 with step of 0.5. While
    the positions of the small terminating points
    stem from the restriction on the sound speed at the configuration center,
    the larger black terminating points located at $\mathcal{R} = 1$ express
    the fact that the configuration radius cannot be extended behind the
    static radius. With $\lambda$ increasing, the curves for higher $n$
    gradually vanish.}
\end{minipage}
\end{figure*}

In order to illustrate clearly the role of the cosmological constant (vacuum
energy) in the character of the GRPs, it is instructive to relate the
extension of the polytropic spheres to the so called static radius of their
external spacetime~\cite{Stu:1983:BULAI:}. The static radius is determined by
the formula~\cite{Stu-Hle:1999:PHYSR4:,Stu-Sch:2011:JCAP:CCMagOnCloud}
\begin{equation*}
  r_{\mathrm{s}} = \left(\frac{3 r_{\mathrm{g}}}{2 \Lambda}\right)^{1/3}\,.
\end{equation*}
At the static radius, the gravitational attraction of the central mass source
(i.e., the galaxy and its halo) is just balanced by the cosmic repulsion. The
static radius defines the region of gravitational
binding~\cite{Stu-Hle:1999:PHYSR4:}, and it should be stressed that the region
of strong cosmological-constant repulsion effects starts behind the static
radius where the cosmic repulsive acceleration prevails the gravitational
attraction and the cosmic expansion
occurs~\cite{Stu:2005:MODPLA:,Stu-Sch:2011:JCAP:CCMagOnCloud,Far-LaL-Pra:2015:JCAP:TurnarRadAccUni,Far:2016:PHYDARU:TurnarRadModGr}.
Using the quantities characterizing the spherical polytropes, we can express
the static radius of the external spacetime of the polytropes in the form
\begin{equation*}
  r_{\mathrm{s}} = \frac{3 v(\xi_{1})}{2 \lambda}^{1/3}
    \left[\frac{\sigma (n+1) c^{2}}{4 \pi G \rhocent}\right]^{1/2} 
             =  \mathcal{L} \frac{3 v(\xi_{1})}{2 \lambda}^{1/3}\,.
\end{equation*}
Then we can introduce a dimensionless `cosmologically' modified radius, i.e.,
the radius expressed in units of the static radius
\begin{equation*}
  \mathcal{R} = \frac{R}{r_{\mathrm{s}}}
    = \frac{\xi_{1} (2 \lambda)^{1/3}}{[3 v(\xi_{1})]^{1/3}}
\end{equation*}
reflecting the role of the cosmic repulsion in the character of the general
relativistic polytropic spheres. It is clear that this role is growing with
the cosmologically modified radius increasing, but the modified radius does
not depend on the central density $\rhocent$ explicitly, but only implicitly
due to magnitude of the cosmological parameter. We can see immediately that
the cosmological parameter $\lambda$ is the most relevant one, however, for
relatively large values of the polytropic index ($n \geq 3$) and the
relativistic parameter ($\sigma > 0.6$), the dimensionless radius $\xi_{1}$
can grow substantially. The results of the numerical calculations are given in
Fig.\,\ref{SigR2rs}. The limits on the dimensionless radius of the polytrope
expressed in units of the static radius of the external spacetime are given in
terms of the functions $\mathcal{R}(\sigma,n,\lambda)$ considered for the
characteristic values of the cosmological parameter $\lambda$ and the
polytropic index $n$ that were used for deduction of GRP global
characteristics. The restrictions have the upper limit at the ratio
$\mathcal{R}=1$ and become stronger with increasing value of $\lambda$ and
increasing value of $n$. For $\lambda = 10^{-12}$, the upper limit of
$\mathcal{R}=1$ is relevant for polytropes with $n=3.5,4$ from the considered
values of $n$, for $\lambda = 10^{-6}$, also the $n=3$ polytropes can reach
the upper limit of $\mathcal{R}=1$, while for $\lambda = 10^{-3}$ the polytrope
$n=2.5$ reaches the limit of $\mathcal{R}=1$ too, but the polytropes with $n=4$
are completely forbidden for such high value of $\lambda$. Of course, the
range of allowed values of the relativistic parameter $\sigma$ for a polytrope
with fixed index $n$ decreases with increasing parameter $\lambda$.

In the special case of uniform energy density spheres (GRPs with the
polytropic index $n=0$), the cosmologically modified dimensionless radius
reads
\begin{equation}
  \mathcal{R} = (2\lambda)^{1/3} .
\end{equation}
For the $n=0$ GRPs the vacuum energy parameter must satisfy the condition
$\lambda < 1/2$---we see immediately that extension of the $n=0$ polytropes
cannot exceed the static radius of the external spacetime. The same statement
holds for GRPs with any value of the polytropic index $n$ as demonstrated by
numerical calculations presented above. We have to stress that increasing
value of the vacuum energy parameter $\lambda$ means decreasing central
density of the polytrope, if the vacuum energy is assumed to be fixed by the
cosmological tests.

The analysis demonstrates that extension of the low-density polytropes
strongly increases with decreasing central density related to the cosmic
repulsion by the cosmological parameter $\lambda$. The rate of the polytropic
extension decrease depends strongly on the polytropic index $n$ and the
relativistic parameter $\sigma$. We can observe in Fig.\,\ref{SigXi1} that
there can be even two branches of the polytropes with fixed polytropic index
$n > 3$.

We can summarize that the results of the numerical analysis of the extension
of the low-density polytropic spheres, where the role of the cosmological
parameter $\lambda$ is relevant, imply that extension of the GRPs cannot
exceed the static radius of their external spacetime. This is a demonstration
of the fact that the gravitationally bound systems are limited by the static
radius, indicated for the first time in~\cite{Stu-Hle:1999:PHYSR4:}.

The cosmologically modified polytrope radius is the most representative
quantity when we relate the polytropes to the most extended objects on the
cosmic scales. On the other hand, in the opposite extreme, related to the most
compact objects, we have to use as the proper measure the dimensionless radius
related to the gravitational radius of the object, given by
\begin{equation}
  \mathcal{R_{\mathrm{g}}} = \frac{R}{r_{\mathrm{g}}} = \mathcal{C}^{-1}\,.
\end{equation}
For very compact objects with large central energy, the role of the
cosmological constant is quite negligible because the cosmological parameter
$\lambda$ has to be extremely low.

\section{\label{halos}Models of Galactic halo}

Finally, we present some comments on the possibility to model galactic dark
matter halos by the GRPs. For these purposes, it is useful to express the
length and mass scales of the relativistic polytropes in the form adjusted to
the astrophysically relevant, galactic conditions. Therefore we express the
length scales~(\ref{scalesLM}) in the form
\begin{align}
  \mathcal{L} &= 1.06\,\frac{[\sigma (n+1)]^{1/2}}%
    {\rhocent^{1/2}}\,(100\,\mathrm{kpc})\,,\nobadge\\
  \mathcal{M} &= 2.22\,\frac{[\sigma (n+1)]^{3/2}}%
    {\rhocent^{1/2}}\,(10^{18}\,\msun)\,,\nobadge
\end{align}
$\rhocent$ to be substituted in units of $10^{-20}\,\mathrm{g/cm^{3}}$. The
length scale of galactic halos related to typical galaxies, similar to the
Milky Way galaxy, is estimated to be $100$--$200\,\mathrm{kpc}$, while the
estimated mass of the halo is considered to be about
$(1$--$5)\times 10^{12}\,\msun$. Of course, in the case of extremely large and
massive galaxies and galaxy clusters, the extension of the halo can be up to
$1\,\mathrm{Mpc}$, and the halo mass could be as large as
$10^{15}\,\msun$~\cite{Zio:2005:MONNR:HDECygX1}.

The polytropic spheres with given mass and length scales are determined by the
solution of the structure equations given by the radial coordinate
$\xi_{1}(n, \sigma, \lambda)$ and the related mass parameter
$v_{1}(n,\sigma, \lambda)$. Generally, the exact solutions can strongly modify
the length and mass scales, however, for low values of the parameter $n$ and
non-relativistic dark matter with $\sigma\ll 1$, the length and mass scales
are decisive. Then we can obtain the GRP extension and mass in agreement with
the galactic halo estimates for $\sigma (n+1) < 10^{-4}$. However, the central
density of such polytropes have to be very small.

\begin{figure*}[t]
\begin{minipage}{0.48\linewidth}
\centering
\includegraphics[width=\linewidth]{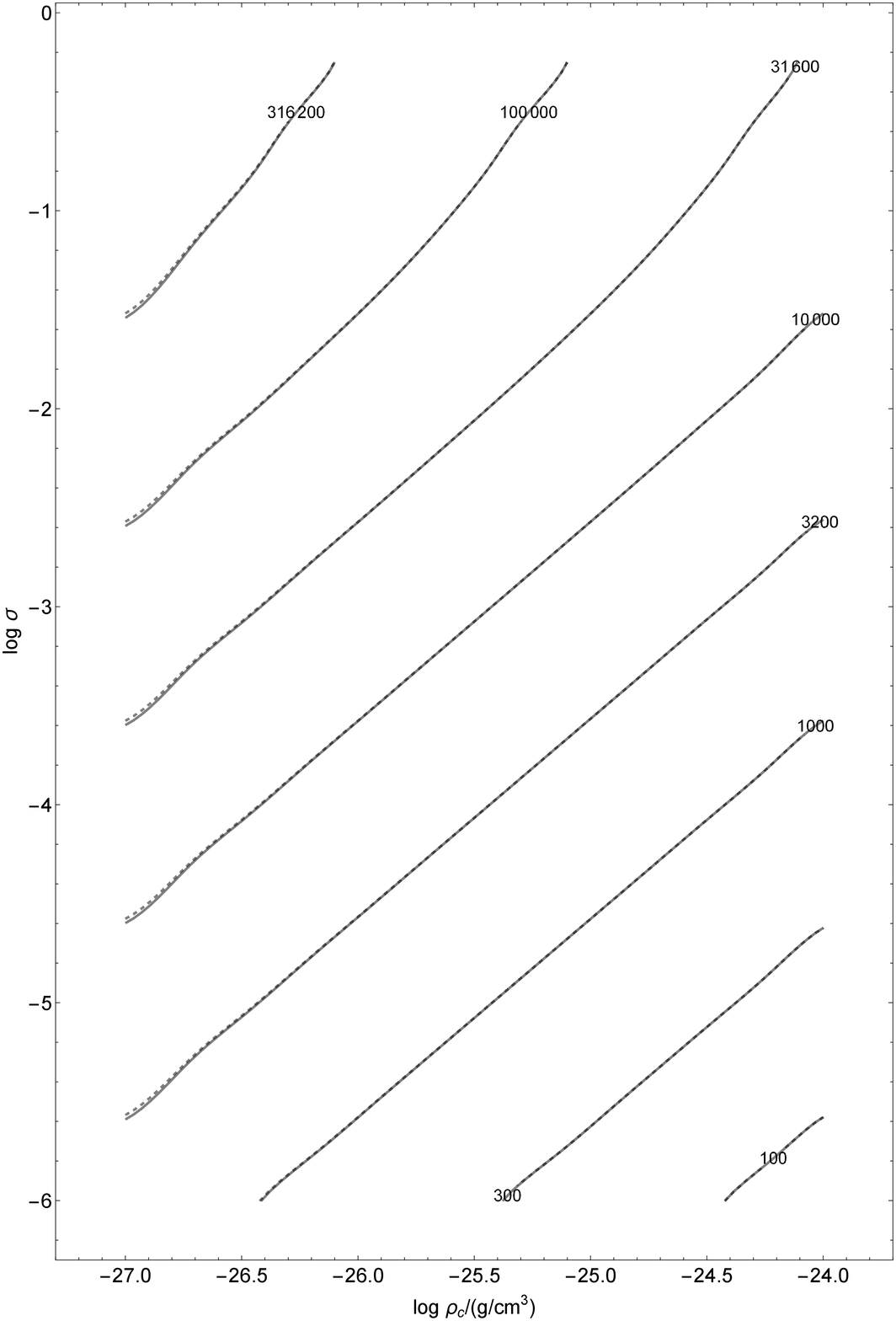}
\end{minipage}\hfill%
\begin{minipage}{0.48\linewidth}
\centering
\includegraphics[width=\linewidth]{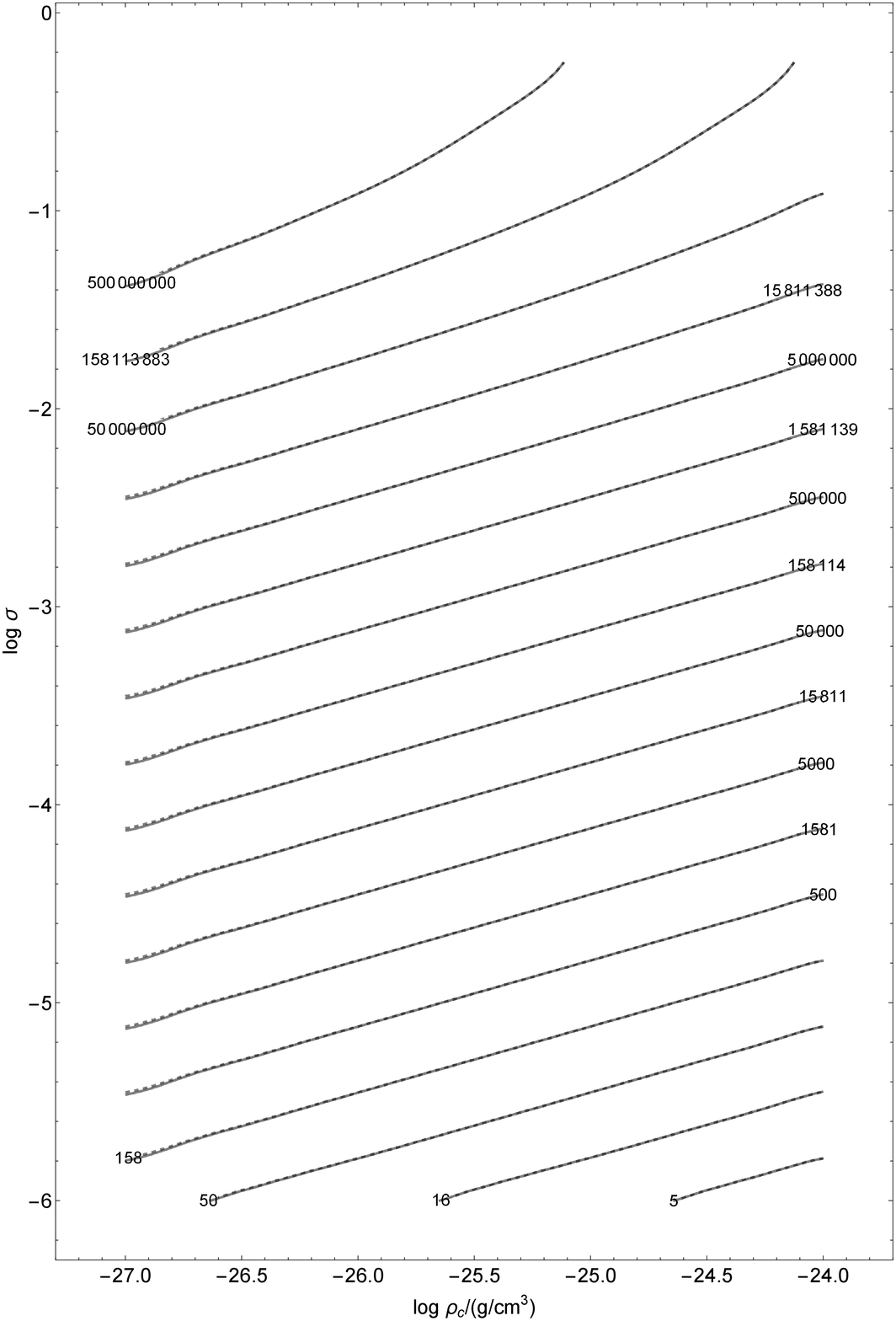}
\end{minipage}
\par\vspace{1.5\baselineskip}\par
\caption{\label{MapN15ov}Comparison of contours $R=\mathrm{const}$ ($R$ in
  kpc, left) and $M=\mathrm{const}$ ($M$ in $10^{12}\,\msun$, right) related
  to the observationally established cosmological constant
  (solid)
  and vanishing cosmological constant
  (dashed) in the parameter space $\sigma$-$\rhocent$ for polytropes with
  $n=1.5$. The presence of cosmological constant increases the radius up to a
  percent and mass up to a few percents for a fixed $\sigma$ and $\rhocent$,
  particularly for small central densities $\rhocent$.}
\end{figure*}

Detailed analysis of possible matching of the GRP extension and mass to the
CDM halo extension and mass is planned for a future paper. Here, using the
numerical methods, we give an insight into the role of the observationally
given cosmological constant on the fitting of the length and mass of the
polytropes to the astrophysically relevant values for a concrete polytrope
model with parameter $n=3/2$ corresponding to the non-relativistic gas. In
Fig.\,\ref{MapN15ov}, the constant values of the polytrope extension $R$ (and
mass $M$) are given as the functions of the parameters $\sigma$ and $\rhocent$
that gives also the parameter $\lambda$. The contours related to the
cosmological constant with observationally given value of
$\Lambda \sim 10^{-56}\,\mathrm{cm^{-2}}$ are related to those corresponding
to $\Lambda=0$.
For the $n=3/2$ polytropes, the influence of the cosmological constant is
strongest for small values of $\sigma$, being on the level of 1~percent for
the extension and 3~percent for the mass of the polytrope.

\section{\label{conc}Concluding remarks}

We have constructed fully general relativistic models of polytropic spheres
immersed in the spacetime with the relict repulsive cosmological constant
$\Lambda=1.3 \times 10^{-56}\,\mathrm{cm^{-2}}$, indicated by wide variety of
recent cosmological tests. The polytropic spheres are characterized by three
dimensionless parameters, namely by the polytropic index $n$, the relativistic
parameter $\sigma=\pcent/\rhocent$ reflecting the role of the (special)
relativistic effects in their structure, and the cosmological parameter
$\lambda=\rho_{\mathrm{vac}}/\rhocent$ reflecting the role of the vacuum
energy density (the repulsive cosmological constant) in their structure. We
have demonstrated that in dependence on the polytropic index $n$, and the
relativistic parameter $\sigma$, the GRPs are not allowed for the values of
the cosmological parameter $\lambda > \lambda_{\mathrm{crit}}(n,\sigma)$. The
value of $\lambda_{\mathrm{crit}}$ increases with the polytropic index
increasing and the relativistic parameter decreasing for the index $n \leq 3$,
while it exhibits more complex behavior for $n=3.5, 4$, when it can grow with
increasing $\sigma$. There exist even some singular states for $n>3.34$
polytropes, when the solutions are not allowed for special values of the
relativistic parameter~\cite{Nil-Ugl:2000:ANNPH1:GRStarsPolyEoS}. For example,
we have found that the $n=3.5$ polytropes are forbidden for one specific value
of $\sigma_{\mathrm{f}}=0.314$, while for the $n=4$ polytropes, there are two
forbidden values of the relativistic parameter $\sigma_{\mathrm{f}1}=0.1503$,
$\sigma_{\mathrm{f}2}=0.338$. For these special values of $\sigma$, extension
of the polytropes diverges. However, the cosmological constant naturally cuts
off these divergent radii, as radius of the polytropic configurations cannot
exceed the static radius of the external spacetime, determined by the combined
effect of the cosmological constant and the mass of the polytrope.

The length and mass scales of the GRPs with fixed polytropic index $n$ are
characterized by the central density $\rhocent$ and the relativistic parameter
$\sigma$. Both of them grow with the central density decreasing and the
relativistic parameter increasing. The real extension of the polytropic
spheres is
influenced by the cosmological parameter $\lambda$, if it is high enough (and
the central density is low enough, allowing for sufficiently high values of
$\lambda$).

Since the dependence of mass and length scales on the central density
$\rhocent$ is of the same character, while it has inverse character in the
case of the relativistic parameter $\sigma$, we can find, for any value of the
polytropic index $n$, the parameters $\rhocent$ and $\sigma$ determining a
polytropic sphere with prescribed values of the radius $R$ and mass $M$. Of
course, it can be done in the region of allowed values of the cosmological
parameter $\lambda$.

Adjusting properly the central density of the polytrope, we are able to
simulate properties of astrophysical objects in wide range, starting in the
region of extremely compact (neutron or strange) stars for extremely high
central densities, through the standard stars and stellar clusters and
finishing in the region of extremely extended low density polytropic
structures that could represent large cold dark matter halos. We demonstrate
that both the extension and mass of the most extended polytropic spheres, when
the role of the cosmological constant has to be very important putting even
strong limits on the extension of such structures, can be in agreement with
the data restricting dark matter halos -- their extension and mass have to be
$\sim 100\,\mathrm{kpc}$ and $\sim 10^{12}\,\msun$ for galaxies of the type of
the Milky Way, going up to $\sim 1\,\mathrm{Mpc}$ and $\sim 10^{14}\,\msun$
for the largest galaxies~\cite{Zio:2005:NUOC2:GalCollObj} or even larger
radius and mass for galaxy clusters. It is interesting that the polytropic
spheres can be relevant also in the framework of the so called little
inflation~\cite{Lin:1985:NEWSCI:UniInfOutChaos} related to the first order
phase transition of quark gluon plasma to the hadron phase at non-negligible
baryon number~\cite{Boe-Til:2010:PHYRL:LitInf} that implies existence of dark
matter halos of mass $M_{\mathrm{cluster}} \sim 10^{6}\,\msun$ relevant for
the physics of globular clusters and emergence of first
stars~\cite{Boe-Til:2012:PHYSR4:LitInf}.

We have demonstrated that extension of the GRPs cannot exceed the so called
static radius of their external spacetimes. Such result supports idea of the
static radius (or turn-around radius) representing an extension limit on
gravitationally bounded configurations in the expanding universe governed by
the cosmological
constant~\cite{Stu:1983:BULAI:,Stu-Hle:1999:PHYSR4:,Stu-Sch:2011:JCAP:CCMagOnCloud}.


In objects with the central density large enough, representing all the cases
of compact objects, stars, and star clusters, the role of the repulsive
cosmological constant is clearly quite negligible, since the cosmological
parameter $\lambda$ is extremely small in such situations due to high central
densities. On the other hand, the numerical analysis shows that the relict
repulsive cosmological constant has a relevant influence on the structure of
GRPs when the length scale $\mathcal{L}$ becomes comparable with the
cosmological length scale $\sim \Lambda^{-1/3}$. It is clear that
$\mathcal{L}$ increases with $\sigma$ increasing and $\rhocent$ decreasing,
thus we can expect strong role of $\Lambda$ in very low density polytropic
configurations. The influence of the relict cosmological constant can be also
amplified for polytropes with the polytropic index high enough. For example,
in the case of $n > 3.5$, the influence of $\lambda > 0$ can lead to an
instability of static polytropic configurations found in the case of
$\lambda = 0$.

The stability of the general relativistic polytropic spheres has been shortly
discussed in~\cite{Stu-Hle-2:2005:RAGtime6and7:CrossRef}. In the case of
uniform density spheres detailed discussion can be found also
in~\cite{Boh-Har:2005:PHYSR4:DynInsFluSph}. We shall discuss the stability of
the polytropes in detail in a future paper.

We also plan to study the influence of the repulsive cosmological constant on
the so called adiabatic fluid spheres, generalizing thus the results of
\cite{Too:1965:ASTRJ2:}, where the equation of state is considered in the more
``popular'' form that can be directly related to the perfect-gas equation of
state, $p = K \rho^{\gamma}_{g}$, with $\rho_{g}$ being the rest-mass density
of gas, and $\gamma$ being the adiabatic index. We can expect that for the
non-relativistic gas, when $\rho \sim \rho_g$, the adiabatic spheres will be
of similar character as the GRPs considered in the present paper, but
significant differences occur for relativistic gas. The adiabatic spheres are
governed by two structure equations with three parameters of the same meaning
as those related to the polytropic spheres. However, the adiabatic structure
equations are more complex in comparison to the polytrope structure equations;
e.g., they do not alow for existence of special solutions determined by
elementary functions, as is the case of the $n=0$ polytropes.

\begin{acknowledgments}
  The~authors acknowledge the Institutional support of the Silesian
  University. One of the authors (ZS) acknowledges the Albert Einstein Center
  for Gravitation and Astrophysics supported by the~Czech Science Foundation
  grant No.~14-37086G. Authors JN and ZS thank the Silesian University grant
  SGS/14/2016.
\end{acknowledgments}

\bibliography{mrabbrev,abbs,refs,pubs,pros}

\providecommand{\uv}[1]{\glqq#1\grqq}
  \makeatletter\providecommand{\BibTeX}{B\hbox{\check@mathfonts
  \fontsize\sf@size\z@ \math@fontsfalse\selectfont
  IB}\TeX}\makeatother\providecommand{\cheatsort}[1]{}
  \providecommand*{\angdeg}{^{\circ}} \providecommand*{\upalpha}{^{\alpha}}
  \providecommand*{\eV}[1][]{\ensuremath{\mathrm{#1e\kern-.10em V}}}
  \let\origemph\emph \renewcommand{\emph}[1]{\origemph{\ #1}} \let\u\v
  \providecommand{\csty}[3]{\textbf{\sffamily [\hspace{0.08em}Cited by (since
  #2) #1: #3\hspace{0.1em}]}}
\begin{thebibliography}{151}%
\makeatletter
\providecommand \@ifxundefined [1]{%
 \@ifx{#1\undefined}
}%
\providecommand \@ifnum [1]{%
 \ifnum #1\expandafter \@firstoftwo
 \else \expandafter \@secondoftwo
 \fi
}%
\providecommand \@ifx [1]{%
 \ifx #1\expandafter \@firstoftwo
 \else \expandafter \@secondoftwo
 \fi
}%
\providecommand \natexlab [1]{#1}%
\providecommand \enquote  [1]{``#1''}%
\providecommand \bibnamefont  [1]{#1}%
\providecommand \bibfnamefont [1]{#1}%
\providecommand \citenamefont [1]{#1}%
\providecommand \href@noop [0]{\@secondoftwo}%
\providecommand \href [0]{\begingroup \@sanitize@url \@href}%
\providecommand \@href[1]{\@@startlink{#1}\@@href}%
\providecommand \@@href[1]{\endgroup#1\@@endlink}%
\providecommand \@sanitize@url [0]{\catcode `\\12\catcode `\$12\catcode
  `\&12\catcode `\#12\catcode `\^12\catcode `\_12\catcode `\%12\relax}%
\providecommand \@@startlink[1]{}%
\providecommand \@@endlink[0]{}%
\providecommand \url  [0]{\begingroup\@sanitize@url \@url }%
\providecommand \@url [1]{\endgroup\@href {#1}{\urlprefix }}%
\providecommand \urlprefix  [0]{URL }%
\providecommand \Eprint [0]{\href }%
\providecommand \doibase [0]{http://dx.doi.org/}%
\providecommand \selectlanguage [0]{\@gobble}%
\providecommand \bibinfo  [0]{\@secondoftwo}%
\providecommand \bibfield  [0]{\@secondoftwo}%
\providecommand \translation [1]{[#1]}%
\providecommand \BibitemOpen [0]{}%
\providecommand \bibitemStop [0]{}%
\providecommand \bibitemNoStop [0]{.\EOS\space}%
\providecommand \EOS [0]{\spacefactor3000\relax}%
\providecommand \BibitemShut  [1]{\csname bibitem#1\endcsname}%
\let\auto@bib@innerbib\@empty
\bibitem [{\citenamefont {Linde}(1990)}]{Lin:1990:InfCos:}%
  \BibitemOpen
  \bibfield  {author} {\bibinfo {author} {\bibfnamefont {A.~D.}\ \bibnamefont
  {Linde}},\ }\href@noop {} {\emph {\bibinfo {title} {Particle Physics and
  Inflationary Cosmology}}}\ (\bibinfo  {publisher} {Gordon and Breach},\
  \bibinfo {address} {New York},\ \bibinfo {year} {1990})\BibitemShut {NoStop}%
\bibitem [{\citenamefont {Krauss}\ and\ \citenamefont
  {Turner}(1995)}]{Kra-Tur:1995:GENRG2:}%
  \BibitemOpen
  \bibfield  {author} {\bibinfo {author} {\bibfnamefont {L.~M.}\ \bibnamefont
  {Krauss}}\ and\ \bibinfo {author} {\bibfnamefont {M.~S.}\ \bibnamefont
  {Turner}},\ }\href {\doibase 10.1007/BF02108229} {\bibfield  {journal}
  {\bibinfo  {journal} {Gen. Relativity Gravitation}\ }\textbf {\bibinfo
  {volume} {27}},\ \bibinfo {pages} {1137} (\bibinfo {year}
  {1995})}\BibitemShut {NoStop}%
\bibitem [{\citenamefont {Ostriker}\ and\ \citenamefont
  {Steinhardt}(1995)}]{Ost-Ste:1995:NATURE:}%
  \BibitemOpen
  \bibfield  {author} {\bibinfo {author} {\bibfnamefont {J.~P.}\ \bibnamefont
  {Ostriker}}\ and\ \bibinfo {author} {\bibfnamefont {P.~J.}\ \bibnamefont
  {Steinhardt}},\ }\href {\doibase 10.1038/377600a0} {\bibfield  {journal}
  {\bibinfo  {journal} {Nature}\ }\textbf {\bibinfo {volume} {377}},\ \bibinfo
  {pages} {600} (\bibinfo {year} {1995})}\BibitemShut {NoStop}%
\bibitem [{\citenamefont {Krauss}(1998)}]{Kra:1998:ASTRJ2:}%
  \BibitemOpen
  \bibfield  {author} {\bibinfo {author} {\bibfnamefont {L.~M.}\ \bibnamefont
  {Krauss}},\ }\href
  {http://cdsaas.u-strasbg.fr:2001/ApJ/journal/issues/ApJ/v501n2/37388/37388.html}
  {\bibfield  {journal} {\bibinfo  {journal} {Astrophys. J.}\ }\textbf
  {\bibinfo {volume} {501}},\ \bibinfo {pages} {461} (\bibinfo {year}
  {1998})}\BibitemShut {NoStop}%
\bibitem [{\citenamefont {Bahcall}\ \emph {et~al.}(1999)\citenamefont
  {Bahcall}, \citenamefont {Ostriker}, \citenamefont {Perlmutter},\ and\
  \citenamefont {Steinhardt}}]{Bah-etal:1999:SCIEN:}%
  \BibitemOpen
  \bibfield  {author} {\bibinfo {author} {\bibfnamefont {N.}~\bibnamefont
  {Bahcall}}, \bibinfo {author} {\bibfnamefont {J.~P.}\ \bibnamefont
  {Ostriker}}, \bibinfo {author} {\bibfnamefont {S.}~\bibnamefont
  {Perlmutter}}, \ and\ \bibinfo {author} {\bibfnamefont {P.~J.}\ \bibnamefont
  {Steinhardt}},\ }\href@noop {} {\bibfield  {journal} {\bibinfo  {journal}
  {Science}\ }\textbf {\bibinfo {volume} {284}},\ \bibinfo {pages} {1481}
  (\bibinfo {year} {1999})}\BibitemShut {NoStop}%
\bibitem [{\citenamefont {Caldwell}\ \emph {et~al.}(1998)\citenamefont
  {Caldwell}, \citenamefont {Dave},\ and\ \citenamefont
  {Steinhardt}}]{Cal-Dav-Ste:1998:PHYRL:}%
  \BibitemOpen
  \bibfield  {author} {\bibinfo {author} {\bibfnamefont {R.~R.}\ \bibnamefont
  {Caldwell}}, \bibinfo {author} {\bibfnamefont {R.}~\bibnamefont {Dave}}, \
  and\ \bibinfo {author} {\bibfnamefont {P.~J.}\ \bibnamefont {Steinhardt}},\
  }\href {http://ojps.aip.org/journal_cgi/dbt?KEY=PRLTAO&Volume=80&Issue=8}
  {\bibfield  {journal} {\bibinfo  {journal} {Phys. Rev. Lett.}\ }\textbf
  {\bibinfo {volume} {80}},\ \bibinfo {pages} {1582} (\bibinfo {year}
  {1998})}\BibitemShut {NoStop}%
\bibitem [{\citenamefont {Armendariz-Picon}\ \emph {et~al.}(2000)\citenamefont
  {Armendariz-Picon}, \citenamefont {Mukhanov},\ and\ \citenamefont
  {Steinhardt}}]{ArP-Muk-Ste:2000:PHYRL:}%
  \BibitemOpen
  \bibfield  {author} {\bibinfo {author} {\bibfnamefont {C.}~\bibnamefont
  {Armendariz-Picon}}, \bibinfo {author} {\bibfnamefont {V.}~\bibnamefont
  {Mukhanov}}, \ and\ \bibinfo {author} {\bibfnamefont {P.~J.}\ \bibnamefont
  {Steinhardt}},\ }\href
  {http://ojps.aip.org/journal_cgi/dbt?KEY=PRLTAO&Volume=85&Issue=21}
  {\bibfield  {journal} {\bibinfo  {journal} {Phys. Rev. Lett.}\ }\textbf
  {\bibinfo {volume} {85}},\ \bibinfo {pages} {4438} (\bibinfo {year}
  {2000})}\BibitemShut {NoStop}%
\bibitem [{\citenamefont {Wang}\ \emph {et~al.}(2000)\citenamefont {Wang},
  \citenamefont {Caldwell}, \citenamefont {Ostriker},\ and\ \citenamefont
  {Steinhardt}}]{Wan-etal:2000:ASTRJ2:}%
  \BibitemOpen
  \bibfield  {author} {\bibinfo {author} {\bibfnamefont {L.}~\bibnamefont
  {Wang}}, \bibinfo {author} {\bibfnamefont {R.~R.}\ \bibnamefont {Caldwell}},
  \bibinfo {author} {\bibfnamefont {J.~P.}\ \bibnamefont {Ostriker}}, \ and\
  \bibinfo {author} {\bibfnamefont {P.~J.}\ \bibnamefont {Steinhardt}},\ }\href
  {http://www.journals.uchicago.edu/doi/abs/10.1086/308331} {\bibfield
  {journal} {\bibinfo  {journal} {Astrophys. J.}\ }\textbf {\bibinfo {volume}
  {530}},\ \bibinfo {pages} {17} (\bibinfo {year} {2000})}\BibitemShut
  {NoStop}%
\bibitem [{\citenamefont {Spergel}\ \emph {et~al.}(2007)\citenamefont
  {Spergel}, \citenamefont {Bean}, \citenamefont {Dore}, \citenamefont {Nolta},
  \citenamefont {Bennett}, \citenamefont {Dunkley}, \citenamefont {Hinshaw},
  \citenamefont {Jarosik}, \citenamefont {Komatsu}, \citenamefont {Page},
  \citenamefont {Peiris}, \citenamefont {Verde}, \citenamefont {Halpern},
  \citenamefont {Hill}, \citenamefont {Kogut}, \citenamefont {Limon},
  \citenamefont {Meyer}, \citenamefont {Odegard}, \citenamefont {Tucker},
  \citenamefont {Weiland}, \citenamefont {Wollack},\ and\ \citenamefont
  {Wright}}]{Spe-etal:2007:ASTJS:3yrWMAP}%
  \BibitemOpen
  \bibfield  {author} {\bibinfo {author} {\bibfnamefont {D.~N.}\ \bibnamefont
  {Spergel}}, \bibinfo {author} {\bibfnamefont {R.}~\bibnamefont {Bean}},
  \bibinfo {author} {\bibfnamefont {O.}~\bibnamefont {Dore}}, \bibinfo {author}
  {\bibfnamefont {M.~R.}\ \bibnamefont {Nolta}}, \bibinfo {author}
  {\bibfnamefont {C.~L.}\ \bibnamefont {Bennett}}, \bibinfo {author}
  {\bibfnamefont {J.}~\bibnamefont {Dunkley}}, \bibinfo {author} {\bibfnamefont
  {G.}~\bibnamefont {Hinshaw}}, \bibinfo {author} {\bibfnamefont
  {N.}~\bibnamefont {Jarosik}}, \bibinfo {author} {\bibfnamefont
  {E.}~\bibnamefont {Komatsu}}, \bibinfo {author} {\bibfnamefont
  {L.}~\bibnamefont {Page}}, \bibinfo {author} {\bibfnamefont {H.~V.}\
  \bibnamefont {Peiris}}, \bibinfo {author} {\bibfnamefont {L.}~\bibnamefont
  {Verde}}, \bibinfo {author} {\bibfnamefont {M.}~\bibnamefont {Halpern}},
  \bibinfo {author} {\bibfnamefont {R.~S.}\ \bibnamefont {Hill}}, \bibinfo
  {author} {\bibfnamefont {A.}~\bibnamefont {Kogut}}, \bibinfo {author}
  {\bibfnamefont {M.}~\bibnamefont {Limon}}, \bibinfo {author} {\bibfnamefont
  {S.~S.}\ \bibnamefont {Meyer}}, \bibinfo {author} {\bibfnamefont
  {N.}~\bibnamefont {Odegard}}, \bibinfo {author} {\bibfnamefont {G.~S.}\
  \bibnamefont {Tucker}}, \bibinfo {author} {\bibfnamefont {J.~L.}\
  \bibnamefont {Weiland}}, \bibinfo {author} {\bibfnamefont {E.}~\bibnamefont
  {Wollack}}, \ and\ \bibinfo {author} {\bibfnamefont {E.~L.}\ \bibnamefont
  {Wright}},\ }\href {\doibase 10.1086/513700} {\bibfield  {journal} {\bibinfo
  {journal} {Astrophys. J. Suppl.}\ }\textbf {\bibinfo {volume} {170}},\
  \bibinfo {pages} {377} (\bibinfo {year} {2007})}\BibitemShut {NoStop}%
\bibitem [{\citenamefont {Riess}\ \emph {et~al.}(2004)\citenamefont {Riess}
  \emph {et~al.}}]{Rie-etal:2004:ASTRJ2:}%
  \BibitemOpen
  \bibfield  {author} {\bibinfo {author} {\bibfnamefont {A.~G.}\ \bibnamefont
  {Riess}} \emph {et~al.},\ }\href@noop {} {\bibfield  {journal} {\bibinfo
  {journal} {Astrophys. J.}\ }\textbf {\bibinfo {volume} {123}},\ \bibinfo
  {pages} {145} (\bibinfo {year} {2004})},\ \Eprint
  {http://arxiv.org/abs/astro-ph/0402512} {astro-ph/0402512} \BibitemShut
  {NoStop}%
\bibitem [{\citenamefont {Caldwell}\ and\ \citenamefont
  {Kamionkowski}(2009)}]{Cal-Kam:2009:NATURE:CosDarkMat}%
  \BibitemOpen
  \bibfield  {author} {\bibinfo {author} {\bibfnamefont {R.}~\bibnamefont
  {Caldwell}}\ and\ \bibinfo {author} {\bibfnamefont {M.}~\bibnamefont
  {Kamionkowski}},\ }\href {\doibase 10.1038/458587a} {\bibfield  {journal}
  {\bibinfo  {journal} {Nature}\ }\textbf {\bibinfo {volume} {458}},\ \bibinfo
  {pages} {587} (\bibinfo {year} {2009})}\BibitemShut {NoStop}%
\bibitem [{\citenamefont {Adami}\ \emph {et~al.}(2013)\citenamefont {Adami},
  \citenamefont {Durret}, \citenamefont {Guennou},\ and\ \citenamefont
  {Da~Rocha}}]{Ada-etal:2013:ASTRA:DifLiYoClG}%
  \BibitemOpen
  \bibfield  {author} {\bibinfo {author} {\bibfnamefont {C.}~\bibnamefont
  {Adami}}, \bibinfo {author} {\bibfnamefont {F.}~\bibnamefont {Durret}},
  \bibinfo {author} {\bibfnamefont {L.}~\bibnamefont {Guennou}}, \ and\
  \bibinfo {author} {\bibfnamefont {C.}~\bibnamefont {Da~Rocha}},\ }\href
  {\doibase 10.1051/0004-6361/201220282} {\bibfield  {journal} {\bibinfo
  {journal} {Astronomy and Astrophysics}\ }\textbf {\bibinfo {volume} {551}},\
  \bibinfo {pages} {A20 (7~pages)} (\bibinfo {year} {2013})}\BibitemShut
  {NoStop}%
\bibitem [{\citenamefont {{Planck Collaboration}}\ \emph
  {et~al.}(2014)\citenamefont {{Planck Collaboration}}, \citenamefont {Ade}
  \emph {et~al.}}]{Ade-etal:2014:ASTRA:PlanckXII}%
  \BibitemOpen
  \bibfield  {author} {\bibinfo {author} {\bibnamefont {{Planck
  Collaboration}}}, \bibinfo {author} {\bibfnamefont {P.~A.~R.}\ \bibnamefont
  {Ade}},  \emph {et~al.},\ }\href {\doibase 10.1051/0004-6361/201321580}
  {\bibfield  {journal} {\bibinfo  {journal} {Astronomy and Astrophysics}\
  }\textbf {\bibinfo {volume} {571}},\ \bibinfo {pages} {A12} (\bibinfo {year}
  {2014})}\BibitemShut {NoStop}%
\bibitem [{\citenamefont {Misner}\ \emph {et~al.}(1973)\citenamefont {Misner},
  \citenamefont {Thorne},\ and\ \citenamefont
  {Wheeler}}]{Mis-Tho-Whe:1973:Gra:}%
  \BibitemOpen
  \bibfield  {author} {\bibinfo {author} {\bibfnamefont {C.~W.}\ \bibnamefont
  {Misner}}, \bibinfo {author} {\bibfnamefont {K.~S.}\ \bibnamefont {Thorne}},
  \ and\ \bibinfo {author} {\bibfnamefont {J.~A.}\ \bibnamefont {Wheeler}},\
  }\href@noop {} {\emph {\bibinfo {title} {Gravitation}}}\ (\bibinfo
  {publisher} {W. H. Freeman and Co},\ \bibinfo {address} {New York, San
  Francisco},\ \bibinfo {year} {1973})\BibitemShut {NoStop}%
\bibitem [{\citenamefont {Stuchl{\'i}k}(1983)}]{Stu:1983:BULAI:}%
  \BibitemOpen
  \bibfield  {author} {\bibinfo {author} {\bibfnamefont {Z.}~\bibnamefont
  {Stuchl{\'i}k}},\ }\href@noop {} {\bibfield  {journal} {\bibinfo  {journal}
  {Bull. Astronom. Inst. Czechoslovakia}\ }\textbf {\bibinfo {volume} {34}},\
  \bibinfo {pages} {129} (\bibinfo {year} {1983})}\BibitemShut {NoStop}%
\bibitem [{\citenamefont {Stuchl{\'i}k}(1984)}]{Stu:1984:BULAI:}%
  \BibitemOpen
  \bibfield  {author} {\bibinfo {author} {\bibfnamefont {Z.}~\bibnamefont
  {Stuchl{\'i}k}},\ }\href@noop {} {\bibfield  {journal} {\bibinfo  {journal}
  {Bull. Astronom. Inst. Czechoslovakia}\ }\textbf {\bibinfo {volume} {35}},\
  \bibinfo {pages} {205} (\bibinfo {year} {1984})}\BibitemShut {NoStop}%
\bibitem [{\citenamefont {Uzan}\ \emph {et~al.}(2011)\citenamefont {Uzan},
  \citenamefont {Ellis},\ and\ \citenamefont
  {Larena}}]{Uza-Ell-Lar:2011:GENRG2:2MassExp}%
  \BibitemOpen
  \bibfield  {author} {\bibinfo {author} {\bibfnamefont {J.-P.}\ \bibnamefont
  {Uzan}}, \bibinfo {author} {\bibfnamefont {G.~F.~R.}\ \bibnamefont {Ellis}},
  \ and\ \bibinfo {author} {\bibfnamefont {J.}~\bibnamefont {Larena}},\ }\href
  {\doibase 10.1007/s10714-010-1081-6} {\bibfield  {journal} {\bibinfo
  {journal} {Gen. Relativity Gravitation}\ }\textbf {\bibinfo {volume} {43}},\
  \bibinfo {pages} {191} (\bibinfo {year} {2011})}\BibitemShut {NoStop}%
\bibitem [{\citenamefont {Grenon}\ and\ \citenamefont
  {Lake}(2010)}]{Gre-Lak:2010:PHYSR4:SwissCheeseMass}%
  \BibitemOpen
  \bibfield  {author} {\bibinfo {author} {\bibfnamefont {C.}~\bibnamefont
  {Grenon}}\ and\ \bibinfo {author} {\bibfnamefont {K.}~\bibnamefont {Lake}},\
  }\href {\doibase 10.1103/PhysRevD.81.023501} {\bibfield  {journal} {\bibinfo
  {journal} {Phys. Rev. D}\ }\textbf {\bibinfo {volume} {81}},\ \bibinfo
  {pages} {023501 [10~pages]} (\bibinfo {year} {2010})}\BibitemShut {NoStop}%
\bibitem [{\citenamefont {Fleury}\ \emph {et~al.}(2013)\citenamefont {Fleury},
  \citenamefont {Dupuy},\ and\ \citenamefont
  {Uzan}}]{Fle-Dup-Uza:2013:PHYSR4:HubbleNonhomUn}%
  \BibitemOpen
  \bibfield  {author} {\bibinfo {author} {\bibfnamefont {P.}~\bibnamefont
  {Fleury}}, \bibinfo {author} {\bibfnamefont {H.}~\bibnamefont {Dupuy}}, \
  and\ \bibinfo {author} {\bibfnamefont {J.-P.}\ \bibnamefont {Uzan}},\ }\href
  {\doibase 10.1103/PhysRevD.87.123526} {\bibfield  {journal} {\bibinfo
  {journal} {Phys. Rev. D}\ }\textbf {\bibinfo {volume} {87}},\ \bibinfo
  {pages} {123526} (\bibinfo {year} {2013})},\ \Eprint
  {http://arxiv.org/abs/1302.5308} {1302.5308} \BibitemShut {NoStop}%
\bibitem [{\citenamefont {Firouzjaee}\ and\ \citenamefont
  {Feghhi}(2016)}]{Fir-Feg:2016:arXiv160805491:PtMassCBH}%
  \BibitemOpen
  \bibfield  {author} {\bibinfo {author} {\bibfnamefont {J.~T.}\ \bibnamefont
  {Firouzjaee}}\ and\ \bibinfo {author} {\bibfnamefont {T.}~\bibnamefont
  {Feghhi}},\ }\href {http://arxiv.org/abs/1608.05491} {{\selectlanguage
  {english}\enquote {\bibinfo {title} {{Point mass Cosmological Black
  Holes}},}\ }} (\bibinfo {year} {2016}),\ \Eprint
  {http://arxiv.org/abs/1608.05491} {arXiv:1608.05491} \BibitemShut {NoStop}%
\bibitem [{\citenamefont {McVittie}(1933)}]{McV:1933:MONNR:MassPartExpUn}%
  \BibitemOpen
  \bibfield  {author} {\bibinfo {author} {\bibfnamefont {G.~C.}\ \bibnamefont
  {McVittie}},\ }\href {\doibase 10.1093/mnras/93.5.325} {\bibfield  {journal}
  {\bibinfo  {journal} {Monthly Notices Roy. Astronom. Soc.}\ }\textbf
  {\bibinfo {volume} {93}},\ \bibinfo {pages} {325} (\bibinfo {year}
  {1933})}\BibitemShut {NoStop}%
\bibitem [{\citenamefont {Nolan}(1998)}]{Nol:1998:PHYSR4:PtMassIsotUn}%
  \BibitemOpen
  \bibfield  {author} {\bibinfo {author} {\bibfnamefont {B.~C.}\ \bibnamefont
  {Nolan}},\ }\href {\doibase 10.1103/PhysRevD.58.064006} {\bibfield  {journal}
  {\bibinfo  {journal} {Phys. Rev. D}\ }\textbf {\bibinfo {volume} {58}},\
  \bibinfo {eid} {064006} (\bibinfo {year} {1998})},\ \Eprint
  {http://arxiv.org/abs/gr-qc/9805041} {gr-qc/9805041} \BibitemShut {NoStop}%
\bibitem [{\citenamefont {Nandra}\ \emph
  {et~al.}(2012{\natexlab{a}})\citenamefont {Nandra}, \citenamefont {Lasenby},\
  and\ \citenamefont {Hobson}}]{Nan-Las-Hob:2012:MONNR:ExpUnMassOb}%
  \BibitemOpen
  \bibfield  {author} {\bibinfo {author} {\bibfnamefont {R.}~\bibnamefont
  {Nandra}}, \bibinfo {author} {\bibfnamefont {A.~N.}\ \bibnamefont {Lasenby}},
  \ and\ \bibinfo {author} {\bibfnamefont {M.~P.}\ \bibnamefont {Hobson}},\
  }\href {\doibase 10.1111/j.1365-2966.2012.20617.x} {\bibfield  {journal}
  {\bibinfo  {journal} {Monthly Notices Roy. Astronom. Soc.}\ }\textbf
  {\bibinfo {volume} {422}},\ \bibinfo {pages} {2945} (\bibinfo {year}
  {2012}{\natexlab{a}})},\ \Eprint {http://arxiv.org/abs/1104.4458}
  {arXiv:1104.4458 [gr-qc]} \BibitemShut {NoStop}%
\bibitem [{\citenamefont {Nandra}\ \emph
  {et~al.}(2012{\natexlab{b}})\citenamefont {Nandra}, \citenamefont {Lasenby},\
  and\ \citenamefont {Hobson}}]{Nan-Las-Hob:2012:MONNR:MassObExpUn}%
  \BibitemOpen
  \bibfield  {author} {\bibinfo {author} {\bibfnamefont {R.}~\bibnamefont
  {Nandra}}, \bibinfo {author} {\bibfnamefont {A.~N.}\ \bibnamefont {Lasenby}},
  \ and\ \bibinfo {author} {\bibfnamefont {M.~P.}\ \bibnamefont {Hobson}},\
  }\href {\doibase 10.1111/j.1365-2966.2012.20618.x} {\bibfield  {journal}
  {\bibinfo  {journal} {Monthly Notices Roy. Astronom. Soc.}\ }\textbf
  {\bibinfo {volume} {422}},\ \bibinfo {pages} {2931} (\bibinfo {year}
  {2012}{\natexlab{b}})},\ \Eprint {http://arxiv.org/abs/1104.4447}
  {arXiv:1104.4447 [gr-qc]} \BibitemShut {NoStop}%
\bibitem [{\citenamefont {Kaloper}\ \emph {et~al.}(2010)\citenamefont
  {Kaloper}, \citenamefont {Kleban},\ and\ \citenamefont
  {Martin}}]{Kal-Kle-Mar:2010:PHYSR4:McVittieLegacy}%
  \BibitemOpen
  \bibfield  {author} {\bibinfo {author} {\bibfnamefont {N.}~\bibnamefont
  {Kaloper}}, \bibinfo {author} {\bibfnamefont {M.}~\bibnamefont {Kleban}}, \
  and\ \bibinfo {author} {\bibfnamefont {D.}~\bibnamefont {Martin}},\ }\href
  {\doibase 10.1103/PhysRevD.81.104044} {\bibfield  {journal} {\bibinfo
  {journal} {Phys. Rev. D}\ }\textbf {\bibinfo {volume} {81}},\ \bibinfo
  {pages} {104044} (\bibinfo {year} {2010})}\BibitemShut {NoStop}%
\bibitem [{\citenamefont {Lake}\ and\ \citenamefont
  {Abdelqader}(2011)}]{Lak-Abd:2011:PHYSR4:McVittieLegacy}%
  \BibitemOpen
  \bibfield  {author} {\bibinfo {author} {\bibfnamefont {K.}~\bibnamefont
  {Lake}}\ and\ \bibinfo {author} {\bibfnamefont {M.}~\bibnamefont
  {Abdelqader}},\ }\href {\doibase 10.1103/PhysRevD.84.044045} {\bibfield
  {journal} {\bibinfo  {journal} {Phys. Rev. D}\ }\textbf {\bibinfo {volume}
  {84}},\ \bibinfo {pages} {044045} (\bibinfo {year} {2011})},\ \Eprint
  {http://arxiv.org/abs/1106.3666} {1106.3666} \BibitemShut {NoStop}%
\bibitem [{\citenamefont {da~Silva}\ \emph {et~al.}(2013)\citenamefont
  {da~Silva}, \citenamefont {Fontanini},\ and\ \citenamefont
  {Guariento}}]{Sil-Fon-Gua:2013:PHYSR4:McVittieST}%
  \BibitemOpen
  \bibfield  {author} {\bibinfo {author} {\bibfnamefont {A.~M.}\ \bibnamefont
  {da~Silva}}, \bibinfo {author} {\bibfnamefont {M.}~\bibnamefont {Fontanini}},
  \ and\ \bibinfo {author} {\bibfnamefont {D.~C.}\ \bibnamefont {Guariento}},\
  }\href {\doibase 10.1103/PhysRevD.87.064030} {\bibfield  {journal} {\bibinfo
  {journal} {Phys. Rev. D}\ }\textbf {\bibinfo {volume} {87}},\ \bibinfo
  {pages} {064030} (\bibinfo {year} {2013})}\BibitemShut {NoStop}%
\bibitem [{\citenamefont {Nolan}(2014)}]{Nol:2014:1408.0044:MvVitOrb}%
  \BibitemOpen
  \bibfield  {author} {\bibinfo {author} {\bibfnamefont {B.~C.}\ \bibnamefont
  {Nolan}},\ }\href {\doibase 10.1088/0264-9381/31/23/235008} {\bibfield
  {journal} {\bibinfo  {journal} {Classical Quantum Gravity}\ }\textbf
  {\bibinfo {volume} {31}},\ \bibinfo {pages} {235008} (\bibinfo {year}
  {2014})},\ \Eprint {http://arxiv.org/abs/1408.0044} {1408.0044} \BibitemShut
  {NoStop}%
\bibitem [{\citenamefont {Stuchl{\'i}k}(2005)}]{Stu:2005:MODPLA:}%
  \BibitemOpen
  \bibfield  {author} {\bibinfo {author} {\bibfnamefont {Z.}~\bibnamefont
  {Stuchl{\'i}k}},\ }\href@noop {} {\bibfield  {journal} {\bibinfo  {journal}
  {Modern Phys. Lett. A}\ }\textbf {\bibinfo {volume} {20}},\ \bibinfo {pages}
  {561} (\bibinfo {year} {2005})}\BibitemShut {NoStop}%
\bibitem [{\citenamefont {Kottler}(1918)}]{Kot:1918:ANNPH2:PhyBasEinsGr}%
  \BibitemOpen
  \bibfield  {author} {\bibinfo {author} {\bibfnamefont {F.}~\bibnamefont
  {Kottler}},\ }\href@noop {} {\bibfield  {journal} {\bibinfo  {journal} {Ann.
  Physik}\ }\textbf {\bibinfo {volume} {56}},\ \bibinfo {pages} {401} (\bibinfo
  {year} {1918})},\ \bibinfo {note} {abstract Number: A1918-01213}\BibitemShut
  {NoStop}%
\bibitem [{\citenamefont {Stuchl{\'i}k}\ and\ \citenamefont
  {Hled{\'i}k}(1999)}]{Stu-Hle:1999:PHYSR4:}%
  \BibitemOpen
  \bibfield  {author} {\bibinfo {author} {\bibfnamefont {Z.}~\bibnamefont
  {Stuchl{\'i}k}}\ and\ \bibinfo {author} {\bibfnamefont {S.}~\bibnamefont
  {Hled{\'i}k}},\ }\href {\doibase 10.1103/PhysRevD.60.044006} {\bibfield
  {journal} {\bibinfo  {journal} {Phys. Rev. D}\ }\textbf {\bibinfo {volume}
  {60}},\ \bibinfo {pages} {044006 (15~pages)} (\bibinfo {year}
  {1999})}\BibitemShut {NoStop}%
\bibitem [{\citenamefont {Stuchl{\'i}k}(2000)}]{Stu:2000:ACTPS2:}%
  \BibitemOpen
  \bibfield  {author} {\bibinfo {author} {\bibfnamefont {Z.}~\bibnamefont
  {Stuchl{\'i}k}},\ }\href {http://adsabs.harvard.edu/abs/2008arXiv0803.2530S}
  {\bibfield  {journal} {\bibinfo  {journal} {Acta Phys. Slovaca}\ }\textbf
  {\bibinfo {volume} {50}},\ \bibinfo {pages} {219} (\bibinfo {year} {2000})},\
  \Eprint {http://arxiv.org/abs/0803.2530} {0803.2530} \BibitemShut {NoStop}%
\bibitem [{\citenamefont {B{\"o}hmer}(2004{\natexlab{a}})}]{Boh:2004:GENRG2:}%
  \BibitemOpen
  \bibfield  {author} {\bibinfo {author} {\bibfnamefont {C.~G.}\ \bibnamefont
  {B{\"o}hmer}},\ }\href {\doibase 10.1023/B:GERG.0000018088.69051.3b}
  {\bibfield  {journal} {\bibinfo  {journal} {Gen. Relativity Gravitation}\
  }\textbf {\bibinfo {volume} {36}},\ \bibinfo {pages} {1039} (\bibinfo {year}
  {2004}{\natexlab{a}})},\ \Eprint {http://arxiv.org/abs/gr-qc/0312027}
  {gr-qc/0312027} \BibitemShut {NoStop}%
\bibitem [{\citenamefont {Carter}(1973)}]{Car:1973:BlaHol:}%
  \BibitemOpen
  \bibfield  {author} {\bibinfo {author} {\bibfnamefont {B.}~\bibnamefont
  {Carter}},\ }in\ \href@noop {} {\emph {\bibinfo {booktitle} {{Black
  Holes}}}},\ \bibinfo {editor} {edited by\ \bibinfo {editor} {\bibfnamefont
  {C.~D.}\ \bibnamefont {Witt}}\ and\ \bibinfo {editor} {\bibfnamefont
  {B.~S.~D.}\ \bibnamefont {Witt}}}\ (\bibinfo  {publisher} {Gordon and
  Breach},\ \bibinfo {address} {New York--London--Paris},\ \bibinfo {year}
  {1973})\ pp.\ \bibinfo {pages} {57--214}\BibitemShut {NoStop}%
\bibitem [{\citenamefont {Stuchl{\'i}k}(1990)}]{Stu:1990:BULAI:}%
  \BibitemOpen
  \bibfield  {author} {\bibinfo {author} {\bibfnamefont {Z.}~\bibnamefont
  {Stuchl{\'i}k}},\ }\href@noop {} {\bibfield  {journal} {\bibinfo  {journal}
  {Bull. Astronom. Inst. Czechoslovakia}\ }\textbf {\bibinfo {volume} {41}},\
  \bibinfo {pages} {341} (\bibinfo {year} {1990})}\BibitemShut {NoStop}%
\bibitem [{\citenamefont {Stuchl{\'i}k}\ and\ \citenamefont
  {Calvani}(1991)}]{Stu-Cal:1991:GENRG2:}%
  \BibitemOpen
  \bibfield  {author} {\bibinfo {author} {\bibfnamefont {Z.}~\bibnamefont
  {Stuchl{\'i}k}}\ and\ \bibinfo {author} {\bibfnamefont {M.}~\bibnamefont
  {Calvani}},\ }\href {\doibase 10.1007/BF00758012} {\bibfield  {journal}
  {\bibinfo  {journal} {Gen. Relativity Gravitation}\ }\textbf {\bibinfo
  {volume} {23}},\ \bibinfo {pages} {507} (\bibinfo {year} {1991})}\BibitemShut
  {NoStop}%
\bibitem [{\citenamefont {Stuchl{\'i}k}\ and\ \citenamefont
  {Hled{\'i}k}(2000)}]{Stu-Hle:2000:CLAQG:}%
  \BibitemOpen
  \bibfield  {author} {\bibinfo {author} {\bibfnamefont {Z.}~\bibnamefont
  {Stuchl{\'i}k}}\ and\ \bibinfo {author} {\bibfnamefont {S.}~\bibnamefont
  {Hled{\'i}k}},\ }\href {\doibase 10.1088/0264-9381/17/21/312} {\bibfield
  {journal} {\bibinfo  {journal} {Classical Quantum Gravity}\ }\textbf
  {\bibinfo {volume} {17}},\ \bibinfo {pages} {4541} (\bibinfo {year}
  {2000})},\ \Eprint {http://arxiv.org/abs/0803.2539} {0803.2539} \BibitemShut
  {NoStop}%
\bibitem [{\citenamefont {Lake}(2002)}]{Lak:2002:PHYSR4:BendLiCC}%
  \BibitemOpen
  \bibfield  {author} {\bibinfo {author} {\bibfnamefont {K.}~\bibnamefont
  {Lake}},\ }\href {\doibase 10.1103/PhysRevD.65.087301} {\bibfield  {journal}
  {\bibinfo  {journal} {Phys. Rev. D}\ }\textbf {\bibinfo {volume} {65}},\
  \bibinfo {pages} {087301} (\bibinfo {year} {2002})}\BibitemShut {NoStop}%
\bibitem [{\citenamefont {Bakala}\ \emph {et~al.}(2007)\citenamefont {Bakala},
  \citenamefont {Čermák}, \citenamefont {Hled{\'i}k}, \citenamefont
  {Stuchl{\'i}k},\ and\ \citenamefont {Truparov\'{a}}}]{Bak-etal:2007:CEURJP:}%
  \BibitemOpen
  \bibfield  {author} {\bibinfo {author} {\bibfnamefont {P.}~\bibnamefont
  {Bakala}}, \bibinfo {author} {\bibfnamefont {P.}~\bibnamefont {Čermák}},
  \bibinfo {author} {\bibfnamefont {S.}~\bibnamefont {Hled{\'i}k}}, \bibinfo
  {author} {\bibfnamefont {Z.}~\bibnamefont {Stuchl{\'i}k}}, \ and\ \bibinfo
  {author} {\bibfnamefont {K.}~\bibnamefont {Truparov\'{a}}},\ }\href {\doibase
  10.2478/s11534-007-0033-6} {\bibfield  {journal} {\bibinfo  {journal}
  {Central European J. Phys.}\ }\textbf {\bibinfo {volume} {5}},\ \bibinfo
  {pages} {599} (\bibinfo {year} {2007})},\ \Eprint
  {http://arxiv.org/abs/0709.4274} {0709.4274} \BibitemShut {NoStop}%
\bibitem [{\citenamefont {Sereno}(2008)}]{Ser:2008:PHYSR4:CCLens}%
  \BibitemOpen
  \bibfield  {author} {\bibinfo {author} {\bibfnamefont {M.}~\bibnamefont
  {Sereno}},\ }\href@noop {} {\bibfield  {journal} {\bibinfo  {journal} {Phys.
  Rev. D}\ }\textbf {\bibinfo {volume} {77}},\ \bibinfo {pages} {043004}
  (\bibinfo {year} {2008})},\ \Eprint {http://arxiv.org/abs/0711.1802}
  {0711.1802} \BibitemShut {NoStop}%
\bibitem [{\citenamefont {Müller}(2008)}]{Mul:2008:GENRG2:FallSchBH}%
  \BibitemOpen
  \bibfield  {author} {\bibinfo {author} {\bibfnamefont {T.}~\bibnamefont
  {Müller}},\ }\href {\doibase 10.1007/s10714-008-0623-7} {\bibfield  {journal}
  {\bibinfo  {journal} {Gen. Relativity Gravitation}\ ,\ \bibinfo {pages} {56}}
  (\bibinfo {year} {2008})}\BibitemShut {NoStop}%
\bibitem [{\citenamefont {Sch{\"{u}}cker}\ and\ \citenamefont
  {Zaimen}(2008)}]{Sch-Zai:2008:0801.3776:CCTimeDelay}%
  \BibitemOpen
  \bibfield  {author} {\bibinfo {author} {\bibfnamefont {T.}~\bibnamefont
  {Sch{\"{u}}cker}}\ and\ \bibinfo {author} {\bibfnamefont {N.}~\bibnamefont
  {Zaimen}},\ }\href {\doibase 10.1051/0004-6361:200809449} {\bibfield
  {journal} {\bibinfo  {journal} {Astronomy and Astrophysics}\ }\textbf
  {\bibinfo {volume} {484}},\ \bibinfo {pages} {103} (\bibinfo {year}
  {2008})},\ \Eprint {http://arxiv.org/abs/0801.3776} {0801.3776} \BibitemShut
  {NoStop}%
\bibitem [{\citenamefont {Villanueva}\ \emph {et~al.}(2012)\citenamefont
  {Villanueva}, \citenamefont {Saavedra}, \citenamefont {Olivares},\ and\
  \citenamefont {Cruz}}]{Vil-etal:2012:ASTSS:PhMoChgAdS}%
  \BibitemOpen
  \bibfield  {author} {\bibinfo {author} {\bibfnamefont {J.~R.}\ \bibnamefont
  {Villanueva}}, \bibinfo {author} {\bibfnamefont {J.}~\bibnamefont
  {Saavedra}}, \bibinfo {author} {\bibfnamefont {M.}~\bibnamefont {Olivares}},
  \ and\ \bibinfo {author} {\bibfnamefont {N.}~\bibnamefont {Cruz}},\ }\href
  {\doibase 10.1007/s10509-012-1333-x} {\bibfield  {journal} {\bibinfo
  {journal} {Astrophys. and Space Sci.}\ }\textbf {\bibinfo {volume} {344}},\
  \bibinfo {pages} {437} (\bibinfo {year} {2012})}\BibitemShut {NoStop}%
\bibitem [{\citenamefont {Zhao}\ and\ \citenamefont
  {Tang}(2015)}]{Zha-Tan:2015:PHYSR4:GraLensSdS}%
  \BibitemOpen
  \bibfield  {author} {\bibinfo {author} {\bibfnamefont {F.}~\bibnamefont
  {Zhao}}\ and\ \bibinfo {author} {\bibfnamefont {J.}~\bibnamefont {Tang}},\
  }\href {\doibase 10.1103/PhysRevD.92.083011} {\bibfield  {journal} {\bibinfo
  {journal} {Phys. Rev. D}\ }\textbf {\bibinfo {volume} {92}},\ \bibinfo
  {pages} {083011} (\bibinfo {year} {2015})}\BibitemShut {NoStop}%
\bibitem [{\citenamefont {Zhao}\ \emph {et~al.}(2016)\citenamefont {Zhao},
  \citenamefont {Tang},\ and\ \citenamefont
  {He}}]{Zha-Tan-He:2016:PHYSR4:GraLensRNdS}%
  \BibitemOpen
  \bibfield  {author} {\bibinfo {author} {\bibfnamefont {F.}~\bibnamefont
  {Zhao}}, \bibinfo {author} {\bibfnamefont {J.}~\bibnamefont {Tang}}, \ and\
  \bibinfo {author} {\bibfnamefont {F.}~\bibnamefont {He}},\ }\href {\doibase
  10.1103/PhysRevD.93.123017} {\bibfield  {journal} {\bibinfo  {journal} {Phys.
  Rev. D}\ }\textbf {\bibinfo {volume} {93}},\ \bibinfo {pages} {123017}
  (\bibinfo {year} {2016})}\BibitemShut {NoStop}%
\bibitem [{\citenamefont {Stuchl{\'i}k}\ and\ \citenamefont
  {Hled{\'i}k}(2002)}]{Stu-Hle:2002:ACTPS2:}%
  \BibitemOpen
  \bibfield  {author} {\bibinfo {author} {\bibfnamefont {Z.}~\bibnamefont
  {Stuchl{\'i}k}}\ and\ \bibinfo {author} {\bibfnamefont {S.}~\bibnamefont
  {Hled{\'i}k}},\ }\href {http://www.acta.sav.sk/acta02/no5/} {\bibfield
  {journal} {\bibinfo  {journal} {Acta Phys. Slovaca}\ }\textbf {\bibinfo
  {volume} {52}},\ \bibinfo {pages} {363} (\bibinfo {year} {2002})},\ \bibinfo
  {note} {erratum notice can be found at
  \url{http://www.acta.sav.sk/acta02/no6/}},\ \Eprint
  {http://arxiv.org/abs/0803.2685} {0803.2685} \BibitemShut {NoStop}%
\bibitem [{\citenamefont {Stuchl{\'i}k}\ and\ \citenamefont
  {Slan\'y}(2004)}]{Stu-Sla:2004:PHYSR4:}%
  \BibitemOpen
  \bibfield  {author} {\bibinfo {author} {\bibfnamefont {Z.}~\bibnamefont
  {Stuchl{\'i}k}}\ and\ \bibinfo {author} {\bibfnamefont {P.}~\bibnamefont
  {Slan\'y}},\ }\href@noop {} {\bibfield  {journal} {\bibinfo  {journal} {Phys.
  Rev. D}\ }\textbf {\bibinfo {volume} {69}},\ \bibinfo {pages} {064001}
  (\bibinfo {year} {2004})},\ \Eprint {http://arxiv.org/abs/gr-qc/0307049}
  {gr-qc/0307049} \BibitemShut {NoStop}%
\bibitem [{\citenamefont {Kraniotis}(2004)}]{Kra:2004:CLAQG:}%
  \BibitemOpen
  \bibfield  {author} {\bibinfo {author} {\bibfnamefont {G.~V.}\ \bibnamefont
  {Kraniotis}},\ }\href@noop {} {\bibfield  {journal} {\bibinfo  {journal}
  {Classical Quantum Gravity}\ }\textbf {\bibinfo {volume} {21}},\ \bibinfo
  {pages} {4743} (\bibinfo {year} {2004})}\BibitemShut {NoStop}%
\bibitem [{\citenamefont {Kraniotis}(2005)}]{Kra:2005:DARK:CCPerPrec}%
  \BibitemOpen
  \bibfield  {author} {\bibinfo {author} {\bibfnamefont {G.~V.}\ \bibnamefont
  {Kraniotis}},\ }in\ \href@noop {} {\emph {\bibinfo {booktitle} {{Dark matter
  in astro- and particle physics. Proceedings of the International Conference
  DARK 2004, College Station, Texas, USA, 3--9 October, 2004}}}},\ \bibinfo
  {editor} {edited by\ \bibinfo {editor} {\bibfnamefont {H.~V.}\ \bibnamefont
  {Klapdor-Kleingrothaus}}\ and\ \bibinfo {editor} {\bibfnamefont
  {R.}~\bibnamefont {Arnowitt}}}\ (\bibinfo  {publisher} {Springer},\ \bibinfo
  {address} {Berlin},\ \bibinfo {year} {2005})\ pp.\ \bibinfo {pages}
  {469--479}\BibitemShut {NoStop}%
\bibitem [{\citenamefont {Kraniotis}(2007)}]{Kra:2007:CLAQG:Periapsis}%
  \BibitemOpen
  \bibfield  {author} {\bibinfo {author} {\bibfnamefont {G.~V.}\ \bibnamefont
  {Kraniotis}},\ }\href@noop {} {\bibfield  {journal} {\bibinfo  {journal}
  {Classical Quantum Gravity}\ }\textbf {\bibinfo {volume} {24}},\ \bibinfo
  {pages} {1775} (\bibinfo {year} {2007})},\ \Eprint
  {http://arxiv.org/abs/gr-qc/0602056} {gr-qc/0602056} \BibitemShut {NoStop}%
\bibitem [{\citenamefont {Cruz}\ \emph {et~al.}(2005)\citenamefont {Cruz},
  \citenamefont {Olivares},\ and\ \citenamefont
  {Villanueva}}]{Cru-Oli-Vil:2005:CLAQG:GeoSdSBH}%
  \BibitemOpen
  \bibfield  {author} {\bibinfo {author} {\bibfnamefont {N.}~\bibnamefont
  {Cruz}}, \bibinfo {author} {\bibfnamefont {M.}~\bibnamefont {Olivares}}, \
  and\ \bibinfo {author} {\bibfnamefont {J.~R.}\ \bibnamefont {Villanueva}},\
  }\href {\doibase 10.1088/0264-9381/22/6/016} {\bibfield  {journal} {\bibinfo
  {journal} {Classical Quantum Gravity}\ }\textbf {\bibinfo {volume} {22}},\
  \bibinfo {pages} {1167} (\bibinfo {year} {2005})}\BibitemShut {NoStop}%
\bibitem [{\citenamefont {Kagramanova}\ \emph {et~al.}(2006)\citenamefont
  {Kagramanova}, \citenamefont {Kunz},\ and\ \citenamefont
  {Lammerzahl}}]{Kag-Kun-Lam:2006:PHYLB:SolarSdS}%
  \BibitemOpen
  \bibfield  {author} {\bibinfo {author} {\bibfnamefont {V.}~\bibnamefont
  {Kagramanova}}, \bibinfo {author} {\bibfnamefont {J.}~\bibnamefont {Kunz}}, \
  and\ \bibinfo {author} {\bibfnamefont {C.}~\bibnamefont {Lammerzahl}},\
  }\href {\doibase 10.1016/j.physletb.2006.01.069} {\bibfield  {journal}
  {\bibinfo  {journal} {Phys. Lett. B}\ }\textbf {\bibinfo {volume} {634}},\
  \bibinfo {pages} {465} (\bibinfo {year} {2006})}\BibitemShut {NoStop}%
\bibitem [{\citenamefont {Aliev}(2007)}]{Ali:2007:PHYSR4:EMPropKadS}%
  \BibitemOpen
  \bibfield  {author} {\bibinfo {author} {\bibfnamefont {A.~N.}\ \bibnamefont
  {Aliev}},\ }\href {\doibase 10.1103/PhysRevD.75.084041} {\bibfield  {journal}
  {\bibinfo  {journal} {Phys. Rev. D}\ }\textbf {\bibinfo {volume} {75}},\
  \bibinfo {pages} {084041} (\bibinfo {year} {2007})}\BibitemShut {NoStop}%
\bibitem [{\citenamefont {Chen}\ and\ \citenamefont
  {Wang}(2008)}]{Che:2008:CHINPB:DkEnGeoMorSchw}%
  \BibitemOpen
  \bibfield  {author} {\bibinfo {author} {\bibfnamefont {J.-H.}\ \bibnamefont
  {Chen}}\ and\ \bibinfo {author} {\bibfnamefont {Y.-J.}\ \bibnamefont
  {Wang}},\ }\href {\doibase 10.1088/1674-1056/17/4/006} {\bibfield  {journal}
  {\bibinfo  {journal} {Chinese Physics~B}\ }\textbf {\bibinfo {volume} {17}},\
  \bibinfo {pages} {1184} (\bibinfo {year} {2008})}\BibitemShut {NoStop}%
\bibitem [{\citenamefont {Iorio}(2009)}]{Ior:2009:NEWASTR:CCDGPGrav}%
  \BibitemOpen
  \bibfield  {author} {\bibinfo {author} {\bibfnamefont {L.}~\bibnamefont
  {Iorio}},\ }\href {\doibase 10.1016/j.newast.2008.08.002} {\bibfield
  {journal} {\bibinfo  {journal} {New Astronomy}\ }\textbf {\bibinfo {volume}
  {14}},\ \bibinfo {pages} {196} (\bibinfo {year} {2009})}\BibitemShut
  {NoStop}%
\bibitem [{\citenamefont {Hackmann}\ \emph {et~al.}(2010)\citenamefont
  {Hackmann}, \citenamefont {Hartmann}, \citenamefont {L{\" a}mmerzahl},\ and\
  \citenamefont {Sirimachan}}]{Hac-etal:2010:PHYSR4:KerrBHCoStr}%
  \BibitemOpen
  \bibfield  {author} {\bibinfo {author} {\bibfnamefont {E.}~\bibnamefont
  {Hackmann}}, \bibinfo {author} {\bibfnamefont {B.}~\bibnamefont {Hartmann}},
  \bibinfo {author} {\bibfnamefont {C.}~\bibnamefont {L{\" a}mmerzahl}}, \ and\
  \bibinfo {author} {\bibfnamefont {P.}~\bibnamefont {Sirimachan}},\ }\href
  {\doibase 10.1103/PhysRevD.82.044024} {\bibfield  {journal} {\bibinfo
  {journal} {Phys. Rev. D}\ }\textbf {\bibinfo {volume} {82}},\ \bibinfo
  {pages} {044024} (\bibinfo {year} {2010})}\BibitemShut {NoStop}%
\bibitem [{\citenamefont {Olivares}\ \emph {et~al.}(2011)\citenamefont
  {Olivares}, \citenamefont {Saavedra}, \citenamefont {Leiva},\ and\
  \citenamefont {Villanueva}}]{Oli-etal:2011:MODPLA:ChaParRNadS}%
  \BibitemOpen
  \bibfield  {author} {\bibinfo {author} {\bibfnamefont {M.}~\bibnamefont
  {Olivares}}, \bibinfo {author} {\bibfnamefont {J.}~\bibnamefont {Saavedra}},
  \bibinfo {author} {\bibfnamefont {C.}~\bibnamefont {Leiva}}, \ and\ \bibinfo
  {author} {\bibfnamefont {J.~R.}\ \bibnamefont {Villanueva}},\ }\href
  {\doibase 10.1142/S0217732311037261} {\bibfield  {journal} {\bibinfo
  {journal} {Modern Phys. Lett. A}\ }\textbf {\bibinfo {volume} {26}},\
  \bibinfo {pages} {2923} (\bibinfo {year} {2011})}\BibitemShut {NoStop}%
\bibitem [{\citenamefont {Chavanis}\ and\ \citenamefont
  {Harko}(2012)}]{Cha-Har:2012:PHYSR4:BEConGRStar}%
  \BibitemOpen
  \bibfield  {author} {\bibinfo {author} {\bibfnamefont {P.-H.}\ \bibnamefont
  {Chavanis}}\ and\ \bibinfo {author} {\bibfnamefont {T.}~\bibnamefont
  {Harko}},\ }\href {\doibase 10.1103/PhysRevD.86.064011} {\bibfield  {journal}
  {\bibinfo  {journal} {Phys. Rev. D}\ }\textbf {\bibinfo {volume} {86}},\
  \bibinfo {pages} {064011} (\bibinfo {year} {2012})}\BibitemShut {NoStop}%
\bibitem [{\citenamefont {Chauvineau}\ and\ \citenamefont
  {Regimbau}(2012)}]{Cha-Reg:2012:PHYSR4:LocEffCC}%
  \BibitemOpen
  \bibfield  {author} {\bibinfo {author} {\bibfnamefont {B.}~\bibnamefont
  {Chauvineau}}\ and\ \bibinfo {author} {\bibfnamefont {T.}~\bibnamefont
  {Regimbau}},\ }\href {\doibase 10.1103/PhysRevD.85.067302} {\bibfield
  {journal} {\bibinfo  {journal} {Phys. Rev. D}\ }\textbf {\bibinfo {volume}
  {85}},\ \bibinfo {pages} {067302} (\bibinfo {year} {2012})}\BibitemShut
  {NoStop}%
\bibitem [{\citenamefont {Zou}\ \emph {et~al.}(2014)\citenamefont {Zou},
  \citenamefont {Li},\ and\ \citenamefont {Li}}]{Zou-Li-Li:2014:INTJMD:TOVSdS}%
  \BibitemOpen
  \bibfield  {author} {\bibinfo {author} {\bibfnamefont {L.}~\bibnamefont
  {Zou}}, \bibinfo {author} {\bibfnamefont {F.-Y.}\ \bibnamefont {Li}}, \ and\
  \bibinfo {author} {\bibfnamefont {T.}~\bibnamefont {Li}},\ }\href {\doibase
  10.1142/S0218271814500163} {\bibfield  {journal} {\bibinfo  {journal}
  {Internat. J. Modern Phys. D}\ }\textbf {\bibinfo {volume} {23}},\ \bibinfo
  {pages} {1450016} (\bibinfo {year} {2014})}\BibitemShut {NoStop}%
\bibitem [{\citenamefont {Sarkar}\ \emph {et~al.}(2014)\citenamefont {Sarkar},
  \citenamefont {Ghosh},\ and\ \citenamefont
  {Bhadra}}]{Sar-Gho-Bha:2014:PHYSR4:NewtAnalogSdS}%
  \BibitemOpen
  \bibfield  {author} {\bibinfo {author} {\bibfnamefont {T.}~\bibnamefont
  {Sarkar}}, \bibinfo {author} {\bibfnamefont {S.}~\bibnamefont {Ghosh}}, \
  and\ \bibinfo {author} {\bibfnamefont {A.}~\bibnamefont {Bhadra}},\ }\href
  {\doibase 10.1103/PhysRevD.90.063008} {\bibfield  {journal} {\bibinfo
  {journal} {Phys. Rev. D}\ }\textbf {\bibinfo {volume} {90}},\ \bibinfo
  {pages} {063008} (\bibinfo {year} {2014})}\BibitemShut {NoStop}%
\bibitem [{\citenamefont {Gonz{\'{a}}lez}\ \emph {et~al.}(2015)\citenamefont
  {Gonz{\'{a}}lez}, \citenamefont {Olivares},\ and\ \citenamefont
  {V{\'{a}}squez}}]{Gon-Oli-Vas:2015:EPHYJC:4DAdS}%
  \BibitemOpen
  \bibfield  {author} {\bibinfo {author} {\bibfnamefont {P.~A.}\ \bibnamefont
  {Gonz{\'{a}}lez}}, \bibinfo {author} {\bibfnamefont {M.}~\bibnamefont
  {Olivares}}, \ and\ \bibinfo {author} {\bibfnamefont {Y.}~\bibnamefont
  {V{\'{a}}squez}},\ }\href {\doibase 10.1140/epjc/s10052-015-3690-4}
  {\bibfield  {journal} {\bibinfo  {journal} {European Physical Journal C}\
  }\textbf {\bibinfo {volume} {75}},\ \bibinfo {pages} {464} (\bibinfo {year}
  {2015})},\ \Eprint {http://arxiv.org/abs/1507.03610} {arXiv:1507.03610}
  \BibitemShut {NoStop}%
\bibitem [{\citenamefont {Maciel}\ \emph {et~al.}(2015)\citenamefont {Maciel},
  \citenamefont {Guariento},\ and\ \citenamefont
  {Molina}}]{Mac-Gua-Mol:2015:PHYSR4:CBHandWH}%
  \BibitemOpen
  \bibfield  {author} {\bibinfo {author} {\bibfnamefont {A.}~\bibnamefont
  {Maciel}}, \bibinfo {author} {\bibfnamefont {D.~C.}\ \bibnamefont
  {Guariento}}, \ and\ \bibinfo {author} {\bibfnamefont {C.}~\bibnamefont
  {Molina}},\ }\href {\doibase 10.1103/PhysRevD.91.084043} {\bibfield
  {journal} {\bibinfo  {journal} {Phys. Rev. D}\ }\textbf {\bibinfo {volume}
  {91}},\ \bibinfo {pages} {084043} (\bibinfo {year} {2015})}\BibitemShut
  {NoStop}%
\bibitem [{\citenamefont {Kunst}\ \emph {et~al.}(2015)\citenamefont {Kunst},
  \citenamefont {Perlick},\ and\ \citenamefont
  {L{\"{a}}mmerzahl}}]{Kun-Per-Lam:2015:PHYSR4:IsofreqPairSdS}%
  \BibitemOpen
  \bibfield  {author} {\bibinfo {author} {\bibfnamefont {D.}~\bibnamefont
  {Kunst}}, \bibinfo {author} {\bibfnamefont {V.}~\bibnamefont {Perlick}}, \
  and\ \bibinfo {author} {\bibfnamefont {C.}~\bibnamefont {L{\"{a}}mmerzahl}},\
  }\href {\doibase 10.1103/PhysRevD.92.024029} {\bibfield  {journal} {\bibinfo
  {journal} {Phys. Rev. D}\ }\textbf {\bibinfo {volume} {92}},\ \bibinfo
  {pages} {024029} (\bibinfo {year} {2015})}\BibitemShut {NoStop}%
\bibitem [{\citenamefont {Zakharov}(2015)}]{Zak:2015:JASPASN:AltToSmBH}%
  \BibitemOpen
  \bibfield  {author} {\bibinfo {author} {\bibfnamefont {A.~F.}\ \bibnamefont
  {Zakharov}},\ }\href {\doibase 10.1007/s12036-015-9345-x} {\bibfield
  {journal} {\bibinfo  {journal} {Journal of Astrophysics and Astronomy}\
  }\textbf {\bibinfo {volume} {36}},\ \bibinfo {pages} {0} (\bibinfo {year}
  {2015})}\BibitemShut {NoStop}%
\bibitem [{\citenamefont
  {Arraut}(2015{\natexlab{a}})}]{Arr:2015:IJMPD:OnBHAltThGraNonlin}%
  \BibitemOpen
  \bibfield  {author} {\bibinfo {author} {\bibfnamefont {I.}~\bibnamefont
  {Arraut}},\ }\href {\doibase 10.1142/S0218271815500224} {\bibfield  {journal}
  {\bibinfo  {journal} {Internat. J. Modern Phys. D}\ }\textbf {\bibinfo
  {volume} {24}},\ \bibinfo {pages} {1550022} (\bibinfo {year}
  {2015}{\natexlab{a}})}\BibitemShut {NoStop}%
\bibitem [{\citenamefont
  {Arraut}(2015{\natexlab{b}})}]{Arr:2015:EURLE:OnApparLoss}%
  \BibitemOpen
  \bibfield  {author} {\bibinfo {author} {\bibfnamefont {I.}~\bibnamefont
  {Arraut}},\ }\href {\doibase 10.1209/0295-5075/109/10002} {\bibfield
  {journal} {\bibinfo  {journal} {Europhys. Lett.}\ }\textbf {\bibinfo {volume}
  {109}},\ \bibinfo {pages} {10002} (\bibinfo {year}
  {2015}{\natexlab{b}})}\BibitemShut {NoStop}%
\bibitem [{\citenamefont {Sporea}\ and\ \citenamefont
  {Borowiec}(2016)}]{Spo-Bor:2016:INTJMD:GBfactorsSdS}%
  \BibitemOpen
  \bibfield  {author} {\bibinfo {author} {\bibfnamefont {C.~A.}\ \bibnamefont
  {Sporea}}\ and\ \bibinfo {author} {\bibfnamefont {A.}~\bibnamefont
  {Borowiec}},\ }\href {\doibase 10.1142/S0218271816500437} {\bibfield
  {journal} {\bibinfo  {journal} {Internat. J. Modern Phys. D}\ }\textbf
  {\bibinfo {volume} {25}},\ \bibinfo {pages} {1650043} (\bibinfo {year}
  {2016})},\ \Eprint {http://arxiv.org/abs/1509.00831} {arXiv:1509.00831}
  \BibitemShut {NoStop}%
\bibitem [{\citenamefont {Jacobson}\ and\ \citenamefont
  {Sotiriou}(2009)}]{Jac-Sot:2009:PHYSR4:StrDynSpiBH}%
  \BibitemOpen
  \bibfield  {author} {\bibinfo {author} {\bibfnamefont {T.}~\bibnamefont
  {Jacobson}}\ and\ \bibinfo {author} {\bibfnamefont {T.~P.}\ \bibnamefont
  {Sotiriou}},\ }\href {\doibase 10.1103/PhysRevD.79.065029} {\bibfield
  {journal} {\bibinfo  {journal} {Phys. Rev. D}\ }\textbf {\bibinfo {volume}
  {79}},\ \bibinfo {pages} {065029} (\bibinfo {year} {2009})}\BibitemShut
  {NoStop}%
\bibitem [{\citenamefont {Kolo\v{s}}\ and\ \citenamefont
  {Stuchl{\'i}k}(2010)}]{Kol-Stu:2010:PHYSR4:CurCarStrLoops}%
  \BibitemOpen
  \bibfield  {author} {\bibinfo {author} {\bibfnamefont {M.}~\bibnamefont
  {Kolo\v{s}}}\ and\ \bibinfo {author} {\bibfnamefont {Z.}~\bibnamefont
  {Stuchl{\'i}k}},\ }\href {\doibase 10.1103/PhysRevD.82.125012} {\bibfield
  {journal} {\bibinfo  {journal} {Phys. Rev. D}\ }\textbf {\bibinfo {volume}
  {82}},\ \bibinfo {pages} {125012 (21 pages)} (\bibinfo {year}
  {2010})}\BibitemShut {NoStop}%
\bibitem [{\citenamefont {Gu}\ and\ \citenamefont
  {Cheng}(2007)}]{Gu-Cheng:2007:GENRG2:CircLoopKdS}%
  \BibitemOpen
  \bibfield  {author} {\bibinfo {author} {\bibfnamefont {Z.}~\bibnamefont
  {Gu}}\ and\ \bibinfo {author} {\bibfnamefont {H.}~\bibnamefont {Cheng}},\
  }\href {\doibase 10.1007/s10714-006-0329-7} {\bibfield  {journal} {\bibinfo
  {journal} {Gen. Relativity Gravitation}\ }\textbf {\bibinfo {volume} {39}},\
  \bibinfo {pages} {1} (\bibinfo {year} {2007})}\BibitemShut {NoStop}%
\bibitem [{\citenamefont {Wang}\ and\ \citenamefont
  {Cheng}(2012)}]{Wan-Che:2012:PHYLB:CirLoopPerTens}%
  \BibitemOpen
  \bibfield  {author} {\bibinfo {author} {\bibfnamefont {L.}~\bibnamefont
  {Wang}}\ and\ \bibinfo {author} {\bibfnamefont {H.}~\bibnamefont {Cheng}},\
  }\href {\doibase http://dx.doi.org/10.1016/j.physletb.2012.05.034} {\bibfield
   {journal} {\bibinfo  {journal} {Phys. Lett. B}\ }\textbf {\bibinfo {volume}
  {713}},\ \bibinfo {pages} {59} (\bibinfo {year} {2012})}\BibitemShut
  {NoStop}%
\bibitem [{\citenamefont {Stuchl{\'i}k}\ and\ \citenamefont
  {Kolo\v{s}}(2012{\natexlab{a}})}]{Stu-Kol:2012:PHYSR4:AccStringLoops}%
  \BibitemOpen
  \bibfield  {author} {\bibinfo {author} {\bibfnamefont {Z.}~\bibnamefont
  {Stuchl{\'i}k}}\ and\ \bibinfo {author} {\bibfnamefont {M.}~\bibnamefont
  {Kolo\v{s}}},\ }\href {\doibase 10.1103/PhysRevD.85.065022} {\bibfield
  {journal} {\bibinfo  {journal} {Phys. Rev. D}\ }\textbf {\bibinfo {volume}
  {85}},\ \bibinfo {pages} {065022 [13~pages]} (\bibinfo {year}
  {2012}{\natexlab{a}})}\BibitemShut {NoStop}%
\bibitem [{\citenamefont {Stuchl{\'i}k}\ and\ \citenamefont
  {Kolo\v{s}}(2012{\natexlab{b}})}]{Stu-Kol:2012:JCAP:StringLoops}%
  \BibitemOpen
  \bibfield  {author} {\bibinfo {author} {\bibfnamefont {Z.}~\bibnamefont
  {Stuchl{\'i}k}}\ and\ \bibinfo {author} {\bibfnamefont {M.}~\bibnamefont
  {Kolo\v{s}}},\ }\href {\doibase 10.1088/1475-7516/2012/10/008} {\bibfield
  {journal} {\bibinfo  {journal} {Journal of Cosmology and Astroparticle
  Physics}\ }\textbf {\bibinfo {volume} {2012}},\ \bibinfo {pages} {008}
  (\bibinfo {year} {2012}{\natexlab{b}})}\BibitemShut {NoStop}%
\bibitem [{\citenamefont {Stuchl{\'i}k}\ and\ \citenamefont
  {Kolo\v{s}}(2014)}]{Stu-Kol:2014:PHYSR4:KerrStrLoopOsc}%
  \BibitemOpen
  \bibfield  {author} {\bibinfo {author} {\bibfnamefont {Z.}~\bibnamefont
  {Stuchl{\'i}k}}\ and\ \bibinfo {author} {\bibfnamefont {M.}~\bibnamefont
  {Kolo\v{s}}},\ }\href {\doibase 10.1103/PhysRevD.89.065007} {\bibfield
  {journal} {\bibinfo  {journal} {Phys. Rev. D}\ }\textbf {\bibinfo {volume}
  {89}},\ \bibinfo {eid} {065007} (\bibinfo {year} {2014})},\ \Eprint
  {http://arxiv.org/abs/1403.2748} {1403.2748 [astro-ph.HE]} \BibitemShut
  {NoStop}%
\bibitem [{\citenamefont {M{\"u}ller}\ and\ \citenamefont
  {Aschenbach}(2007)}]{Mul-Asc:2007:CLAQG:VelProKadSBH}%
  \BibitemOpen
  \bibfield  {author} {\bibinfo {author} {\bibfnamefont {A.}~\bibnamefont
  {M{\"u}ller}}\ and\ \bibinfo {author} {\bibfnamefont {B.}~\bibnamefont
  {Aschenbach}},\ }\href {\doibase 10.1088/0264-9381/24/10/009} {\bibfield
  {journal} {\bibinfo  {journal} {Classical Quantum Gravity}\ }\textbf
  {\bibinfo {volume} {24}},\ \bibinfo {pages} {2637} (\bibinfo {year}
  {2007})},\ \Eprint {http://arxiv.org/abs/0704.3963} {0704.3963} \BibitemShut
  {NoStop}%
\bibitem [{\citenamefont {Slan\'y}\ and\ \citenamefont
  {Stuchl{\'i}k}(2008)}]{Sla-Stu:2008:CLAQG:CmtNoMonKadS}%
  \BibitemOpen
  \bibfield  {author} {\bibinfo {author} {\bibfnamefont {P.}~\bibnamefont
  {Slan\'y}}\ and\ \bibinfo {author} {\bibfnamefont {Z.}~\bibnamefont
  {Stuchl{\'i}k}},\ }\href {\doibase 10.1088/0264-9381/25/3/038001} {\bibfield
  {journal} {\bibinfo  {journal} {Classical Quantum Gravity}\ }\textbf
  {\bibinfo {volume} {25}},\ \bibinfo {pages} {038001} (\bibinfo {year}
  {2008})}\BibitemShut {NoStop}%
\bibitem [{\citenamefont {Stuchl{\'i}k}\ \emph
  {et~al.}(2000{\natexlab{a}})\citenamefont {Stuchl{\'i}k}, \citenamefont
  {Slan\'y},\ and\ \citenamefont {Hled{\'i}k}}]{Stu-Sla-Hle:2000:ASTRA:}%
  \BibitemOpen
  \bibfield  {author} {\bibinfo {author} {\bibfnamefont {Z.}~\bibnamefont
  {Stuchl{\'i}k}}, \bibinfo {author} {\bibfnamefont {P.}~\bibnamefont
  {Slan\'y}}, \ and\ \bibinfo {author} {\bibfnamefont {S.}~\bibnamefont
  {Hled{\'i}k}},\ }\href {http://adsabs.harvard.edu/abs/2000A\%26A...363..425S}
  {\bibfield  {journal} {\bibinfo  {journal} {Astronomy and Astrophysics}\
  }\textbf {\bibinfo {volume} {363}},\ \bibinfo {pages} {425} (\bibinfo {year}
  {2000}{\natexlab{a}})}\BibitemShut {NoStop}%
\bibitem [{\citenamefont {Slan\'y}\ and\ \citenamefont
  {Stuchl{\'i}k}(2005)}]{Sla-Stu:2005:CLAQG:}%
  \BibitemOpen
  \bibfield  {author} {\bibinfo {author} {\bibfnamefont {P.}~\bibnamefont
  {Slan\'y}}\ and\ \bibinfo {author} {\bibfnamefont {Z.}~\bibnamefont
  {Stuchl{\'i}k}},\ }\href
  {http://www.iop.org/EJ/abstract/-search=44947255.2/0264-9381/22/17/019}
  {\bibfield  {journal} {\bibinfo  {journal} {Classical Quantum Gravity}\
  }\textbf {\bibinfo {volume} {22}},\ \bibinfo {pages} {3623} (\bibinfo {year}
  {2005})}\BibitemShut {NoStop}%
\bibitem [{\citenamefont {Rezzolla}\ \emph {et~al.}(2003)\citenamefont
  {Rezzolla}, \citenamefont {Zanotti},\ and\ \citenamefont
  {Font}}]{Rez-Zan-Fon:2003:ASTRA:}%
  \BibitemOpen
  \bibfield  {author} {\bibinfo {author} {\bibfnamefont {L.}~\bibnamefont
  {Rezzolla}}, \bibinfo {author} {\bibfnamefont {O.}~\bibnamefont {Zanotti}}, \
  and\ \bibinfo {author} {\bibfnamefont {J.~A.}\ \bibnamefont {Font}},\ }\href
  {\doibase 10.1051/0004-6361:20031457} {\bibfield  {journal} {\bibinfo
  {journal} {Astronomy and Astrophysics}\ }\textbf {\bibinfo {volume} {412}},\
  \bibinfo {pages} {603} (\bibinfo {year} {2003})}\BibitemShut {NoStop}%
\bibitem [{\citenamefont {Aschenbach}(2008)}]{Asc:2008:CHIAA:MassSpinBHQPO}%
  \BibitemOpen
  \bibfield  {author} {\bibinfo {author} {\bibfnamefont {B.}~\bibnamefont
  {Aschenbach}},\ }\href@noop {} {\bibfield  {journal} {\bibinfo  {journal}
  {Chinese Astronom. Astrophys.}\ }\textbf {\bibinfo {volume} {8}},\ \bibinfo
  {pages} {291} (\bibinfo {year} {2008})},\ \bibinfo {note} {7th International
  Workshop on Multifrequency Behaviour of High Energy Cosmic Sources, Vulcano,
  Italy, May~28--Jun~02, 2007}\BibitemShut {NoStop}%
\bibitem [{\citenamefont {Stuchl{\'i}k}\ \emph {et~al.}(2005)\citenamefont
  {Stuchl{\'i}k}, \citenamefont {Slan\'y}, \citenamefont {T{\" o}r{\" o}k},\
  and\ \citenamefont {Abramowicz}}]{Stu-etal:2005:PHYSR4:AschenUnexpTopo}%
  \BibitemOpen
  \bibfield  {author} {\bibinfo {author} {\bibfnamefont {Z.}~\bibnamefont
  {Stuchl{\'i}k}}, \bibinfo {author} {\bibfnamefont {P.}~\bibnamefont
  {Slan\'y}}, \bibinfo {author} {\bibfnamefont {G.}~\bibnamefont {T{\" o}r{\"
  o}k}}, \ and\ \bibinfo {author} {\bibfnamefont {M.~A.}\ \bibnamefont
  {Abramowicz}},\ }\href {\doibase 10.1103/PhysRevD.71.024037} {\bibfield
  {journal} {\bibinfo  {journal} {Phys. Rev. D}\ }\textbf {\bibinfo {volume}
  {71}},\ \bibinfo {pages} {024037} (\bibinfo {year} {2005})}\BibitemShut
  {NoStop}%
\bibitem [{\citenamefont {Kučáková}\ \emph {et~al.}(2011)\citenamefont
  {Kučáková}, \citenamefont {Slan\'y},\ and\ \citenamefont
  {Stuchl{\'i}k}}]{Kuc-Sla-Stu:2011:JCAP:ToroPerFlRNadS}%
  \BibitemOpen
  \bibfield  {author} {\bibinfo {author} {\bibfnamefont {H.}~\bibnamefont
  {Kučáková}}, \bibinfo {author} {\bibfnamefont {P.}~\bibnamefont {Slan\'y}}, \
  and\ \bibinfo {author} {\bibfnamefont {Z.}~\bibnamefont {Stuchl{\'i}k}},\
  }\href {\doibase 10.1088/1475-7516/2011/01/033} {\bibfield  {journal}
  {\bibinfo  {journal} {Journal of Cosmology and Astroparticle Physics}\
  }\textbf {\bibinfo {volume} {2011}},\ \bibinfo {pages} {033} (\bibinfo {year}
  {2011})}\BibitemShut {NoStop}%
\bibitem [{\citenamefont {Perez}\ \emph {et~al.}(2013)\citenamefont {Perez},
  \citenamefont {Romero},\ and\ \citenamefont
  {Perez~Bergliaffa}}]{Per-Rom-PeB:2013:ASTRA:AccDiBHModGra}%
  \BibitemOpen
  \bibfield  {author} {\bibinfo {author} {\bibfnamefont {D.}~\bibnamefont
  {Perez}}, \bibinfo {author} {\bibfnamefont {G.~E.}\ \bibnamefont {Romero}}, \
  and\ \bibinfo {author} {\bibfnamefont {S.~E.}\ \bibnamefont
  {Perez~Bergliaffa}},\ }\href {\doibase 10.1051/0004-6361/201220378}
  {\bibfield  {journal} {\bibinfo  {journal} {Astronomy and Astrophysics}\
  }\textbf {\bibinfo {volume} {551}},\ \bibinfo {pages} {A4} (\bibinfo {year}
  {2013})}\BibitemShut {NoStop}%
\bibitem [{\citenamefont {Chakraborty}(2015)}]{Cha:2015:CLAQG:EqConfArndBH}%
  \BibitemOpen
  \bibfield  {author} {\bibinfo {author} {\bibfnamefont {S.}~\bibnamefont
  {Chakraborty}},\ }\href {\doibase 10.1088/0264-9381/32/7/075007} {\bibfield
  {journal} {\bibinfo  {journal} {Classical Quantum Gravity}\ }\textbf
  {\bibinfo {volume} {32}},\ \bibinfo {pages} {075007} (\bibinfo {year}
  {2015})}\BibitemShut {NoStop}%
\bibitem [{\citenamefont {Pugliese}\ and\ \citenamefont
  {Stuchl{\'i}k}(2015)}]{Pug-Stu:2015:ASTJS:RingedAccDiskEq}%
  \BibitemOpen
  \bibfield  {author} {\bibinfo {author} {\bibfnamefont {D.}~\bibnamefont
  {Pugliese}}\ and\ \bibinfo {author} {\bibfnamefont {Z.}~\bibnamefont
  {Stuchl{\'i}k}},\ }\href {\doibase 10.1088/0067-0049/221/2/25} {\bibfield
  {journal} {\bibinfo  {journal} {Astrophys. J. Suppl.}\ }\textbf {\bibinfo
  {volume} {221}},\ \bibinfo {pages} {25} (\bibinfo {year} {2015})}\BibitemShut
  {NoStop}%
\bibitem [{\citenamefont {Pugliese}\ and\ \citenamefont
  {Stuchl{\'i}k}(2016)}]{Pug-Stu:2016:ASTJS:RingAccDiskInstab}%
  \BibitemOpen
  \bibfield  {author} {\bibinfo {author} {\bibfnamefont {D.}~\bibnamefont
  {Pugliese}}\ and\ \bibinfo {author} {\bibfnamefont {Z.}~\bibnamefont
  {Stuchl{\'i}k}},\ }\href {\doibase 10.3847/0067-0049/223/2/27} {\bibfield
  {journal} {\bibinfo  {journal} {Astrophys. J. Suppl.}\ }\textbf {\bibinfo
  {volume} {223}},\ \bibinfo {pages} {27} (\bibinfo {year} {2016})}\BibitemShut
  {NoStop}%
\bibitem [{\citenamefont {Karkowski}\ and\ \citenamefont
  {Malec}(2013)}]{Kar-Mal:2013:PHYSR4:BondiAccCBH}%
  \BibitemOpen
  \bibfield  {author} {\bibinfo {author} {\bibfnamefont {J.}~\bibnamefont
  {Karkowski}}\ and\ \bibinfo {author} {\bibfnamefont {E.}~\bibnamefont
  {Malec}},\ }\href {\doibase 10.1103/PhysRevD.87.044007} {\bibfield  {journal}
  {\bibinfo  {journal} {Phys. Rev. D}\ }\textbf {\bibinfo {volume} {87}},\
  \bibinfo {pages} {044007} (\bibinfo {year} {2013})}\BibitemShut {NoStop}%
\bibitem [{\citenamefont {Mach}\ and\ \citenamefont
  {Malec}(2013)}]{Mac-Mal:2013:PHYSR4:StabBondiAccSadS}%
  \BibitemOpen
  \bibfield  {author} {\bibinfo {author} {\bibfnamefont {P.}~\bibnamefont
  {Mach}}\ and\ \bibinfo {author} {\bibfnamefont {E.}~\bibnamefont {Malec}},\
  }\href {\doibase 10.1103/PhysRevD.88.084055} {\bibfield  {journal} {\bibinfo
  {journal} {Phys. Rev. D}\ }\textbf {\bibinfo {volume} {88}},\ \bibinfo
  {pages} {084055} (\bibinfo {year} {2013})}\BibitemShut {NoStop}%
\bibitem [{\citenamefont {Mach}\ \emph {et~al.}(2013)\citenamefont {Mach},
  \citenamefont {Malec},\ and\ \citenamefont
  {Karkowski}}]{Mac-Mal-Kar:2013:PHYSR4:SphSteadAccFlow}%
  \BibitemOpen
  \bibfield  {author} {\bibinfo {author} {\bibfnamefont {P.}~\bibnamefont
  {Mach}}, \bibinfo {author} {\bibfnamefont {E.}~\bibnamefont {Malec}}, \ and\
  \bibinfo {author} {\bibfnamefont {J.}~\bibnamefont {Karkowski}},\ }\href
  {\doibase 10.1103/PhysRevD.88.084056} {\bibfield  {journal} {\bibinfo
  {journal} {Phys. Rev. D}\ }\textbf {\bibinfo {volume} {88}},\ \bibinfo
  {pages} {084056} (\bibinfo {year} {2013})}\BibitemShut {NoStop}%
\bibitem [{\citenamefont {Mach}(2015)}]{Mac:2015:PHYSR4:HomoclinAccrSadS}%
  \BibitemOpen
  \bibfield  {author} {\bibinfo {author} {\bibfnamefont {P.}~\bibnamefont
  {Mach}},\ }\href {\doibase 10.1103/PhysRevD.91.084016} {\bibfield  {journal}
  {\bibinfo  {journal} {Phys. Rev. D}\ }\textbf {\bibinfo {volume} {91}},\
  \bibinfo {pages} {084016} (\bibinfo {year} {2015})}\BibitemShut {NoStop}%
\bibitem [{\citenamefont {Ficek}(2015)}]{Fic:2015:CLAQG:BondiTypeAcc}%
  \BibitemOpen
  \bibfield  {author} {\bibinfo {author} {\bibfnamefont {F.}~\bibnamefont
  {Ficek}},\ }\href {\doibase 10.1088/0264-9381/32/23/235008} {\bibfield
  {journal} {\bibinfo  {journal} {Classical Quantum Gravity}\ }\textbf
  {\bibinfo {volume} {32}},\ \bibinfo {pages} {235008} (\bibinfo {year}
  {2015})}\BibitemShut {NoStop}%
\bibitem [{\citenamefont {Stuchl{\'i}k}\ and\ \citenamefont
  {Kov\'{a}\v{r}}(2008)}]{Stu-Kov:2008:INTJMD:PsNewtSdS}%
  \BibitemOpen
  \bibfield  {author} {\bibinfo {author} {\bibfnamefont {Z.}~\bibnamefont
  {Stuchl{\'i}k}}\ and\ \bibinfo {author} {\bibfnamefont {J.}~\bibnamefont
  {Kov\'{a}\v{r}}},\ }\href {http://arxiv.org/abs/0803.3641} {\bibfield
  {journal} {\bibinfo  {journal} {INTJMD}\ }\textbf {\bibinfo {volume} {17}},\
  \bibinfo {pages} {2089} (\bibinfo {year} {2008})},\ \Eprint
  {http://arxiv.org/abs/0803.3641} {0803.3641} \BibitemShut {NoStop}%
\bibitem [{\citenamefont {Stuchl{\'i}k}\ \emph
  {et~al.}(2009{\natexlab{a}})\citenamefont {Stuchl{\'i}k}, \citenamefont
  {Slan\'y},\ and\ \citenamefont
  {Kov\'{a}\v{r}}}]{Stu-Sla-Kov:2009:CLAQG:PseNewSdS}%
  \BibitemOpen
  \bibfield  {author} {\bibinfo {author} {\bibfnamefont {Z.}~\bibnamefont
  {Stuchl{\'i}k}}, \bibinfo {author} {\bibfnamefont {P.}~\bibnamefont
  {Slan\'y}}, \ and\ \bibinfo {author} {\bibfnamefont {J.}~\bibnamefont
  {Kov\'{a}\v{r}}},\ }\href {\doibase 10.1088/0264-9381/26/21/215013}
  {\bibfield  {journal} {\bibinfo  {journal} {Classical Quantum Gravity}\
  }\textbf {\bibinfo {volume} {26}},\ \bibinfo {pages} {215013 (34~pp)}
  (\bibinfo {year} {2009}{\natexlab{a}})},\ \Eprint
  {http://arxiv.org/abs/0910.3184v1} {0910.3184v1} \BibitemShut {NoStop}%
\bibitem [{\citenamefont {Stuchl{\'i}k}\ and\ \citenamefont
  {Schee}(2011)}]{Stu-Sch:2011:JCAP:CCMagOnCloud}%
  \BibitemOpen
  \bibfield  {author} {\bibinfo {author} {\bibfnamefont {Z.}~\bibnamefont
  {Stuchl{\'i}k}}\ and\ \bibinfo {author} {\bibfnamefont {J.}~\bibnamefont
  {Schee}},\ }\href {\doibase 10.1088/1475-7516/2011/09/018} {\bibfield
  {journal} {\bibinfo  {journal} {Journal of Cosmology and Astroparticle
  Physics}\ }\textbf {\bibinfo {volume} {9}},\ \bibinfo {eid} {018} (\bibinfo
  {year} {2011})}\BibitemShut {NoStop}%
\bibitem [{\citenamefont {Schee}\ \emph {et~al.}(2013)\citenamefont {Schee},
  \citenamefont {Stuchl{\'i}k},\ and\ \citenamefont
  {Petrásek}}]{Sch-Stu-Pet:2013:JCAP:MOND}%
  \BibitemOpen
  \bibfield  {author} {\bibinfo {author} {\bibfnamefont {J.}~\bibnamefont
  {Schee}}, \bibinfo {author} {\bibfnamefont {Z.}~\bibnamefont {Stuchl{\'i}k}},
  \ and\ \bibinfo {author} {\bibfnamefont {M.}~\bibnamefont {Petrásek}},\
  }\href {\doibase 10.1088/1475-7516/2013/12/026} {\bibfield  {journal}
  {\bibinfo  {journal} {Journal of Cosmology and Astroparticle Physics}\
  }\textbf {\bibinfo {volume} {12}},\ \bibinfo {eid} {026} (\bibinfo {year}
  {2013})},\ \Eprint {http://arxiv.org/abs/1312.0817} {1312.0817} \BibitemShut
  {NoStop}%
\bibitem [{\citenamefont {Stuchl{\'i}k}\ and\ \citenamefont
  {Schee}(2012{\natexlab{a}})}]{Stu-Sch:2012:INTJMD:GRvsPsNewtMagClou}%
  \BibitemOpen
  \bibfield  {author} {\bibinfo {author} {\bibfnamefont {Z.}~\bibnamefont
  {Stuchl{\'i}k}}\ and\ \bibinfo {author} {\bibfnamefont {J.}~\bibnamefont
  {Schee}},\ }\href {\doibase 10.1142/S0218271812500319} {\bibfield  {journal}
  {\bibinfo  {journal} {Internat. J. Modern Phys. D}\ }\textbf {\bibinfo
  {volume} {21}},\ \bibinfo {pages} {1250031} (\bibinfo {year}
  {2012}{\natexlab{a}})}\BibitemShut {NoStop}%
\bibitem [{\citenamefont {Gimon}\ and\ \citenamefont
  {Ho{\v{r}}ava}(2004)}]{Gim-Hor:2004:hep-th0405019:GodHolo}%
  \BibitemOpen
  \bibfield  {author} {\bibinfo {author} {\bibfnamefont {E.~G.}\ \bibnamefont
  {Gimon}}\ and\ \bibinfo {author} {\bibfnamefont {P.}~\bibnamefont
  {Ho{\v{r}}ava}},\ }\href@noop {} {\enquote {\bibinfo {title} {{Over-Rotating
  Black Holes, G{\"o}del Holography and the Hypertube}},}\ } (\bibinfo {year}
  {2004}),\ \Eprint {http://arxiv.org/abs/hep-th/0405019v1} {hep-th/0405019v1}
  \BibitemShut {NoStop}%
\bibitem [{\citenamefont {Boyda}\ \emph {et~al.}(2003)\citenamefont {Boyda},
  \citenamefont {Ganguli}, \citenamefont {Ho{\v{r}}ava},\ and\ \citenamefont
  {Varadarajan}}]{Boy-etal:2003:PHYSR4:HoloProtChron}%
  \BibitemOpen
  \bibfield  {author} {\bibinfo {author} {\bibfnamefont {E.~K.}\ \bibnamefont
  {Boyda}}, \bibinfo {author} {\bibfnamefont {S.}~\bibnamefont {Ganguli}},
  \bibinfo {author} {\bibfnamefont {P.}~\bibnamefont {Ho{\v{r}}ava}}, \ and\
  \bibinfo {author} {\bibfnamefont {U.}~\bibnamefont {Varadarajan}},\ }\href
  {\doibase 10.1103/PhysRevD.67.106003} {\bibfield  {journal} {\bibinfo
  {journal} {Phys. Rev. D}\ }\textbf {\bibinfo {volume} {67}},\ \bibinfo
  {pages} {106003} (\bibinfo {year} {2003})},\ \Eprint
  {http://arxiv.org/abs/hep-th/0212087v2} {hep-th/0212087v2} \BibitemShut
  {NoStop}%
\bibitem [{\citenamefont {Gimon}\ and\ \citenamefont
  {Hořava}(2009)}]{Gim-Hor:2009:PHYLB:AstVioSignStr}%
  \BibitemOpen
  \bibfield  {author} {\bibinfo {author} {\bibfnamefont {E.~G.}\ \bibnamefont
  {Gimon}}\ and\ \bibinfo {author} {\bibfnamefont {P.}~\bibnamefont {Hořava}},\
  }\href@noop {} {\bibfield  {journal} {\bibinfo  {journal} {Phys. Lett. B}\
  }\textbf {\bibinfo {volume} {672}},\ \bibinfo {pages} {299} (\bibinfo {year}
  {2009})},\ \Eprint {http://arxiv.org/abs/0706.2873v1} {0706.2873v1}
  \BibitemShut {NoStop}%
\bibitem [{\citenamefont {Stuchl{\'i}k}\ and\ \citenamefont
  {Schee}(2012{\natexlab{b}})}]{Stu-Sch:2012:CLAQG:ObsPrimKerrSS}%
  \BibitemOpen
  \bibfield  {author} {\bibinfo {author} {\bibfnamefont {Z.}~\bibnamefont
  {Stuchl{\'i}k}}\ and\ \bibinfo {author} {\bibfnamefont {J.}~\bibnamefont
  {Schee}},\ }\href {\doibase 10.1088/0264-9381/29/6/065002} {\bibfield
  {journal} {\bibinfo  {journal} {Classical Quantum Gravity}\ }\textbf
  {\bibinfo {volume} {29}},\ \bibinfo {pages} {065002} (\bibinfo {year}
  {2012}{\natexlab{b}})}\BibitemShut {NoStop}%
\bibitem [{\citenamefont {de~Felice}(1974)}]{deFel:1974:ASTRA:}%
  \BibitemOpen
  \bibfield  {author} {\bibinfo {author} {\bibfnamefont {F.}~\bibnamefont
  {de~Felice}},\ }\href {http://adsabs.harvard.edu/article_service.html}
  {\bibfield  {journal} {\bibinfo  {journal} {Astronomy and Astrophysics}\
  }\textbf {\bibinfo {volume} {34}},\ \bibinfo {pages} {15} (\bibinfo {year}
  {1974})}\BibitemShut {NoStop}%
\bibitem [{\citenamefont {Cunningham}(1975)}]{Cun:1975:ASTRJ2:}%
  \BibitemOpen
  \bibfield  {author} {\bibinfo {author} {\bibfnamefont {C.~T.}\ \bibnamefont
  {Cunningham}},\ }\href@noop {} {\bibfield  {journal} {\bibinfo  {journal}
  {Astrophys. J.}\ }\textbf {\bibinfo {volume} {202}},\ \bibinfo {pages} {788}
  (\bibinfo {year} {1975})}\BibitemShut {NoStop}%
\bibitem [{\citenamefont {de~Felice}(1978)}]{deFel:1978:NATURE:InstabNS}%
  \BibitemOpen
  \bibfield  {author} {\bibinfo {author} {\bibfnamefont {F.}~\bibnamefont
  {de~Felice}},\ }\href {\doibase 10.1038/273429a0} {\bibfield  {journal}
  {\bibinfo  {journal} {Nature}\ }\textbf {\bibinfo {volume} {273}},\ \bibinfo
  {pages} {429} (\bibinfo {year} {1978})}\BibitemShut {NoStop}%
\bibitem [{\citenamefont {Stuchl{\'i}k}(1980)}]{Stu:1980:BULAI:}%
  \BibitemOpen
  \bibfield  {author} {\bibinfo {author} {\bibfnamefont {Z.}~\bibnamefont
  {Stuchl{\'i}k}},\ }\href@noop {} {\bibfield  {journal} {\bibinfo  {journal}
  {Bull. Astronom. Inst. Czechoslovakia}\ }\textbf {\bibinfo {volume} {31}},\
  \bibinfo {pages} {129} (\bibinfo {year} {1980})}\BibitemShut {NoStop}%
\bibitem [{\citenamefont {Stuchl{\'i}k}\ \emph {et~al.}(2011)\citenamefont
  {Stuchl{\'i}k}, \citenamefont {Hled{\'i}k},\ and\ \citenamefont
  {Truparov\'{a}}}]{Stu-Hle-Tru:2011:CLAQG:}%
  \BibitemOpen
  \bibfield  {author} {\bibinfo {author} {\bibfnamefont {Z.}~\bibnamefont
  {Stuchl{\'i}k}}, \bibinfo {author} {\bibfnamefont {S.}~\bibnamefont
  {Hled{\'i}k}}, \ and\ \bibinfo {author} {\bibfnamefont {K.}~\bibnamefont
  {Truparov\'{a}}},\ }\href {\doibase 10.1088/0264-9381/28/15/155017}
  {\bibfield  {journal} {\bibinfo  {journal} {Classical Quantum Gravity}\
  }\textbf {\bibinfo {volume} {28}} (\bibinfo {year} {2011}),\
  10.1088/0264-9381/28/15/155017}\BibitemShut {NoStop}%
\bibitem [{\citenamefont {Hioki}\ and\ \citenamefont
  {Maeda}(2009)}]{Hio-Mae:2009:PHYSR4:KerrSpinMeas}%
  \BibitemOpen
  \bibfield  {author} {\bibinfo {author} {\bibfnamefont {K.}~\bibnamefont
  {Hioki}}\ and\ \bibinfo {author} {\bibfnamefont {K.-i.}\ \bibnamefont
  {Maeda}},\ }\href {\doibase 10.1103/PhysRevD.80.024042} {\bibfield  {journal}
  {\bibinfo  {journal} {Phys. Rev. D}\ }\textbf {\bibinfo {volume} {80}},\
  \bibinfo {pages} {024042 (9~pages)} (\bibinfo {year} {2009})},\ \Eprint
  {http://arxiv.org/abs/0904.3575v3} {0904.3575v3} \BibitemShut {NoStop}%
\bibitem [{\citenamefont {Stuchl{\'i}k}\ and\ \citenamefont
  {Schee}(2010)}]{Stu-Sch:2010:CLAQG:AppKepDiOrKerrSSp}%
  \BibitemOpen
  \bibfield  {author} {\bibinfo {author} {\bibfnamefont {Z.}~\bibnamefont
  {Stuchl{\'i}k}}\ and\ \bibinfo {author} {\bibfnamefont {J.}~\bibnamefont
  {Schee}},\ }\href {\doibase 10.1088/0264-9381/27/21/215017} {\bibfield
  {journal} {\bibinfo  {journal} {Classical Quantum Gravity}\ }\textbf
  {\bibinfo {volume} {27}},\ \bibinfo {pages} {215017 (39~pages)} (\bibinfo
  {year} {2010})},\ \Eprint {http://arxiv.org/abs/1101.3569} {1101.3569}
  \BibitemShut {NoStop}%
\bibitem [{\citenamefont {Stuchl{\'i}k}\ and\ \citenamefont
  {Schee}(2012{\natexlab{c}})}]{Stu-Sch:2012:CLAQG:CntRKerrSSpi}%
  \BibitemOpen
  \bibfield  {author} {\bibinfo {author} {\bibfnamefont {Z.}~\bibnamefont
  {Stuchl{\'i}k}}\ and\ \bibinfo {author} {\bibfnamefont {J.}~\bibnamefont
  {Schee}},\ }\href {\doibase 10.1088/0264-9381/29/2/025008} {\bibfield
  {journal} {\bibinfo  {journal} {Classical Quantum Gravity}\ }\textbf
  {\bibinfo {volume} {29}},\ \bibinfo {pages} {025008} (\bibinfo {year}
  {2012}{\natexlab{c}})}\BibitemShut {NoStop}%
\bibitem [{\citenamefont {Stuchl{\'i}k}\ and\ \citenamefont
  {Schee}(2013)}]{Stu-Sch:2013:CLAQG:UHEKerrGeo}%
  \BibitemOpen
  \bibfield  {author} {\bibinfo {author} {\bibfnamefont {Z.}~\bibnamefont
  {Stuchl{\'i}k}}\ and\ \bibinfo {author} {\bibfnamefont {J.}~\bibnamefont
  {Schee}},\ }\href {\doibase 10.1088/0264-9381/30/7/075012} {\bibfield
  {journal} {\bibinfo  {journal} {Classical Quantum Gravity}\ }\textbf
  {\bibinfo {volume} {30}},\ \bibinfo {pages} {075012} (\bibinfo {year}
  {2013})}\BibitemShut {NoStop}%
\bibitem [{\citenamefont {Bosma}(1981)}]{Bos:1981:ASTRJ1:21cmSpiGal}%
  \BibitemOpen
  \bibfield  {author} {\bibinfo {author} {\bibfnamefont {A.}~\bibnamefont
  {Bosma}},\ }\href {\doibase 10.1086/113062} {\bibfield  {journal} {\bibinfo
  {journal} {Astronom. J.}\ }\textbf {\bibinfo {volume} {86}},\ \bibinfo
  {pages} {1791} (\bibinfo {year} {1981})}\BibitemShut {NoStop}%
\bibitem [{\citenamefont {Rubin}(1982)}]{Rub:1982:HIContNorGal:SaSbScGal}%
  \BibitemOpen
  \bibfield  {author} {\bibinfo {author} {\bibfnamefont {V.~C.}\ \bibnamefont
  {Rubin}},\ }in\ \href@noop {} {{\selectlanguage {english}\emph {\bibinfo
  {booktitle} {{Comparative HI Content of Normal Galaxies, Proceedings of the
  Workshop}}}}},\ \bibinfo {editor} {edited by\ \bibinfo {editor} {\bibnamefont
  {Knudsen}}}\ (\bibinfo {year} {1982})\ p.~\bibinfo {pages} {42},\ \bibinfo
  {note} {1982chcn.conf...42R}\BibitemShut {NoStop}%
\bibitem [{\citenamefont {Binney}\ and\ \citenamefont
  {Tremaine}(1988)}]{Bin-Tre:1988:GalacDynam:}%
  \BibitemOpen
  \bibfield  {author} {\bibinfo {author} {\bibfnamefont {J.}~\bibnamefont
  {Binney}}\ and\ \bibinfo {author} {\bibfnamefont {S.}~\bibnamefont
  {Tremaine}},\ }\href@noop {} {{\selectlanguage {english}\emph {\bibinfo
  {title} {{Galactic Dynamics}}}}},\ {Princeton Series in Astrophysics}\
  (\bibinfo  {publisher} {Princeton University Press},\ \bibinfo {address}
  {Princeton},\ \bibinfo {year} {1988})\ p.\ \bibinfo {pages} {755}\BibitemShut
  {NoStop}%
\bibitem [{\citenamefont {Iorio}(2010)}]{Ior:2010:MONNR:GalOrbMoDarkMat}%
  \BibitemOpen
  \bibfield  {author} {\bibinfo {author} {\bibfnamefont {L.}~\bibnamefont
  {Iorio}},\ }\href {\doibase 10.1111/j.1365-2966.2009.15811.x} {\bibfield
  {journal} {\bibinfo  {journal} {Monthly Notices Roy. Astronom. Soc.}\
  }\textbf {\bibinfo {volume} {401}},\ \bibinfo {pages} {2012} (\bibinfo {year}
  {2010})}\BibitemShut {NoStop}%
\bibitem [{\citenamefont {Navarro}\ \emph {et~al.}(1997)\citenamefont
  {Navarro}, \citenamefont {Frenk},\ and\ \citenamefont
  {White}}]{Nav-Fre-Whi:1997:ASTRJ2:UniDeProHiCl}%
  \BibitemOpen
  \bibfield  {author} {\bibinfo {author} {\bibfnamefont {J.~F.}\ \bibnamefont
  {Navarro}}, \bibinfo {author} {\bibfnamefont {C.~S.}\ \bibnamefont {Frenk}},
  \ and\ \bibinfo {author} {\bibfnamefont {S.~D.~M.}\ \bibnamefont {White}},\
  }\href {\doibase 10.1086/304888} {\bibfield  {journal} {\bibinfo  {journal}
  {Astrophys. J.}\ }\textbf {\bibinfo {volume} {490}},\ \bibinfo {pages} {493}
  (\bibinfo {year} {1997})}\BibitemShut {NoStop}%
\bibitem [{\citenamefont {Cremaschini}\ and\ \citenamefont
  {Stuchl{\'i}k}(2013)}]{Cre-Stu:2013:INTJMD:KinThGravSys}%
  \BibitemOpen
  \bibfield  {author} {\bibinfo {author} {\bibfnamefont {C.}~\bibnamefont
  {Cremaschini}}\ and\ \bibinfo {author} {\bibfnamefont {Z.}~\bibnamefont
  {Stuchl{\'i}k}},\ }\href {\doibase 10.1142/S0218271813500776} {\bibfield
  {journal} {\bibinfo  {journal} {Internat. J. Modern Phys. D}\ }\textbf
  {\bibinfo {volume} {22}},\ \bibinfo {eid} {1350077} (\bibinfo {year}
  {2013})}\BibitemShut {NoStop}%
\bibitem [{\citenamefont {Weinberg}(2008)}]{Wei:2008:Cosmology:1}%
  \BibitemOpen
  \bibfield  {author} {\bibinfo {author} {\bibfnamefont {S.}~\bibnamefont
  {Weinberg}},\ }\href@noop {} {{\selectlanguage {english}\emph {\bibinfo
  {title} {{Cosmology}}}}},\ \bibinfo {edition} {1st}\ ed.\ (\bibinfo
  {publisher} {Oxford University Press},\ \bibinfo {address} {New York},\
  \bibinfo {year} {2008})\ p.\ \bibinfo {pages} {544}\BibitemShut {NoStop}%
\bibitem [{\citenamefont {Tooper}(1964)}]{Too:1964:ASTRJ2:}%
  \BibitemOpen
  \bibfield  {author} {\bibinfo {author} {\bibfnamefont {R.~F.}\ \bibnamefont
  {Tooper}},\ }\href@noop {} {\bibfield  {journal} {\bibinfo  {journal}
  {Astrophys. J.}\ }\textbf {\bibinfo {volume} {140}},\ \bibinfo {pages} {434}
  (\bibinfo {year} {1964})}\BibitemShut {NoStop}%
\bibitem [{\citenamefont {Börner}(1993)}]{Bor:1993:EarlyUniv:}%
  \BibitemOpen
  \bibfield  {author} {\bibinfo {author} {\bibfnamefont {G.}~\bibnamefont
  {Börner}},\ }\href@noop {} {\emph {\bibinfo {title} {{The Early Universe}}}}\
  (\bibinfo  {publisher} {Springer-Verlag},\ \bibinfo {address}
  {Berlin--Heidelberg--New York},\ \bibinfo {year} {1993})\BibitemShut
  {NoStop}%
\bibitem [{\citenamefont {Kolb}\ and\ \citenamefont
  {Turner}(1990)}]{Kol-Tur:1990:EarUni:}%
  \BibitemOpen
  \bibfield  {author} {\bibinfo {author} {\bibfnamefont {E.~W.}\ \bibnamefont
  {Kolb}}\ and\ \bibinfo {author} {\bibfnamefont {M.~S.}\ \bibnamefont
  {Turner}},\ }\href@noop {} {\emph {\bibinfo {title} {{The Early Universe}}}}\
  (\bibinfo  {publisher} {Addison-Wesley},\ \bibinfo {address} {Redwood City,
  California},\ \bibinfo {year} {1990})\ \bibinfo {note} {the Advanced Book
  Program}\BibitemShut {NoStop}%
\bibitem [{\citenamefont {Shapiro}\ and\ \citenamefont
  {Teukolsky}(1983)}]{Sha-Teu:1983:BHWDNS:}%
  \BibitemOpen
  \bibfield  {author} {\bibinfo {author} {\bibfnamefont {S.~L.}\ \bibnamefont
  {Shapiro}}\ and\ \bibinfo {author} {\bibfnamefont {S.~A.}\ \bibnamefont
  {Teukolsky}},\ }\href@noop {} {\emph {\bibinfo {title} {{Black Holes, White
  Dwarfs and Neutron Stars: The Physics of Compact Objects}}}}\ (\bibinfo
  {publisher} {John Wiley \&{} Sons},\ \bibinfo {address} {New York},\ \bibinfo
  {year} {1983})\ p.\ \bibinfo {pages} {672}\BibitemShut {NoStop}%
\bibitem [{\citenamefont {Stuchl{\'i}k}\ \emph {et~al.}(2001)\citenamefont
  {Stuchl{\'i}k}, \citenamefont {Hled{\'i}k}, \citenamefont {\v{S}olt\'{e}s},\
  and\ \citenamefont {{\O}stgaard}}]{Stu-etal:2001:PHYSR4:}%
  \BibitemOpen
  \bibfield  {author} {\bibinfo {author} {\bibfnamefont {Z.}~\bibnamefont
  {Stuchl{\'i}k}}, \bibinfo {author} {\bibfnamefont {S.}~\bibnamefont
  {Hled{\'i}k}}, \bibinfo {author} {\bibfnamefont {J.}~\bibnamefont
  {\v{S}olt\'{e}s}}, \ and\ \bibinfo {author} {\bibfnamefont {E.}~\bibnamefont
  {{\O}stgaard}},\ }\href {\doibase 10.1103/PhysRevD.64.044004} {\bibfield
  {journal} {\bibinfo  {journal} {Phys. Rev. D}\ }\textbf {\bibinfo {volume}
  {64}},\ \bibinfo {pages} {044004 (17~pages)} (\bibinfo {year}
  {2001})}\BibitemShut {NoStop}%
\bibitem [{\citenamefont {Nilsson}\ and\ \citenamefont
  {Uggla}(2000{\natexlab{a}})}]{Nil-Ugg:2000:ANNPH1:GRStarPoEqSt}%
  \BibitemOpen
  \bibfield  {author} {\bibinfo {author} {\bibfnamefont {U.~S.}\ \bibnamefont
  {Nilsson}}\ and\ \bibinfo {author} {\bibfnamefont {C.}~\bibnamefont
  {Uggla}},\ }in\  \cite{Nil-Ugl:2000:ANNPH1:GRStarsPolyEoS},\ pp.\ \bibinfo
  {pages} {292--319},\ \Eprint {http://arxiv.org/abs/gr-qc/0002022}
  {gr-qc/0002022} \BibitemShut {NoStop}%
\bibitem [{\citenamefont {B{\" o}hmer}\ and\ \citenamefont
  {Fodor}(2008)}]{Boe-Fod:2008:PHYSR4:}%
  \BibitemOpen
  \bibfield  {author} {\bibinfo {author} {\bibfnamefont {C.~G.}\ \bibnamefont
  {B{\" o}hmer}}\ and\ \bibinfo {author} {\bibfnamefont {G.}~\bibnamefont
  {Fodor}},\ }\href {\doibase 10.1103/PhysRevD.77.064008} {\bibfield  {journal}
  {\bibinfo  {journal} {Phys. Rev. D}\ }\textbf {\bibinfo {volume} {77}},\
  \bibinfo {pages} {064008} (\bibinfo {year} {2008})}\BibitemShut {NoStop}%
\bibitem [{\citenamefont {Abramowicz}(1990)}]{Abr:1990:MONNR:}%
  \BibitemOpen
  \bibfield  {author} {\bibinfo {author} {\bibfnamefont {M.~A.}\ \bibnamefont
  {Abramowicz}},\ }\href
  {http://adsabs.harvard.edu/cgi-bin/nph-bib_query?bibcode=1990MNRAS.245..733A&db_key=AST}
  {\bibfield  {journal} {\bibinfo  {journal} {Monthly Notices Roy. Astronom.
  Soc.}\ }\textbf {\bibinfo {volume} {245}},\ \bibinfo {pages} {733} (\bibinfo
  {year} {1990})}\BibitemShut {NoStop}%
\bibitem [{\citenamefont {Stuchl{\'i}k}\ \emph
  {et~al.}(2000{\natexlab{b}})\citenamefont {Stuchl{\'i}k}, \citenamefont
  {Hled{\'i}k},\ and\ \citenamefont {Jur{\'a}{\v
  n}}}]{Stu-Hle-Jur:2000:CLAQG:}%
  \BibitemOpen
  \bibfield  {author} {\bibinfo {author} {\bibfnamefont {Z.}~\bibnamefont
  {Stuchl{\'i}k}}, \bibinfo {author} {\bibfnamefont {S.}~\bibnamefont
  {Hled{\'i}k}}, \ and\ \bibinfo {author} {\bibfnamefont {J.}~\bibnamefont
  {Jur{\'a}{\v n}}},\ }\href {\doibase 10.1088/0264-9381/17/14/307} {\bibfield
  {journal} {\bibinfo  {journal} {Classical Quantum Gravity}\ }\textbf
  {\bibinfo {volume} {17}},\ \bibinfo {pages} {2691} (\bibinfo {year}
  {2000}{\natexlab{b}})},\ \Eprint {http://arxiv.org/abs/0803.2533} {0803.2533}
  \BibitemShut {NoStop}%
\bibitem [{\citenamefont {Hladík}\ and\ \citenamefont
  {Stuchl{\'i}k}(2011)}]{Hla-Stu:2011:JCAP:}%
  \BibitemOpen
  \bibfield  {author} {\bibinfo {author} {\bibfnamefont {J.}~\bibnamefont
  {Hladík}}\ and\ \bibinfo {author} {\bibfnamefont {Z.}~\bibnamefont
  {Stuchl{\'i}k}},\ }\href {\doibase 10.1088/1475-7516/2011/07/012} {\bibfield
  {journal} {\bibinfo  {journal} {Journal of Cosmology and Astroparticle
  Physics}\ }\textbf {\bibinfo {volume} {2011}},\ \bibinfo {pages} {012}
  (\bibinfo {year} {2011})}\BibitemShut {NoStop}%
\bibitem [{\citenamefont {Ziolkowski}(2008)}]{Zio:2008:CHIAA:MassBHU}%
  \BibitemOpen
  \bibfield  {author} {\bibinfo {author} {\bibfnamefont {J.}~\bibnamefont
  {Ziolkowski}},\ }\href@noop {} {\bibfield  {journal} {\bibinfo  {journal}
  {Chinese Astronom. Astrophys.}\ }\textbf {\bibinfo {volume} {8}},\ \bibinfo
  {pages} {273} (\bibinfo {year} {2008})},\ \bibinfo {note} {{7th International
  Workshop on Multifrequency Behaviour of High Energy Cosmic Sources, Vulcano,
  Italy, May~28--Jun~02, 2007}}\BibitemShut {NoStop}%
\bibitem [{\citenamefont {Arraut}(2013)}]{Ara:2013:MODPLA:PropGwAsymdS}%
  \BibitemOpen
  \bibfield  {author} {\bibinfo {author} {\bibfnamefont {I.}~\bibnamefont
  {Arraut}},\ }\href {\doibase 10.1142/S0217732313500193} {\bibfield  {journal}
  {\bibinfo  {journal} {Modern Phys. Lett. A}\ }\textbf {\bibinfo {volume}
  {28}} (\bibinfo {year} {2013}),\ 10.1142/S0217732313500193}\BibitemShut
  {NoStop}%
\bibitem [{\citenamefont {Arraut}(2014)}]{Ara:2014:1406.2571:KomarMass}%
  \BibitemOpen
  \bibfield  {author} {\bibinfo {author} {\bibfnamefont {I.}~\bibnamefont
  {Arraut}},\ }\href {http://arxiv.org/abs/1406.2571} {\bibfield  {journal}
  {\bibinfo  {journal} {ArXiv e-prints}\ } (\bibinfo {year} {2014})},\ \Eprint
  {http://arxiv.org/abs/1406.2571} {arXiv:1406.2571 [gr-qc]} \BibitemShut
  {NoStop}%
\bibitem [{\citenamefont {Landau}\ and\ \citenamefont
  {Lifshitz}(1987)}]{Lan-Lif:1987:FluidMech:}%
  \BibitemOpen
  \bibfield  {author} {\bibinfo {author} {\bibfnamefont {L.~D.}\ \bibnamefont
  {Landau}}\ and\ \bibinfo {author} {\bibfnamefont {E.~M.}\ \bibnamefont
  {Lifshitz}},\ }\href@noop {} {\emph {\bibinfo {title} {Fluid Mechanics}}}\
  (\bibinfo  {publisher} {Pergamon Press},\ \bibinfo {address} {Oxford, Great
  Britain},\ \bibinfo {year} {1987})\BibitemShut {NoStop}%
\bibitem [{\citenamefont {Abramowicz}\ \emph {et~al.}(1988)\citenamefont
  {Abramowicz}, \citenamefont {Carter},\ and\ \citenamefont
  {Lasota}}]{Abr-Car-Las:1988:GENRG2:}%
  \BibitemOpen
  \bibfield  {author} {\bibinfo {author} {\bibfnamefont {M.~A.}\ \bibnamefont
  {Abramowicz}}, \bibinfo {author} {\bibfnamefont {B.}~\bibnamefont {Carter}},
  \ and\ \bibinfo {author} {\bibfnamefont {J.}~\bibnamefont {Lasota}},\
  }\href@noop {} {\bibfield  {journal} {\bibinfo  {journal} {Gen. Relativity
  Gravitation}\ }\textbf {\bibinfo {volume} {20}},\ \bibinfo {pages} {1173}
  (\bibinfo {year} {1988})}\BibitemShut {NoStop}%
\bibitem [{\citenamefont {Abramowicz}\ \emph {et~al.}(1993)\citenamefont
  {Abramowicz}, \citenamefont {Miller},\ and\ \citenamefont
  {Stuchl{\'i}k}}]{Abr-Mil-Stu:1993:PHYSR4:}%
  \BibitemOpen
  \bibfield  {author} {\bibinfo {author} {\bibfnamefont {M.~A.}\ \bibnamefont
  {Abramowicz}}, \bibinfo {author} {\bibfnamefont {J.~C.}\ \bibnamefont
  {Miller}}, \ and\ \bibinfo {author} {\bibfnamefont {Z.}~\bibnamefont
  {Stuchl{\'i}k}},\ }\href {http://link.aps.org/abstract/PRD/v47/p1440}
  {\bibfield  {journal} {\bibinfo  {journal} {Phys. Rev. D}\ }\textbf {\bibinfo
  {volume} {47}},\ \bibinfo {pages} {1440} (\bibinfo {year}
  {1993})}\BibitemShut {NoStop}%
\bibitem [{\citenamefont {Kov\'{a}\v{r}}\ and\ \citenamefont
  {Stuchl{\'i}k}(2007)}]{Kov-Stu:2007:CLAQG:}%
  \BibitemOpen
  \bibfield  {author} {\bibinfo {author} {\bibfnamefont {J.}~\bibnamefont
  {Kov\'{a}\v{r}}}\ and\ \bibinfo {author} {\bibfnamefont {Z.}~\bibnamefont
  {Stuchl{\'i}k}},\ }\href@noop {} {\bibfield  {journal} {\bibinfo  {journal}
  {Classical Quantum Gravity}\ }\textbf {\bibinfo {volume} {24}},\ \bibinfo
  {pages} {565} (\bibinfo {year} {2007})},\ \Eprint
  {http://arxiv.org/abs/gr-qc/0701028} {gr-qc/0701028} \BibitemShut {NoStop}%
\bibitem [{\citenamefont {Semerák}(1995)}]{Sem:1995:NUOC2:}%
  \BibitemOpen
  \bibfield  {author} {\bibinfo {author} {\bibfnamefont {O.}~\bibnamefont
  {Semerák}},\ }\href@noop {} {\bibfield  {journal} {\bibinfo  {journal} {Nuovo
  Cimento B}\ }\textbf {\bibinfo {volume} {110}},\ \bibinfo {pages} {973}
  (\bibinfo {year} {1995})}\BibitemShut {NoStop}%
\bibitem [{\citenamefont {Hled{\'i}k}(2002)}]{Hle:2002:JB60:}%
  \BibitemOpen
  \bibfield  {author} {\bibinfo {author} {\bibfnamefont {S.}~\bibnamefont
  {Hled{\'i}k}},\ }in\ \href@noop {} {\emph {\bibinfo {booktitle} {Gravitation:
  {F}ollowing the {P}rague {I}nspiration ({A} {V}olume in {C}elebration of the
  60th {B}irthday of {J}iří {B}ičák)}}},\ \bibinfo {editor} {edited by\
  \bibinfo {editor} {\bibfnamefont {O.}~\bibnamefont {Semer\'{a}k}}, \bibinfo
  {editor} {\bibfnamefont {J.}~\bibnamefont {Podolsk\'{y}}}, \ and\ \bibinfo
  {editor} {\bibfnamefont {M.}~\bibnamefont {Žofka}}}\ (\bibinfo  {publisher}
  {World Scientific},\ \bibinfo {address} {New Jersey, London, Singapore, Hong
  Kong},\ \bibinfo {year} {2002})\ pp.\ \bibinfo {pages} {161--192}\BibitemShut
  {NoStop}%
\bibitem [{\citenamefont {Stuchl{\'i}k}\ \emph {et~al.}(2012)\citenamefont
  {Stuchl{\'i}k}, \citenamefont {Hladík}, \citenamefont {Urbanec},\ and\
  \citenamefont {T{\" o}r{\" o}k}}]{Stu-etal:2012:GENRG2:NeutrinoTrap}%
  \BibitemOpen
  \bibfield  {author} {\bibinfo {author} {\bibfnamefont {Z.}~\bibnamefont
  {Stuchl{\'i}k}}, \bibinfo {author} {\bibfnamefont {J.}~\bibnamefont
  {Hladík}}, \bibinfo {author} {\bibfnamefont {M.}~\bibnamefont {Urbanec}}, \
  and\ \bibinfo {author} {\bibfnamefont {G.}~\bibnamefont {T{\" o}r{\" o}k}},\
  }\href {\doibase 10.1007/s10714-012-1346-3} {\bibfield  {journal} {\bibinfo
  {journal} {Gen. Relativity Gravitation}\ }\textbf {\bibinfo {volume} {44}},\
  \bibinfo {pages} {1393} (\bibinfo {year} {2012})}\BibitemShut {NoStop}%
\bibitem [{\citenamefont {Abramowicz}\ \emph {et~al.}(1995)\citenamefont
  {Abramowicz}, \citenamefont {Nurowski},\ and\ \citenamefont
  {Wex}}]{Abr-Nur-Wex:1995:CLAQG:}%
  \BibitemOpen
  \bibfield  {author} {\bibinfo {author} {\bibfnamefont {M.~A.}\ \bibnamefont
  {Abramowicz}}, \bibinfo {author} {\bibfnamefont {P.}~\bibnamefont
  {Nurowski}}, \ and\ \bibinfo {author} {\bibfnamefont {N.}~\bibnamefont
  {Wex}},\ }\href {http://www.iop.org/EJ/abstract/0264-9381/12/6/012}
  {\bibfield  {journal} {\bibinfo  {journal} {Classical Quantum Gravity}\
  }\textbf {\bibinfo {volume} {12}},\ \bibinfo {pages} {1467} (\bibinfo {year}
  {1995})}\BibitemShut {NoStop}%
\bibitem [{\citenamefont {B{\"o}hmer}(2004{\natexlab{b}})}]{Boh:2004:GENRQC:}%
  \BibitemOpen
  \bibfield  {author} {\bibinfo {author} {\bibfnamefont {C.~G.}\ \bibnamefont
  {B{\"o}hmer}},\ }\href@noop {} {\bibfield  {journal} {\bibinfo  {journal}
  {General Relativity and Quantum Cosmology}\ ,\ \bibinfo {pages} {1219}}
  (\bibinfo {year} {2004}{\natexlab{b}})},\ \Eprint
  {http://arxiv.org/abs/gr-qc/0409030} {gr-qc/0409030} \BibitemShut {NoStop}%
\bibitem [{\citenamefont {Stuchl{\'i}k}\ \emph
  {et~al.}(2009{\natexlab{b}})\citenamefont {Stuchl{\'i}k}, \citenamefont {T{\"
  o}r{\" o}k}, \citenamefont {Hled{\'i}k},\ and\ \citenamefont
  {Urbanec}}]{Stu-etal:2009:CLAQG:NeuTrap1Eff}%
  \BibitemOpen
  \bibfield  {author} {\bibinfo {author} {\bibfnamefont {Z.}~\bibnamefont
  {Stuchl{\'i}k}}, \bibinfo {author} {\bibfnamefont {G.}~\bibnamefont {T{\"
  o}r{\" o}k}}, \bibinfo {author} {\bibfnamefont {S.}~\bibnamefont
  {Hled{\'i}k}}, \ and\ \bibinfo {author} {\bibfnamefont {M.}~\bibnamefont
  {Urbanec}},\ }\href {\doibase 10.1088/0264-9381/26/3/035003} {\bibfield
  {journal} {\bibinfo  {journal} {Classical Quantum Gravity}\ }\textbf
  {\bibinfo {volume} {26}},\ \bibinfo {pages} {035003 (17~pp)} (\bibinfo {year}
  {2009}{\natexlab{b}})}\BibitemShut {NoStop}%
\bibitem [{\citenamefont {Nilsson}\ and\ \citenamefont
  {Uggla}(2000{\natexlab{b}})}]{Nil-Ugl:2000:ANNPH1:GRStarsPolyEoS}%
  \BibitemOpen
  \bibfield  {author} {\bibinfo {author} {\bibfnamefont {U.~S.}\ \bibnamefont
  {Nilsson}}\ and\ \bibinfo {author} {\bibfnamefont {C.}~\bibnamefont
  {Uggla}},\ }\href {\doibase 10.1006/aphy.2000.6090} {\bibfield  {journal}
  {\bibinfo  {journal} {Ann. Physics}\ }\textbf {\bibinfo {volume} {286}},\
  \bibinfo {pages} {292} (\bibinfo {year} {2000}{\natexlab{b}})},\ \Eprint
  {http://arxiv.org/abs/gr-qc/0002022} {gr-qc/0002022} \BibitemShut {NoStop}%
\bibitem [{\citenamefont {Faraoni}\ \emph {et~al.}(2015)\citenamefont
  {Faraoni}, \citenamefont {Lapierre-L{\'{e}}onard},\ and\ \citenamefont
  {Prain}}]{Far-LaL-Pra:2015:JCAP:TurnarRadAccUni}%
  \BibitemOpen
  \bibfield  {author} {\bibinfo {author} {\bibfnamefont {V.}~\bibnamefont
  {Faraoni}}, \bibinfo {author} {\bibfnamefont {M.}~\bibnamefont
  {Lapierre-L{\'{e}}onard}}, \ and\ \bibinfo {author} {\bibfnamefont
  {A.}~\bibnamefont {Prain}},\ }\href {\doibase 10.1088/1475-7516/2015/10/013}
  {\bibfield  {journal} {\bibinfo  {journal} {Journal of Cosmology and
  Astroparticle Physics}\ }\textbf {\bibinfo {volume} {2015}},\ \bibinfo
  {pages} {013} (\bibinfo {year} {2015})}\BibitemShut {NoStop}%
\bibitem [{\citenamefont {Faraoni}(2016)}]{Far:2016:PHYDARU:TurnarRadModGr}%
  \BibitemOpen
  \bibfield  {author} {\bibinfo {author} {\bibfnamefont {V.}~\bibnamefont
  {Faraoni}},\ }\href {\doibase 10.1016/j.dark.2015.11.001} {\bibfield
  {journal} {\bibinfo  {journal} {Physics of the Dark Universe}\ }\textbf
  {\bibinfo {volume} {11}},\ \bibinfo {pages} {11} (\bibinfo {year} {2016})},\
  \Eprint {http://arxiv.org/abs/1508.00475} {arXiv:1508.00475} \BibitemShut
  {NoStop}%
\bibitem [{\citenamefont
  {Ziolkowski}(2005{\natexlab{a}})}]{Zio:2005:MONNR:HDECygX1}%
  \BibitemOpen
  \bibfield  {author} {\bibinfo {author} {\bibfnamefont {J.}~\bibnamefont
  {Ziolkowski}},\ }\href {\doibase 10.1111/j.1365-2966.2005.08796.x} {\bibfield
   {journal} {\bibinfo  {journal} {Monthly Notices Roy. Astronom. Soc.}\
  }\textbf {\bibinfo {volume} {358}},\ \bibinfo {pages} {851} (\bibinfo {year}
  {2005}{\natexlab{a}})}\BibitemShut {NoStop}%
\bibitem [{\citenamefont
  {Ziolkowski}(2005{\natexlab{b}})}]{Zio:2005:NUOC2:GalCollObj}%
  \BibitemOpen
  \bibfield  {author} {\bibinfo {author} {\bibfnamefont {J.}~\bibnamefont
  {Ziolkowski}},\ }\href {\doibase 10.1393/ncb/i2005-10118-0} {\bibfield
  {journal} {\bibinfo  {journal} {Nuovo Cimento della Societa Italiana di
  fisica B~-- General physics relativity astronomy and mathematical physics and
  methods}\ }\textbf {\bibinfo {volume} {120}},\ \bibinfo {pages} {757}
  (\bibinfo {year} {2005}{\natexlab{b}})},\ \bibinfo {note} {{Vulcano Workshop
  2004 on Frontier Objects in Astrophysics, Vulcano, Italy, May 24--29,
  2004}}\BibitemShut {NoStop}%
\bibitem [{\citenamefont {Linde}(1985)}]{Lin:1985:NEWSCI:UniInfOutChaos}%
  \BibitemOpen
  \bibfield  {author} {\bibinfo {author} {\bibfnamefont {A.}~\bibnamefont
  {Linde}},\ }\href@noop {} {\bibfield  {journal} {\bibinfo  {journal} {New
  Scientist}\ }\textbf {\bibinfo {volume} {105}},\ \bibinfo {pages} {14}
  (\bibinfo {year} {1985})}\BibitemShut {NoStop}%
\bibitem [{\citenamefont {Boeckel}\ and\ \citenamefont
  {Schaffner-Bielich}(2010)}]{Boe-Til:2010:PHYRL:LitInf}%
  \BibitemOpen
  \bibfield  {author} {\bibinfo {author} {\bibfnamefont {T.}~\bibnamefont
  {Boeckel}}\ and\ \bibinfo {author} {\bibfnamefont {J.}~\bibnamefont
  {Schaffner-Bielich}},\ }\href {\doibase 10.1103/PhysRevLett.105.041301}
  {\bibfield  {journal} {\bibinfo  {journal} {Phys. Rev. Lett.}\ }\textbf
  {\bibinfo {volume} {105}},\ \bibinfo {pages} {041301} (\bibinfo {year}
  {2010})}\BibitemShut {NoStop}%
\bibitem [{\citenamefont {Boeckel}\ and\ \citenamefont
  {Schaffner-Bielich}(2012)}]{Boe-Til:2012:PHYSR4:LitInf}%
  \BibitemOpen
  \bibfield  {author} {\bibinfo {author} {\bibfnamefont {T.}~\bibnamefont
  {Boeckel}}\ and\ \bibinfo {author} {\bibfnamefont {J.}~\bibnamefont
  {Schaffner-Bielich}},\ }\href {\doibase 10.1103/PhysRevD.85.103506}
  {\bibfield  {journal} {\bibinfo  {journal} {Phys. Rev. D}\ }\textbf {\bibinfo
  {volume} {85}},\ \bibinfo {pages} {103506} (\bibinfo {year}
  {2012})}\BibitemShut {NoStop}%
\bibitem [{\citenamefont {Stuchl{\'i}k}\ and\ \citenamefont
  {Hled{\'i}k}(2005)}]{Stu-Hle-2:2005:RAGtime6and7:CrossRef}%
  \BibitemOpen
  \bibfield  {author} {\bibinfo {author} {\bibfnamefont {Z.}~\bibnamefont
  {Stuchl{\'i}k}}\ and\ \bibinfo {author} {\bibfnamefont {S.}~\bibnamefont
  {Hled{\'i}k}},\ }in\ \href@noop {} {\emph {\bibinfo {booktitle} {Proceedings
  of RAGtime 6/7: Workshops on black holes and neutron stars, Opava,
  16--18/18--20 September 2004/2005}}},\ \bibinfo {editor} {edited by\ \bibinfo
  {editor} {\bibfnamefont {S.}~\bibnamefont {Hled{\'i}k}}\ and\ \bibinfo
  {editor} {\bibfnamefont {Z.}~\bibnamefont {Stuchl{\'i}k}}}\ (\bibinfo
  {publisher} {Silesian University in Opava},\ \bibinfo {address} {Opava},\
  \bibinfo {year} {2005})\ pp.\ \bibinfo {pages} {209--222}\BibitemShut
  {NoStop}%
\bibitem [{\citenamefont {B{\"o}hmer}\ and\ \citenamefont
  {Harko}(2005)}]{Boh-Har:2005:PHYSR4:DynInsFluSph}%
  \BibitemOpen
  \bibfield  {author} {\bibinfo {author} {\bibfnamefont {C.~G.}\ \bibnamefont
  {B{\"o}hmer}}\ and\ \bibinfo {author} {\bibfnamefont {T.}~\bibnamefont
  {Harko}},\ }\href {\doibase 10.1103/PhysRevD.71.084026} {\bibfield  {journal}
  {\bibinfo  {journal} {Phys. Rev. D}\ }\textbf {\bibinfo {volume} {71}},\
  \bibinfo {pages} {084026} (\bibinfo {year} {2005})}\BibitemShut {NoStop}%
\bibitem [{\citenamefont {Tooper}(1965)}]{Too:1965:ASTRJ2:}%
  \BibitemOpen
  \bibfield  {author} {\bibinfo {author} {\bibfnamefont {R.~F.}\ \bibnamefont
  {Tooper}},\ }\href@noop {} {\bibfield  {journal} {\bibinfo  {journal}
  {Astrophys. J.}\ }\textbf {\bibinfo {volume} {142}},\ \bibinfo {pages} {1541}
  (\bibinfo {year} {1965})}\BibitemShut {NoStop}%
\end{thebibliography}%

\end{document}